\documentclass[11pt,a4paper]{article}

\usepackage[a4paper]{geometry}
\usepackage{amssymb,latexsym,amsmath,amsfonts,amsthm,mathtools,stix2}
\usepackage{graphicx}
\usepackage[percent]{overpic}
\usepackage{float}
\usepackage{epstopdf}
\usepackage{tikz}
\usetikzlibrary{arrows.meta,decorations.markings}
\usepackage{xcolor}
\usepackage{fullpage}
\usepackage{comment,verbatim}
\usepackage[title,titletoc]{appendix}
\usepackage[
    colorlinks=true,
    citecolor=blue,
    linkcolor=black,
    urlcolor=black
]{hyperref}

\DeclareMathOperator{\diag}{diag}

\newcommand{\dd}{\,\mathrm d}
\newcommand{\ii}{\mathrm i}
\newcommand{\Ree}{\operatorname{Re}}
\newcommand{\Imm}{\operatorname{Im}}

\newtheorem{theorem}{Theorem}[section]
\newtheorem{lemma}[theorem]{Lemma}
\newtheorem{RHP}{Riemann--Hilbert problem}[section]
\newtheorem{proposition}[theorem]{Proposition}
\newtheorem{definition}[theorem]{Definition}
\newtheorem{assumption}[theorem]{Assumption}
\newtheorem{remark}[theorem]{Remark}

\numberwithin{equation}{section}
\allowdisplaybreaks
\title{\bfseries
The focusing coupled nonlinear Schr\"odinger equation with parallel nonzero boundary conditions: long--time asmptotics  
}
\date{}

\author{\hspace{0.6 cm}{Haibing Zhang$^{a}$, Xianguo Geng$^{a,b}$\footnote{\footnotesize
 Corresponding author. {\sl Email address}: xggeng@zzu.edu.cn} , Bo Xue$^{a}$,  Minxin Jia$^{a}$}\\
\leftline{\hspace{0.6 cm}{\small{\sl $^{a}$ School of Mathematics and Statistics, Zhengzhou University, 100 Kexue Road, Zhengzhou, }}}\\
\leftline{\hspace{0.6 cm}{\small{\sl \quad Henan 450001, People's Republic of China}}}\\
\leftline{\hspace{0.6 cm}{\small{\sl $^{b}$ School of Mathematics and Statistics, North China University of Water Resources }}}\\
\leftline{\hspace{0.6 cm}{\small{\sl \quad and Electric Power, Zhengzhou, Henan 450011, People's Republic of China}}}\\}

\begin{document}
\maketitle
\begin{abstract}
We study the long-time behavior of solutions of the focusing
coupled nonlinear Schr\"odinger equations with parallel constant
nonzero boundary conditions in the solitonless setting.
We distinguish three types of asymptotic regions: plane-wave
regions, modulated elliptic-wave regions, and Painlev\'e-II
transition regions.
In each region, we derive the leading asymptotic term and
the first correction, together with explicit error bounds.
Our results show that, although localized perturbations may
have components both parallel and orthogonal to the background,
the leading asymptotic term is always parallel to the background
vector. In particularly, in the elliptic-wave regions, the leading order is given by a
scalar modulated elliptic wave multiplying the background vector,  while the component of the solution orthogonal to the
background is \(O(t^{-1/2})\). This is consistent with the linear stability analysis:
the background is modulationally unstable to parallel
perturbations but stable to orthogonal perturbations.
The proof is based on applying the Deift--Zhou nonlinear
steepest-descent method, together with a \(G\)-function mechanism,
to the associated \(3\times3\) Riemann--Hilbert problem.
\end{abstract}

\tableofcontents

\section{Introduction}

\subsection{Background}

We study the long-time behavior of solutions to the focusing coupled
nonlinear Schr\"odinger (CNLS) equations~\cite{Manakov1974,BZ1970,CP1999},
also known as the Manakov system,
\begin{equation}\label{eq:manakov-nzbc}
\begin{cases}
 \mathrm{i}q_{1,t}+q_{1,xx}
 +2\bigl(|q_1|^2+|q_2|^2-q_0^2\bigr)q_1=0,\\[2mm]
 \mathrm{i}q_{2,t}+q_{2,xx}
 +2\bigl(|q_1|^2+|q_2|^2-q_0^2\bigr)q_2=0,
\end{cases}
\end{equation}
subject to constant nonzero boundary conditions (NZBCs)
\begin{equation}
 \lim_{x\to\pm\infty}\mathbf q(x,t)
 =\mathbf q_\pm=\mathbf q_0e^{\mathrm{i}\vartheta_\pm},
 \label{E:NZBCs}
\end{equation}
where \(\mathbf q=(q_1,q_2)^{\mathsf T}\),
\(\lVert\mathbf q_0\rVert=q_0>0\), and
\(\vartheta_\pm\in[0,2\pi)\).
Here, \(\lVert\cdot\rVert\) denotes the Euclidean norm.
The boundary vectors have the same polarization and may differ
only in phase. We therefore refer to \eqref{E:NZBCs} as
parallel NZBCs. Compared with the standard form of the focusing CNLS equations~\cite{Manakov1974},
\eqref{eq:manakov-nzbc} contains the additional linear term
\(-2q_0^2q_j\), \(j=1,2\). This term was added so that the asymptotic values of the potential are independent of time.
The gauge transformation
\(\mathbf Q=e^{2\mathrm{i}q_0^2t}\mathbf q\) removes it and
recovers the standard form.

The CNLS equations \eqref{eq:manakov-nzbc} is an integrable
two-component extension of the well-known nonlinear Schr\"odinger
equation. The system can arise when two propagating
fields are nonlinearly coupled.  Models of this type are known to appear in the study of 
polarization modes in birefringent optical
fibers~\cite{A1987,A1989,BDW1987,FrisquetEtAl2015}, multicomponent
Bose--Einstein condensates~\cite{CP1999,QZL2019}, and crossing
water-wave trains~\cite{OnoratoOsborneSerio2006}.
From a mathematical perspective, the system \eqref{eq:manakov-nzbc}
is a basic example of a completely integrable system.
More precisely, it can be written as the compatibility condition
for the following \(3\times3\) Lax pair~\cite{Manakov1974,KrausBiondiniKovacic2015}:
\begin{equation}
 \boldsymbol\phi_x=X\boldsymbol\phi,
 \qquad
 \boldsymbol\phi_t=T\boldsymbol\phi,
 \label{eq:lax-pair-k}
\end{equation}
where
\begin{align}
 X&=-\mathrm ikJ+ \mathcal Q, \qquad
 T=2\mathrm ik^2J-\mathrm iJ(\mathcal{Q}_x-\mathcal{Q}^2-q_0^2I_{3 \times 3})-2k\mathcal{Q},
 \label{eq:lax-matrices-k}\\
 J&=\begin{pmatrix}1&\mathbf0^{\mathsf T}\\[1pt]\mathbf0&-I_{2 \times 2}\end{pmatrix}, \qquad
\mathcal{Q}=\begin{pmatrix}0&-\mathbf q^\dagger\\ \mathbf q&0_{2\times2}\end{pmatrix}.
 \nonumber
\end{align}
Consequently, many methods from the theory of integrable systems
have been applied to this system, for example inverse scattering transform~(IST) for decaying initial
data~\cite{Manakov1974,Liu2019},  Hirota's bilinear method~\cite{RadhakrishnanLakshmananHietarinta1997,KannaLakshmanan2001},
Darboux transformations ~\cite{Wright2003,LingGuoZhao2014,LingZhaoGuo2015},
Algebro-geometric methods~\cite{ElginEnolskiIts2007,Kalla2011}, the Kadomtsev--Petviashvili (KP) hierarchy reduction method~\cite{YangYang2023}, and the dressing methods~\cite{GelashRaskovalov2023}.  The integrable structure of the system has also been used
to establish rigorous nonlinear stability results for
localized solutions with zero boundary conditions; see for example~\cite{LingPelinovskySu2026,LingSu2027}.

The nonzero background~\eqref{E:NZBCs} of the focusing CNLS
equations has attracted considerable attention because it
supports a rich variety of exact nonlinear waves.
Examples include dark--bright vector solitons, vector breathers
of Kuznetsov--Ma and Akhmediev type~\cite{MM2013,
KrausBiondiniKovacic2015,GelashRaskovalov2023},
vector rogue waves, and semirational vector
solutions~\cite{BaronioDegasperisConfortiWabnitz2012,
MuQinGrimshaw2015}.
A systematic framework for studying solutions with these NZBCs
is provided by the IST developed by Kraus, Biondini, and
Kova\v{c}i\v{c}~\cite{KrausBiondiniKovacic2015}.
Their analysis treats both the direct and inverse scattering
problems and yields exact soliton solutions and several
families of vector breathers from discrete scattering data
in the reflectionless case. 
Beyond these exact solutions, the IST provides a spectral
framework for studying the long--time evolution of localized perturbations
of the background. In modern formulations of the IST, the inverse problem is usually formulated as a Riemann--Hilbert~(RH) problem. In the pioneering paper \cite{pz93}, Deift and Zhou introduced the nonlinear steepest descent method for oscillatory RH problems, thereby establishing the long-time asymptotic behavior of solutions to the modified KdV equation. Since then, the nonlinear steepest descent method  and its extensions have become a fundamental tool for studying long-time evolution of integrable systems. This method has also proved effective for studying long-time
asymptotics in problems with nonzero backgrounds, including
the Toda shock problem~\cite{EgorovaMichorTeschl2018,
EgorovaMichorPryimakTeschl2023},
the KdV equation with step-like initial
data~\cite{EgorovaGladkaKotlyarovTeschl2013,
EgorovaPiorkowskiTeschl2024},
and the scalar focusing NLS equation with step-like
oscillatory backgrounds~\cite{BoutetDeMonvelLenellsShepelsky2021,
BoutetDeMonvelLenellsShepelsky2022}. However, although the IST was given  in  ~\cite{KrausBiondiniKovacic2015}, the corresponding  nonlinear steepest descent analysis has not been established yet. This problem was left as open direction in Refs.\cite{KrausBiondiniKovacic2015,P2023}. This is one of the motivation for this work.

As in the scalar case, the nonzero background~\eqref{E:NZBCs}
of the focusing CNLS equations exhibits long-wave modulational
instability, known as Benjamin--Feir instability in the context
of water waves~\cite{BF1967}.
In the vector setting, however, polarization introduces an
additional degree of freedom.
Even when the boundary vectors are parallel, a localized
perturbation may have components both parallel and orthogonal
to the background.
The linear stability analysis in Section~\ref{S:lsN} shows
that the background is modulationally unstable to parallel
perturbations but linearly stable to orthogonal perturbations.
The parallel component has the same unstable dispersion branch
as the scalar focusing NLS equation with
NZBCs~\cite{BiondiniMantzavinos2016}.
This linear analysis describes the early stage of instability,
during which unstable long-wave perturbations grow exponentially.
Once their amplitudes become comparable to the background
amplitude, nonlinear effects become important and the linear
approximation is no longer adequate.
The subsequent evolution is usually referred to as the
nonlinear stage of modulational instability.
This raises the question of how the different linear stability
properties of the two polarization components are reflected
in the nonlinear stage of modulational instability.
In particular, we ask whether the orthogonal component
contributes to the leading long-time behavior of the solution.

Characterizing the nonlinear stage of modulational instability is an important and challenging problem, even in the scalar case.  Recently, Biondini and Mantzavinos made great progress in this
direction~\cite{BiondiniMantzavinos2016,BiondiniMantzavinos2017}.
They realized that combining the IST method  with the Deift--Zhou nonlinear
steepest descent method could provide
an effective approach to understanding the nonlinear
development of modulational instability.
They then applied this idea to the focusing NLS equation
with NZBCs~(namely, setting \(q_2\equiv0\) in~\eqref{eq:manakov-nzbc}--\eqref{E:NZBCs}.); see Ref.~\cite{BiondiniMantzavinos2017}.
Their results show that a broad class of localized perturbations
of a constant background share a common long-time asymptotic
structure. In particular, the envelope
of the leading modulated elliptic waves is independent of
the initial perturbation. In this sense, the long-time
asymptotic state of the nonlinear stage of modulational
instability is \emph{universal}.
Building on these results, Biondini, Li, Mantzavinos, and
Trillo~\cite{BiondiniLiMantzavinosTrillo2018} carried out
analytical and numerical studies of several other NLS-type
models. These include models with power-law and saturable
nonlinearities, a nonlocal model for thermal media, the
dispersion-managed NLS equation, and the Ablowitz--Ladik
equation, an integrable discrete counterpart of the NLS
equation. Their findings provided evidence that this qualitative
universality is shared by a range of integrable and
nonintegrable NLS-type models.
It is therefore natural to ask whether this universality
also holds for the focusing CNLS equations with parallel
constant NZBCs.

The aim of this paper is to address these questions by deriving
precise long-time asymptotic formulas for solutions
of~\eqref{eq:manakov-nzbc}--\eqref{E:NZBCs}.
We work in the solitonless setting under suitable assumptions.
Our strategy is to apply the Deift--Zhou nonlinear
steepest-descent method to the associated \(3\times3\) RH
problem formulated through the
IST~\cite{KrausBiondiniKovacic2015}.
The analysis is inspired by the work of Biondini and
Mantzavinos on the scalar focusing NLS
equation with NZBCs~\cite{BLM2021,BiondiniMantzavinos2017}.
However, the vector setting introduces new features and
technical difficulties.
The main highlights of this  work are mainly reflected in the following aspects:

\begin{itemize}
\item[$\bullet$]
In the vector case, we need to carry out the steepest-descent
analysis for a \(3\times3\) matrix RH problem, in contrast
to the \(2\times2\) problem in the scalar
case~\cite{BiondiniMantzavinos2017}.
Although the jump matrix on the branch cut \(B\) has an
embedded \(2\times2\) structure, the real-axis jump is ``truly'' $3 \times 3$; see \eqref{eq:CR-k}.
The analysis is therefore more involved than in the scalar case.
Another important feature is the need to introduce a
\(G\)-function transformation in the steepest-descent analysis.
Implementing this transformation also presents additional
difficulties in the \(3\times3\) setting.

\item[$\bullet$]
We derive not only the leading asymptotic term but also
an explicit expression for the first correction, together
with a bound on the remaining error.
In the scalar analysis of~\cite{BiondiniMantzavinos2017},
this correction is included in the error term.
Computing it requires a detailed analysis of the local
models near several distinguished spectral points.
A particularly delicate step is the analysis of a nonstandard
parabolic-cylinder model near \(k_0\) in Section~\ref{subsubsec:elliptic-local-k0}.
Reducing this local RH problem to the standard
parabolic-cylinder model is a nontrivial part of the analysis.

\item[$\bullet$]
We also study the transition regions between the plane-wave
and elliptic-wave regions, which were not treated
in~\cite{BiondiniMantzavinos2017}.
We show that the first asymptotic correction in these
regions is related to the increasing tritronqu\'ee solution
of an inhomogeneous Painlev\'e-II equation.
\end{itemize}
\subsection{Linear stability of the constant background}
\label{S:lsN}

We first examine the linear stability of the constant background.
This analysis can serve as a reference point for the nonlinear
long-time asymptotics derived below.

Since the two limiting backgrounds differ
only by a constant phase, it therefore suffices to consider the background $\mathbf q_+$. Let \(\mathbf e=\mathbf q_+/q_0\in\mathbb C^2\), then $\mathbf q_+=q_0\mathbf e$. Choose a unit vector
\(\mathbf e^\perp\) satisfying
\(\mathbf e^\dagger\mathbf e^\perp=0\). Consider
the perturbation expansion
\begin{equation}
 \mathbf q(x,t)
 =q_0\mathbf e
 +\epsilon\bigl(u(x,t)\mathbf e+v(x,t)\mathbf e^\perp\bigr)
 +O(\epsilon^2),
 \label{eq:intro-linear-perturbation}
\end{equation}
where \(\epsilon\) is a small real parameter and \(\{u,v \}\) 
are complex-valued functions. 
Using \eqref{eq:intro-linear-perturbation}, we obtain the following  perturbation expansion
\begin{align} \label{E:annnn}
|q_1|^2+|q_2|^2
 =q_0^2+\epsilon q_0(u+\overline u)+O(\epsilon^2).
\end{align}
Then we substitute \eqref{eq:intro-linear-perturbation} and \eqref{E:annnn} into
\eqref{eq:manakov-nzbc} and collect the terms of order
\(\epsilon\). This gives
\begin{equation}
 \mathrm{i}u_t+u_{xx}+2q_0^2(u+\overline u)=0,
 \qquad
 \mathrm{i}v_t+v_{xx}=0.
 \label{eq:intro-linearized-system}
\end{equation}
Thus, the perturbations parallel and orthogonal to the background
decouple at the linear level. The parallel component satisfies
the linearized scalar focusing NLS equation, while the orthogonal
component satisfies the free Schr\"odinger equation.

Following the scalar analysis in
\cite[Section~6.1]{BiondiniFagerstrom2015}, we separate
the real and imaginary parts by writing
\[
 u=\frac12(r+\mathrm{i}s),
 \qquad
 v=\frac12(a+\mathrm{i}b),
\]
where \(r,s,a,b\) are real-valued functions.
Then \eqref{eq:intro-linearized-system} becomes
\begin{equation}
 \begin{aligned}
  r_t+s_{xx}&=0,
  &\qquad s_t-r_{xx}-4q_0^2r&=0,\\
  a_t+b_{xx}&=0,
  &\qquad b_t-a_{xx}&=0.
 \end{aligned}
 \label{eq:ls-real-system}
\end{equation}
These systems may be solved with Fourier transforms.
Introduce  the convention
\[
 \widehat f(K,t)
 =\int_{\mathbb R}f(x,t)e^{-\mathrm{i}Kx}\,\mathrm{d}x,
 \qquad K\in\mathbb R.
\]
It then follows from  \eqref{eq:ls-real-system} that 
 \begin{align} \label{eq:intro-dispersion-relation}
  \partial_t
  \begin{pmatrix}
   \widehat r\\
   \widehat s
  \end{pmatrix}
  =
  \begin{pmatrix}
   0&K^2\\
   4q_0^2-K^2&0
  \end{pmatrix}
  \begin{pmatrix}
   \widehat r\\
   \widehat s
  \end{pmatrix},
  \qquad
  \partial_t
  \begin{pmatrix}
   \widehat a\\
   \widehat b
  \end{pmatrix}
  =
  \begin{pmatrix}
   0&K^2\\
   -K^2&0
  \end{pmatrix}
  \begin{pmatrix}
   \widehat a\\
   \widehat b
  \end{pmatrix}.
 \end{align}
 For each \(K\), the eigenvalues of the two coefficient matrices
determine the temporal growth  of the Fourier
components.
These eigenvalues are \(\pm\mathrm{i}\Omega_{\mathrm I}\)
and \(\pm\mathrm{i}\Omega_{\mathrm{II}}\), respectively, where
$
 \Omega_{\mathrm{I}}=K\sqrt{K^2-4q_0^2}
$ and $
 \Omega_{\mathrm{II}}=K^2
$.

The first system describes perturbations parallel to the
background.
For \(0<|K|<2q_0\), \(\Omega_{\mathrm I}\) is purely imaginary,
so the corresponding matrix has a positive real eigenvalue.
Parallel perturbations can therefore grow exponentially,
and the background is modulationally unstable in this
polarization.
The second system in \eqref{eq:intro-dispersion-relation} describes orthogonal perturbations.
Since \(\Omega_{\mathrm{II}}=K^2\) is real for every \(K\),
the orthogonal component has no exponentially growing
Fourier components and is spectrally stable.
Hence, exponentially growing linear perturbations are
confined to the background direction.

\subsection{Main results}

In this subsection, we state our main results. We begin by recalling
the IST  for the initial-value
problem~\eqref{eq:manakov-nzbc}--\eqref{E:NZBCs}, developed in
\cite{KrausBiondiniKovacic2015}.
A key feature of the scattering analysis with NZBCs is the
appearance of multivalued functions in the construction of the
Jost solutions. In the present setting, this construction involves
the function \(\lambda(k)\), which satisfies
\(
 \lambda(k)^2=k^2+q_0^2
\).
This relation naturally defines a two-sheeted Riemann surface
with branch points at \(k=\pm\mathrm{i}q_0\).
To simplify the analysis, the authors in \cite{KrausBiondiniKovacic2015} introduce the 
uniformization variable
$
 \mathrm z=k+\lambda(k)
$,
with inverse transformation
\[
 k=\frac12\left(\mathrm z-\frac{q_0^2}{\mathrm z}\right),
 \qquad
 \lambda=\frac12\left(\mathrm z+\frac{q_0^2}{\mathrm z}\right),
 \qquad \mathrm z\ne0.
\]
Their scattering analysis is consequently carried out in the
\(\mathrm z\)-plane. For the subsequent steepest-descent analysis, however,
it is more convenient to work directly in the spectral \(k\)-plane.
We therefore express the corresponding results
of~\cite{KrausBiondiniKovacic2015} in the spectral \(k\)-plane.

\medskip
\noindent
\textbf{\textit{Direct scattering.}}
The direct scattering transform associates the initial potential
\(\mathbf q(x,0)\) with its scattering data. In general, these
data have two components: (i) continuous scattering data, encoded
by reflection coefficients defined on the continuous spectrum,
and (ii) discrete scattering data, consisting of discrete
eigenvalues and their associated norming constants. The latter
determine the soliton contributions to the solution.
In the present setting, the continuous spectrum of the direct scattering
problem consists of those values of \(k\) for which
\(\lambda(k)\) is real. Thus,  the continuous spectrum is 
\[
 \Sigma=\mathbb R\cup[-\mathrm{i}q_0,\mathrm{i}q_0].
\]
To define the reflection coefficients, let \(A(k)=(a_{ij}(k))\)
denote the scattering matrix in~\eqref{eq:A-integral-k}. On
\(\Sigma\setminus\{\pm\mathrm{i}q_0\}\), the reflection
coefficients are defined  by
\begin{equation}
 \rho_1(k)=\frac{a_{21}(k)}{a_{11}(k)},
 \qquad
 \rho_2(k)=\frac{a_{31}(k)}{a_{11}(k)},
 \qquad
 \rho_3(k)=\frac{a_{32}(k)}{a_{22}(k)}.
 \label{eq:intro-reflection-coefficients}
\end{equation}

The scattering coefficient \(a_{11}(k)\) may have 
square-root singularities at the branch points
\(\pm\mathrm{i}q_0\); see \eqref{E:arsssh}.
In this paper, we assume that the leading singular coefficient
at each branch point is nonzero.
This assumption simplifies several technical arguments,
particularly the verification that the Beals--Coifman matrix
\(M\), defined by~\eqref{eq:RHP-unknown-k}, remains bounded
near the branch points.

For simplicity, we also assume that the discrete scattering
data are absent and thus work in the solitonless setting.
Accordingly, we impose the following assumptions.

\begin{assumption}[Spectral assumptions]
\label{ass:intro-no-discrete-spectrum}
The scattering data generated by the initial potential
\(\mathbf q(x,0)\) satisfy the following conditions:
\begin{enumerate}
 \item[(i)] \textit{Generic behavior at the branch points.}
We assume that
  \[
  \lim_{k\to\pm\mathrm{i}q_0}
  (k\mp\mathrm{i}q_0)^{1/2}a_{11}(k)
  \ne 0,\qquad \lim_{k\to\pm\mathrm{i}q_0} a_{22}(k) \ne 0.
 \]

 \item[(ii)] \textit{Absence of discrete spectrum.} We assume that
 the analytic scattering coefficients \(a_{11}\) and
 \(a_{22}\) have no zeros in their respective domains
 of analyticity.
\end{enumerate}
\end{assumption}

\begin{remark} 
This solitonless assumption simplifies the inverse problem by removing the
pole conditions associated with discrete eigenvalues. It also
allows us to focus on the contribution of the continuous spectrum
to modulational instability. For the scalar focusing NLS equation,
Biondini and Fagerstrom~\cite{BiondiniFagerstrom2015} showed that
\emph{solitons cannot be the mechanism that mediates modulational
instability}. Instead, the instability arises from certain portions
of the continuous spectrum of the associated scattering problem.
These results show that modulational instability can occur without
solitons. We therefore adopt the solitonless assumption to simplify
the analysis.
\end{remark}
Thus, under the assumptions above, we have  established
the direct scattering map
\[
 \mathbf q(x,0)\longmapsto
\big\{\rho_1(k),\rho_2(k),\rho_3(k) \big\}.
\]

\medskip
\noindent
\textbf{\textit{Inverse scattering.}}
The inverse scattering transform recovers the potential $\mathbf q(x,t)$ from
the scattering data.  This proceeds in two
steps: the scattering data first determine a matrix RH problem,
and the potential is then reconstructed from the large-\(k\)
expansion of its solution.

To obtain an RH formulation, one typically uses the Jost solutions
and the analytic diagonal entries of the scattering matrix to
construct the so-called Beals--Coifman simultaneous solution of
the Lax pair. Analyzing this sectionally meromorphic matrix-valued
function yields the jump relations, pole conditions, and
normalization. These
conditions specify the associated RH problem.
In our solitonless setting, no pole conditions are required, so
the inverse problem involves only the continuous scattering data
\(\{\rho_j\}_{j=1}^3\) generated by the initial data. This leads to
the following \(3\times3\) RH problem.

\begin{RHP}
\label{rhp:M-solitonless}
Let $\Sigma^\circ=\Sigma\setminus\{0,\pm\mathrm{i}q_0\}$.
For each fixed \((x,t)\), find a \(3\times3\) matrix-valued function
\(M(x,t,k)\)  satisfying the following
conditions.
\begin{enumerate}
\item[(i)] \textit{Analyticity.}
The function \(M(x,t,\cdot)\) is analytic in
\(\mathbb C\setminus\Sigma\) and admits continuous nontangential
boundary values \(M^\pm(x,t,k)\) on \(\Sigma^\circ\).

\item[(ii)] \textit{Jump relations.}
The boundary values satisfy
\begin{equation}
 M^+(x,t,k)=M^-(x,t,k)V(x,t,k),
 \qquad k\in\Sigma^\circ,
 \label{eq:M-RHP-jump}
\end{equation}
where jump matrix $V(x,t,k)$ is given by \eqref{eq:piecewise-jump-k}.

\item[(iii)] \textit{Normalization.}
As \(k\to\infty\),
\begin{equation}
 M(x,t,k)
 =M_\infty+\frac{M_1(x,t)}{k}+O(k^{-2}),
 \label{eq:M-RHP-normalization}
\end{equation}
where
\[
 M_\infty=
\begin{pmatrix}
  1&0&0\\[6pt]
\mathbf  0
   &\dfrac{\mathbf q_+^\perp}{q_0}
   &\dfrac{\mathbf q_+}{q_0}
 \end{pmatrix}.
\]

\item[(iv)]
\textit{Local boundedness.}
The matrix \(M(x,t,k)\) remains bounded as \(k\) approaches
\(0\) or \(\pm\mathrm{i}q_0\).
\end{enumerate}
\end{RHP}

We next state the following reconstruction theorem, whose proof is given
in Section~\ref{sec:direct-scattering-k}.

\begin{theorem}
\label{thm:reconstruction}
Let \(M(x,t,k)\) be the solution of the Riemann--Hilbert
problem~\ref{rhp:M-solitonless} associated with the scattering
data generated by the initial potential \(\mathbf q(x,0)\).
Then the solution of the initial-value
problem~\eqref{eq:manakov-nzbc}--\eqref{E:NZBCs} is reconstructed by
\begin{equation}
 \mathbf q(x,t)
 =-2\mathrm{i}
 \begin{pmatrix}
 (M_1(x,t))_{21}\\[1mm]
 (M_1(x,t))_{31}
 \end{pmatrix}.
 \label{eq:reconstruction}
\end{equation}
\end{theorem}


\medskip
\noindent
\textbf{\textit{Long-time asymptotic results.}}
The reconstruction theorem shows that the long-time behavior of
the solution of the initial-value problem can be determined by
analyzing the RH problem~\ref{rhp:M-solitonless}.
The central task of the present work is therefore to derive the
long-time asymptotics of this \(3\times3\) matrix RH problem
using the Deift--Zhou nonlinear steepest-descent method.  The main idea of this method is to carry out a sequence of exact
transformations that reduce the original problem to a small-norm
RH problem as \(t\to\infty\). We now present the long-time asymptotic results.

\begin{figure}
\centering
\begin{tikzpicture}[
    x=.18cm,
    y=1.05cm,
    >={Stealth[length=2.5mm,width=1.7mm]},
    line cap=round,
    line join=round,
    axis/.style={black,line width=1.1pt},
    boundary/.style={black!65,line width=.65pt},
    critical/.style={
        black!75,
        line width=.8pt,
        dash pattern=on 3.5pt off 3pt
    },
    regionlabel/.style={font=\large,inner sep=2pt}
]

\definecolor{pwfill}{RGB}{217,229,244}
\definecolor{ewfill}{RGB}{207,233,220}
\definecolor{trfill}{RGB}{248,219,188}

\def\tmax{5.2}
\def\xmax{35}
\pgfmathsetmacro{\speed}{4*sqrt(2)}
\def\bandwidth{2.6}

\begin{scope}
    \clip (-\xmax,0) rectangle (\xmax,\tmax);

    \fill[pwfill]
        (-\xmax,0) rectangle (\xmax,\tmax);

    \fill[ewfill]
        (0,0)
        -- ({-\speed*\tmax},\tmax)
        -- ({\speed*\tmax},\tmax)
        -- cycle;

    \foreach \sgn in {-1,1} {
        \fill[trfill]
            plot[domain=0:\tmax,samples=100,variable=\s]
                ({\sgn*(\speed*\s
                   +\bandwidth*(sqrt(1+\s)-1))},\s)
            --
            plot[domain=\tmax:0,samples=100,variable=\s]
                ({\sgn*(\speed*\s
                   -\bandwidth*(sqrt(1+\s)-1))},\s)
            -- cycle;

        \foreach \side in {-1,1} {
            \draw[boundary]
                plot[domain=0:\tmax,samples=100,variable=\s]
                    ({\sgn*(\speed*\s
                       +\side*\bandwidth*(sqrt(1+\s)-1))},\s);
        }

        \draw[critical]
            (0,0)--({\sgn*\speed*\tmax},\tmax);
    }
\end{scope}

\draw[axis,->] (-35.5,0)--(36.8,0)
    node[right=4pt] {$x$};
\draw[axis,->] (0,0)--(0,5.65)
    node[above=4pt] {$t$};

\node[regionlabel] at (-27.5,1.45)
    {$\mathrm P_{\mathrm{II}}$};
\node[regionlabel] at (27.5,1.45)
    {$\mathrm P_{\mathrm I}$};

\node[regionlabel] at (-8.2,3.25)
    {$\mathrm E_{\mathrm I}$};
\node[regionlabel] at (8.2,3.25)
    {$\mathrm E_{\mathrm{II}}$};

\node[regionlabel,fill=trfill,inner sep=.5pt]
    at ({-\speed*3.7},3.7)
    {$\mathcal T_{\mathrm{II}}$};
\node[regionlabel,fill=trfill,inner sep=.5pt]
    at ({\speed*3.7},3.7)
    {$\mathcal T_{\mathrm I}$};

\end{tikzpicture}
\caption{Different asymptotic regions given in Definition \ref{D:asy}}
\label{fig:asymptotic-regions}
\end{figure}
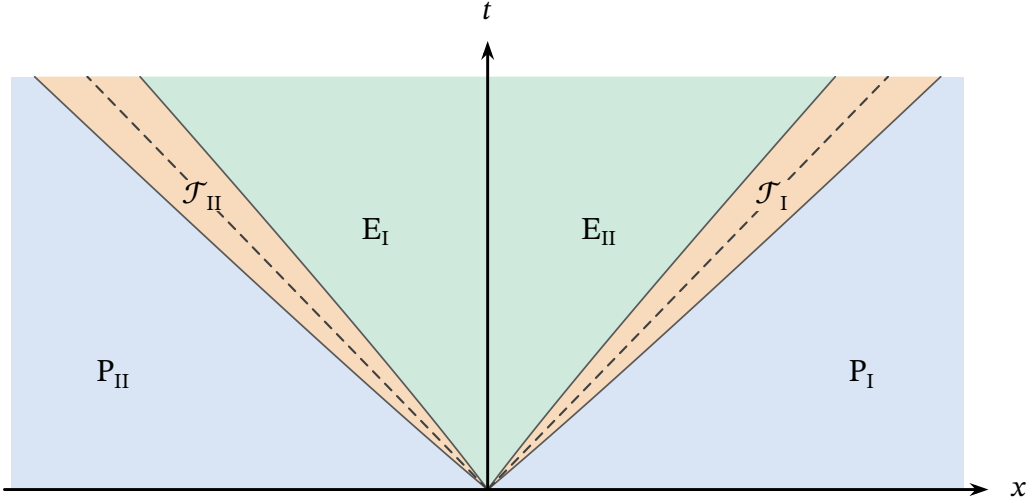

The Deift--Zhou steepest-descent analysis shows that the solution
of the RH problem exhibits qualitatively different long-time
behavior in different regions of the \((x,t)\)-plane.
In terms of the self-similar variable \(\xi=x/t\), we distinguish
two outer plane-wave regions, two central modulated elliptic-wave
regions, and two Painlev\'e-type transition regions near the
boundaries separating the plane-wave and elliptic-wave regions. These regions are defined below; see Figure~\ref{fig:asymptotic-regions}
for an illustration.

\begin{definition} \label{D:asy}
Let \(\xi=x/t\), and fix a constant \(C>0\).
We define the following asymptotic regions:
\begin{enumerate}
\item[(i)] The two plane-wave regions:
$
 \mathrm{P}_{\mathrm{I}}
 =\{(x,t): \xi>4\sqrt{2}q_0\},\quad
 \mathrm{P}_{\mathrm{II}}
 =\{(x,t): \xi<-4\sqrt{2}q_0\}.
$

\item[(ii)] The two elliptic-wave regions:
$$
 \mathrm{E}_{\mathrm{I}}
 =\{(x,t):  -4\sqrt{2}q_0<\xi<0\},\quad
 \mathrm{E}_{\mathrm{II}}
 =\{(x,t):  0<\xi<4\sqrt{2}q_0\}.
$$

\item[(iii)] The two Painlev\'e-type transition regions:
\begin{align*}
 \mathcal{T}_{\mathrm{I}}
 =\{(x,t):
       |\xi-4\sqrt{2}q_0|\leq Ct^{-2/3}\},\quad
 \mathcal{T}_{\mathrm{II}}
 =\{(x,t):
       |\xi+4\sqrt{2}q_0|\leq Ct^{-2/3}\}.
\end{align*}
\end{enumerate}
\end{definition}

The regions defined above are determined by the qualitative
behavior of the phase functions, which in turn leads to different
strategies for contour deformation in the steepest-descent
analysis of the RH problem. Although the \(3\times3\) matrix structure requires us to analyze
three phase functions simultaneously, the resulting division into
asymptotic regions agrees with that of the scalar focusing NLS
problem~\cite{BiondiniMantzavinos2017}.
An interesting feature of the vector setting, however, is that
the asymptotic formulas reveal a clear geometric structure.
On the other hand,  since the focusing CNLS system is invariant under \(x\mapsto-x\),
it suffices to analyze one region from each of the three pairs.
We choose \(\mathrm{P}_{\mathrm{I}}\), \(\mathrm{E}_{\mathrm{I}}\),
and \(\mathcal{T}_{\mathrm{I}}\) and give precise asymptotic
formulas in these regions.

The following theorem describes the long-time asymptotics
in the plane-wave region.

\begin{theorem}[Plane-wave asymptotics]
\label{thm:long-time-two-stationary}
Let \(\mathbf q(x,t)\) be a solution of the focusing CNLS
equations~\eqref{eq:manakov-nzbc} with the nonzero boundary
conditions~\eqref{E:NZBCs}, and assume that its initial data
satisfy Assumptions~\ref{ass:intro-no-discrete-spectrum}
and~\ref{A:intii}.
For any compact subset \(\mathcal K_+\) of
\((4\sqrt{2}q_0,\infty)\), the solution admits the following
asymptotic expansion, uniformly for \(\xi\in\mathcal K_+\):
\begin{equation}
 \mathbf q(x,t)
 =\mathbf q_+e^{-2\mathrm{i}g_\infty(\xi)}
 +\frac{1}{\sqrt t}\,\mathbf q_{\mathrm{rad}}(x,t)
 +O(t^{-1}\log t),
 \qquad t\to\infty,
 \label{eq:final-asymptotics}
\end{equation}
where the phase \(g_\infty\) is defined by~\eqref{eq:g-infinity},
and the radiative term is given by
\begin{equation}
 \mathbf q_{\mathrm{rad}}(x,t)
 =-\frac{2\mathrm{i}}{q_0}\left(
 \widetilde{\mathcal E}^{(1)}_{21}
 e^{-2\mathrm{i}g_\infty}\mathbf q_+
 +\widetilde{\mathcal E}^{(1)}_{31}
 e^{-\mathrm{i}g_\infty}\mathbf q_+^\perp
 \right).
 \label{eq:intro-plane-radiation}
\end{equation}
The matrix \(\widetilde{\mathcal E}^{(1)}\) is given explicitly
in~\eqref{eq:Etilde-explicit}.
\end{theorem}

The asymptotic formula shows that the leading term is a scalar
multiple of the background vector and therefore lies entirely
in the background direction. The subleading term decomposes
into components parallel and orthogonal to the background.

We next present the results in the elliptic-wave region.
The preceding theorem concerns the plane-wave region on the
\(\xi>0\) side. For comparison, we now consider the elliptic-wave
region on the \(\xi<0\) side, namely \(\mathrm{E}_{\mathrm{I}}\).
The main difference on this side is the reversal of the sign
structure of the phase functions. We therefore begin with a
diagonal rescaling of the RH unknown to adjust the factorizations
of the jump matrices. This transformation modifies the jumps
on \(\Sigma\) and replaces the reflection coefficients
in~\eqref{eq:intro-reflection-coefficients} with
\begin{equation}
 r_1(k)=\frac{a_{13}(k)}{a_{11}(k)},
 \qquad
 r_2 (k)=\frac{a_{21}(k)}{a_{22}(k)},
 \qquad
 r_3(k)=\frac{a_{23}(k)}{a_{22}(k)}.
 \label{eq:intro-rescaled-reflection-coefficients}
\end{equation}

Another interesting feature of this region is the need for a
\(G\)-function transformation, whose construction is closely related
to the following nonlinear modulation system:
\begin{subequations}
\label{eq:elliptic-modulation-system-1}
\begin{equation}
\operatorname{Re}\alpha
=
-k_0+\frac{\xi}{4},
\qquad
\operatorname{Im}\alpha
=
\sqrt{
2k_0^2-\frac{\xi k_0}{2}+q_0^2
}.
\label{eq:elliptic-modulation-algebraic-1}
\end{equation}
\begin{equation}
\int_{-\mathrm{i}q_0}^{\mathrm{i}q_0}
\sqrt{
\frac{
\left(z+k_0-\frac{\xi}{4}\right)^2
+2k_0^2-\frac{\xi k_0}{2}+q_0^2
}{
z^2+q_0^2
}
}
(z-k_0)\,\mathrm{d}z
=
0.
\label{eq:elliptic-modulation-moment-1}
\end{equation}
\end{subequations}
The same modulation system also appears in the scalar focusing
NLS problem~\cite{BiondiniMantzavinos2017}. In particular, Biondini and Mantzavinos
showed in~\cite[Lemma~5.4]{BiondiniMantzavinos2017} that
the integral equation~\eqref{eq:elliptic-modulation-moment-1}
uniquely determines \(k_0(\xi)\) for every
\(\xi\in(-4\sqrt{2}q_0,0)\).
The algebraic relations~\eqref{eq:elliptic-modulation-algebraic-1}
then uniquely determine \(\alpha(\xi)\).
Thus, the modulation system~\eqref{eq:elliptic-modulation-system-1}
defines the two points \(k_0(\xi)\) and \(\alpha(\xi)\)
throughout the elliptic-wave region \(\mathrm{E}_{\mathrm{I}}\).
These points depend only on \(q_0\) and \(\xi\) and are independent
of the initial perturbation.

We are now ready to state the asymptotic theorem for the
elliptic-wave region.

\begin{theorem}[Elliptic-wave asymptotics]
\label{thm:long-EW}
Assume that the initial data \(\mathbf q(x,0)\) satisfy Assumptions~\ref{ass:intro-no-discrete-spectrum}
and~\ref{A:intii},  and let
\(\mathcal K_-\) be any compact subset of
\((-4\sqrt{2}q_0,0)\).
Assume further that
\begin{equation}
 r_1(k_0)\neq0,
 \qquad r_1(\alpha)\neq0,
 \qquad \xi\in\mathcal K_-.
 \label{eq:intro-elliptic-genericity}
\end{equation}
Then the solution $\mathbf q(x,t)$ admits the following asymptotic expansion,
uniformly for \(\xi\in\mathcal K_-\):
\begin{equation}
 \mathbf q(x,t)
 =\mathbf q^{\mathrm{ell}}(x,t)
 +t^{-1/2}\mathbf q_{\mathrm{rad}}^{\mathrm{ell}}(x,t)
 +O(t^{-1}\log t),
 \qquad t\to\infty,
 \label{eq:intro-elliptic-asymptotics}
\end{equation}
where the leading term is a modulated elliptic wave given by
\[
\begin{aligned}
 \mathbf q^{\mathrm{ell}}(x,t)
 =\frac{q_0+\Imm \alpha}{q_0}
 e^{-2\mathrm{i}(G_\infty t+\widehat g_\infty)}
 \frac{\vartheta\!\left(\frac{\Omega t-\omega}{2\pi}
                    +2\nu_\infty-\frac14\right)\vartheta(-\frac12)}
      {\vartheta\!\left(\frac{\Omega t-\omega}{2\pi}-\frac14\right)
       \vartheta(2\nu_\infty-\frac12)}
 \mathbf q_-.
\end{aligned}
\]
Here, the quantities
\(\Omega,\omega,G_\infty,\widehat g_\infty,\nu_\infty\)
and the genus-one theta function \(\vartheta\) are defined in
Section~\ref{sec:elliptic-wave}.
The radiative term is given by
\begin{equation}
 \mathbf q_{\mathrm{rad}}^{\mathrm{ell}}(x,t)
 =-\frac{2\mathrm{i}}{q_0}\left(
 \widecheck{\mathcal E}_{21}
 e^{-2\mathrm{i}(G_\infty t+\widehat g_\infty)}\mathbf q_-
 +\widecheck{\mathcal E}_{31}
 e^{-\mathrm{i}(G_\infty t+\widehat g_\infty)}\mathbf q_-^\perp
 \right),
 \label{eq:intro-elliptic-radiation}
\end{equation}
where the matrix \(\widecheck{\mathcal E}\) is given explicitly
in~\eqref{eq:elliptic-scaled-error-explicit}.
\end{theorem}

\begin{remark}
It can be observed that at leading order, the modulated
elliptic wave lies entirely in the background direction.
This structure is consistent with the linear stability analysis,
which shows that perturbations parallel to the background are
modulationally unstable through the same mechanism as in the
scalar focusing NLS equation.
The modulated elliptic wave can be viewed as the nonlinear state
that develops as this instability saturates.
\end{remark}

\begin{remark}
The leading asymptotic term is obtained from a \(3\times3\)
outer RH model constructed by embedding the \(2\times2\)
outer model for the scalar focusing NLS equation~\cite{BiondiniMantzavinos2017}.
Consequently, the scalar factor in our leading term has the
same theta-function structure as in the scalar problem.
Therefore, the  envelope of the leading elliptic
wave depends only on \(q_0\) and \(\xi\) and is independent of
the initial perturbation. 
In this sense, the nonlinear stage of modulational instability
also exhibits \textit{universality} for the constant parallel
backgrounds considered here.
\end{remark}

The asymptotic formulas for the plane-wave and elliptic-wave
regions are valid in the interiors of these regions, but they
are not uniform as \(\xi\) approaches the critical values
\(\pm4\sqrt{2}q_0\). A separate analysis is therefore needed
to describe the long-time behavior in the transition regions
near these critical rays. The transition regions also exhibit interesting asymptotic
behavior. We show that the first correction to the leading
term in these regions can be expressed in terms of a
distinguished Painlev\'e transcendent.
We introduce this special function below.

Define the critical spectral point \(k_c=q_0/\sqrt{2}\).
A double-scaling analysis near \(k_c\) leads to a local RH model
that can be transformed into the standard Jimbo--Miwa
Painlev\'e-II model studied systematically in~\cite{Miller}.
We define
\begin{equation}
 \widetilde\nu_1
 =-\frac{1}{2\pi}\log\bigl(1+|\rho_2(k_c)|^2\bigr).
 \label{eq:intro-transition-nu}
\end{equation}
The first correction to the leading term can then be expressed
in terms of the increasing tritronqu\'ee solution \(Q(y)\)
of the inhomogeneous Painlev\'e-II equation
\begin{equation}
 Q''+\frac23yQ-2Q^3
 +\frac23\mathrm{i}\widetilde\nu_1-\frac13=0,
 \label{eq:intro-PII}
\end{equation}
characterized by the asymptotic condition
\begin{equation}
 Q(y)=\mathrm{i}\left(-\frac y3\right)^{1/2}
 -\left(\frac14-\frac{\mathrm{i} \widetilde{\nu}_1}{2}\right)\frac1y
 +O(|y|^{-5/2}),
 \qquad |y|\to\infty, \quad |\arg(-y)|<2\pi/3.
 \label{eq:PII-Q-asymptotics-int}
\end{equation}
This distinguished solution also appears in the study of
infinite-order rogue waves by Bilman, Ling, and
Miller~\cite{BilmanLingMiller2020}.

We can now state the long-time asymptotics in the transition
region \(\mathcal{T}_{\mathrm{I}}\).

\begin{theorem}[Painlev\'e-II asymptotics]
\label{thm:tr-final}
Assume that the initial data \(\mathbf q(x,0)\) satisfy Assumptions~\ref{ass:intro-no-discrete-spectrum}
and~\ref{A:intii}.   Then the solution $\mathbf q(x,t)$ admits the following asymptotic expansion,
uniformly for 
\((x,t)\in\mathcal T_{\mathrm I}\):
\begin{equation}
 \mathbf q(x,t)
 =
 \left(1+t^{-1/3}q_P(x,t)\right)e^{-2\mathrm{i}\widetilde g_\infty}\mathbf q_+
 +O(t^{-1/2}),
 \qquad t\to\infty,
 \label{eq:tr-final-asymptotics}
\end{equation}
where the phase \(\widetilde g_\infty\) is defined
by~\eqref{eq:tr-g-infty}, and the bounded correction
\(q_P(x,t)\) is given by
\begin{equation*}
 q_P(x,t)
 =-\frac{\mathrm{i}\left(
 \kappa_1\mathcal V^*(y)
 +\kappa_2\mathcal V(y)
 +\kappa_3\widehat{\mathcal V}(y)
 \right)}
 {q_0\left(\frac{\sqrt6}{9q_0}\right)^{1/3}}, \qquad y=-\frac{2}{\sqrt3}\left(\frac{8\sqrt6}{9q_0}\right)^{-1/3}
 t^{2/3}(\xi-4\sqrt2q_0).
\end{equation*}
The coefficients \(\{  \kappa_j \}_{j=1}^3\) are given in~\eqref{kappaj123}.
The functions \(\mathcal V\) and \(\widehat{\mathcal V}\)
are expressed in terms of the increasing tritronqu\'ee solution
\(Q(y)\) of the inhomogeneous Painlev\'e-II
equation~\eqref{eq:intro-PII} through
\eqref{E:defmathcalV} and~\eqref{E:hatmathcalV}, respectively.
\end{theorem}

\begin{remark}
We briefly explain the \(O(t^{-1/2})\) remainder.
The contribution at this order comes from the parabolic cylinder
local model near \(k_d=7q_0/(4\sqrt{2})\).
The corresponding coefficient can be computed explicitly from
the asymptotic expansion of this local model.
We omit this calculation for simplicity, as our main focus
is the leading correction of order \(t^{-1/3}\).
\end{remark}

\begin{remark}
An interesting feature of the transition asymptotics is that
the Painlev\'e-type transition occurs only in the background
direction. More precisely, the correction of order \(t^{-1/3}\)
is a scalar multiple of the background vector.
\end{remark}

\subsection{Outline of the paper}
The paper is organized as follows.

\begin{itemize}
\item[$\bullet$]
Section~\ref{sec:direct-scattering-k} develops the direct and
inverse scattering formulation on the spectral \(k\)-plane.
We formulate the \(3\times3\) RH problem and derive the
reconstruction formula.

\item[$\bullet$]
In Section~\ref{sec:steepest-descent}, we carry out the
Deift--Zhou steepest-descent analysis in the plane-wave
region \(\mathrm P_{\mathrm I}\).
We perform a sequence of transformations, and construct a genus-zero outer parametrix together
with the required local parametrices. These constructions
allow us to obtain a small-norm
RH problem. The proof of
Theorem~\ref{thm:long-time-two-stationary} is completed in
Section~\ref{subsec:reconstruction-asymptotics}.

\item[$\bullet$]
Section~\ref{sec:elliptic-wave} studies the long-time
asymptotics in the elliptic-wave region $\mathrm E_{\mathrm{I}}$.
The steepest-descent analysis is more involved here.
We introduce a \(G\)-function transformation that leads
to a genus-one outer model. We also construct Airy local
parametrices and a nonstandard parabolic-cylinder parametrix
near distinguished spectral points.
The analysis is again reduced to a small-norm RH problem,
and the proof of Theorem~\ref{thm:long-EW} is completed in
Section~\ref{subsec:elliptic-reconstruction}.

\item[$\bullet$]
In Section~\ref{sec:positive-transition}, we derive the
Painlev\'e-II asymptotics in the transition region
\(\mathcal T_{\mathrm I}\).
We carry out a double-scaling analysis near the critical
point \(k_c\) and construct a suitable Painlev\'e-II local
parametrix. As in the other regions, the remaining analysis
is reduced to a small-norm RH problem.
The proof of Theorem~\ref{thm:tr-final} is completed in
Section~\ref{S:poggP}.

\item[$\bullet$]
The appendices present the parabolic-cylinder, Airy, and
Painlev\'e-II model problems used in the analysis.
\end{itemize}

\medskip
\noindent
\textbf{\textit{Notation.}}
We use the following notation throughout the paper. The symbols
\(\mathbb C_\pm\) denote the  upper and lower half-planes. 
 The norm in the classical space $L^p(\mathbb R)$, $1 \le p \le \infty$, is written as $\|\cdot \|_{L^p}$. For a $3 \times 3$
matrix $A$, we also define  $\|A \|_{L^p}=\mathrm{max}_{i,j=1,2,3} \| A_{ij}\|_{L^p}$.
Given a function \(f\) and an oriented contour, we denote
its boundary values from the left and right sides of the
contour by \(f^+\) and \(f^-\), respectively.  For a scalar or matrix quantity,
\(f^*\) denotes entrywise complex conjugation. For a complex-valued function \(f(k)\), we write
\(\bar{f}(k)=f^*(k^*)\). The symbol \(\dagger\) denotes the Hermitian
transpose. For a column vector
\(\mathbf v=(v_1,v_2)^{\mathsf T}\), we define
\[
 \mathbf v^\perp=(v_2^*,-v_1^*)^{\mathsf T}.
\]
The symbol \(C\) denotes a positive constant whose
value may change from line to line.  We write \(a\lesssim b\) to mean that \(a\leq Cb\)
for some constant \(C>0\).

\section
{Riemann--Hilbert formulation on the  \(k\)-plane}
\label{sec:direct-scattering-k}
In~\cite{KrausBiondiniKovacic2015}, the IST for the focusing
CNLS equations with NZBCs was developed using a uniformization
variable. For the subsequent Deift--Zhou steepest-descent
analysis, however, it is more convenient to formulate both
the direct and inverse scattering problems in the original
spectral \(k\)-plane.
As we shall see, the analysis closely parallels that in
the uniformization variable, although some results need
to be reformulated.   We  omit some of the proofs for brevity and refer the reader to \cite{KrausBiondiniKovacic2015} for the corresponding arguments.

\subsection{Jost eigenfunctions and the scattering matrix}

Recall that the focusing
CNLS equations~\eqref{eq:manakov-nzbc} is the compatibility condition  of  Lax pair \eqref{eq:lax-pair-k}.
Indeed, the zero-curvature condition
\(X_t-T_x+[X,T]=0\) is equivalent to
\eqref{eq:manakov-nzbc}.
Let \(Q_\pm=\lim_{x\to\pm\infty}Q(x,t)\) and
\(X_\pm=-\mathrm ikJ+Q_\pm\).  The eigenvalues of \(X_\pm\) are
\(-\mathrm i\lambda\), \(\mathrm ik\), and \(\mathrm i\lambda\),
where
\begin{equation}
 \lambda(k)=\sqrt{k^2+q_0^2}.
 \label{eq:lambda-k}
\end{equation}
We take
\[
 B=[-\mathrm{i}q_0,\mathrm{i}q_0]
\]
as the branch cut and choose the branch of \(\lambda\)
that is analytic in \(\mathbb C\setminus B\) and satisfies
\(\lambda(k)\sim k\) as \(k\to\infty\).  
We orient the branch cut \(B\) upward and, for \(k\in B\) away from
the branch points, define \(\lambda(k)\) as its boundary value from the
right-hand side of \(B\), namely,
\(
\lambda(k):=\lambda^{-}(k)
\).
Since the two boundary values differ by a sign, it follows that
\[
\lambda^{+}(k)=-\lambda^{-}(k)=-\lambda(k),
\qquad k\in B.
\]

It's easy to verify that the matrices
\begin{equation}
 E_\pm(k)=
 \begin{pmatrix}
  1&0&-\frac{\mathrm i(k-\lambda)}{q_0}\\[6pt]
  \frac{\mathrm i(\lambda-k)}{q_0^2}\mathbf q_\pm
   &\frac{\mathbf q_\pm^\perp}{q_0}
   &\frac{\mathbf q_\pm}{q_0}
 \end{pmatrix}
 \label{eq:Epm-k}
\end{equation}
simultaneously diagonalize the limiting Lax matrices:
\begin{align*}
 E_\pm^{-1}X_\pm E_\pm=\mathrm i\Lambda,  \qquad
 E_\pm^{-1}T_\pm E_\pm=-\mathrm i\Omega,
\end{align*}
where
\begin{align}
 \mathrm i\Lambda=\operatorname{diag}
 (-\mathrm i\lambda,\mathrm ik,\mathrm i\lambda), \qquad 
 -\mathrm i\Omega=\operatorname{diag}
 \bigl(2\mathrm ik\lambda,-\mathrm i(k^2+\lambda^2),
 -2\mathrm ik\lambda\bigr).
 \nonumber
\end{align}
Moreover, we can define the function $d(k)$ by
\begin{equation}
 d(k)=\det E_\pm(k)=\frac{2\lambda}{k+\lambda}.
 \label{eq:d-k}
\end{equation}

Define \(\Sigma=\mathbb R \cup B\). For $k \in \Sigma \setminus \{\pm \ii q_0 \}$, let
\(\boldsymbol\phi_\pm(x,t,k)\) denote the simultaneous solutions of
\eqref{eq:lax-pair-k} normalized by
\begin{equation}
 \boldsymbol\phi_\pm(x,t,k)
 =E_\pm(k)e^{\Theta(x,t,k)}+o(1),
 \qquad x\to\pm\infty,
 \label{eq:Jost-normalization}
\end{equation}
where  the diagonal phase is given by
\begin{equation*}
 \Theta(x,t,k)=\mathrm i\Lambda x-\mathrm i\Omega t
 =\operatorname{diag}(\theta_1,\theta_2,\theta_3),
\end{equation*}
with
\begin{equation}
 \theta_1=-\mathrm i\lambda x+2\mathrm ik\lambda t,
 \qquad
 \theta_2=\mathrm ikx-\mathrm i(k^2+\lambda^2)t,
\qquad
\theta_3=-\theta_1. \label{eq:theta-k}
\end{equation}
It is convenient to remove the background oscillations by setting
\(\mu_\pm=\boldsymbol\phi_\pm e^{-\Theta}\). Let
\(\Delta Q_\pm=Q-Q_\pm\).  It is straightforward to verify that \(\mu_{\pm}\) satisfy the following Volterra integral equations:
\begin{equation}
 \begin{split}
 \mu_\pm(x,t,k)=E_\pm(k)
 +\int_{\pm\infty}^{x}
 E_\pm e^{\mathrm i(x-y)\Lambda}E_\pm^{-1}
 \Delta Q_\pm(y,t)\mu_\pm(y,t,k)
 e^{-\mathrm i(x-y)\Lambda}\,\mathrm dy .
 \end{split}
 \label{eq:Volterra-k}
\end{equation}

\begin{proposition}
\label{prop:Jost-analyticity-k}

Suppose that, for each fixed \(t\), there exists
\(a\in\mathbb R\) such that
\[
 \mathbf q(\cdot,t)-\mathbf q_-\in L^1(-\infty,a),
 \qquad
 \mathbf q(\cdot,t)-\mathbf q_+\in L^1(a,\infty).
\]
Then  \(\mu_\pm(x,t,k)\) are well defined for\( k \in \Sigma\setminus \{ \pm \ii q_0  \} \).
Moreover, the columns \(\mu_{-,1}\) and \(\mu_{+,2}\)
admit analytic continuations to \(\mathbb C_+\setminus B\),
whereas \(\mu_{-,2}\) and \(\mu_{+,1}\) admit analytic
continuations to \(\mathbb C_-\setminus B\).
No analogous analytic continuation is guaranteed for the
third columns \(\mu_{\pm,3}\).

\end{proposition}

\begin{proof}
Writing the Volterra integral equations \eqref{eq:Volterra-k} columnwise,
the conjugation by \(e^{\mathrm i(x-y)\Lambda}\) gives rise to exponential
terms whose growth or decay depends on the signs of the imaginary parts
of the corresponding phase functions.  For \(\mu_{-,1}\) and \(\mu_{+,2}\),
these exponential factors remain bounded along the  integration intervals when
\(k\in\mathbb C_+\setminus B\), whereas for
\(\mu_{-,2}\) and \(\mu_{+,1}\) they remain bounded when
\(k\in\mathbb C_-\setminus B\).
Together with the \(L^1\)-decay assumptions on
\(\mathbf q-\mathbf q_\pm\), these bounds imply that the corresponding
Neumann series converge absolutely and locally uniformly with respect
to \(k\) on the indicated domains. Since each term of the Neumann
series is analytic in \(k\), the resulting Jost columns admit analytic
continuations to the stated regions.
\end{proof}

\begin{remark}
The corresponding analyticity result in the uniformized \(z\)-plane is given in
\cite[Theorem~2.1]{KrausBiondiniKovacic2015}. The regions above are the projections,
onto the selected \(k\)-sheet, of the corresponding analyticity domains in the \(z\)-plane.
\end{remark}

For $k \in \Sigma\setminus
\{\pm\mathrm iq_0\}$, Abel's identity gives
\begin{equation}
 \det\mu_\pm =  \det E_{\pm}(k)=d(k),
 \qquad
 \det\boldsymbol\phi_\pm
 =d(k)e^{\theta_2(x,t,k)}.
 \label{eq:Jost-determinants}
\end{equation}
 Away from the branch points,
\(d(k)\ne0\), so both Jost matrices are fundamental solutions.
Therefore, there exists a scattering matrix \(A(k)=(a_{ij}(k))\) independent of \(x\) and \(t\) such that
\begin{equation}
 \boldsymbol\phi_-(x,t,k)=\boldsymbol\phi_+(x,t,k)A(k), \qquad k \in \Sigma\setminus
\{\pm\mathrm iq_0\}.
 \label{eq:scattering-relation-k}
\end{equation}
  Taking determinants in
\eqref{eq:scattering-relation-k} yields \(\det A=1\).
Since the scattering matrix is independent of \(t\), we set \(t=0\) and
suppress the time dependence. 
Then the scattering matrix admits the integral representation
\begin{equation}
\begin{aligned}
A(k)
={}&
\int_{0}^{\infty}
e^{-\mathrm{i}y\Lambda(k)}
E_+^{-1}(k)
\bigl[Q(y)-Q_+\bigr]
\boldsymbol\phi_-(y,k)\,dy
\\
&\quad
+E_+^{-1}(k)E_-(k)
\left[
I+
\int_{-\infty}^{0}
e^{-\mathrm{i}y\Lambda(k)}
E_-^{-1}(k)
\bigl[Q(y)-Q_-\bigr]\boldsymbol\phi_-(y,k)\,dy
\right].
\label{eq:A-integral-k}
\end{aligned}
\end{equation}
Write \(A^{-1}(k)=(b_{ij}(k))\). It follows from the integral
representations of the scattering coefficients that \(a_{11}(k)\)
and \(b_{22}(k)\) admit analytic continuations to
\(\mathbb C_+\setminus B\), whereas \(a_{22}(k)\) and \(b_{11}(k)\)
admit analytic continuations to \(\mathbb C_-\setminus B\).

We next investigate the symmetry properties of the Jost solutions and
the scattering matrix. We begin with the symmetry induced by complex
conjugation of the spectral parameter, \(k\mapsto k^*\). The
complex-conjugation reduction of the Lax pair yields the following
result.

\begin{proposition}
\label{prop:conjugation-k}
Define
\(
\mathbf C(k):=\operatorname{diag}\bigl(d(k),1,d(k)\bigr).
\)
Then, for all \(k\) for which the relevant quantities are defined, the
Jost solutions satisfy
\begin{equation}
\bigl(\boldsymbol{\phi}_{\pm}^{\dagger}(x,t,k^*)\bigr)^{-1}\mathbf C(k)
=
\boldsymbol{\phi}_{\pm}(x,t,k),
\label{eq:Jost-conjugation-k}
\end{equation}
and the scattering matrix obeys
\begin{equation}
A^{\dagger}(k^*)
=
\mathbf C(k)A^{-1}(k) \mathbf C^{-1}(k).
\label{eq:scattering-conjugation-k}
\end{equation}
\end{proposition}

\begin{proof}
The reduction \(Q^{\dagger}=-Q\) implies that the coefficient matrices
of the Lax pair satisfy
\[
X^{\dagger}(x,t,k^*)=-X(x,t,k),
\qquad
T^{\dagger}(x,t,k^*)=-T(x,t,k).
\]
Consequently,
\(
\bigl(\boldsymbol{\phi}_{\pm}^{\dagger}(x,t,k^*)\bigr)^{-1}
\)
is a fundamental solution of the Lax pair at \(k\). Comparing its
asymptotic behavior with the normalization of
\(\boldsymbol{\phi}_{\pm}(x,t,k)\) in
\eqref{eq:Jost-normalization}, and then using uniqueness, gives
\eqref{eq:Jost-conjugation-k}.

Taking the Hermitian conjugate of the scattering relation at \(k^*\)
and using \eqref{eq:Jost-conjugation-k}, we obtain
\[
A(k)
=
\mathbf C^{-1}(k)\bigl(A^{\dagger}(k^*)\bigr)^{-1}\mathbf C(k),
\]
which is equivalent to \eqref{eq:scattering-conjugation-k}.
\end{proof}

The second symmetry corresponds to the involution
\(z\mapsto\widehat z\) introduced
in~\cite{KrausBiondiniKovacic2015}, which exchanges the two
sheets of the associated Riemann surface.
Since we fix a single branch of \(\lambda(k)\) throughout
this work, the corresponding symmetry in the \(k\)-plane
relates the boundary values on opposite sides of the
branch cut \(B\).
Indeed, crossing \(B\) interchanges the two sheets, or equivalently,
changes \(\lambda\) into \(-\lambda\). Comparing the boundary values of
the first and third columns of \eqref{eq:Epm-k}, and using the
uniqueness of the normalized Jost solutions, we obtain
\begin{equation} \label{eq:cut-symmetry-k}
\begin{cases}
\boldsymbol{\phi}_{-,1}^{+}
=
-\frac{\mathrm{i}\bigl(k+\lambda \bigr)}{q_0}
 \boldsymbol{\phi}_{-,3}^{-},\\
\boldsymbol{\phi}_{-,3}^{+}
=
-\frac{\mathrm{i}\bigl(k+\lambda\bigr)}{q_0}
 \boldsymbol{\phi}_{-,1}^{-},
\end{cases}
\quad
 k\in B_+, \qquad
\begin{cases}
\boldsymbol{\phi}_{+,1}^{+}
=
-\frac{\mathrm{i}\bigl(k+\lambda\bigr)}{q_0}
 \boldsymbol{\phi}_{+,3}^{-},\\
\boldsymbol{\phi}_{+,3}^{+}
=
-\frac{\mathrm{i}\bigl(k+\lambda\bigr)}{q_0}
 \boldsymbol{\phi}_{+,1}^{-},
\end{cases}
\quad k\in B_-,
\end{equation}
where $B_{\pm}=\mathbb C_{\pm } \cap B$.  We remind the reader that, by convention, \(\lambda(k)=\lambda^{-}(k)\) here.

We next investigate the relations between the boundary values of the scattering matrix on the two sides of the branch cut $B$. The following lemma summarizes these relations.

\begin{lemma}
\label{lem:scattering-cut-symmetry-k}
For $k\in B\setminus\{\pm\mathrm{i}q_0\}$, the boundary values of the scattering coefficients satisfy
\begin{align}
    a_{11}^{\pm}(k)&=a_{33}^{\mp}(k),
    &
    a_{22}^{\pm}(k)&=a_{22}^{\mp}(k),
    &
    a_{13}^{\pm}(k)&=a_{31}^{\mp}(k),
    \label{eq:scattering-cut-symmetry-diagonal}
    \\[2mm]
    a_{21}^{+}(k)
    &=
    -\frac{\mathrm{i}(k+\lambda)}{q_0}\,a_{23}^{-}(k),
    &
    a_{12}^{+}(k)
    &=
    \frac{\mathrm{i}q_0}{k+\lambda}\,a_{32}^{-}(k),
    \label{eq:scattering-cut-symmetry-first}
    \\[2mm]
    a_{23}^{+}(k)
    &=
    -\frac{\mathrm{i}(k+\lambda)}{q_0}\,a_{21}^{-}(k),
    &
    a_{32}^{+}(k)
    &=
    \frac{\mathrm{i}q_0}{k+\lambda}\,a_{12}^{-}(k).
    \label{eq:scattering-cut-symmetry-second}
\end{align}
\end{lemma}

\begin{proof}
We prove the first identity; the remaining identities follow in the same manner. A direct calculation gives the following Wronskian representation:
\begin{equation}
    a_{11}(k)
    =
    \frac{1}{d(k)e^{\theta_2(x,t,k)}}
    W\bigl[
        \phi_{-,1}(x,t,k),
        \phi_{+,2}(x,t,k),
        \phi_{+,3}(x,t,k)
    \bigr],
    \label{eq:a11-Wronskian-k}
\end{equation}
where $W[\cdot]$ denotes the standard Wronskian determinant.
Using
\(
    d^{+}(k)
    =
    \frac{(k+\lambda)^2}{q_0^2}\,d^{-}(k)
\)
together with the column symmetries in \eqref{eq:cut-symmetry-k}, we obtain
\begin{align*}
    a_{11}^{+}(k)
    &=
    \frac{1}{d^{+}(k)e^{\theta_2(x,t,k)}}
    W\bigl[
        \phi_{-,1}^{+},
        \phi_{+,2}^{+},
        \phi_{+,3}^{+}
    \bigr]
    \\
    &=
    \frac{q_0^2}{(k+\lambda)^2}
    \frac{1}{d^{-}(k)e^{\theta_2(x,t,k)}}
    \left(
        -\frac{\bigl[-\mathrm{i}(k+\lambda)\bigr]^2}{q_0^2}
    \right)
    W\bigl[
        \phi_{+,1}^{-},
        \phi_{+,2}^{-},
        \phi_{-,3}^{-}
    \bigr]
    \\
    &=a_{33}^{-}(k).
\end{align*}
This proves the first identity. All the other identities are obtained analogously.
\end{proof}

\begin{remark}
\label{sheetchageB}
The same symmetry relations hold for the entries of the inverse scattering matrix $A^{-1}=(b_{ij})$. More precisely, each $a_{ij}$ in the identities above may be replaced by the corresponding entry $b_{ij}$.
\end{remark}

It's remains to consider the large-\(k\) behavior of the modified Jost
eigenfunctions $\mu_{\pm }$. The corresponding analysis in the uniformization
variable \(z\) was carried out in \cite{KrausBiondiniKovacic2015}.  Since, on  our \(k \)-plane,
\[
z=k+\lambda(k)=2k+O(k^{-1}),
\qquad k\to\infty,
\]
the following results are obtained directly by expressing their
\(z\)-plane asymptotics in terms of \(k\). We therefore omit the proof.
If \(\mathbf q(x,t)\) is sufficiently smooth and approaches
its background value sufficiently rapidly as \(x\to \pm \infty\),
the Volterra integral equations for the modified Jost
eigenfunctions yield
\begin{equation}
\mu_{\pm,1}(x,t,k)
=
\begin{pmatrix}
1\\[1mm]
\frac{\mathrm{i}}{2k}\mathbf{q}(x,t)
\end{pmatrix}+\frac{1}{k}
\begin{pmatrix}
\star\\[1mm]
\mathbf{0}
\end{pmatrix}
+O(k^{-2}),
\qquad k\to\infty.
\label{eq:mu1-large-k}
\end{equation}
Here, \(\star\) denotes a function of \(x\) and \(t\)
whose explicit expression is omitted for convenience.
Similarly, the third columns satisfy
\begin{equation}
\mu_{\pm,3}(x,t,k)
=
\begin{pmatrix}
0\\[2mm]
\frac{\mathbf{q}_{\pm}}{q_0}
\end{pmatrix} 
+O(k^{-1}),
\qquad k\to\infty.
\label{eq:mu3-large-k}
\end{equation}
We next determine the corresponding asymptotic behavior of the
diagonal scattering coefficients. Recall that the scattering matrix
\(A(k)=(a_{ij}(k))\) is defined by
\[
\phi_{-}(x,t,k)=\phi_{+}(x,t,k)A(k).
\]
Combining the large-\(k\) expansions of the Jost eigenfunctions with
this scattering relation gives
\begin{equation}\label{eq:a22-large-k-phase}
\begin{aligned}
a_{11}(k)=1+O(k^{-1}),
\qquad k\to\infty,
\\
a_{22}(k)
=
\frac{\mathbf{q}_{-}^{\dagger}\mathbf{q}_{+}}{q_0^2}
+O(k^{-1}),
\qquad k\to\infty.
\end{aligned}
\end{equation}
Moreover, since the boundary values are parallel and satisfy
\(
\mathbf{q}_{+}
=
\mathrm{e}^{\mathrm{i}(\theta_{+}-\theta_{-})}\mathbf{q}_{-}
\),
 we have
\begin{equation}
a_{22}(k)
=
\mathrm{e}^{\mathrm{i}(\theta_{+}-\theta_{-})}
+O(k^{-1}),
\qquad k\to\infty.
\end{equation}
Thus, in contrast to \(a_{11}(k)\), the coefficient \(a_{22}(k)\)
does not in general tend to \(1\).

\subsection{Adjoint eigenfunctions}
Compared with the inverse scattering analysis of integrable systems admitting a $2\times2$ Lax pair, the $3\times3$ case is considerably more involved, as it requires an additional analysis of the adjoint Lax pair. This analysis allows us to use the adjoint eigenfunctions to construct auxiliary eigenfunctions with the required analyticity, thereby supplying the missing analytic columns needed to formulate the RH problem. The adjoint Lax pair is defined as follows:
\begin{equation}
 \widetilde{\boldsymbol\phi}_x
 =\widetilde X\widetilde{\boldsymbol\phi},
 \qquad
 \widetilde{\boldsymbol\phi}_t
 =\widetilde T\widetilde{\boldsymbol\phi},
 \label{eq:adjoint-Lax-k}
\end{equation}
where 
$$
 \widetilde X(x,t,k)=X^*(x,t,k^*),\quad
 \widetilde T(x,t,k)=T^*(x,t,k^*).
$$
The following lemma, proved in~\cite{KrausBiondiniKovacic2015}, provides the basis for our construction of the auxiliary eigenfunctions.
\begin{lemma}[Cross-product construction]
\label{lem:cross-product-k}
If \(\widetilde{\mathbf v}\) and \(\widetilde{\mathbf w}\) solve
\eqref{eq:adjoint-Lax-k}, then
\begin{equation}
 \mathbf u(x,t,k)
 =e^{\theta_2(x,t,k)}
 \bigl(\widetilde{\mathbf v}(x,t,k)
 \times\widetilde{\mathbf w}(x,t,k)\bigr)
 \label{eq:cross-product-k}
\end{equation}
solves the original Lax pair \eqref{eq:lax-pair-k}.
\end{lemma}

In analogy with the definition of the Jost solutions, we define the adjoint Jost solutions through the following asymptotic normalization conditions:
\begin{equation}
    \widetilde{\boldsymbol\phi}_\pm(x,t,k)
    =
    E_\pm^*(k^*)e^{-\Theta(x,t,k)}+o(1),
    \qquad x\to\pm\infty,
    \label{eq:adjoint-Jost-k}
\end{equation}
and introduce the modified adjoint Jost eigenfunctions
\(
    \widetilde{\mu}_\pm
    =
    \widetilde{\boldsymbol\phi}_\pm e^{\Theta}
\).
The reversal of the signs in the exponential factors changes the analyticity properties of the corresponding columns. More precisely, $\widetilde{\mu}_{-,2}$ and $\widetilde{\mu}_{+,1}$ admit analytic continuations to $\mathbb{C}_+\setminus B$, whereas $\widetilde{\mu}_{-,1}$ and $\widetilde{\mu}_{+,2}$ admit analytic continuations to $\mathbb{C}_-\setminus B$.

In fact, a direct relation between the Jost solutions and their adjoint counterparts can be readily established. The asymptotic normalizations as $x\to\pm\infty$, together with Lemma~\ref{lem:cross-product-k}, imply that, for each cyclic permutation $(j,\ell,m)$ of $(1,2,3)$,
\begin{equation}
    \phi_{\pm,j}
    =
    \frac{e^{\theta_2}}{\mathcal{J}_j}
    \bigl(
        \widetilde{\phi}_{\pm,\ell}
        \times
        \widetilde{\phi}_{\pm,m}
    \bigr),
    \label{eq:cyclic-cross-product-k}
\end{equation}
where $\mathcal{J}_1=\mathcal{J}_3=1$ and $\mathcal{J}_2=d(k)$.


We are now in a position to introduce the auxiliary eigenfunctions.

\begin{lemma}[Auxiliary analytic eigenfunctions]
\label{lem:auxiliary-k}
Define
\begin{equation} \label{eq:auxiliary-k}
\begin{aligned}
 \chi_1(x,t,k)=e^{\theta_2(x,t,k)}
 \bigl(\widetilde\phi_{+,1}(x,t,k)\times\widetilde\phi_{-,2}(x,t,k)\bigr),
 \\
 \chi_2(x,t,k)=e^{\theta_2(x,t,k)}
 \bigl(\widetilde\phi_{-,1}(x,t,k)\times\widetilde\phi_{+,2}(x,t,k)\bigr).
\end{aligned}
\end{equation}
Then \(\chi_1\) is analytic in \(\mathbb C_+\setminus B\), whereas
\(\chi_2\) is analytic in \(\mathbb C_-\setminus B\). For $k \in \Sigma$,
\begin{align}
 \chi_1(x,t,k)
 &=a_{11}\phi_{-,3}(x,t,k)-a_{13}\phi_{-,1}(x,t,k)
  =b_{22}\phi_{+,3}(x,t,k)-b_{23}\phi_{+,2}(x,t,k),
 \label{eq:chi1-decomposition-k}\\
 \chi_2(x,t,k)
 &=b_{11}\phi_{+,3}(x,t,k)-b_{13}\phi_{+,1}(x,t,k)
  =a_{22}\phi_{-,3}(x,t,k)-a_{23}\phi_{-,2}(x,t,k).
 \label{eq:chi2-decomposition-k}
\end{align}
\end{lemma}

\begin{proof}
The analyticity of $\chi_1$ and $\chi_2$ follows directly from the analyticity properties of the adjoint Jost columns established above. 
To obtain the decompositions, insert the adjoint scattering relation into one factor of each cross product and then apply \eqref{eq:cyclic-cross-product-k}. Alternating bilinearity of the cross product leaves precisely the two terms displayed in each identity.
\end{proof}
Furthermore, it follows from~\cite[Lemma~3.11]{KrausBiondiniKovacic2015} that, as $k\to\infty$,
\begin{equation} \label{E:chi12asy}
\begin{aligned}
\chi_1(x,t,k)e^{-\theta_3(x,t,k)}
&=
\begin{pmatrix}
0\\[1mm]
\dfrac{\mathbf q_-}{q_0}
\end{pmatrix}
+O(k^{-1}),
\qquad k\in\mathbb C_+\setminus B,
\\
\chi_2(x,t,k)e^{-\theta_3(x,t,k)}
&=
\begin{pmatrix}
0\\[1mm]
\dfrac{\mathbf q_+}{q_0}
\end{pmatrix}
+O(k^{-1}),
\qquad k\in\mathbb C_-\setminus B.
\end{aligned}
\end{equation}

\subsection{The Riemann--Hilbert problem}

We now assemble the columns needed to formulate the inverse problem. Before doing so,  we impose the  spectral assumption \ref{ass:intro-no-discrete-spectrum} and restrict our attention to the solitonless case.

We then define a sectionally analytic function on $\mathbb{C}\setminus B$ as follows:
\begin{equation}
 M(x,t,k)=
 \begin{cases}
 \displaystyle
  \left(\dfrac{\phi_{-,1}}{a_{11}d},\ \phi_{+,2},\
        \dfrac{\chi_1}{b_{22}}\right)e^{-\Theta},
      &k\in\mathbb C_+\setminus B,\\[11pt]
 \displaystyle
  \left(\phi_{+,1},\ \dfrac{\phi_{-,2}}{a_{22}},\
        \dfrac{\chi_2}{b_{11}d}\right)e^{-\Theta},
      &k\in\mathbb C_-\setminus B.
 \end{cases}
 \label{eq:RHP-unknown-k}
\end{equation}

\begin{remark}
In~\cite{KrausBiondiniKovacic2015}, the two-sheeted Riemann surface is uniformized by the variable $z$, and the RH problem is formulated on the resulting $z$-plane. The $z$-plane is divided into four domains of analyticity, with a separate analytic matrix-valued function defined in each domain. In the present work, by contrast, we formulate the RH problem on a single sheet of the Riemann surface. It therefore suffices to divide this sheet into two domains of analyticity and define the RH unknown separately in each domain.
\end{remark}

\begin{remark}
The two factors $d^{-1}$ are introduced to normalize $M$ so that $\det M=1$.
\end{remark}

Set
\[
    \Sigma^\circ
    :=
    \Sigma\setminus\{0,\pm\mathrm{i}q_0\}.
\]
We now describe the jump relations satisfied on $\Sigma^\circ$ by the matrix $M$ defined in \eqref{eq:RHP-unknown-k}. The corresponding jump matrices are expressed in terms of the following three reflection coefficients:
\begin{equation}
     \rho_1(k)=\frac{a_{21}(k)}{a_{11}(k)},
 \qquad
 \rho_2(k)=\frac{a_{31}(k)}{a_{11}(k)},
 \qquad
 \rho_3(k)=\frac{a_{32}(k)}{a_{22}(k)}.
    \label{eq:reflection-coefficients-k}
\end{equation}
Recall that
\(
 B_{\pm}=B\cap\mathbb C_{\pm}
\).
The explicit
jump relation is as follows. For convenience, we may suppress the dependence on \(x\) and \(t\).

\begin{proposition}
\label{prop:explicit-jumps-k}
The boundary values of \eqref{eq:RHP-unknown-k} on $\Sigma^\circ$  satisfy
\begin{equation}
 M^+(x,t,k)=M^-(x,t,k)V(x,t,k),
 \qquad k\in\Sigma^\circ,
 \label{eq:RHP-jump-k}
\end{equation}
where
\begin{equation}
 V(x,t,k)=
 \begin{cases}
  V_{\mathbb R}(x,t,k),&k\in\mathbb R\setminus\{0\},\\
  V_{B_+}(x,t,k),&k\in B_+,\\
  V_{B_-}(x,t,k),&k\in B_-.
 \end{cases}
 \label{eq:piecewise-jump-k}
\end{equation}
On the real axis,
\begin{equation}
 \label{eq:CR-k}
 V_{\mathbb R}(k)=e^{\Theta(k)}
 \begin{pmatrix}
 \dfrac{1}{d(k)}
 \left(1+\dfrac{|\rho_1(k)|^2}{d(k)}+|\rho_2(k)|^2\right)
 &\dfrac{\rho_1^*(k)}{d(k)}
 &\rho_2^*(k)-\rho_1^*(k)\rho_3^*(k)\\[7pt]
 \dfrac{\rho_1(k)}{d(k)}
 &1
 &-d(k)\rho_3^*(k)\\[5pt]
 \rho_2(k)-\rho_1(k)\rho_3(k)
 &-d(k)\rho_3(k)
 &d(k)\left(1+d(k)|\rho_3(k)|^2\right)
 \end{pmatrix}
 e^{-\Theta(k)}.
\end{equation}
On the two halves of the cut,
\begin{equation}
\label{eq:VB-k}
\begin{aligned}
 V_{B_+}(k)
 &=
 \begin{pmatrix}
 -\dfrac{\mathrm{i}(\lambda-k)}{q_0}
   \rho_2^+(k)e^{2\theta_1(k)}
 &0&-\dfrac{2\mathrm{i}\lambda}{q_0}\\[7pt]
 0&1&0\\[5pt]
 -\dfrac{\mathrm{i}q_0}{2\lambda}
   \left(1-\rho_2^+(k)\rho_2^-(k)\right)
 &0&\dfrac{\mathrm{i}(k+\lambda)}{q_0}
   \rho_2^-(k)e^{2\theta_3(k)}
 \end{pmatrix},
 \\[4mm]
 V_{B_-}(k)
 &=
 \begin{pmatrix}
 -\dfrac{\mathrm{i}(k+\lambda)}{q_0}
   \bar\rho_2^-(k)e^{2\theta_1(k)}
 &0&-\dfrac{\mathrm{i}q_0}{2\lambda}
   \left(1-\bar\rho_2^+(k)\bar\rho_2^-(k)\right)\\[7pt]
 0&1&0\\[5pt]
 -\dfrac{2\mathrm{i}\lambda}{q_0}
 &0&\dfrac{\mathrm{i}(\lambda-k)}{q_0}
   \bar\rho_2^+(k)e^{2\theta_3(k)}
 \end{pmatrix}.
\end{aligned}
\end{equation}
\end{proposition}

\begin{proof}

\emph{Jump on \(\mathbb R\).}
Combining the scattering relation with
\eqref{eq:chi2-decomposition-k}, we obtain
\begin{align} \label{E:wjkln}
 \phi_{+,3}(k)
 =d(k)\frac{\chi_2 (k)}{b_{11}(k)d(k)}+\rho_2^*(k)\phi_{+,1}(k),
 \qquad k\in\mathbb R.
\end{align}
Using the scattering relation once more, together with
\eqref{E:wjkln}, gives
\[
 \phi_{+,2}(k)
 =\frac{\rho_1^*(k)}{d(k)}\phi_{+,1}(k)
  +\frac{\phi_{-,2}(k)}{a_{22}(k)}
  -d(k)\rho_3(k)\frac{\chi_2(k)}{b_{11}(k)d(k)}.
\]
The identity above is equivalent to the relation for the
second column in~\eqref{eq:RHP-jump-k} on the real axis.
On the other hand, the scattering relation and 
\eqref{eq:chi1-decomposition-k} yield
\begin{align}
 \frac{\phi_{-,1}(k)}{a_{11}(k)d(k)}
 &=\frac{1}{d(k)}
   \left(
      \phi_{+,1}(k)+\rho_1(k)\phi_{+,2}(k)
      +\rho_2(k)\phi_{+,3}(k)
   \right),
 \label{eq:real-jump-proof-1}\\
 \frac{\chi_1(k)}{b_{22}(k)}
 &=\phi_{+,3}(k)-d(k)\rho_3^*(k)\phi_{+,2}(k).
 \label{eq:real-jump-proof-2}
\end{align}
Substituting the preceding expressions for \(\phi_{+,2}\)
and \(\phi_{+,3}\) into~\eqref{eq:real-jump-proof-1}
and~\eqref{eq:real-jump-proof-2} yields two identities.
These are equivalent to the relations for the first and
third columns in~\eqref{eq:RHP-jump-k} on the real axis,
respectively.  This establishes \eqref{eq:CR-k}.

\emph{Jump on \(B_+\).}
The second column $\phi_{+,2}$ is unchanged under sheet interchange; hence
\(
    \phi_{+,2}^+=\phi_{+,2}^-
\) for $k \in B_+$.
Since $A^{-1}=(b_{ij})$ and $\det A=1$, the cofactor formula gives
\(
    b_{22}=a_{11}a_{33}-a_{13}a_{31}
\).
It therefore follows from
Lemma~\ref{lem:scattering-cut-symmetry-k} that
\begin{equation}
    \frac{b_{22}^-(k)}{a_{11}^- (k) a_{11}^+(k)}
    =
    1-\rho_2^+(k)\rho_2^-(k).
    \label{eq:Bplus-cofactor-k}
\end{equation}
Combining \eqref{eq:cut-symmetry-k}, \eqref{eq:Bplus-cofactor-k} with
\(
    \chi_1
    =
    a_{11}\phi_{-,3}-a_{13}\phi_{-,1}
\), we obtain
\begin{align}
    \frac{\phi_{-,1}^+(k)}{a_{11}^+(k) d^+(k)}
    &=
    -\frac{\mathrm{i}(k+\lambda)}{q_0d^+ (k)}
    \left[
        \rho_2^+(k) \frac{\phi_{-,1}^-(k)}{a_{11}^-(k)}
        +(1-\rho_2^+(k)\rho_2^-(k))
        \frac{\chi_1^- (k)}{b_{22}^-(k)}
    \right],
   \nonumber\\
    \frac{\chi_1^+ (k)}{b_{22}^+(k)}
    &=
    -\frac{\mathrm{i}(k+\lambda)}{q_0}
    \left[
        d^- (k) \frac{\phi_{-,1}^- (k) }{a_{11}^-(k) d^-(k)}
        -\rho_2^-(k) \frac{\chi_1^-(k) }{b_{22}^-(k)}
    \right].
    \nonumber
\end{align}
Finally, using the scalar identities
\begin{equation}
    -\frac{\mathrm{i}(k+\lambda)}{q_0}\frac{d^-(k)}{d^+(k)}
    =
    -\frac{\mathrm{i}(\lambda-k)}{q_0},
    \qquad
    -\frac{\mathrm{i}(k+\lambda)}{q_0d^+ (k)}
    =
    -\frac{\mathrm{i}q_0}{2\lambda},
    \qquad
    -\frac{\mathrm{i}(k+\lambda)d^- (k)}{q_0}
    =
    -\frac{2\mathrm{i}\lambda}{q_0}
    \label{eq:cut-scalar-identities-k}
\end{equation}
we obtain a relation equivalent to the jump condition
for \(M\) on \(B_+\).

The jump relation on \(B_-\) is obtained by a similar calculation.
\end{proof}
\begin{remark}
It is straightforward to verify that
\(
    \det V_{\mathbb R}
    =
    \det V_{B_+}
    =
    \det V_{B_-}
    =
    1.
\)
\end{remark}

We next examine the behavior of the matrix-valued function
\(M\), defined by~\eqref{eq:RHP-unknown-k}, near the branch
points \(\pm\mathrm{i}q_0\).
We first recall the corresponding behavior of the Jost
eigenfunctions and the analytic scattering coefficients. The analysis in~\cite[Section~3.4]{KrausBiondiniKovacic2015}
shows that the modified Jost eigenfunctions are bounded
at the branch points. Moreover,
It follows by~\cite[Eq.~(3.16)]{KrausBiondiniKovacic2015} that the scattering matrix \(A(k)\) may have 
square-root singularities at the branch points:
\begin{equation} \label{E:arsssh}
A(k)
=
\frac{1}{(k\mp \mathrm{i}q_0)^{1/2}}
\begin{pmatrix}
\mathfrak{a}_{\pm} & \mathfrak{b}_{\pm} & \pm\mathfrak{a}_{\pm}\\
0 & 0 & 0\\
\mp\mathfrak{a}_{\pm} & \mp\mathfrak{b}_{\pm} & -\mathfrak{a}_{\pm}
\end{pmatrix}
+O(1),
\qquad k\to\pm\mathrm{i}q_0.
\end{equation}
Throughout this paper, we restrict our attention to the
generic case in which \(\mathfrak{a}_{\pm} \ne 0 \); see Assumption~\ref{ass:intro-no-discrete-spectrum}.
Moreover, the auxiliary eigenfunctions are also bounded, as follows
from their cross-product representations. Since
the function \(d(k)\) vanishes like \((k\pm \ii q_0)^{1/2}\)
at either branch point, 
\(
 M(x,t,k)=O(1)
\) as  $k\to\pm\mathrm{i}q_0$.

\begin{remark}
Under Assumption~\ref{ass:intro-no-discrete-spectrum},   it follows from~\eqref{E:arsssh} that as $ k \to \pm \ii q_0$,
 \begin{align} \label{E:rhojlhdy}
\rho_1(k)=O(k\mp \mathrm{i}q_0)^{1/2}), \qquad 
\rho_2(k)=\mp1+O\!\left(|k\mp\mathrm{i}q_0|^{1/2}\right), \qquad
\rho_3(k)=O(k\mp \mathrm{i}q_0)^{-1/2}).
\end{align}
These expansions ensure that all entries of \(V_{B_+}\)
and \(V_{B_-}\) remain bounded near \(\mathrm{i}q_0\)
and \(-\mathrm{i}q_0\), respectively.
\end{remark}

The boundedness of \(M(x,t,k)\) as \(k\to0\) is immediate. It remains to  examine the asymptotic behavior of the matrix-valued function $M$ as $k\to\infty$. Combining \eqref{eq:mu1-large-k}, \eqref{eq:mu3-large-k},  \eqref{eq:a22-large-k-phase}, and \eqref{E:chi12asy} with a direct calculation, we obtain
\begin{equation}
M_{\infty}:= \lim_{k  \to \infty} M(x,t,k)=
\begin{pmatrix}
  1&0&0\\[6pt]
\mathbf  0
   &\dfrac{\mathbf q_+^\perp}{q_0}
   &\dfrac{\mathbf q_+}{q_0}
 \end{pmatrix}.
\label{eq:M-infinity}
\end{equation}
A standard Liouville argument shows that RH
problem~\ref{rhp:M-solitonless} has at most one solution.
Since the matrix \(M\) defined by~\eqref{eq:RHP-unknown-k}
satisfies all the conditions of this RH problem, it is
the unique solution.  The coefficient of $k^{-1}$ in \eqref{eq:mu1-large-k} then yields the  reconstruction  formula \eqref{eq:reconstruction}.    This proves  Theorem \ref{thm:reconstruction}.

To facilitate the subsequent Deift--Zhou steepest-descent analysis,
we normalize the Riemann--Hilbert problem for $M$. Introduce
\begin{equation} \label{E:tildesigma}
    N(x,t,k)
    =
    \tilde{\sigma}^{-1}M_{\infty}^{-1}M(x,t,k)\tilde{\sigma},
    \qquad
   \tilde{ \sigma}=
    \begin{pmatrix}
        1&0&0\\
        0&0&1\\
        0&1&0
    \end{pmatrix}.
\end{equation}
Then $N(x,t,k)$ satisfies the following Riemann--Hilbert problem.

\begin{RHP}
Find a $3\times3$ matrix-valued function $N(x,t,k)$ satisfying the
following conditions.

\begin{enumerate}
\item[(i)]
The function $N(x,t,\cdot)$ is analytic for
\(
    k\in\mathbb{C}\setminus\Sigma^\circ
\),
and admits non-tangential boundary values
$N^\pm(x,t,k)$ on $\Sigma^\circ$. These boundary values satisfy the
jump condition
\begin{equation}
    N^+(x,t,k)
    =
    N^-(x,t,k)\widetilde V(x,t,k),
    \qquad k\in\Sigma^\circ,
\end{equation}
where $\widetilde V(x,t,k)= \tilde{ \sigma}^{-1} V(x,t,k) \tilde{ \sigma}$.
\item[(ii)]
As $k\to\infty$,
\begin{equation}
    N(x,t,k)
    =
I+O\!\left(\frac{1}{k}\right).
\end{equation}

\item[(iii)]
The matrix \(N(x,t,k)\) remains bounded as \(k\) approaches
\(0\) or \(\pm\mathrm{i}q_0\).
\end{enumerate}
\end{RHP}

To facilitate the matrix factorizations in the following sections, we write
\(\widetilde V(x,t,k)\) explicitly as
\begin{equation}
    \widetilde V(x,t,k)
    =
    \begin{cases}
        \widetilde V_{\mathbb R}(x,t,k),
            & k\in\mathbb R,\\[1mm]
        \widetilde V_{B_+}(x,t,k),
            & k\in B_+,\\[1mm]
        \widetilde V_{B_-}(x,t,k),
            & k\in B_-.
    \end{cases}
\end{equation}
Define $\theta_{ij}=\theta_i-\theta_j$. Then $\widetilde V_{\mathbb R}(x,t,k)$ is given by
\begin{equation}
\widetilde V_{\mathbb R}(k)
=
\begin{pmatrix}
\frac{1}{d(k)}
\left(
1+\frac{|\rho_1(k)|^2}{d(k)}+|\rho_2(k)|^2
\right)
&
\bigl(\rho_2(k)-\rho_1(k)\rho_3(k)\bigr)^*
e^{\theta_{13}(k)}
&
\frac{\rho_1^*(k)}{d(k)}e^{\theta_{12}(k)}
\\[3mm]
\bigl(\rho_2(k)-\rho_1(k)\rho_3(k)\bigr)
e^{\theta_{31}(k)}
&
d(k)\left(1+d(k)|\rho_3(k)|^2\right)
&
-d(k)\rho_3(k)e^{\theta_{32}(k)}
\\[3mm]
\frac{\rho_1(k)}{d(k)}e^{\theta_{21}(k)}
&
-d(k)\rho_3^*(k)e^{\theta_{23}(k)}
&
1
\end{pmatrix}.
\end{equation}
On the two halves of the cut, we have 
\begin{align}
\widetilde V_{B_+}(x,t,k)
&=
\begin{pmatrix}
-\frac{\mathrm{i}(\lambda-k)}{q_0}
\rho_2^+(k)e^{2\theta_1(x,t,k)}
&
-\frac{2\mathrm{i}\lambda}{q_0}
&
0
\\[3mm]
-\frac{\mathrm{i}q_0}{2\lambda}
\left(1-\rho_2^+(k)\rho_2^-(k)\right)
&
\frac{\mathrm{i}(k+\lambda)}{q_0}
\rho_2^-(k)e^{2\theta_3(x,t,k)}
&
0
\\[3mm]
0&0&1
\end{pmatrix},
\label{eq:Vtilde-Bplus-k}
\\[4mm]
\widetilde V_{B_-}(x,t,k)
&=
\begin{pmatrix}
-\frac{\mathrm{i}(\lambda+k)}{q_0}
\bar{\rho}_2^-(k)e^{2\theta_1(x,t,k)}
&
-\frac{\mathrm{i}q_0}{2\lambda}
\left(1-\bar{\rho}_2^+(k)\bar{\rho}_2^-(k)\right)
&
0
\\[3mm]
-\frac{2\mathrm{i}\lambda}{q_0}
&
\frac{\mathrm{i}(\lambda-k)}{q_0}
\bar{\rho}_2^+(k)e^{2\theta_3(x,t,k)}
&
0
\\[3mm]
0&0&1
\end{pmatrix}.
\label{eq:Vtilde-Bminus-k}
\end{align}

In the following sections, we apply the Deift--Zhou nonlinear
steepest-descent method to the RH problem for \(N\).
The Deift--Zhou analysis is greatly simplified if the reflection
coefficients admit analytic continuations off the continuous
spectrum \(\Sigma\).  This can be achieved by imposing the
following  assumption on the initial data.

\begin{assumption}
\label{A:intii}
We assume that there exists \(L>0\) such that
\[
 \mathbf q(x,0)=\mathbf q_-,
 \qquad x<-L
 \quad\text{and}\quad
 \mathbf q(x,0)=\mathbf q_+,
 \qquad x>L.
\]
\end{assumption}

We make the following comment on this assumption.

\begin{remark}
The compact-support assumption is stronger than is strictly
necessary for analytic continuation.
For example, in the scalar setting considered
in~\cite{BiondiniMantzavinos2017}, exponential convergence
of the initial data to the corresponding backgrounds is
sufficient to guarantee analyticity of the reflection
coefficient in a neighborhood of the continuous spectrum.
In the present \(3\times3\) setting, however, the corresponding
estimates for the Neumann series are less straightforward.
We therefore adopt the stronger compact-support assumption,
under which the required analyticity follows directly
from~\eqref{eq:A-integral-k}.

For initial data satisfying weaker decay conditions, one may
instead use rational approximations of the reflection
coefficients~\cite{pz93} or the \(\bar\partial\) extension
of the Deift--Zhou method~\cite{BJM2018}.
However, adapting either approach to the present problem
would require nontrivial additional work.
Nonetheless, we regard the compact-support assumption as a
technical simplification rather than an essential restriction
and expect the resulting asymptotic description to capture
the behavior of solutions under more general initial data.
\end{remark}

\section{Long-time asymptotics in the plane-wave region}
\label{sec:steepest-descent}

The main idea of the Deift--Zhou nonlinear steepest descent method is to reduce the original RH problem, through a sequence of transformations, to  a  error RH  problem. A key step is to deform the jump contour, a procedure
we refer to as lens opening. The lens contours are chosen according to the signature tables of the relevant phase functions, so that the oscillatory entries of the jump matrices can be analytically continued into regions where they decay exponentially. We therefore begin by examining the properties of these phase functions. In this section, We restrict our analysis to the plane-wave region $\mathrm{P}_{\mathrm{I}}$:
\(
    \xi>4\sqrt{2}\,q_0
\).  Let  \(\mathcal K_+\)  be  any compact subset of
\((4\sqrt{2}q_0,\infty)\).  For convenience, we may suppress the dependence on
\(x\), \(t\), and \(k\) in the remainder of the paper.

Recall that
\(
    \theta_{ij}=\theta_i-\theta_j
\).
The stationary points of the three relevant phase functions $\theta_{12}$, $\theta_{13}$ and $\theta_{23}$ are described as follows.

\begin{enumerate}
\item The phase function $\theta_{13}$ has two distinct real stationary points:
\begin{equation} \label{E:expreofkwk0}
    k_w
    =
    \frac{\xi-\sqrt{\xi^2-32q_0^2}}{8},
    \qquad
    k_0
    =
    \frac{\xi+\sqrt{\xi^2-32q_0^2}}{8}.
\end{equation}

\item The phase function $\theta_{12}$ has a unique real stationary point:
\begin{equation}
    k_1
    =
    \frac{\xi}{4}-\frac{q_0^2}{\xi}.
\end{equation}

\item The phase function $\theta_{23}$ has no real stationary points.
\end{enumerate}

\begin{figure}[H]
 \centering
 \setlength{\tabcolsep}{2pt}
 \begin{tabular}{ccc}
  \includegraphics[width=0.315\textwidth]
    {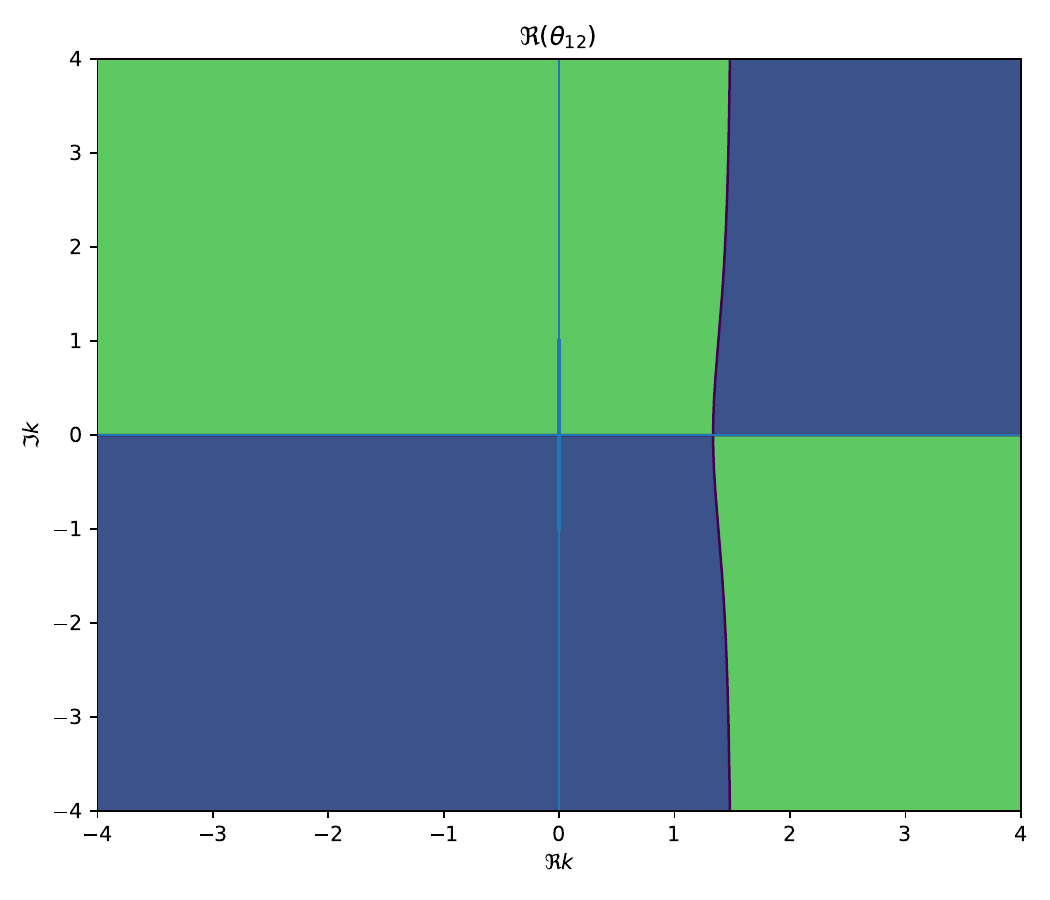}
  &
  \includegraphics[width=0.315\textwidth]
    {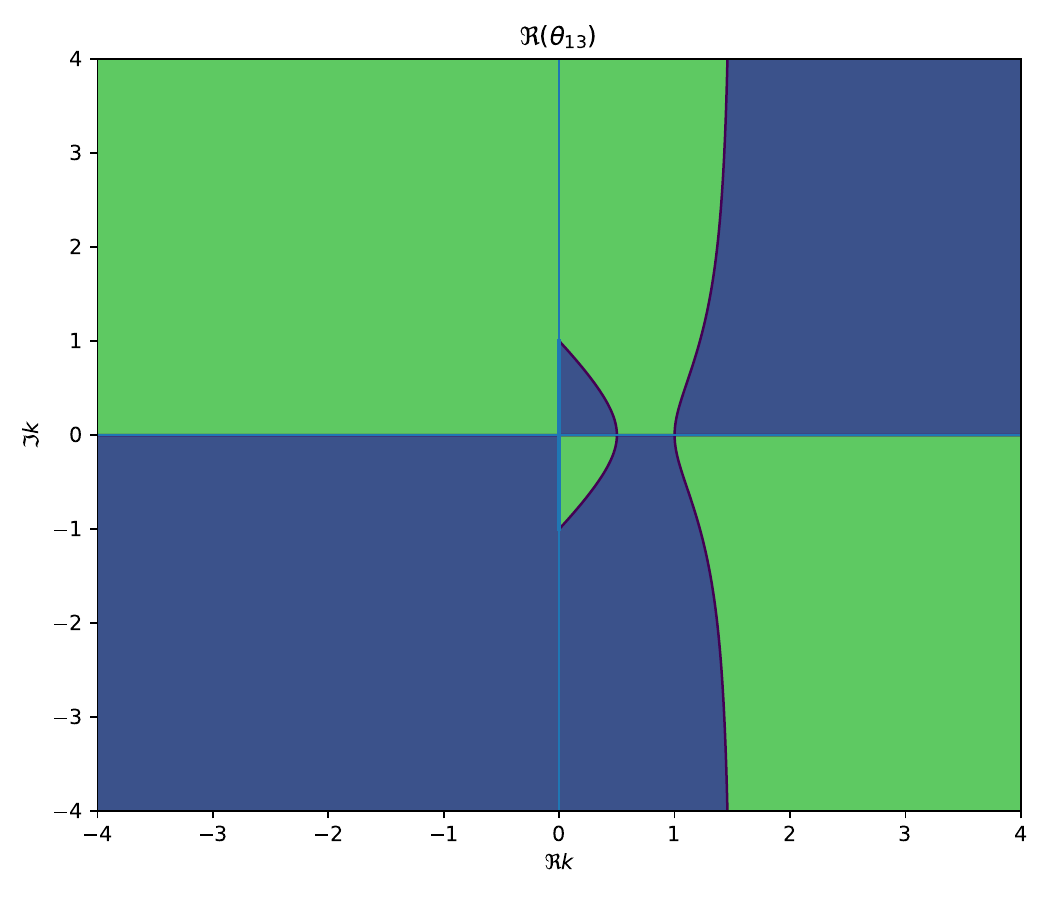}
  &
  \includegraphics[width=0.315\textwidth]
    {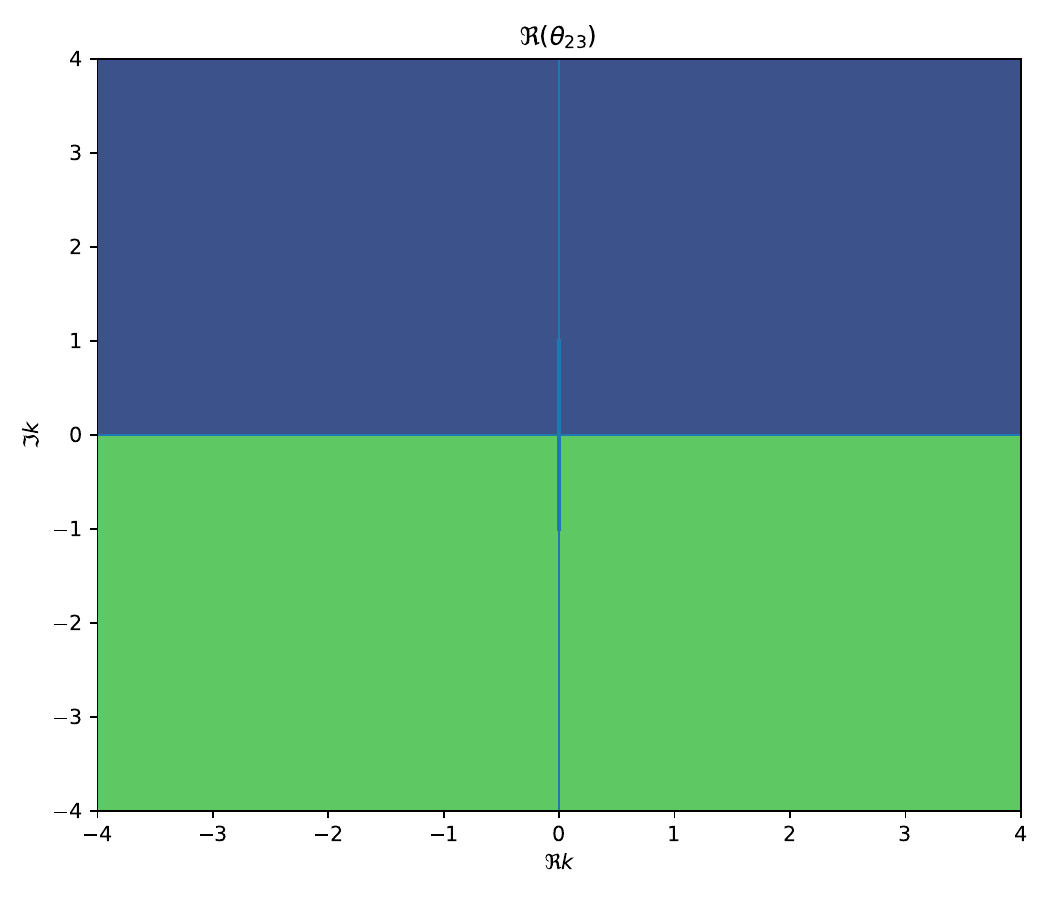}
 \end{tabular}
 \caption{Signature tables for
\(\operatorname{Re}\theta_{12}\),
\(\operatorname{Re}\theta_{13}\), and
\(\operatorname{Re}\theta_{23}\) (from left to right),
with \(q_0=1\) and \(\xi=6\).
The green and blue regions correspond to
\(\operatorname{Re}\theta_{ij}>0\) and
\(\operatorname{Re}\theta_{ij}<0\), respectively.}
 \label{fig:phase-signature-charts}
\end{figure}

For $\xi>4\sqrt{2}\,q_0$, it is straightforward to verify that
\(
    0<k_w<k_0<k_1
\).
The signature charts of the three phase functions in this regime are shown in Figure~\ref{fig:phase-signature-charts}.

\subsection{Opening of lenses: first transformation}
\label{subsec:first-lens}
We next perform the lens-opening transformation. The key step is to factorize the jump matrices in accordance with the signature table of the relevant phase functions, so that the oscillatory factors can be continued off the real axis into regions where they decay exponentially. In view of the configuration of the stationary points when $\xi>4\sqrt{2}\,q_0$~(see Fig~\ref{fig:phase-signature-charts}), different factorizations of the real-axis jump matrix $\widetilde V_{\mathbb R}$ are required on the three intervals
\[
    (-\infty,k_0),\qquad
    (k_0,k_1),\qquad
    (k_1,+\infty).
\]
We now present the corresponding factorizations, which can be verified by direct calculation.

 On $(-\infty,k_0)$ we use the exact
four-factor decomposition
\begin{equation}
 \widetilde V_{\mathbb R}
 =V_U^{(1)}V_{r,U}^{(1)}V_{r,L}^{(1)}V_L^{(1)},
 \label{eq:left-four-factorization}
\end{equation}
with
\begin{subequations}\label{eq:left-four-factors}
\begin{align}
 V_U^{(1)}&=
 \begin{pmatrix}
  d^{-1}& \bar \rho_2e^{\theta_{13}}&0\\
  0&d&0\\0&0&1
 \end{pmatrix},
 &
 V_L^{(1)}&=
 \begin{pmatrix}
  d^{-1}&0&0\\
  \rho_2e^{\theta_{31}}&d&0\\0&0&1
 \end{pmatrix},\\[2mm]
 V_{r,U}^{(1)}&=
 \begin{pmatrix}
  d&0&(\bar \rho_1+d \bar {\rho}_2\rho_3)e^{\theta_{12}}\\
  0&1&-\rho_3e^{\theta_{32}}\\0&0&1
 \end{pmatrix},
 &
 V_{r,L}^{(1)}&=
 \begin{pmatrix}
  1&0&0\\
  0&d^{-1}&0\\
  (\rho_1+d\rho_2 \bar{\rho}_3)e^{\theta_{21}}
    &-\bar \rho_3 e^{\theta_{23}}&1
 \end{pmatrix}.
\end{align}
\end{subequations}
On the interval $(k_0,k_1)$, the jump matrix admits the following factorization:
\begin{equation}
 \widetilde V_{\mathbb R}=V_a^{(1)}V_b^{(1)}V_c^{(1)},
 \qquad k\in(k_0,k_1),
 \label{eq:middle-factorization}
\end{equation}
where
\begin{subequations}\label{eq:middle-factors}
\begin{align}
 V_a^{(1)}&=
 \begin{pmatrix}
  1&0&d^{-1}\bar \rho_1e^{\theta_{12}}\\
  0&d&-d\rho_3e^{\theta_{32}}\\
  0&0&1
 \end{pmatrix},\quad
 V_b^{(1)}=
 \begin{pmatrix}
  1+\rho_2  \bar \rho_2&\bar \rho_2 e^{\theta_{13}}&0\\
  \rho_2 e^{\theta_{31}}&1&0\\
  0&0&1
 \end{pmatrix},\quad
 V_c^{(1)}&=
 \begin{pmatrix}
  d^{-1}&0&0\\
  0&1&0\\
  d^{-1}\rho_1e^{\theta_{21}}&-d \bar \rho_3 e^{\theta_{23}}&1
 \end{pmatrix}.
\end{align}
\end{subequations}
Finally, on $(k_1,\infty)$ we use
\begin{equation}
 \widetilde V_{\mathbb R}=V_d^{(1)}V_e^{(1)}V_f^{(1)},
 \qquad k\in(k_1,\infty),
 \label{eq:right-factorization}
\end{equation}
where
\begin{subequations}\label{eq:right-factors}
\begin{align}
 V_d^{(1)}&=
 \begin{pmatrix}
  1&0&0\\0&d&-d\rho_3e^{\theta_{32}}\\0&0&1
 \end{pmatrix},\\
 V_e^{(1)}&=
 \begin{pmatrix}
  1+d^{-1}\bar \rho_1 \rho_1+\rho_2 \bar \rho_2
    &\bar \rho_2 e^{\theta_{13}}&d^{-1} \bar \rho_1 e^{\theta_{12}}\\
  \rho_2e^{\theta_{31}}&1&0\\
  \rho_1e^{\theta_{21}}&0&1
 \end{pmatrix},\\
 V_f^{(1)}&=
 \begin{pmatrix}
  d^{-1}&0&0\\0&1&0\\0&-d \bar \rho_3 e^{\theta_{23}}&1
 \end{pmatrix}.
\end{align}
\end{subequations}

\begin{remark}
Recall that
\(
    \bar{\rho}_j(k):=\rho_j^*(k^*)
\).
This convention ensures that the conjugated quantities admit the required analytic continuations.
\end{remark}

The factorizations above constitute only the first stage of the lens-opening procedure. At this stage, we analytically continue off the real axis only those matrix entries whose exponential factors already have the desired sign structure. To treat the remaining off-diagonal entries, whose phases have the opposite sign structure, we introduce the conjugating matrix functions $\Delta_1$ and $\Delta_2$; see section~\ref{subsec:scalar-conjugations}. These functions modify the factorization so that the corresponding exponential factors have the required decay properties and can then be continued into the appropriate regions.

Alternatively, one may adopt a more direct lens-opening strategy. One may first introduce $\Delta_1$ and $\Delta_2$, and then factorize the jump matrix on each of the three intervals so that the exponential terms in all three off-diagonal entries of every factor decay in the corresponding lens regions. All the lenses can then be opened in a single step. This approach produces a sequence of transformations equivalent to that adopted here. We prefer the present two-step procedure because it makes the role of the signature structure more transparent.

We also introduce the following two diagonal matrices:
\begin{equation}
    V_g^{(1)}
    =
    \operatorname{diag}(d^{-1},1,1),
    \qquad
    V_h^{(1)}
    =
    \operatorname{diag}(1,d,1).
    \label{eq:first-h}
\end{equation}
 Some of the jump matrices produced by the lens-opening transformation contain nontrivial diagonal entries, and these matrices are used to remove the corresponding diagonal factors.

\begin{figure}[htbp]
\centering

\begin{tikzpicture}[
    x=0.92cm,
    y=0.92cm,
    >=Stealth,
    contour/.style={
        black,
        line width=2pt,
        line cap=round,
        line join=round
    },
    axis/.style={
        black,
        dashed,
        line width=1.50pt,
        line cap=round
    },
    branch/.style={
        black,
        line width=2.2pt,
        line cap=round
    },
    raylabel/.style={
        font=\small,
        fill=white,
        inner sep=1.0pt
    },
        dir/.style={
        postaction={decorate},
        decoration={
            markings,
            mark=at position .56 with {
                \arrow[scale=0.8]{>}
            }
        }
    }
]

\draw[axis] (-6,0)--(6,0);

\draw[contour,dir] (-6,1.8)--(0,1.8);
\draw[contour,dir] (0,1.8)--(6,1.8);
\draw[contour,dir] (-6,-1.8)--(0,-1.8);
\draw[contour,dir] (0,-1.8)--(6,-1.8);

\draw[contour,dir] (-2.30,1.80)--(-0.50,0);
\draw[contour,dir] (-2.30,-1.80)--(-0.50,0);

\draw[contour,dir] (-0.50,0)--(2.50,0);

\draw[contour,dir] (0.70,1.80)--(2.50,0);
\draw[contour,dir] (0.70,-1.80)--(2.50,0);

\draw[contour,dir] (2.50,0)--(6,0);

\draw[branch,dir] (-4.00,-1.25)--(-4.00,1.25);

\fill[black] (-0.50,0) circle (2.8pt);
\fill[black] (2.50,0) circle (2.8pt);

\node[below=6pt] at (-0.50,0) {$k_0$};
\node[below=6pt] at (2.50,0) {$k_1$};

\node at (-3.30, 0.88) {$D_1$};
\node at (-3.30,-0.88) {$D_2$};

\node at (0.2, 1.00) {$D_3$};
\node at (0.2,-1.00) {$D_4$};

\node at (3.85, 0.82) {$D_5$};
\node at (3.85,-0.82) {$D_6$};

\node at (0.20, 2.45) {$D_7$};
\node at (0.20,-2.45) {$D_8$};

\node[raylabel] at (-4.42,0.5) {$B$};

%

\node[raylabel] at (-2.02, 0.98) {$1$};
\node[raylabel] at (-2.02,-0.98) {$2$};

\node[raylabel] at (2.05, 0.98) {$3$};
\node[raylabel] at (2.05,-0.98) {$4$};

\node[raylabel] at (0.5,0.28) {$5$};
\node[raylabel] at (4.55,0.28) {$6$};

\end{tikzpicture}

\caption{The jump countor  $\Sigma^{(1)}$ and the regions
$D_1,\ldots,D_8$.}
\label{fig:first-deformation}
\end{figure}
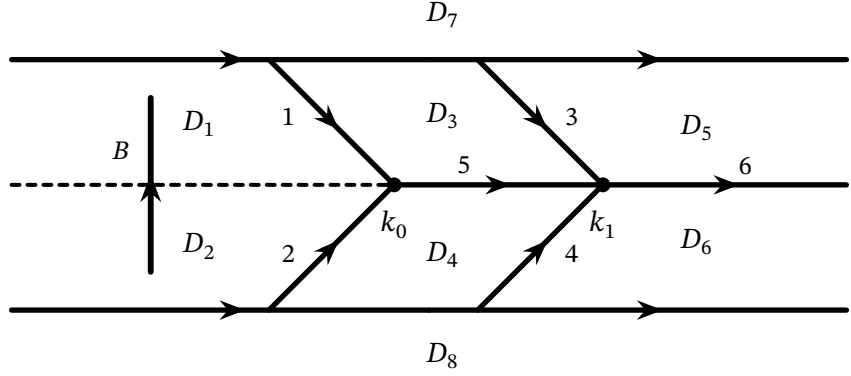

We are now ready to introduce the first transformation, defined as follows:
\begin{equation}
 N^{(1)}(x,t,k)=N(x,t,k)G_1(k),
 \label{eq:first-transformation}
\end{equation}
where
\begin{equation}
G_1(k)=
\begin{cases}
 (V_L^{(1)})^{-1}(V_{r,L}^{(1)})^{-1},&k\in D_1,\\
 V_U^{(1)}V_{r,U}^{(1)},&k\in D_2,\\
 (V_c^{(1)})^{-1},&k\in D_3,\\
 V_a^{(1)},&k\in D_4,\\
 (V_f^{(1)})^{-1},&k\in D_5,\\
 V_d^{(1)},&k\in D_6,\\
 (V_g^{(1)})^{-1},&k\in D_7,\\
 V_h^{(1)},&k\in D_8.
\end{cases}
\label{eq:G1-piecewise}
\end{equation}
All matrices in this definition have been given explicitly in
\eqref{eq:left-four-factors}, \eqref{eq:middle-factors},
\eqref{eq:right-factors}, and \eqref{eq:first-h}.   The first deformed contour and its eight complementary regions $\{D_{j}\}_{j=1}^8$ are shown in
Figure~\ref{fig:first-deformation}.

\begin{remark}
Using \eqref{E:rhojlhdy} and the expressions for \(G_1(k)\)
in \(D_1\) and \(D_2\), we see that \(G_1(k)\) remains bounded
near the branch points \(\pm\mathrm{i}q_0\).
\end{remark}

The new contour is denoted by $\Sigma^{(1)}$ .  The jump matrices on the contour segments labeled $1,\ldots,6$ in Figure~\ref{fig:first-deformation} are given by
\begin{subequations}\label{eq:first-principal-jumps}
\begin{align}
 V_1^{(1)}&=
 \begin{pmatrix}
  1&0&0\\ \rho_2e^{\theta_{31}}&1&0\\ 0&0&1
 \end{pmatrix},
 &V_2^{(1)}&=
 \begin{pmatrix}
  1&\bar{\rho}_2e^{\theta_{13}}&0\\ 0&1&0\\ 0&0&1
 \end{pmatrix},\\[2mm]
 V_3^{(1)}&=
 \begin{pmatrix}
  1&0&0\\ 0&1&0\\ \rho_1e^{\theta_{21}}&0&1
 \end{pmatrix},
 &V_4^{(1)}&=
 \begin{pmatrix}
  1&0&d^{-1} \bar{\rho}_1e^{\theta_{12}}\\ 0&1&0\\ 0&0&1
 \end{pmatrix},\\[2mm]
 V_5^{(1)}&=V_b^{(1)},
 &V_6^{(1)}&=V_e^{(1)}.
\end{align}
\end{subequations}
Here, $V_j^{(1)}$ denotes the jump matrix on the contour segment labeled $j$. The same convention will be used for the jump matrices arising from subsequent transformations.
The sign structure of the phase functions shows that the jump matrices
on all unlabeled contour segments in Figure~\ref{fig:first-deformation}
approach the identity matrix exponentially fast as \(t\to\infty\).
We therefore omit their explicit expressions.

It remains to identify the jump on the branch cut $B$.  Before the deformation,
the jump matrices on the upper and lower parts of the branch cut can be factorized as
\begin{equation}
 \widetilde V_{B_+}=(V_L^{(1)-})^{-1}V_sV_L^{(1)+},
 \qquad
 \widetilde V_{B_-}=V_U^{(1)-}V_t(V_U^{(1)+})^{-1},
 \label{eq:branch-pre-factorization}
\end{equation}
where the matrices $V_s$ and $V_t$ are given by
\begin{equation}
 V_s=
 \begin{pmatrix}
  0&-\frac{\mathrm i(k+\lambda)}{q_0d^+}&0\\[2mm]
  -\frac{\mathrm i(k+\lambda)}{q_0}d^-&0&0\\
  0&0&1
 \end{pmatrix},
 \qquad
 V_t=
 \begin{pmatrix}
  0&-\frac{\mathrm i(k+\lambda)}{q_0}d^-&0\\[2mm]
  -\frac{\mathrm i(k+\lambda)}{q_0d^+}&0&0\\
  0&0&1
 \end{pmatrix}.
 \label{eq:Vs-Vt}
\end{equation}
Here $d^\pm$ are the boundary values on the two sides of $B$.  Consequently, direct matrix multiplication on $B_+$ gives
\begin{equation}
 V_{B_+}^{(1)}
 =V_{r,L}^{(1)-}V_L^{(1)-}\widetilde V_{B_+}
  (V_L^{(1)+})^{-1}(V_{r,L}^{(1)+})^{-1}
 =
 \begin{pmatrix}
  0&-\frac{\mathrm i(k+\lambda)}{q_0}&0\\[2mm]
  -\frac{\mathrm i(k+\lambda)}{q_0}&0&0\\[2mm]
  f_1&f_2&1
 \end{pmatrix},
 \label{eq:first-branch-jump-with-remainders}
\end{equation}
where the two entries $f_1$ and $f_2$ are given by
\begin{align}
 f_1&=\left[-\frac{b_{23}^-}{b_{22}^-}
              \left(-\frac{\mathrm i(k+\lambda)}{q_0}\right)
              +\frac{b_{21}^+}{b_{22}^+}\right]e^{\theta_{23}},
 \nonumber\\
 f_2&=\left[-\frac{b_{21}^-}{b_{22}^-}
              \left(-\frac{\mathrm i(k+\lambda)}{q_0}\right)
              +\frac{b_{23}^+}{b_{22}^+}\right]e^{\theta_{21}}.
 \label{eq:branch-remainders}
\end{align}
Applying the sheet-interchange symmetry stated in Remark~\ref{sheetchageB}, we immediately obtain
\(
    f_1=f_2=0
\).
Therefore
\begin{equation}
V_{B_+}^{(1)}=
 \begin{pmatrix}
  0&-\frac{\mathrm i(k+\lambda)}{q_0}&0\\[2mm]
  -\frac{\mathrm i(k+\lambda)}{q_0}&0&0\\
  0&0&1
 \end{pmatrix}.
 \label{eq:first-branch-jump}
\end{equation}
An analogous calculation on $B_-$ yields
\[
    V_{B_-}^{(1)}
    =
    V_{B_+}^{(1)}
    =:
    V_B^{(1)}.
\]

\subsection{Scalar conjugations and further contour deformations}
\label{subsec:scalar-conjugations}
We next introduce a scalar conjugation function $\delta_1(k)$. This function will be used to define a matrix transformation that yields a new factorization of the jump matrix on $(k_0,\infty)$. We define $\delta_1$ as the solution of the following scalar RH problem: $\delta_1$ is analytic in $\mathbb C\setminus[k_0,\infty)$ and satisfies
\begin{equation*}
    \delta_1^+(k)
    =
    \frac{\delta_1^-(k)}{1+|\rho_2(k)|^2},
    \qquad k\in(k_0,\infty);
    \qquad
    \delta_1(k)\to1,
    \quad k\to\infty.
\end{equation*}
By the Sokhotski--Plemelj formula, the solution of the scalar RH problem
 is given by
\begin{equation}
 \delta_1(k)=
 \exp\left\{-\frac{1}{2\pi\mathrm i}
 \int_{k_0}^{\infty}
 \frac{\log(1+|\rho_2(s)|^2)}{s-k}\,\mathrm ds\right\},
 \qquad k\in\mathbb C\setminus[k_0,\infty).
 \label{eq:delta1-solution}
\end{equation}
Set
\begin{equation}
 \Delta_1(k)=\operatorname{diag}(\delta_1,\delta_1^{-1},1).
 \label{eq:Delta1}
\end{equation}
Then the second transformation is defined by
\begin{equation}
    N^{(2)}(k)=N^{(1)}(k)\Delta_1(k),
    \qquad k\in\mathbb C.
    \label{eq:second-transformation}
\end{equation}
We denote the jump contour for $N^{(2)}$ by $\Sigma^{(2)}$. This transformation introduces no new contours, so we have
\(
    \Sigma^{(2)}=\Sigma^{(1)}
\).
The corresponding jump matrix is denoted by $V^{(2)}$. Because $\Delta_1$ is analytic away from $[k_0,\infty)$, the jump matrices on all contour segments outside this interval are obtained by conjugation:
\begin{equation}
    V^{(2)}(k)
    =
    \Delta_1(k)^{-1}V^{(1)}(k)\Delta_1(k).
    \label{eq:delta1-conjugation}
\end{equation}
On $[k_0,\infty)$, however, the distinct boundary values of $\Delta_1$ lead to more involved jump matrices. We now give the corresponding expressions for $V^{(2)}$ and introduce the factorizations required for the subsequent contour deformation.

Using the jump condition satisfied by $\Delta_1$, a direct calculation gives, on $(k_0,k_1)$,
\begin{equation}
    V_5^{(2)}
    =
    V_a^{(2)}V_b^{(2)},
    \label{eq:V5-second-factorization}
\end{equation}
where
\begin{align}\label{eq:VaVb-second}
    V_a^{(2)}
    =
    \begin{pmatrix}
        1&0&0\\
        \dfrac{\rho_2}{1+\rho_2\bar\rho_2}
        (\delta_1^-)^2e^{\theta_{31}}&1&0\\
        0&0&1
    \end{pmatrix},
    \qquad
    V_b^{(2)}
    =
    \begin{pmatrix}
        1&
        \dfrac{\bar\rho_2}{1+\rho_2\bar\rho_2}
        (\delta_1^+)^{-2}e^{\theta_{13}}&0\\
        0&1&0\\
        0&0&1
    \end{pmatrix}.
\end{align}
On $(k_1,\infty)$, the jump matrix $V_6^{(1)}$ has a more complicated structure. Conjugation by the two boundary values of $\Delta_1$ leads to the expression
\begin{equation}
    V_6^{(2)}
    =
    V_c^{(2)}V_d^{(2)}.
    \label{eq:V6-cd-factorization}
\end{equation}
where
\begin{subequations}\label{eq:VcVd-second}
\begin{align}
 V_c^{(2)}&=
 \begin{pmatrix}
  1&0&d^{-1}(\delta_{1}^-)^{-1} \bar \rho_1e^{\theta_{12}}\\
  \dfrac{\rho_2}{1+\rho_2 \bar \rho_2}(\delta_{1}^-)^{2}e^{\theta_{31}}
    &1&0\\
  0&0&1
 \end{pmatrix},\\
 V_d^{(2)}&=
 \begin{pmatrix}
  1&\dfrac{\bar \rho_2}{1+\rho_2 \bar \rho_2}
       (\delta_{1}^+)^{-2}e^{\theta_{13}}&0\\
  0&1&0\\
  \rho_1 (\delta_{1}^+)e^{\theta_{21}}&0&1
 \end{pmatrix}.
\end{align}
\end{subequations}
Although this factorization appears simple, it is not compatible with the signature structure of the corresponding phase functions. We therefore refactor the product in \eqref{eq:V6-cd-factorization}. Direct multiplication yields the following exact three-factor decomposition:
\begin{equation}
 V_6^{(2)}=V_e^{(2)}V_g^{(2)}V_f^{(2)},
 \label{eq:V6-egf-factorization}
\end{equation}
with
\begin{subequations}\label{eq:Vegf-second}
\begin{align}
 V_e^{(2)}&=
 \begin{pmatrix}
  1&0&0\\[1mm]
  \dfrac{\rho_2}{1+\rho_2 \bar \rho_2}(\delta_{1}^-)^{2}e^{\theta_{31}}
  &1&-\dfrac{\bar \rho_1 \rho_2}{d(1+\rho_2 \bar \rho_2)}
  \delta_{1}^- e^{\theta_{32}}\\[3mm]
  0&0&1
 \end{pmatrix},\\[2mm]
 V_g^{(2)}&=
 \begin{pmatrix}
  1+\dfrac{d^{-1}|\rho_1|^2}{1+\rho_2 \bar \rho_2}
  &0&d^{-1}\bar \rho_1 (\delta_{1}^-)^{-1}e^{\theta_{12}}\\[3mm]
  0&1&0\\[1mm]
  \rho_1\delta_{1}^+ e^{\theta_{21}}&0&1
 \end{pmatrix},\\[2mm]
 V_f^{(2)}&=
 \begin{pmatrix}
  1&\dfrac{\bar \rho_2}{1+\rho_2 \bar \rho_2}
  (\delta_{1}^+)^{-2}e^{\theta_{13}}&0\\[3mm]
  0&1&0\\[1mm]
  0&-\dfrac{\bar \rho_2 \rho_1}{1+\rho_2 \bar \rho_2}
  (\delta_{1}^+)^{-1}e^{\theta_{23}}&1
 \end{pmatrix}.
\end{align}
\end{subequations}

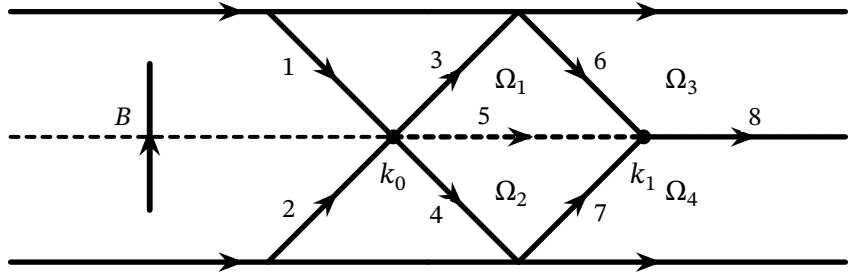
\begin{figure}
\centering

\begin{tikzpicture}[
    x=0.92cm,
    y=0.92cm,
    >=Stealth,
    contour/.style={
        black,
        line width=2pt,
        line cap=round,
        line join=round
    },
    axis/.style={
        black,
        dashed,
        line width=1.50pt,
        line cap=round
    },
    branch/.style={
        black,
        line width=2.2pt,
        line cap=round
    },
    raylabel/.style={
        font=\small,
        fill=white,
        inner sep=2.5pt
    },
     dir/.style={
        postaction={decorate},
        decoration={
            markings,
            mark=at position .56 with {
                \arrow[scale=0.8]{>}
            }
        }
    }
]

\draw[axis] (-6,0)--(6,0);

\draw[contour,dir] (-6,1.8)--(0,1.8);
\draw[contour,dir] (0,1.8)--(6,1.8);
\draw[contour,dir] (-6,-1.8)--(0,-1.8);
\draw[contour,dir] (0,-1.8)--(6,-1.8);

\draw[branch,dir] (-4.0,-1.05)--(-4.0,1.05);

\coordinate (kzero) at (-0.5,0);
\coordinate (kup)   at (1.3,1.8);
\coordinate (kdown) at (1.3,-1.8);
\coordinate (kone)  at (3.1,0);

\draw[contour,dir] (-2.3,1.8)--(kzero);
\draw[contour,dir] (-2.3,-1.8)--(kzero);

\draw[contour,dir] (kzero)--(kup);
\draw[contour,dir] (kzero)--(kdown);

\draw[contour,dir] (kup)--(kone);
\draw[contour,dir] (kdown)--(kone);

\draw[contour,dashed,dir] (kzero)--(kone);
\draw[contour,dir] (kone)--(6,0);

\fill[black] (kzero) circle (2.8pt);
\fill[black] (kone)  circle (2.8pt);

\node[below=6pt] at (kzero) {$k_0$};
\node[below=6pt] at (kone)  {$k_1$};

\node[raylabel] at (-4.38,0.3) {$B$};

\node at (1.20, 0.80) {$\Omega_1$};
\node at (1.20,-0.80) {$\Omega_2$};

\node at (3.65, 0.80) {$\Omega_3$};
\node at (3.65,-0.80) {$\Omega_4$};

\node[raylabel] at (-2.00, 1.02) {$1$};
\node[raylabel] at (-2.00,-1.02) {$2$};

\node[raylabel] at (0.12, 1.12) {$3$};
\node[raylabel] at (0.12,-1.12) {$4$};

\node[raylabel] at (0.80, 0.30) {$5$};

\node[raylabel] at (2.48, 1.10) {$6$};
\node[raylabel] at (2.48,-1.10) {$7$};

\node[raylabel] at (4.70, 0.30) {$8$};

\end{tikzpicture}

\caption{The jump contour $\Sigma^{(3)}$ for $N^{(2)}$ and the regions $\{\Omega_j \}_{j=1}^4$.}

\label{fig:second-deformation}
\end{figure}

We now introduce the third transformation. Let $\Omega_j$, $j=1,\ldots,4$, denote the regions shown in Figure~\ref{fig:second-deformation}, and define
\begin{equation}
    N^{(3)}(k)=N^{(2)}(k)G_3(k),
    \qquad
    G_3(k)=
    \begin{cases}
        (V_b^{(2)}(k))^{-1},&k\in\Omega_1,\\
        V_a^{(2)}(k),&k\in\Omega_2,\\
        (V_f^{(2)}(k))^{-1},&k\in\Omega_3,\\
        V_e^{(2)}(k),&k\in\Omega_4,\\
        I, & \text{elsewhere}.
    \end{cases}
    \label{eq:third-transformation}
\end{equation}
\begin{remark}
The factors appearing in the transformation matrices involve the boundary values $\delta_1^\pm$. When these factors are analytically extended into the lens regions $\Omega_j$, the quantities $\delta_1^\pm$ in their expressions are understood as the corresponding analytic functions $\delta_1(k)$ in the appropriate domains. We adopt this convention throughout the remainder of the analysis without further comment.
\end{remark}

We denote the resulting jump contour by $\Sigma^{(3)}$ and the corresponding jump matrix by $V^{(3)}$. We list below only the jumps associated with the labeled portions of the contour shown in Figure~\ref{fig:second-deformation}:
\begin{subequations}\label{eq:third-jumps}
\begin{align}
 V_1^{(3)}&=
 \begin{pmatrix}
  1&0&0\\ \rho_2\delta_1^2e^{\theta_{31}}&1&0\\ 0&0&1
 \end{pmatrix},
 \qquad V_2^{(3)}=
 \begin{pmatrix}
  1&\bar \rho_2\delta_1^{-2}e^{\theta_{13}}&0\\ 0&1&0\\ 0&0&1
 \end{pmatrix},\\
 V_6^{(3)}&=V_b^{(2)}V_3^{(2)}(V_f^{(2)})^{-1}
  =\begin{pmatrix}
   1&0&0\\ 0&1&0\\ \rho_1\delta_1e^{\theta_{21}}&0&1
  \end{pmatrix},\\[2mm]
 V_7^{(3)}&=(V_e^{(2)})^{-1}V_4^{(2)}V_a^{(2)}
  =\begin{pmatrix}
   1&0&d^{-1} \bar \rho_1\delta_1^{-1}e^{\theta_{12}}\\
   0&1&0\\ 0&0&1
  \end{pmatrix},\\[2mm]
  V_3^{(3)}&=V_b^{(2)},
 \qquad V_4^{(3)}=V_a^{(2)},\\
 V_5^{(3)}&=I,
 \qquad V_8^{(3)}=V_g^{(2)}.
\end{align}
\end{subequations}
Moreover, the present transformation leaves the jump on the branch cut $B$ unchanged; hence,
$
V^{(3)}_B=V^{(2)}_B.
$

At this stage, the only jump that remains on the real axis is the one supported on $[k_1,\infty)$. Our next goal is to deform this jump away from the real axis. To achieve a factorization that is compatible with the sign structure of the phase functions, we introduce an additional scalar conjugation function. Specifically, we define
\begin{equation}
h(k)=1+\frac{d(k)^{-1}|\rho_1(k)|^2}{1+|\rho_2(k)|^2},
\label{eq:h-delta2}
\end{equation}
and let $\delta_2$ denote the solution of the scalar RH problem that is analytic in $\mathbb{C}\setminus [k_1,\infty)$ and satisfies
\begin{equation*}
\delta_{2}^+(k)=\frac{\delta_{2}^-(k)}{h(k)},
\quad k\in[k_1,\infty);
\qquad
\delta_2(k)\to1,\qquad k\to\infty.
\end{equation*}
Thus by the Sokhotski--Plemelj formula, we have
\begin{equation}
 \delta_2(k)=
 \exp\left\{-\frac{1}{2\pi\mathrm i}
 \int_{k_1}^{\infty}\frac{\log h(s)}{s-k}\,\mathrm ds\right\},
 \qquad k\in\mathbb C\setminus[k_1,\infty).
 \label{eq:delta2-solution}
\end{equation}
We introduce the fourth transformation
\begin{equation}
N^{(4)}(k)=N^{(3)}(k)\Delta_2(k),
\qquad k\in\mathbb{C},
\label{eq:fourth-transformation}
\end{equation}
where
$$
\Delta_2(k)=\operatorname{diag}\bigl(\delta_2(k),1,\delta_2(k)^{-1}\bigr).
$$
We denote the resulting jump contour and jump matrix by $\Sigma^{(4)}$ and $V^{(4)}$, respectively. Since this transformation does not alter the contour, we have
$
\Sigma^{(4)}=\Sigma^{(3)}.
$

For all portions of the contour away from $[k_1,\infty)$, the corresponding jumps are conjugated by $\Delta_2$. On the interval $[k_1,\infty)$, we define
\begin{equation}
H(k)=1+|\rho_2(k)|^2+d(k)^{-1}|\rho_1(k)|^2,
\label{eq:H-fourth}
\end{equation}
in terms of which the jump inherited from $V_8^{(3)}=V_g^{(2)}$ admits the factorization
\begin{equation}
V_8^{(4)}
=(\Delta_2^-)^{-1}V_8^{(3)}\Delta_2^+
=V_a^{(4)}V_b^{(4)},
\qquad k\in(k_1,\infty).
\label{eq:V8-fourth-factorization}
\end{equation}
where
\begin{subequations}\label{eq:Vab-fourth}
\begin{align}
 V_a^{(4)}&=
 \begin{pmatrix}
  1&0&0\\ 0&1&0\\
  \dfrac{\rho_1}{1+\rho_2 \bar \rho_2 +d^{-1}\rho_1 \bar \rho_1}\delta_{1}^-(\delta_{2}^-)^{2}e^{\theta_{21}}&0&1
 \end{pmatrix},\\[2mm]
 V_b^{(4)}&=
 \begin{pmatrix}
  1&0&\dfrac{d^{-1} \bar \rho_1}{1+\rho_2 \bar \rho_2 +d^{-1}\rho_1 \bar \rho_1}
  (\delta_{1}^+)^{-1}(\delta_{2}^+)^{-2}e^{\theta_{12}}\\[3mm]
  0&1&0\\ 0&0&1
 \end{pmatrix}.
\end{align}
\end{subequations}

\begin{figure}
\centering

\begin{tikzpicture}[
    x=0.92cm,
    y=0.92cm,
    >=Stealth,
    contour/.style={
        black,
        line width=2pt,
        line cap=round,
        line join=round
    },
    axis/.style={
        black,
        dashed,
        line width=1.50pt,
        line cap=round
    },
    branch/.style={
        black,
        line width=2.2pt,
        line cap=round
    },
    raylabel/.style={
        font=\small,
        fill=white,
        inner sep=2.5pt
    },
     dir/.style={
        postaction={decorate},
        decoration={
            markings,
            mark=at position .56 with {
                \arrow[scale=0.8]{>}
            }
        }
    }
]

\draw[axis] (-6,0)--(6,0);

\draw[contour,dir] (-6,1.8)--(0,1.8);
\draw[contour,dir] (0,1.8)--(6,1.8);

\draw[contour,dir] (-6,-1.8)--(0,-1.8);
\draw[contour,dir] (0,-1.8)--(6,-1.8);

\draw[branch,dir] (-4.0,-1.05)--(-4.0,1.05);

\coordinate (kzero) at (-0.5,0);
\coordinate (kup) at (1.3,1.8);
\coordinate (kdown) at (1.3,-1.8);
\coordinate (kone) at (3.1,0);

\coordinate (krup) at (3.82,0.72);
\coordinate (krdown) at (3.82,-0.72);

\draw[contour,dir] (-2.3,1.8)--(kzero);
\draw[contour,dir] (-2.3,-1.8)--(kzero);

\draw[contour,dir] (kzero)--(kup);
\draw[contour,dir] (kzero)--(kdown);

\draw[contour,dir] (kup)--(kone);
\draw[contour,dir] (kdown)--(kone);


\draw[contour] (kone)--(krup);
\draw[contour] (kone)--(krdown);

\draw[contour,dir] (krup)--(6,0.72);
\draw[contour,dir] (krdown)--(6,-0.72);

\fill[black] (kzero) circle (2.8pt);
\fill[black] (kone) circle (2.8pt);

\node[below=6pt] at (kzero) {$k_0$};
\node[below=6pt] at (kone) {$k_1$};

\node[raylabel] at (-4.38,0.30) {$B$};

\node at (4.68,0.27) {$\mathcal{R}_1$};
\node at (4.68,-0.27) {$\mathcal{R}_2$};

\node[raylabel] at (-2.00,1.02) {$1$};
\node[raylabel] at (-2.00,-1.02) {$2$};

\node[raylabel] at (0.12,1.12) {$3$};
\node[raylabel] at (0.12,-1.12) {$4$};

\node[raylabel] at (2.40,1.20) {$6$};
\node[raylabel] at (2.40,-1.20) {$7$};

\node[raylabel] at (3.72,1.00) {$9$};
\node[raylabel] at (3.72,-1.00) {$10$};

\end{tikzpicture}

\caption{The jump contour $\Sigma^{(5)}$ for $N^{(5)}$ and the regions
$\{\mathcal{R}_j\}_{j=1}^{2}$.}
\label{fig:final-lenses}

\end{figure}
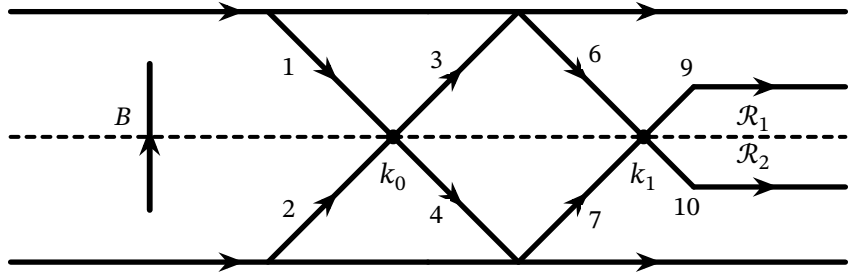

The fifth transformation performs the final lens opening. Its main purpose is to deform the jump on $[k_1,\infty)$ away from the real axis. We define
\begin{equation}
N^{(5)}(k)=N^{(4)}(k)G_5(k),
\qquad
G_5(k)=
\begin{cases}
\bigl(V_b^{(4)}(k)\bigr)^{-1}, & k\in \mathcal R_1,\\
V_a^{(4)}(k), & k\in \mathcal R_2,\\
I, & \text{elsewhere}.
\end{cases}
\end{equation}
The regions $\{\mathcal R_j \}_{j=1}^2$ and the resulting jump contour $\Sigma^{(5)}$ are shown in Figure~\ref{fig:final-lenses}.  The corresponding jump matrix is denoted by $V^{(5)}$. Then we have
\begin{align*}
V^{(5)}_j=V^{(4)}_j, \quad j=1,...,7, \qquad V^{(5)}_8=I, \qquad V^{(5)}_9
=V^{(4)}_b,\qquad V^{(5)}_{10}=V^{(4)}_a, \qquad V^{(5)}_B=V^{(4)}_B.
\end{align*}

\subsection[The g-function conjugation]{The $g$-function conjugation}
\label{subsec:g-conjugation}
A direct calculation gives the following exact expression for the jump matrix $V^{(5)}$ on the branch cut $B$:
\begin{equation}
V^{(5)}_B
=
\begin{pmatrix}
0&
-\frac{\ii(k+\lambda)}{q_0}\delta_1^{-2}\delta_2^{-1}
&0\\[2mm]
-\frac{\ii(k+\lambda)}{q_0}\delta_1^2\delta_2
&0&0\\
0&0&1
\end{pmatrix}.
\end{equation}
We next introduce a transformation that reduces $V^{(5)}_B$ to a constant matrix. The first step is to define a scalar $g$-function. We require $g(\xi,\cdot)$ to be analytic in
$\mathbb C\setminus B$ and to satisfy
\begin{equation*}
 e^{\mathrm i(g^++g^-)(\xi,k)}\delta_1(\xi,k)^2
 \delta_2(\xi,k)=1,\quad k\in B;
 \qquad
 \frac{g(\xi,k)}{\lambda(k)}=\mathcal O(k^{-1}),\qquad k\to\infty.
\end{equation*}
If we define
\begin{equation*}
 f(\xi,s)=\mathrm i\log\bigl(\delta_1(\xi,s)^{-2}
 \delta_2(\xi,s)^{-1}\bigr),\qquad s\in B,
\end{equation*}
 then the Sokhotski--Plemelj formula  gives
\begin{equation}
 g(\xi,k)=\frac{\lambda(k)}{2\pi\mathrm i}
 \int_B\frac{f(\xi,s)}{\lambda(s)(s-k)}\,\mathrm ds.
 \label{eq:g-solution}
\end{equation}
In particular, by a direct calculation we have
\begin{equation}
 g_\infty(\xi):=\lim_{k\to\infty}g(\xi,k)
 =-\frac{1}{2\pi\mathrm i}\int_B
 \frac{f(\xi,s)}{\lambda(s)}\,\mathrm ds.
 \label{eq:g-infinity}
\end{equation}
The reductions $\lambda(k^*)^*=\lambda(k)$ and
$\overline{\delta_j(\xi,k^*)}=\delta_j(\xi,k)^{-1}$ imply
$\overline{g_\infty}=g_\infty$.

Let
\begin{equation}
 \sigma=\operatorname{diag}(1,-1,0),
 \qquad
 D(k)=\operatorname{diag}\bigl(d(k)^{-1/2},d(k)^{-1/2},1\bigr),
 \label{eq:sigma-D-sixth}
\end{equation}
where the branch of $d(k)^{-1/2}$ is chosen so that
$
    d(k)^{-1/2}\to1
    $, $k\to\infty$.  The sixth transformation is defined as follows
\begin{equation}
 N^{(6)}(x,t,k)=e^{-\mathrm i g_\infty\sigma}
 N^{(5)}(x,t,k)D(k)e^{\mathrm i g(\xi,k)\sigma}.
 \label{eq:sixth-transformation}
\end{equation}
The contour is unchanged: $\Sigma^{(6)}=\Sigma^{(5)}$~(see Fig.~\ref{fig:final-lenses}).   By construction, the jump on the fixed branch cut is now the constant matrix
\begin{equation}
 V_B^{(6)}=
 \begin{pmatrix}
  0&-\mathrm i&0\\
  -\mathrm i&0&0\\
  0&0&1
 \end{pmatrix}.
 \label{eq:constant-branch-jump}
\end{equation}
\begin{remark}
Since \(D(k)\) has at most fourth-root singularities at the
branch points \(\pm\mathrm{i}q_0\), the same is true of
\(N^{(6)}\). This explains why we impose the same local
growth condition at these points in the outer model
RH problem~\ref{R:Pewoutmodel}.
\end{remark}

Near $k_0$ the four jumps take the form
\begin{subequations}\label{eq:sixth-jumps-k0}
\begin{alignat}{2}
V_1^{(6)}
&=
\begin{pmatrix}
1&0&0\\
\rho_2\delta_1^2\delta_2e^{2\mathrm{i}g}e^{\theta_{31}}&1&0\\
0&0&1
\end{pmatrix},
&\qquad
V_2^{(6)}
&=
\begin{pmatrix}
1&\bar{\rho}_2\delta_1^{-2}\delta_2^{-1}e^{-2\mathrm{i}g}
e^{\theta_{13}}&0\\
0&1&0\\
0&0&1
\end{pmatrix},
\\[2mm]
V_3^{(6)}
&=
\begin{pmatrix}
1&\dfrac{\bar{\rho}_2}{1+\rho_2\bar{\rho}_2}
\delta_1^{-2}\delta_2^{-1}e^{-2\mathrm{i}g}e^{\theta_{13}}&0\\[3mm]
0&1&0\\
0&0&1
\end{pmatrix},
&\qquad
V_4^{(6)}
&=
\begin{pmatrix}
1&0&0\\
\dfrac{\rho_2}{1+\rho_2\bar{\rho}_2}
\delta_1^2\delta_2e^{2\mathrm{i}g}e^{\theta_{31}}&1&0\\[3mm]
0&0&1
\end{pmatrix}.
\end{alignat}
\end{subequations}
Near $k_1$, the jump matrices on the four rays labeled $6$, $7$, $9$, and $10$ are given explicitly by
\begin{subequations}\label{eq:sixth-jumps-k1}
\begin{alignat}{2}
V_6^{(6)}
&=
\begin{pmatrix}
1&0&0\\
0&1&0\\
d^{-1/2}\rho_1\delta_1\delta_2^{2}e^{\mathrm i g}e^{\theta_{21}}&0&1
\end{pmatrix},
&\qquad
V_7^{(6)}
&=
\begin{pmatrix}
1&0&d^{-1/2}\bar \rho_1\delta_1^{-1}\delta_2^{-2}
e^{-\mathrm i g}e^{\theta_{12}}\\
0&1&0\\
0&0&1
\end{pmatrix},
\\[2mm]
V_9^{(6)}
&=
\begin{pmatrix}
1&0&\dfrac{d^{-1/2}\bar \rho_1\delta_1^{-1}\delta_2^{-2}}
{1+d^{-1}\rho_1\bar \rho_1+\rho_2\bar\rho_2}
e^{-\mathrm i g}e^{\theta_{12}}\\[3mm]
0&1&0\\
0&0&1
\end{pmatrix},
&\qquad
V_{10}^{(6)}
&=
\begin{pmatrix}
1&0&0\\
0&1&0\\
\dfrac{d^{-1/2}\rho_1\delta_1\delta_2^{2}}
{1+d^{-1}\rho_1\bar \rho_1+\rho_2\bar\rho_2}
e^{\mathrm i g}e^{\theta_{21}}&0&1
\end{pmatrix}.
\end{alignat}
\end{subequations}
On all remaining contour segments, which stay away from $k_0$, $k_1$, and the branch cut $B$, the jump matrices converge uniformly and exponentially fast to the identity as $t\to\infty$.

\subsection{Local parametrices near the stationary points}
\label{subsec:local-parametrices}
\begin{figure}[htbp]
\centering
\begin{tikzpicture}[>=Stealth,scale=.95,
 contour/.style={line width=1.25pt,line cap=round,line join=round},
 raylabel/.style={font=\small,fill=white,inner sep=1.6pt},
 dir/.style={postaction={decorate},decoration={markings,
 mark=at position .64 with {\arrow{>}}}}]
 \begin{scope}[xshift=-3.4cm]
  \draw[contour,dir] (-1.5,1.5)--(0,0);
   \draw[contour,dir] (0,0)--(1.5,-1.5);
  \draw[contour,dir] (-1.5,-1.5)--(0,0);
  \draw[contour,dir] (0,0)--(1.5,1.5);
  \fill (0,0) circle (2pt);
  \node[raylabel] at (0,-0.48) {$k_0$};
  \node[raylabel] at (-1.25,0.90) {$1$}; \node[raylabel] at (-1.25,-0.90) {$2$};
  \node[raylabel] at (1.25,0.90) {$3$}; \node[raylabel] at (1.25,-0.90) {$4$};
  \draw[dashed,gray!65,line width=.85pt] (0,0) circle (1.75);
  \node at (0,-2.15) {$D_\epsilon(k_0)$};
 \end{scope}
 \begin{scope}[xshift=3.4cm]
 \draw[contour,dir] (-1.5,1.5)--(0,0);
   \draw[contour,dir] (0,0)--(1.5,-1.5);
  \draw[contour,dir] (-1.5,-1.5)--(0,0);
  \draw[contour,dir] (0,0)--(1.5,1.5);
  \fill (0,0) circle (2pt);
  \node[raylabel] at (0,-0.48) {$k_1$};
  \node[raylabel] at (-1.25,0.90) {$6$}; \node[raylabel] at (-1.25,-0.90) {$7$};
  \node[raylabel] at (1.25,0.90) {$9$}; \node[raylabel] at (1.35,-0.90) {$10$};
  \draw[dashed,gray!65,line width=.85pt] (0,0) circle (1.75);
  \node at (0,-2.15) {$D_\epsilon(k_1)$};
 \end{scope}
\end{tikzpicture}
\caption{The local cross-shaped contours near $k_0$ and $k_1$.}
\label{fig:local-crosses}
\end{figure}
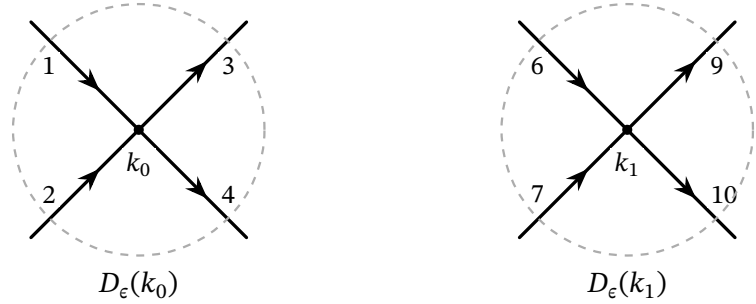
In this subsection, we construct local parametrices in neighborhoods of the stationary points $k_0$ and $k_1$. We first fix $\epsilon>0$ sufficiently small so that the disks
\begin{equation}
    D_\epsilon(k_j)
    =
    \{k\in\mathbb{C}:|k-k_j|<\epsilon\},
    \qquad j=0,1,
    \label{eq:stationary-disks}
\end{equation}
are disjoint and do not intersect the branch cut $B$; see Fig.~\ref{fig:local-crosses}. On the portions of the deformed contour lying outside these disks and away from the branch cut $B$, the jump matrix $V^{(6)}$ converges uniformly and exponentially fast to the identity as $t\to\infty$. Inside each disk, we construct a local parametrix using the standard parabolic-cylinder model described in Appendix~\ref{App:AAA}.
\subsubsection{The  local model near $k_0$}
We first construct the local parametrix near $k_0$. Introduce the local coordinate
\begin{equation}
    \zeta_1
    =
    \sqrt{\gamma_1(\xi)t}\,(k-k_0),  \qquad k \in D_{\epsilon}(k_0),
    \label{eq:zeta1}
\end{equation}
where
$
    \gamma_1(\xi)
    :=
    -\frac{\mathrm{i}}{t}
    \partial_k^2\theta_{13}(x,t,k_0)>0
$ for $\xi > 4\sqrt{2}q_0$. Then, uniformly for $k\in D_\epsilon(k_0)$,
\begin{equation}
 \theta_{13}(x,t,k)=\theta_{13}(x,t,k_0)
 +\frac{\mathrm i}{2}\zeta_1^2+O\bigl(t|k-k_0|^3\bigr).
 \label{eq:theta13-local}
\end{equation}
To separate the endpoint singularity of $\delta_1$ at $k_0$ from its regular part, let
\[
    \log_\pi z
    :=
    \log|z|+\mathrm{i}\arg z,
    \qquad
    \arg z\in(-\pi,\pi).
\]
Define
\begin{align} \label{eq:eta1-local}
    \nu_1
    :=
    -\frac{1}{2\pi}
    \log\bigl(1+|\rho_2(k_0)|^2\bigr),
  \qquad 
    \eta_1(k)
    :=
    \frac{1}{2\pi\mathrm{i}}
    \int_{k_0}^{\infty}
    \log_\pi(s-k)\,
    \mathrm{d}\log\bigl(1+|\rho_2(s)|^2\bigr).
\end{align}
Integrating by parts in \eqref{eq:delta1-solution}  gives
\begin{equation}
    \delta_1(k)
    =
    \mathrm{e}^{\mathrm{i}\nu_1\log_\pi(k_0-k)}
    \mathrm{e}^{\eta_1(k)}.
    \label{eq:delta1-local-form}
\end{equation}
Using the local coordinate \eqref{eq:zeta1} and the logarithmic representation \eqref{eq:delta1-local-form}, the scalar factor appearing in the jump matrices \eqref{eq:sixth-jumps-k0} can therefore be written as
\begin{equation}
    \delta_1^{-2}\delta_2^{-1}\mathrm{e}^{-2\mathrm{i}g}
    =
    \mathrm{e}^{-2\mathrm{i}\nu_1\log_\pi(-\zeta_1)}
    \delta_{1,a}\delta_{1,b}(k), ,  \qquad k \in D_{\epsilon}(k_0),
    \label{eq:delta-combination-k0}
\end{equation}
where the $k$-independent factor $\delta_{1,a}$ and the residual factor $\delta_{1,b}$ are defined by
\begin{align}
    \delta_{1,a}
    &:=
    \bigl(\sqrt{\gamma_1(\xi)t}\bigr)^{2\mathrm{i}\nu_1}
    \mathrm{e}^{-2\eta_1(k_0)}
    \delta_2(k_0)^{-1}
    \mathrm{e}^{-2\mathrm{i}g(k_0)},
    \label{eq:delta1a}\\
    \delta_{1,b}(k)
    &:=
    \frac{\delta_2(k_0)}{\delta_2(k)}
    \mathrm{e}^{-2\mathrm{i}(g(k)-g(k_0))}
    \mathrm{e}^{-2(\eta_1(k)-\eta_1(k_0))}.
    \label{eq:delta1b}
\end{align}
In particular, as $k\to k_0$,
\[
    \delta_{1,b}(k)
    =
    1+O\bigl(|k-k_0|\log|k-k_0|\bigr).
\]

Define the diagonal conjugating matrix
\begin{equation}
    Y_A
    =
    \operatorname{diag}\left(
        \delta_{1,a}^{1/2}e^{\theta_{13}(k_0)/2},
        \delta_{1,a}^{-1/2}e^{-\theta_{13}(k_0)/2},
        1
    \right).
    \label{eq:YA}
\end{equation}
After conjugation by \(Y_A\), the jump matrices satisfy
\(
    Y_A^{-1}V_j^{(6)}Y_A
    \longrightarrow
    V_j^{k_0}
\), $j=1,2,3,4$, as \(t\to\infty\) with \(\zeta_1\) fixed, where the limiting jump matrices
\(V_j^{k_0}\) are given by
\begin{subequations}\label{eq:local-jumps-k0}
\begin{align}
 V_1^{k_0}(\zeta_1)&=
 \begin{pmatrix}
  1&0&0\\
   \mathfrak{q}_0(-\zeta_1)^{2\mathrm i\nu_1}
  e^{-\mathrm i\zeta_1^2/2}&1&0\\
  0&0&1
 \end{pmatrix},\\[2mm]
 V_2^{k_0}(\zeta_1)&=
 \begin{pmatrix}
  1&- \mathfrak{p}_0(-\zeta_1)^{-2\mathrm i\nu_1}
  e^{\mathrm i\zeta_1^2/2}&0\\
  0&1&0\\ 0&0&1
 \end{pmatrix},\\[2mm]
 V_3^{k_0}(\zeta_1)&=
 \begin{pmatrix}
  1&-\dfrac{ \mathfrak{p}_0}{1- \mathfrak{p}_0 \mathfrak{q}_0}
  (-\zeta_1)^{-2\mathrm i\nu_1}e^{\mathrm i\zeta_1^2/2}&0\\[3mm]
  0&1&0\\ 0&0&1
 \end{pmatrix},\\[2mm]
 V_4^{k_0}(\zeta_1)&=
 \begin{pmatrix}
  1&0&0\\
  \dfrac{ \mathfrak{q}_0 }{1- \mathfrak{p}_0  \mathfrak{q}_0}
  (-\zeta_1)^{2\mathrm i\nu_1}e^{-\mathrm i\zeta_1^2/2}&1&0\\[3mm]
  0&0&1
 \end{pmatrix}.
\end{align}
\end{subequations}
with
\begin{equation}
  \mathfrak{p}_0=-\rho_2^*(k_0),\qquad  \mathfrak{q}_0=\rho_2(k_0).
 \label{eq:pc-parameters-k0}
\end{equation}
Accordingly, we seek a matrix-valued function
\(N^{k_0}(x,t,k)\) defined in \(D_\epsilon(k_0)\) and satisfying the
following conditions: $\mathrm{[(i)]}$
\(N^{k_0}\) is analytic in
\(
    D_\epsilon(k_0)
    \setminus
    \bigcup_{j=1}^{4}\Sigma_j^{(6)}
\),
and its boundary values on
\(
    D_\epsilon(k_0)
    \cap
    \bigcup_{j=1}^{4}\Sigma_j^{(6)}
\)
satisfy the jump conditions in \eqref{eq:local-jumps-k0}.
$\mathrm{[(ii)]}$
\(N^{k_0}\) is normalized by
\(
    N^{k_0}(x,t,k)\to I
\), $ \zeta_1\to\infty$.

The solution of this local RH problem can be expressed in terms of the
standard parabolic-cylinder model
\(N^{\mathrm{PC}}(\zeta;p,q)\) introduced in
Appendix~\ref{App:AAA}. More precisely,
\begin{equation}
    N^{k_0}(x,t,k)
    =
    Y_A
    \begin{pmatrix}
        N^{\mathrm{PC}}
        (-\zeta_1(k);\mathfrak{p}_0,\mathfrak{q}_0)
        & \mathbf{0}_{2\times1}\\
        \mathbf{0}_{1\times2} & 1
    \end{pmatrix}
    Y_A^{-1}, \qquad k \in D_\epsilon(k_0).
    \label{eq:local-model-k0}
\end{equation}
It follows directly from Lemma~\ref{L:asyofLemmaPC} that, as
\(\zeta_1\to\infty\),
\begin{equation} \label{E:asyofNko}
    N^{k_0}\bigl(x,t,k(\zeta_1)\bigr)
    =
    I
    +
    \frac{1}{\zeta_1}
    Y_A
      \mathcal{N}_{k_0}^{(1)}
    Y_A^{-1}
    +
    O(\zeta_1^{-2}),
    \end{equation}
    where 
    \begin{equation}
    \mathcal{N}_{k_0}^{(1)}
    =
    \begin{pmatrix}
        0 & -\frac{\sqrt{2\pi}\,e^{-\pi\nu_1/2}e^{\pi\mathrm i/4}}
 {\mathfrak{q}_0\,\Gamma(\mathrm i\nu_1)} &0\\
        -\frac{\sqrt{2\pi}\,e^{-\pi\nu_1/2}e^{-\pi\mathrm i/4}}
 {\mathfrak{p}_0 \,\Gamma(-\mathrm i\nu_1 )}  & 0 &0\\
 0&0&0
    \end{pmatrix}.
    \label{eq:PC-expansion}
\end{equation}

\subsubsection{The local model near $k_1$}

Similarly, in the disk \(D_\epsilon(k_1)\), we introduce the local
conformal coordinate
\begin{equation}
    \zeta_2
    =
    \sqrt{\gamma_2(\xi)t}\,(k-k_1),
    \label{eq:zeta2}
\end{equation}
where
\(
    \gamma_2(\xi)
    =
    -\frac{\mathrm{i}}{t}
    \partial_k^2\theta_{12}(x,t,k_1)
    >0
\).
The Taylor expansion of the phase function $\theta_{12}$ about \(k_1\) then takes the form
\begin{equation}
    \theta_{12}(x,t,k)
    =
    \theta_{12}(x,t,k_1)
    +
    \frac{\mathrm{i}}{2}\zeta_2^2
    +
    O\bigl(t|k-k_1|^3\bigr), \qquad k\in D_{\epsilon}(k_1).
    \label{eq:theta12-local}
\end{equation}
As before, we rewrite \(\delta_2(k)\) in the form
\begin{equation}
    \delta_2(k)
    =
    e^{\mathrm{i}\nu_2\log_\pi(k_1-k)}
    e^{-\eta_2(k)},
    \label{eq:delta2-local-form}
\end{equation}
where
\begin{align}
    \nu_2
    &=
    -\frac{1}{2\pi}\log h(k_1)
    =
    \frac{1}{2\pi}
    \log
    \frac{1+|\rho_2(k_1)|^2}
    {1+d(k_1)^{-1}|\rho_1(k_1)|^2+|\rho_2(k_1)|^2},
    \label{eq:nu2}\\
    \eta_2(k)
    &=
    -\frac{1}{2\pi\mathrm{i}}
    \int_{k_1}^{\infty}
    \log_\pi(s-k)\,
    \mathrm{d}\log h(s).
    \label{eq:eta2-local}
\end{align}
The scalar factor appearing in the \((1,3)\)-entries of the jump
matrices in \eqref{eq:sixth-jumps-k1} can then be written as
\begin{equation}
    \delta_1^{-1}\delta_2^{-2}e^{-\mathrm{i}g}
    =
    e^{-2\mathrm{i}\nu_2\log_\pi(-\zeta_2)}
    \delta_{2,a}
    \begin{cases}
        \bigl(1+|\rho_2(k_1)|^2\bigr)\delta_{2,b}^+(k),
        & k\in\mathbb{C}_+ \cap D_{\epsilon}(k_1),\\[1mm]
        \delta_{2,b}^-(k),
        & k\in\mathbb{C}_-\cap D_{\epsilon}(k_1),
    \end{cases}
    \label{eq:delta-combination-k1}
\end{equation}
where
\begin{align}
 \delta_{2,a}&=\bigl(\sqrt{\gamma_2t}\bigr)^{2\mathrm i\nu_2}
 \delta_1^-(k_1)^{-1}e^{-\mathrm i g(k_1)}e^{2\eta_2(k_1)},
 \label{eq:delta2a}\\
 \delta^{\pm}_{2,b}(k)&=
 \frac{\delta_1^{\pm}(k_1)}{\delta_1(k)}
 e^{-\mathrm i(g(k)-g(k_1))}
 e^{2(\eta_2(k)-\eta_2(k_1))}.
 \label{eq:delta2b}
\end{align}
In particular, as \(k\to k_1\) nontangentially from
\(\mathbb{C}_\pm\),
\begin{equation} \label{E:delta2pmgj}
    \delta_{2,b}^{\pm}(k)
    =
    1
    +
    O\left(
        |k-k_1|\,\bigl|\log|k-k_1|\bigr|
    \right).
\end{equation}

We now introduce a diagonal conjugating matrix \(Y_B\) to reduce the jump matrix near $k_1$
to the standard form of the parabolic-cylinder model. Define
\begin{equation}
    Y_B
    =
    \operatorname{diag}\left(
        \delta_{2,a}^{1/2}e^{\theta_{12}(k_1)/2},
        1,
        \delta_{2,a}^{-1/2}e^{-\theta_{12}(k_1)/2}
    \right).
    \label{eq:YB}
\end{equation}
Then, as \(t\to\infty\) with \(\zeta_2\) fixed, the jump matrices near
\(k_1\) satisfy
\[
    Y_B^{-1}V_j^{(6)}Y_B
    \longrightarrow
    V_j^{k_1},
    \qquad j=6,7,9,10,
\]
where the limiting jump matrices \(V_j^{k_1}\) are given by
\begin{subequations}\label{eq:local-jumps-k1}
\begin{align}
 V_6^{k_1}(\zeta_2)&=
 \begin{pmatrix}
  1&0&0\\ 0&1&0\\
  \mathfrak{q}_1(-\zeta_2)^{2\mathrm i\nu_2}
  e^{-\mathrm i\zeta_2^2/2}&0&1
 \end{pmatrix},\\[2mm]
 V_7^{k_1}(\zeta_2)&=
 \begin{pmatrix}
  1&0&-\mathfrak{p}_1(-\zeta_2)^{-2\mathrm i\nu_2}
  e^{\mathrm i\zeta_2^2/2}\\
  0&1&0\\ 0&0&1
 \end{pmatrix},\\[2mm]
 V_9^{k_1}(\zeta_2)&=
 \begin{pmatrix}
  1&0&-\dfrac{\mathfrak{p}_1}{1-\mathfrak{p}_1 \mathfrak{q}_1}
  (-\zeta_2)^{-2\mathrm i\nu_2}e^{\mathrm i\zeta_2^2/2}\\[3mm]
  0&1&0\\ 0&0&1
 \end{pmatrix},\\[2mm]
 V_{10}^{k_1}(\zeta_2)&=
 \begin{pmatrix}
  1&0&0\\ 0&1&0\\
  \dfrac{\mathfrak{q}_1}{1-\mathfrak{p}_1\mathfrak{q}_1}
  (-\zeta_2)^{2\mathrm i\nu_2}e^{-\mathrm i\zeta_2^2/2}&0&1
 \end{pmatrix},
\end{align}
\end{subequations}
with
\begin{equation}
 \mathfrak{p}_1=-d(k_1)^{-1/2}\rho_1^*(k_1),
 \qquad
  \mathfrak{q}_1=
 \frac{d(k_1)^{-1/2}\rho_1(k_1)}{1+|\rho_2(k_1)|^2}.
 \label{eq:pc-parameters-k1}
\end{equation}
Let \(N^{\mathrm{PC}}(\zeta;p,q)\) denote the solution of the standard
parabolic-cylinder model introduced in Appendix~\ref{App:AAA}, and let
\(N_{ij}^{\mathrm{PC}}(\zeta;p,q)\) denote its \((i,j)\)-entry.
Accordingly, the local parametrix near \(k_1\) is defined by
\[
    N^{k_1}(x,t,k)
    =
    Y_B
    \begin{pmatrix}
        N_{11}^{\mathrm{PC}}
        (-\zeta_2(k);\mathfrak{p}_1,\mathfrak{q}_1)
        & 0
        & N_{12}^{\mathrm{PC}}
        (-\zeta_2(k);\mathfrak{p}_1,\mathfrak{q}_1)
        \\[1mm]
        0 & 1 & 0
        \\[1mm]
        N_{21}^{\mathrm{PC}}
        (-\zeta_2(k);\mathfrak{p}_1,\mathfrak{q}_1)
        & 0
        & N_{22}^{\mathrm{PC}}
        (-\zeta_2(k);\mathfrak{p}_1,\mathfrak{q}_1)
    \end{pmatrix}
    Y_B^{-1}  \qquad k \in D_\epsilon(k_1).
\]
It follows directly from Lemma~\ref{L:asyofLemmaPC} that, as
\(\zeta_2\to\infty\),
\begin{equation} \label{E:asyofNk1}
    N^{k_1}\bigl(x,t,k(\zeta_2)\bigr)
    =
    I
    +
    \frac{1}{\zeta_2}
    Y_B
      \mathcal{N}_{k_1}^{(1)}
    Y_B^{-1}
    +
    O(\zeta_2^{-2}),
    \end{equation}
    where 
    \begin{equation}
    \mathcal{N}_{k_1}^{(1)}
    =
    \begin{pmatrix}
        0 & 0& -\frac{\sqrt{2\pi}\,e^{-\pi\nu_2 /2}e^{\pi\mathrm i/4}}
 {\mathfrak{q}_1\,\Gamma(\mathrm i\nu_2)} \\
  0&0&0\\
        -\frac{\sqrt{2\pi}\,e^{-\pi\nu_2 /2}e^{-\pi\mathrm i/4}}
 {\mathfrak{p}_1 \,\Gamma(-\mathrm i\nu_2 )}  & 0 &0
    \end{pmatrix}.
    \label{eq:PC-expansionzeta2}
\end{equation}

For convenience, we combine the local parametrices at the two stationary
points. Set
\[
    \mathcal{D}
    =
    D_\epsilon(k_0)\cup D_\epsilon(k_1),
\]
and define the combined local parametrix by
\begin{equation}
    N^{\mathrm{loc}}(x,t,k)
    =
    \begin{cases}
        N^{k_0}(x,t,k),
        & k\in D_\epsilon(k_0),\\[1mm]
        N^{k_1}(x,t,k),
        & k\in D_\epsilon(k_1).
    \end{cases}
    \label{eq:combined-local-model}
\end{equation}
Then we have the following lemma.

\begin{lemma}
\label{lem:matching-estimates}
For each fixed \((x,t)\), the local parametrix
\(N^{\mathrm{loc}}(x,t,k)\) defined in
\eqref{eq:combined-local-model} is analytic and bounded in
\(
    \mathcal{D}
    \setminus
   \Sigma^{(6)}
\).
Its boundary values satisfy
\[
    N^{\mathrm{loc},+}(x,t,k)
    =
    N^{\mathrm{loc},-}(x,t,k)
    V^{\mathrm{loc}}(x,t,k),
    \qquad
    k\in\Sigma^{(6)}\cap\mathcal{D}.
\]
As \(t\to\infty\), the jump matrix
\(V^{\mathrm{loc}}\) approximates \(V^{(6)}\) in the sense that
\begin{align}
    \left\|V^{(6)}-V^{\mathrm{loc}}\right\|_
    {L^\infty(\Sigma^{(6)}\cap\mathcal{D})}
    &\lesssim t^{-1/2}\log t,
    \label{eq:local-jump-Linfty-p}\\
    \left\|V^{(6)}-V^{\mathrm{loc}}\right\|_
    {L^1(\Sigma^{(6)}\cap\mathcal{D})}
    &\lesssim t^{-1}\log t.
    \label{eq:local-jump-L1}
\end{align}
Moreover, \(N^{\mathrm{loc}}\) satisfies 
\begin{equation}
    \left\|
        \bigl(N^{\mathrm{loc}}\bigr)^{-1}-I
    \right\|_{L^\infty(\partial\mathcal{D})}
    =
    O(t^{-1/2}), \qquad t \to \infty.
    \label{eq:local-matching-Linfty}
\end{equation}
\end{lemma}

\begin{proof}

We prove the estimates~\eqref{eq:local-jump-Linfty-p} and~\eqref{eq:local-jump-L1} near \(k_1\); the analysis near \(k_0\) is
identical. Consider, for example, the jump on ray \(\Sigma^{(6)}_6 \cap \mathcal{D}\).
Then for \(k \in \Sigma^{(6)}_6 \cap \mathcal{D}\), we have
$$
V^{(6)}-V^{\mathrm{loc}}= Y_B 
\begin{pmatrix}
0&0&0\\
0&0&0\\
f(k)&0&0
\end{pmatrix}
Y_B^{-1},
$$
where $f(k)$ is given  by
\begin{equation}
    f(k)
    =
    \frac{e^{2\mathrm{i}\nu_2\log_\pi(-\zeta_2(k))}
    e^{-\mathrm{i}\zeta_2(k)^2/2}
    }{1+|\rho_2(k_1)|^2}\left(
        d(k)^{-1/2}\rho_1(k)\delta_{2,b}^+(k)^{-1}e^{\theta_{21}-\theta_{21}(x,t,k_1)  + \mathrm{i}\frac{\zeta_2(k)^2}{2} }-d(k_1)^{-1/2}\rho_1(k_1)
    \right).
    \label{eq:representative-f}
\end{equation}
The matrices \(Y_B\) and \(Y_B^{-1}\) are uniformly bounded for
\(\xi \in \mathcal{K}_+\). It therefore suffices to estimate
the scalar function \(f(k)\).
The local Lipschitz continuity of \(d\) and the reflection coefficients
near \(k_1\), together with \eqref{E:delta2pmgj}, yields
\begin{equation}
    \left|\frac{\rho_1}{d^{1/2}}(k)-\frac{\rho_1}{d^{1/2}}(k_1) \right|
    \lesssim
    |k-k_1|,
    \quad
    |\delta_{2,b}^+(k)-1|
    \lesssim
    |k-k_1|\,\bigl|\log|k-k_1|\bigr|,
    \qquad
    k\in\Sigma_6^{(6)}\cap\mathcal{D}.
    \label{eq:lipschitz-local}
\end{equation}
Along \(\Sigma_6^{(6)}\cap\mathcal{D}\), the first exponential factor in
\eqref{eq:representative-f} is uniformly bounded, whereas the second is
bounded by \(e^{-ct|k-k_1|^2}\). Moreover, by \eqref{eq:theta12-local} and the sign structure of
\(\theta_{21}\) on \(\Sigma_6^{(6)}\cap\mathcal{D}\), after choosing
\(\epsilon>0\) sufficiently small, we obtain
\begin{align*}
    &\left|
        e^{-\mathrm{i}\zeta_2(k)^2/2}
        \left(
            e^{
                \theta_{21}(x,t,k)-\theta_{21}(x,t,k_1)
                +\mathrm{i}\zeta_2(k)^2/2
            }
            -1
        \right)
    \right|
    \\
    &\qquad
    =
    \left|
        e^{\theta_{21}(x,t,k)-\theta_{21}(x,t,k_1)}
        -
        e^{-\mathrm{i}\zeta_2(k)^2/2}
    \right|
    \lesssim
    t|k-k_1|^3e^{-ct|k-k_1|^2},
\end{align*}
for \(k\in\Sigma_6^{(6)}\cap\mathcal{D}\).
Consequently,
\begin{equation}
    |f(k)|
    \lesssim
    |k-k_1|
    \left(
        1+\bigl|\log|k-k_1|\bigr|
    \right)
    e^{-ct|k-k_1|^2} + t|k-k_1|^3e^{-ct|k-k_1|^2} 
    \lesssim
    t^{-1/2}\log t.
    \label{eq:f-Linfty}
\end{equation}
This establishes the required \(L^\infty\) estimate for
\(V^{(6)}-V^{\mathrm{loc}}\) on
\(\Sigma_6^{(6)}\cap\mathcal{D}\).  Integrating
the same bound along the ray gives
\begin{align}
    \|f\|_{L^1(\Sigma^{(6)}\cap D_\epsilon(k_1))}
    &\lesssim
    \int_0^\infty
    u\bigl(1+|\log u|\bigr)e^{-ctu^2}\,\mathrm{d}u + t \int_0^\infty
    u^3 e^{-ctu^2}\,\mathrm{d}u
    \nonumber \lesssim
    t^{-1}\log t.
    \label{eq:f-L1}
\end{align}
This proves the corresponding \(L^1\) estimate on
\(\Sigma_6^{(6)}\cap\mathcal{D}\). 
The remaining rays near \(k_1\) can be treat in the same manner, which
establishes the desired estimates in \(D_\epsilon(k_1)\).
Repeating
the argument near \(k_0\) yields
\eqref{eq:local-jump-Linfty-p} and \eqref{eq:local-jump-L1}.

It remains to prove \eqref{eq:local-matching-Linfty}. On
\(\partial D_\epsilon(k_j)\), the corresponding local coordinate
\(\zeta_j\) tends to infinity as \(t\to\infty\). Substituting the
definitions of \(\zeta_1\) and \(\zeta_2\) into
\eqref{E:asyofNko} and \eqref{E:asyofNk1}, respectively, gives
\begin{subequations}
\label{eq:local-inverse-expansions}
\begin{align}
    \bigl(N^{k_0}\bigr)^{-1}(x,t,k)
    &=
    I
    -
    \frac{
        Y_A\mathcal{N}_{k_0}^{(1)}Y_A^{-1}
    }{
        \sqrt{\gamma_1(\xi)t}\,(k-k_0)
    }
    +
    O(t^{-1}),
    &&
    k\in\partial D_\epsilon(k_0),
    \\
    \bigl(N^{k_1}\bigr)^{-1} (x,t,k)
    &=
    I
    -
    \frac{
        Y_B\mathcal{N}_{k_1}^{(1)}Y_B^{-1}
    }{
        \sqrt{\gamma_2(\xi)t}\,(k-k_1)
    }
    +
    O(t^{-1}),
    &&
    k\in\partial D_\epsilon(k_1).
\end{align}
\end{subequations}
These expansions immediately
imply \eqref{eq:local-matching-Linfty}.

\end{proof}

\subsection{The  outer model}
\label{subsec:outer-model}

Away from the neighborhoods of the stationary points, all jump matrices
are asymptotically close to the identity, except for the constant jump on
the branch cut \(B\) given in \eqref{eq:constant-branch-jump}. The leading
global approximation is therefore determined by the following outer-model
RH problem.

\begin{RHP}[Outer-model RH problem] \label{R:Pewoutmodel}
Find a \(3\times3\) matrix-valued function \(N_B(k)\) satisfying the
following conditions:

\begin{enumerate}
\item[(i)]
\(N_B(k)\) is analytic in \(\mathbb{C}\setminus B\).

\item[(ii)]
\(N_B\) admits boundary values on
\(B\setminus\{\pm\mathrm{i}q_0\}\) satisfying
\begin{equation}
N_{B}^+(k)=N_{B}^-(k)V_B^{(6)},\qquad k\in B.
\label{eq:outer-RHP}
\end{equation}

\item[(iii)]
As \(k\to\infty\),
\[
N_B(k)=I+O(k^{-1}).
\]

\item[(iv)]
At the endpoints of the branch cut,
\[
N_B(k)=O\bigl(|k\mp\mathrm{i}q_0|^{-1/4}\bigr),
\qquad k\to\pm\mathrm{i}q_0.
\]
\end{enumerate}
\end{RHP}

This RH problem can be solved explicitly. Introduce the scalar function
\begin{equation}
\Lambda(k)=
\left(\frac{k-\mathrm{i}q_0}{k+\mathrm{i}q_0}\right)^{1/4},
\qquad
\Lambda(k)\to1\quad\text{as}\quad k\to\infty,
\label{eq:Lambda-outer}
\end{equation}
where the branch is chosen so that \(\Lambda\) is analytic in
\(\mathbb{C}\setminus B\). Its boundary values on the oriented branch cut
satisfy
\[
\Lambda^+(k)=\mathrm{i}\Lambda^-(k),\qquad k\in B.
\]

It follows by direct calculation that the solution of the outer-model RH
problem is
\begin{equation}
N_B(k)=\frac{1}{2}
\begin{pmatrix}
\Lambda(k)+\Lambda(k)^{-1}
&-\bigl(\Lambda(k)-\Lambda(k)^{-1}\bigr)&0\\
-\bigl(\Lambda(k)-\Lambda(k)^{-1}\bigr)
&\Lambda(k)+\Lambda(k)^{-1}&0\\
0&0&2
\end{pmatrix}.
\label{eq:outer-model}
\end{equation}
Expanding \(\Lambda(k)\) at infinity gives
\begin{equation}
N_B(k)=I+\frac{1}{k}
\begin{pmatrix}
0&\mathrm{i}q_0/2&0\\
\mathrm{i}q_0/2&0&0\\
0&0&0
\end{pmatrix}
+O(k^{-2}),
\qquad k\to\infty.
\label{eq:outer-expansion}
\end{equation}

\subsection{The small norm RH problem}
\label{subsec:error-problem}
We define the following transformation to obtain a small norm RH problem:
\begin{equation}
\mathcal{E}(x,t,k)=
\begin{cases}
N^{(6)}(x,t,k)N_B(k)^{-1},
&k\in\mathbb{C}\setminus\mathcal{D},\\
N^{(6)}(x,t,k)\bigl(N^{\mathrm{loc}}(x,t,k)\bigr)^{-1}N_B(k)^{-1},
&k\in\mathcal{D}.
\end{cases}
\label{eq:error-definition}
\end{equation}
Its jump contour, denoted by \(\Sigma^{\mathcal{E}}\), is given by
\(
\Sigma^{\mathcal{E}}
=
\left( \Sigma^{(6)} \setminus B\right)\cup\partial\mathcal{D}
\).
The corresponding jump matrix is
\begin{equation}
V^{\mathcal{E}}(k)=
\begin{cases}
N_BV^{(6)}N_B^{-1},
&k\in\Sigma^{(6)}\setminus \left( \overline{\mathcal{D}} \cup B\right),\\
N_B N^{\mathrm{loc},-} V^{(6)}
\bigl(N^{\mathrm{loc},+} \bigr)^{-1}N_B^{-1},
&k\in\Sigma^{(6)}\cap\mathcal{D},\\
N_B\bigl(N^{\mathrm{loc}}\bigr)^{-1}N_B^{-1},
&k\in\partial\mathcal{D}.
\end{cases}
\label{eq:error-jump}
\end{equation}
In particular, right multiplication by \(N_B^{-1}\) cancels the jump
across the branch cut \(B\). The following lemma provides estimates for the jump matrix.

\begin{lemma} \label{L:estiofWe}
Let
\(
w^{\mathcal{E}}(k)=V^{\mathcal{E}}(k)-I
\).
Then, as  \(t\to\infty\), there exists a constant \(c>0\) such that
\begin{subequations}
\label{eq:w-error-estimates}
\begin{align}
\|w^{\mathcal{E}}\|_{L^1\cap L^\infty(\Sigma^{\mathcal{E}}\setminus \overline{\mathcal{D}})}
&\lesssim e^{-ct},
\label{eq:w-error-outside}\\
\|w^{\mathcal{E}}\|_{L^\infty(\partial\mathcal{D})}
&\lesssim t^{-1/2},
\label{eq:w-error-boundary}\\
\|w^{\mathcal{E}}\|_{L^\infty(\Sigma^{\mathcal{E}}\cap\mathcal{D})}
&\lesssim t^{-1/2}\log t,
\label{eq:w-error-local-Linfty}\\
\|w^{\mathcal{E}}\|_{L^1(\Sigma^{\mathcal{E}}\cap\mathcal{D})}
&\lesssim t^{-1}\log t.
\label{eq:w-error-local-L1}
\end{align}
\end{subequations}
\end{lemma}

\begin{proof}
Outside $\mathcal D$, the jump across \(B\) has been canceled
by the outer model, while all remaining jumps are exponentially close to
the identity. 
The uniform boundedness of \(N_B\) and \(N_B^{-1}\) on  $\Sigma^{(6)}\setminus(\mathcal{D} \cup B)$
therefore gives~\eqref{eq:w-error-outside}.
On \(\partial\mathcal{D}\),
the matching estimate \eqref{eq:local-matching-Linfty} then implies~\eqref{eq:w-error-boundary}.

Finally, on \(\Sigma^{\mathcal{E}}\cap\mathcal{D}\), the jump relation
\(N^{\mathrm{loc},+}=N^{\mathrm{loc},-}V^{\mathrm{loc}}\) gives
\[
w^{\mathcal{E}}
=
N_BN_-^{\mathrm{loc}}
\bigl(V^{(6)}-V^{\mathrm{loc}}\bigr)
\bigl(N_+^{\mathrm{loc}}\bigr)^{-1}N_B^{-1}.
\]
All conjugating factors are uniformly bounded on \(\Sigma^{\mathcal{E}}\cap\mathcal{D}\). Hence, \eqref{eq:local-jump-Linfty-p} and
\eqref{eq:local-jump-L1} yield~\eqref{eq:w-error-local-Linfty} and~\eqref{eq:w-error-local-L1}.
\end{proof}

The estimates in \eqref{eq:w-error-estimates} further imply that
\begin{equation*}
\|w^{\mathcal{E}}\|_{L^1(\Sigma^{\mathcal{E}})}
\lesssim t^{-1/2},
\qquad
\|w^{\mathcal{E}}\|_{L^\infty(\Sigma^{\mathcal{E}})}
\lesssim t^{-1/2}\log t.
\end{equation*}
By interpolation, we also obtain
\begin{align}\label{eq:w-L2-Linfty}
\|w^{\mathcal{E}}\|_{L^2(\Sigma^{\mathcal{E}})}  \lesssim 
 \| w^{\mathcal E}\|_{L^1(\Sigma^{\mathcal E})}^{1/2}
 \| w^{\mathcal E}\|_{L^\infty(\Sigma^{\mathcal E})}^{1/2}
 \lesssim t^{-1/2}(\log t)^{1/2}.
\end{align}
Consequently, the operator \(C_{w^{\mathcal{E}}}\), defined by
\[
C_{w^{\mathcal{E}}}(h)
:=
C_-\bigl(hw^{\mathcal{E}}\bigr),
\]
has sufficiently small norm for all sufficiently large \(t\).  Hence, \(I-C_{w^{\mathcal{E}}}\) is invertible on
\(L^2(\Sigma^{\mathcal{E}})\) by a Neumann-series argument.  
Therefore,
\[
\mu^{\mathcal{E}}
=
I+
\bigl(I-C_{w^{\mathcal{E}}}\bigr)^{-1}
C_{w^{\mathcal{E}}}I
\]
is well defined. By the standard theory of  RH problems~\cite{pz93}, the
solution of the error RH problem is given by
\begin{equation}
\mathcal{E}(x,t,k)
=
I+\frac{1}{2\pi\mathrm{i}}
\int_{\Sigma^{\mathcal{E}}}
\frac{\mu^{\mathcal{E}}(x,t,s)w^{\mathcal{E}}(x,t,s)}{s-k}\,\mathrm{d}s,
\qquad k\in\mathbb{C}\setminus\Sigma^{\mathcal{E}}.
\label{eq:error-Cauchy-solution}
\end{equation}
Expanding the Cauchy kernel in \eqref{eq:error-Cauchy-solution} as
\(k\to\infty\), we obtain
\begin{equation}
\mathcal{E}(x,t,k)
=
I+\frac{\mathcal{E}^{(1)}(x,t)}{k}+O(k^{-2}), \qquad 
\mathcal{E}^{(1)}(x,t)
=
-\frac{1}{2\pi\mathrm{i}}
\int_{\Sigma^{\mathcal{E}}}
\mu^{\mathcal{E}}(x,t,s)w^{\mathcal{E}}(x,t,s)\,\mathrm{d}s.
\label{eq:error-first-coefficient}
\end{equation}

\begin{lemma}
\label{lem:error-first-coefficient}

As \(t\to\infty\), the coefficient \(\mathcal{E}^{(1)}\) admits the
asymptotic expansion
\begin{equation}
\begin{aligned}
\mathcal{E}^{(1)}(x,t)
={}&
\frac{
N_B(k_0)Y_A\mathcal{N}_{k_0}^{(1)}Y_A^{-1}N_B(k_0)^{-1}
}{
\sqrt{\gamma_1(\xi)t}
} +
\frac{
N_B(k_1)Y_B\mathcal{N}_{k_1}^{(1)}Y_B^{-1}N_B(k_1)^{-1}
}{
\sqrt{\gamma_2(\xi)t}
}
+O(t^{-1}\log t).
\end{aligned}
\label{eq:E1-residue-formula}
\end{equation}
The error term is uniform for \(\xi \in\mathcal{K}_{+}\).
\end{lemma}

\begin{proof}
The $L^2$ bounds in \eqref{eq:w-L2-Linfty} imply
$\|\mu^{\mathcal{E}}-I\|_{L^2(\Sigma^{\mathcal{E}})}
=O(t^{-1/2}(\log t)^{1/2})$.  Cauchy--Schwarz, together with
\eqref{eq:w-error-estimates}, shows that
\begin{equation}
\begin{aligned}
 \mathcal E^{(1)}(x,t)=&-\frac{1}{2\pi\mathrm i}
 \int_{\partial\mathcal D}w^{\mathcal E}(x,t,s)\,\mathrm ds -\frac{1}{2\pi\mathrm i}
 \int_{\Sigma^{\mathcal E} \setminus \partial\mathcal D}w^{\mathcal E}(x,t,s)\,\mathrm ds-\frac{1}{2\pi\mathrm i}
 \int_{\Sigma^{\mathcal E} }(\mu^{\mathcal{E}}(x,t,s)-I) w^{\mathcal E}(x,t,s) \,\mathrm ds \\
 =&
 -\frac{1}{2\pi\mathrm i}
 \int_{\partial\mathcal D}w^{\mathcal E}(x,t,s)\,\mathrm ds
 +O(t^{-1}\log t), \qquad  t \to \infty.
 \label{eq:E1-circle-integral}
 \end{aligned}
\end{equation}
Substituting
\eqref{eq:local-inverse-expansions} into the remaining circle integrals and
evaluating them by residues gives \eqref{eq:E1-residue-formula}.
\end{proof}

\subsection{Proof of Theorem~\ref{thm:long-time-two-stationary}}
\label{subsec:reconstruction-asymptotics}
We now recover the original RH solution by reversing the six
transformations. In taking the large-\(k\) limit required by the
reconstruction formula, we restrict \(k\) to either \(D_7\) or \(D_8\)
in Figure~\ref{fig:first-deformation}. In these regions, all
lens-opening factors except \(G_1\) are equal to the identity, where
\(G_1\) equals either \(V_g^{(1)}\) or \(V_h^{(1)}\), as defined in
\eqref{eq:first-h}, according to the chosen region. Therefore, the inverse
transformations reduce to
\begin{equation}
M
=
M_\infty\widetilde{\sigma}e^{\mathrm{i}g_\infty\sigma}
\mathcal{E}N_Be^{-\mathrm{i}g\sigma}
D^{-1}\Delta_2^{-1}\Delta_1^{-1}
G_1^{-1}\widetilde{\sigma}^{-1}.
\label{eq:undo-transformations}
\end{equation}
Because the reconstruction formula~\eqref{eq:reconstruction} depends only on the first column of
the \(k^{-1}\) coefficient, we introduce
\begin{equation}
M_1(x,t)
=
\lim_{k\to\infty}k\bigl(M(x,t,k)-M_\infty\bigr),
\qquad
\bigl(M_1\bigr)_{\mathrm{rc}}
=
\begin{pmatrix}
(M_1)_{21}\\
(M_1)_{31}
\end{pmatrix}.
\label{eq:M1-reconstruction-column}
\end{equation}
Then by \eqref{eq:reconstruction}, we obtain 
\begin{align} \label{E:asycgggg}
\mathbf{q}(x,t)=-2 \ii \bigl(M_1\bigr)_{\mathrm{rc}}(x,t).
\end{align}

We next compute
\(\bigl(M_1\bigr)_{\mathrm{rc}}\). 
First, the outer model has the expansion
\begin{equation}
N_B(k)
=
I+\frac{N_B^{(1)}}{k}+O(k^{-2}),
\qquad k\to\infty,\qquad
N_B^{(1)}
=
\begin{pmatrix}
0&\mathrm{i}q_0/2&0\\
\mathrm{i}q_0/2&0&0\\
0&0&0
\end{pmatrix}.
\label{eq:NB1-explicit}
\end{equation}
Therefore, we next determine the contribution from the error matrix $\mathcal E$. Define
\begin{equation}
\widetilde{\mathcal{E}}^{(1)}(x,t)
=
\frac{
N_B(k_0)Y_A\mathcal{N}_{k_0}^{(1)}
Y_A^{-1}N_B(k_0)^{-1}
}{
\sqrt{\gamma_1(\xi)}
}
+
\frac{
N_B(k_1)Y_B\mathcal{N}_{k_1}^{(1)}
Y_B^{-1}N_B(k_1)^{-1}
}{
\sqrt{\gamma_2(\xi)}
}.
\label{eq:Etilde-explicit}
\end{equation}
It follows from \eqref{eq:error-first-coefficient} and
\eqref{eq:E1-residue-formula} that
\begin{equation}
\mathcal{E}(k)
=
I+\frac{1}{k}
\left(
t^{-1/2}\widetilde{\mathcal{E}}^{(1)}
+O(t^{-1}\log t)
\right)
+O(k^{-2}),
\qquad k\to\infty.
\label{eq:E-expii}
\end{equation}
Finally, we note that the coefficients of \(k^{-1}\) in the expansions of \(D^{-1}\),
\(\Delta_2^{-1}\), \(\Delta_1^{-1}\), and \(G_1^{-1}\) are all diagonal.
A direct calculation shows that these
diagonal terms do not contribute to
\(\bigl(M_1\bigr)_{\mathrm{rc}}\).

Combining these observations, substituting the large-\(k\) expansions \eqref{eq:NB1-explicit} and \eqref{eq:E-expii}
into \eqref{eq:undo-transformations}, and collecting the coefficients of
\(k^{-1}\), we obtain
\begin{equation}
\begin{aligned}
\bigl(M_1\bigr)_{\mathrm{rc}}
={}&
\left[
M_\infty\widetilde{\sigma}e^{\mathrm{i}g_\infty\sigma}
\left(
N_B^{(1)}
+t^{-1/2}\widetilde{\mathcal{E}}^{(1)}
\right)
e^{-\mathrm{i}g_\infty\sigma}\widetilde{\sigma}^{-1}
\right]_{\{2,3\}\times\{1\}}+O(t^{-1}\log t).
\end{aligned}
\label{eq:M1-root-column}
\end{equation}
Here, for any \(3\times3\) matrix \(A\),
\[
[A]_{\{2,3\}\times\{1\}}
=
\begin{pmatrix}
A_{21}\\
A_{31}
\end{pmatrix}.
\]
After simplifying the first term on the right-hand side of
\eqref{eq:M1-root-column} and substituting the resulting expression into
the reconstruction formula~\eqref{E:asycgggg}, we obtain
the asserted long-time asymptotic formula~\eqref{eq:final-asymptotics}.

\section{Long-time asymptotics in the elliptic-wave region}
\label{sec:elliptic-wave}

\begin{figure}
 \centering
 \setlength{\tabcolsep}{2pt}
 \begin{tabular}{ccc}
  \includegraphics[width=0.315\textwidth]
    {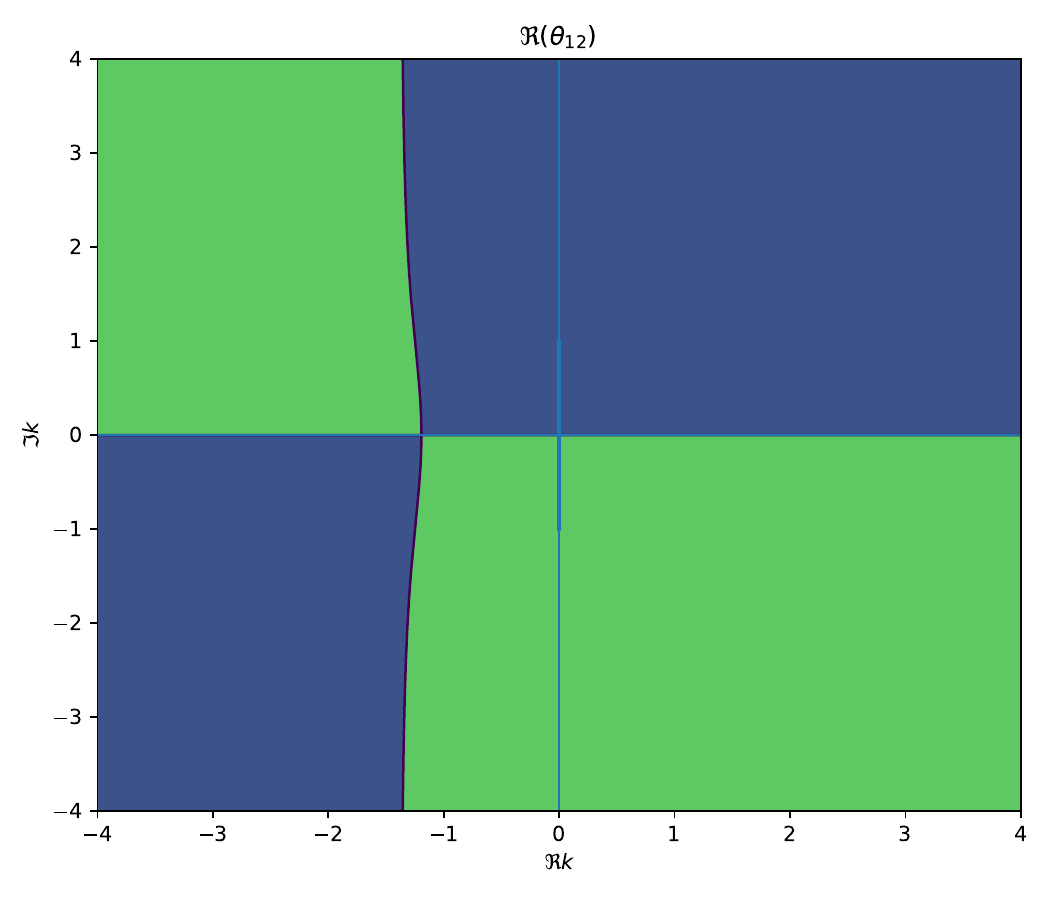}
  &
  \includegraphics[width=0.315\textwidth]
    {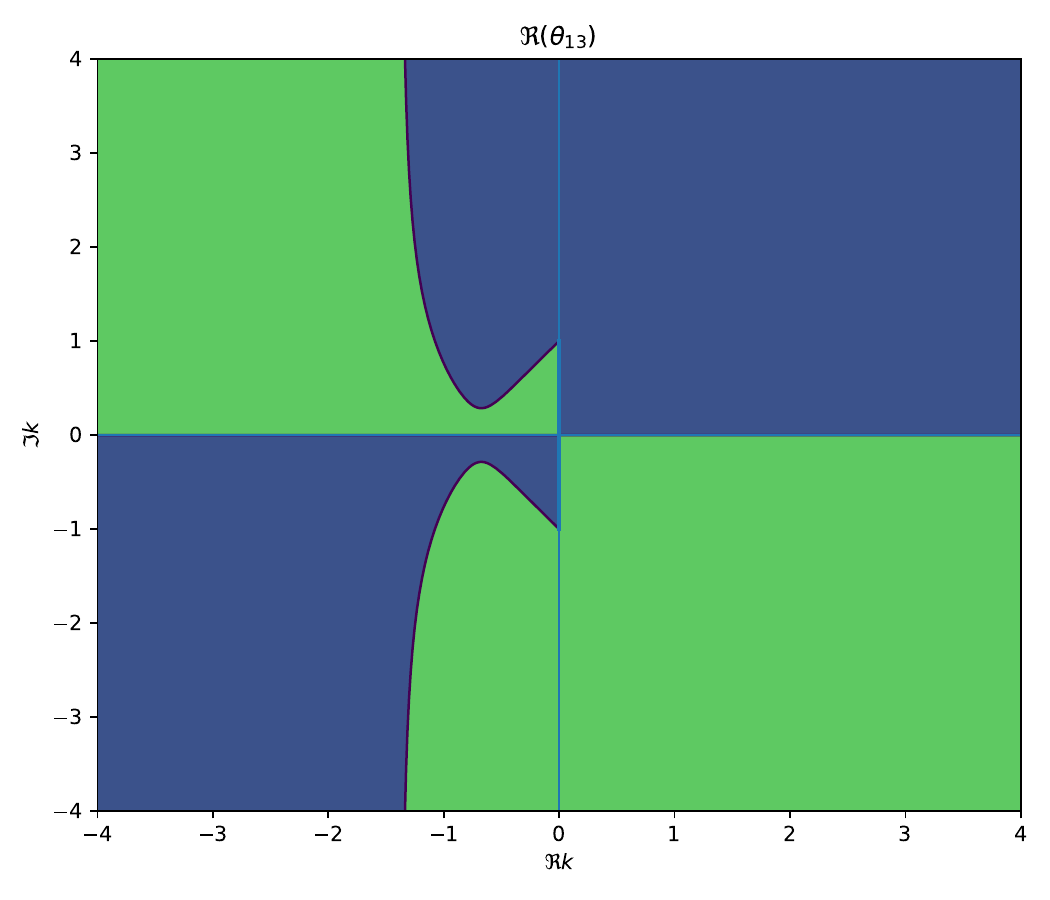}
  &
  \includegraphics[width=0.315\textwidth]
    {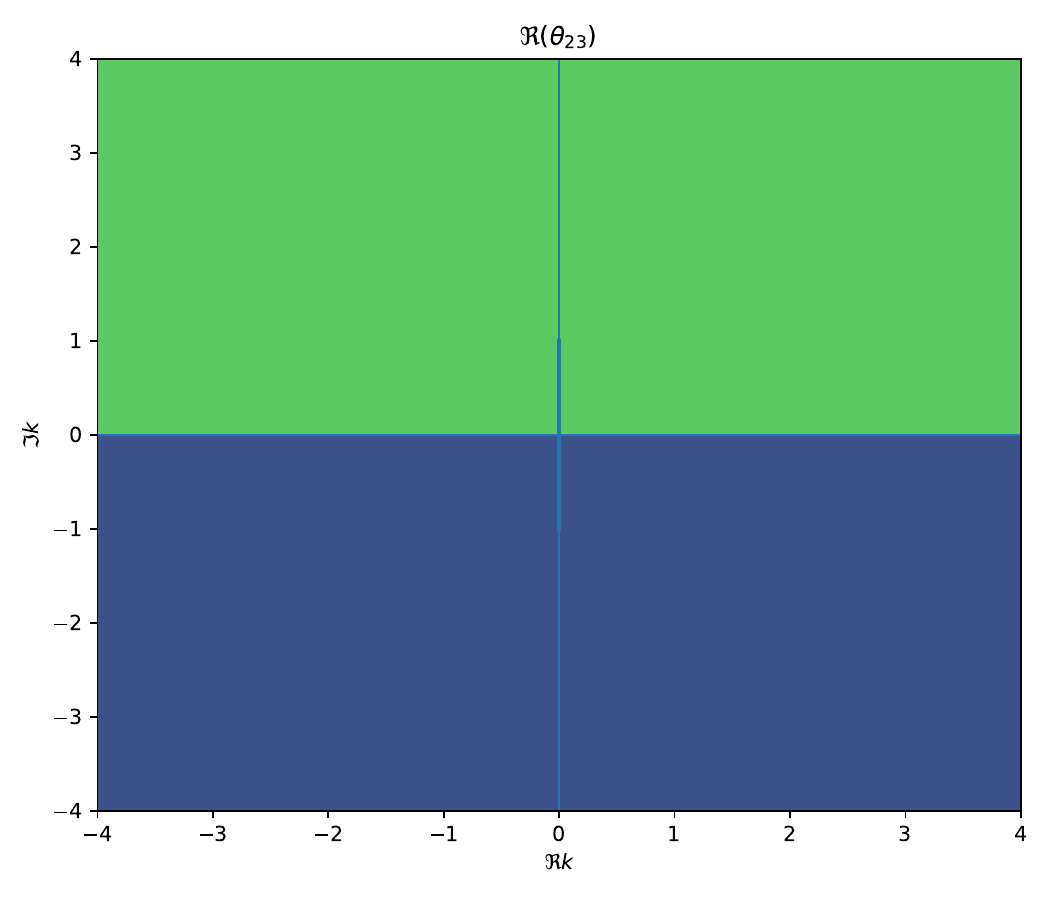}
 \end{tabular}
\caption{Signature tables for
\(\operatorname{Re}\theta_{12}\),
\(\operatorname{Re}\theta_{13}\), and
\(\operatorname{Re}\theta_{23}\) (from left to right),
with \(q_0=1\) and \(\xi=-5.5\).
The green and blue regions correspond to
\(\operatorname{Re}\theta_{ij}>0\) and
\(\operatorname{Re}\theta_{ij}<0\), respectively.}
 \label{fig:phase-signature-charts-E-I}
\end{figure}

We now turn to the modulated elliptic-wave sector
\(\mathrm{E}_{\mathrm{I}}\). Throughout this section, we assume that
\[
\xi\in(-4\sqrt{2}\,q_0,0).
\]
 Let  \(\mathcal K_-\)  be  any compact subset of
\((-4\sqrt{2}\,q_0,0)\)
In this regime, the signature charts of the phase functions
\(\theta_{ij}\) are shown in Figure~\ref{fig:phase-signature-charts-E-I}. A key observation is that the signature structure for \(\xi<0\) is
reversed relative to that for \(\xi>0\). Accordingly, we apply a diagonal
rescaling to the RH problem so that the jump matrices admit the ``opposite''
 factorizations. As shown below, the transformed jump matrices
have precisely the form required for the subsequent analysis.

Define the diagonal rescaling matrix \(S(k)\) by
\begin{equation}
 S(k)=
 \begin{cases}
 \displaystyle
 \diag\!\left(a_{11}(k),\frac{b_{22}(k)}{a_{11}(k)},
                    \frac{1}{b_{22}(k)}\right),&k\in\mathbb C^+ \setminus B,\\[4mm]
 \displaystyle
 \diag\!\left(\frac{1}{b_{11}(k)},\frac{b_{11}(k)}{a_{22}(k)},
                    a_{22}(k)\right),&k\in\mathbb C^- \setminus B.
 \end{cases}
 \label{eq:elliptic-S}
\end{equation}
Using the large-\(k\) asymptotics of the scattering coefficients
\(a_{ij}\) and \(b_{ij}\) established in Section~\ref{sec:direct-scattering-k}, we readily obtain
\begin{equation}
 S_\infty:=\lim_{k\to\infty}S(k)
 =\diag\!\left(1,e^{-\mathrm i(\vartheta_+-\vartheta_-)},
                   e^{\mathrm i(\vartheta_+-\vartheta_-)}\right).
 \label{eq:elliptic-S-infty}
\end{equation}
We now introduce the transformation
\begin{equation}
\widehat{N}(x,t,k)
=
S_\infty^{-1}N(x,t,k)S(k),
\qquad
k\in\mathbb{C}\setminus\Sigma.
\label{eq:elliptic-Nhat}
\end{equation}
The jump contour remains unchanged, but the corresponding jump matrix is
\begin{equation}
\widehat{V}(k)
=
\bigl(S^-(k)\bigr)^{-1}\widetilde{V}(k)S^+(k).
\label{eq:elliptic-Vhat}
\end{equation}
We now derive the explicit form of the jump matrix \(\widehat{V}(k)\).
The diagonal rescaling naturally leads to the following reflection
coefficients:
\begin{equation}
r_1(k)=\frac{a_{13}(k)}{a_{11}(k)},
\qquad
r_2(k)=\frac{a_{21}(k)}{a_{22}(k)},
\qquad
r_3(k)=\frac{a_{23}(k)}{a_{22}(k)}.
\label{eq:elliptic-reflection-coefficients}
\end{equation}
These coefficients should not be confused with the reflection
coefficients \(\rho_j\) introduced in
Section~\ref{sec:steepest-descent}. In what follows, the dependence on \(x,t, k\) is suppressed whenever it is clear from the context. 
On the real axis, the jump matrix can be written as
\begin{equation}
\widehat{V}_{\mathbb{R}}
=
\widehat{V}_L\widehat{V}_U
=
\begin{pmatrix}
1&0&0\\
-d\bar r_1e^{\theta_{31}}&d&0\\
r_2e^{\theta_{21}}&r_3e^{\theta_{23}}&1
\end{pmatrix}
\begin{pmatrix}
d^{-1}&-r_1e^{\theta_{13}}&d^{-1}\bar r_2e^{\theta_{12}}\\
0&1&d^{-1}\bar r_3e^{\theta_{32}}\\
0&0&1
\end{pmatrix}.
\label{eq:elliptic-basic-factorization}
\end{equation}
The jump matrices on \(B_+\) and \(B_-\) are given by

\begin{align} \label{eq:elliptic-initial-cut-jumps}
\widehat{V}_{B_+}
=
\begin{pmatrix}
-\frac{\mathrm{i}(\lambda-k)}{q_0}r_1^-e^{\theta_{13}}
&-\frac{2\mathrm{i}\lambda}{q_0}(1-r_1^-r_1^+)&0\\[3mm]
-\frac{\mathrm{i}q_0}{2\lambda}
&\frac{\mathrm{i}(k+\lambda)}{q_0}r_1^+e^{\theta_{31}}&0\\
0&0&1
\end{pmatrix}, \qquad
\widehat{V}_{B_-}
=
\begin{pmatrix}
-\frac{\mathrm{i}(k+\lambda)}{q_0}\bar r_1^+e^{\theta_{13}}
&-\frac{\mathrm{i}q_0}{2\lambda}&0\\[3mm]
-\frac{2\mathrm{i}\lambda}{q_0}(1-\bar r_1^-\bar r_1^+)
&\frac{\mathrm{i}(\lambda-k)}{q_0}\bar r_1^-e^{\theta_{31}}&0\\
0&0&1
\end{pmatrix}.
\end{align}

\begin{remark}
By \eqref{E:arsssh}, the scattering coefficients \(a_{11}\)
and \(b_{11}\) may have inverse square-root singularities
at the branch points \(\pm\mathrm{i}q_0\).
Consequently, the transformation \eqref{eq:elliptic-Nhat}
may introduce such singularities in the first column of
\(\widehat N\) at \(\mathrm{i}q_0\) and in its second column
at \(-\mathrm{i}q_0\).
These possible singularities are canceled by the next
transformation. Indeed, \eqref{E:arsssh} gives
\[
\begin{aligned}
 \lim_{k\to\mp\mathrm{i}q_0}r_1(k)=\pm1; \qquad
 r_2(k)=O(1),\quad r_3(k)=O(1),
 \qquad k\to\pm\mathrm{i}q_0.
\end{aligned}
\]
Combining these relations with
\(d(k)=O(|k\mp\mathrm{i}q_0|^{1/2})\) as
\(k\to\pm\mathrm{i}q_0\), and the explicit expressions for
\(\widehat G_1\) in \(\mathcal R_1^{(1)}\) and
\(\mathcal R_1^{(2)}\), we verify that
\(\widehat N^{(1)}\), defined by
\eqref{eq:elliptic-first-transformation}, remains bounded
near both branch points.
\end{remark}

\subsection{Opening of lenses: first transformation}
We  observe that the signature structure of the phase
\(\theta_{13}\) in the elliptic-wave sector differs substantially from
that in the plane-wave sector.  In the plane-wave sector, the real stationary points naturally
determine the boundaries of the regions in which the oscillatory factors
decay. In the modulated elliptic-wave sector, however, the original phase
\(\theta_{13}\) has no real stationary points. As shown in
Figure~\ref{fig:phase-signature-charts-E-I}, the zero level curves of
\(\operatorname{Re}\theta_{13}\) emanating from infinity terminate
directly on the branch cut \(B\). Consequently, any deformation toward
\(B\) must cross a region in which the sign of
\(\operatorname{Re}\theta_{13}\) is incompatible with the desired
exponential decay. Thus, the original phase functions alone do not
produce uniformly decaying jumps on the deformed contour.

To overcome this difficulty, we follow the approach of~\cite{BiondiniMantzavinos2017} and
introduce a \(G\)-function later in the analysis. The corresponding
transformation replaces the original phases \(\theta_{ij}\) with modified
phases \(h_{ij}\) having the required signature structure. In particular,
the modified phases \(h_{13}\) and \(h_{12}\) have auxiliary ``stationary
points'' \(k_0\) and \(k_1\), respectively. The subsequent steepest descent
analysis can then be carried out in close analogy with that in section~\ref{sec:steepest-descent}. Although we retain the
notation \(k_0\) and \(k_1\) from the plane-wave analysis, their
definitions in the present sector are different and somewhat more
involved.

We emphasize that the order of the transformations adopted below differs
from the most direct conceptual order. A natural strategy would be to
introduce the \(G\)-function first, open the lenses according to the
signature structure of the modified phases \(h_{ij}\), and then construct
the outer and local models before carrying out the small-norm analysis.
In the present work, by contrast, we first open the lenses along contours
chosen in anticipation of the signature structure of \(h_{ij}\), and
introduce the \(G\)-function only after the lens opening has been
completed. The \(G\)-function transformation then makes the jumps on the
new contours exponentially close to the identity away from the branch cut
and the relevant critical points. The model problems and the small-norm
analysis are introduced afterward. These two orderings lead to the same
final RH problem. Following~\cite{BiondiniMantzavinos2017}, we adopt the latter ordering
because it makes the motivation for introducing the \(G\)-function and
the details of its construction more transparent.

We now begin the lens-opening deformation.  We first define
\(k_0=k_0(\xi)\). 
For \(\xi\in(-4\sqrt{2}\,q_0,0)\), we define
\(k_0=k_0(\xi)\) as the unique solution of the integral
equation~\eqref{eq:elliptic-modulation-moment}.
Its existence and uniqueness are guaranteed by
Lemma~\ref{lem:unique-k0}.
The point \(k_0\) is also the unique real solution of
\(
\frac{\mathrm{d}h_{13}}{\mathrm{d}k}=0
\).
Once \(k_0\) is fixed, the algebraic
relations~\eqref{eq:elliptic-modulation-algebraic}
uniquely determine \(\alpha=\alpha(\xi)\).
We briefly explain the origin of the modulation
system~\eqref{eq:elliptic-modulation-system}.
The parameters \(k_0\) and \(\alpha\) enter the construction
of the modified phase \(h_{13}\), and the system is obtained
by requiring this phase to have the desired properties.
Now we can then define \(k_1\) in terms of \(k_0\) and \(\alpha\).
More precisely, we set
\begin{equation}
k_1
=
\frac{
16|\alpha|^2k_0^2-\xi^2q_0^2
}{
32k_0|\alpha|^2
+32(\operatorname{Re}\alpha)k_0^2
-8\xi q_0^2
}.
\label{eq:elliptic-k1}
\end{equation}
In fact, \(k_1\) is the unique real solution of
\(
\frac{\mathrm{d}h_{12}}{\mathrm{d}k}=0
\).

Moreover, we have the following lemma.

\begin{lemma}\label{lem:elliptic-k1-k0}
For \(\xi\in(-4\sqrt2q_0,0)\), one has \(k_1<k_0\).
\end{lemma}

\begin{proof}
Using the definitions of \(k_0\) and \(\alpha\), rewrite
\(k_1=F_1/F_2\), where
\begin{align}
 F_1={}&k_0^4\left(
 48-16\frac{\xi}{k_0}
 +\frac{\xi^2}{k_0^2 }
 +16\frac{q_0^2}{k_0^2}
 -\frac{q_0^2\xi^{2} }{k_0^4}
 \right),
 \label{eq:elliptic-F1}\\
 F_2={}&k_0^3\left[
 64-24\frac{\xi}{k_0}
 +2 \frac{\xi^2}{k_0^2} 
 +8\frac{q_0^2}{k_0^2}
      \left(4-\frac{\xi}{k_0}\right)
 \right].
 \label{eq:elliptic-F2}
\end{align}
Consequently,
\begin{equation}
 k_1-k_0=\frac{F_1-k_0F_2}{F_2}
 =\frac{\widetilde F_1}{F_2},
 \label{eq:elliptic-k1-minus-k0}
\end{equation}
where
\begin{equation}
 \widetilde F_1=k_0^4\left[
 -16+8\frac{\xi}{k_0}
 -\frac{\xi^2}{k_0^2}
 -8\frac{q_0^2}{k_0^2}
   \left(2-\frac{\xi}{k_0}\right)
 -\frac{q_0^2 \xi^2 }{k_0^4 } 
 \right].
 \label{eq:elliptic-F1-tilde}
\end{equation}
Since \(\Re\alpha<0\), we have
\(0>k_0>\xi/4\), and hence
\(\xi/k_0>4\).  Put
\begin{equation*}
 a=\frac{q_0^2}{\xi^2},
 \quad y=\frac{\xi}{k_0}, \  \  \text{then }
 \quad a>\frac1{32},\quad y>4.
\end{equation*}
Equivalently, \(\widetilde F_1\) and \(F_2\) may now be regarded as
functions of the two variables \((a,y)\).
Direct simplification gives
\begin{equation*}
 \frac{\widetilde F_1}{k_0^4}=-(y-4)^2(ay^2+1)<0,
\end{equation*}
and
\begin{align}
 \frac{F_2}{k_0^3}
 =64-24y+2y^2+8ay^2(4-y) 
 &=2(y-4)(y-8-4ay^2)\nonumber \\
 &<2(y-4)\left(y-8-\frac{y^2}{8}\right)<0.
 \label{eq:elliptic-F2-sign}
\end{align}
Because \(k_0^3<0\), it follows that \(F_2>0\).  Therefore
\(k_1-k_0<0\), as claimed.
\end{proof}

We are now ready to factor the jump matrix on the real axis. As in the
preceding section, we consider separate factorizations on the three
intervals
\(
(-\infty,k_1)\),\( (k_1,k_0)\) and \((k_0,\infty)\).
On \((k_0,\infty)\),  we use
\begin{equation}
 \widehat V_{\mathbb R}
 =\widehat V_L^{(1)}\widehat V_{r,L}^{(1)}
  \widehat V_{r,U}^{(1)}\widehat V_U^{(1)},
 \label{eq:elliptic-right-four-factorization}
\end{equation}
where
\begin{subequations}\label{eq:elliptic-right-four-factors}
\begin{align}
 \widehat V_L^{(1)}&=
 \begin{pmatrix}
 d^{-1}&0&0\\ -\bar r_1e^{\theta_{31}}&d&0\\0&0&1
 \end{pmatrix},
 &
 \widehat V_{r,L}^{(1)}&=
 \begin{pmatrix}
 d&0&0\\0&1&0\\r_2e^{\theta_{21}}&r_3e^{\theta_{23}}&1
 \end{pmatrix},\\[3mm]
 \widehat V_{r,U}^{(1)}&=
 \begin{pmatrix}
 1&0&d^{-1}\bar r_2e^{\theta_{12}}\\
 0&d^{-1}&d^{-1}\bar r_3e^{\theta_{32}}\\0&0&1
 \end{pmatrix},
 &
 \widehat V_U^{(1)}&=
 \begin{pmatrix}
 d^{-1}&-r_1e^{\theta_{13}}&0\\0&d&0\\0&0&1
 \end{pmatrix}.
\end{align}
\end{subequations}
The above factorizations is useful because  we observe that the jump matrices on \(B_+\) and \(B_-\)
admit the factorizations
\begin{align}
 \widehat V_{B_+}
 =
 \left(\widehat V_U^{(1)-}\right)^{-1}
 \widehat V_{B_+}^{(1)}
 \left(\widehat V_U^{(1)}\right)^+,
 \qquad
 \widehat V_{B_-}
 =
 \widehat V_L^{(1) -} \widehat V_{B_-}^{(1)}
 \left(\widehat V_L^{(1)+}\right)^{-1}.
\end{align}
Here
\begin{align}
 \widehat V_{B_+}^{(1)}
 =
 \begin{pmatrix}
 0
 &
 -\frac{\mathrm i(k+\lambda)}{q_0}(d^+)^{-1}
 &
 0
 \\[2mm]
 -\frac{\mathrm i(k+\lambda)}{q_0}d^-
 &
 0
 &
 0
 \\
 0&0&1
 \end{pmatrix},
 \qquad
 \widehat V_{B_-}^{(1)}
 =
 \begin{pmatrix}
 0
 &
 -\frac{\mathrm i(k+\lambda)}{q_0}d^-
 &
 0
 \\[2mm]
 -\frac{\mathrm i(k+\lambda)}{q_0}(d^+)^{-1}
 &
 0
 &
 0
 \\
 0&0&1
 \end{pmatrix}.
\end{align}
Thus once we remove the jump on $(k_0,\infty)$, Then the jump on
$B$ will be much simpler.

On the interval \((k_1,k_0)\), the real-axis jump is decomposed as
$
 \widehat V_{\mathbb R}
 =\widehat V_a^{(1)}\widehat V_b^{(1)}\widehat V_c^{(1)}
$,
where
\begin{align} \label{eq:elliptic-middle-first-factors}
 \widehat V_a^{(1)}=&
 \begin{pmatrix}
 1&0&0\\0&d&0\\(r_2+\bar r_1r_3)e^{\theta_{21}}&r_3e^{\theta_{23}}&1
 \end{pmatrix},\qquad
 \widehat V_b^{(1)}=
 \begin{pmatrix}
 1&-r_1e^{\theta_{13}}&0\\
 -\bar r_1e^{\theta_{31}}&1+r_1\bar r_1&0\\
 0&0&1
 \end{pmatrix},\nonumber \\
 \widehat V_c^{(1)}=&
 \begin{pmatrix}
 d^{-1}&0&d^{-1}(\bar r_2+r_1\bar r_3)e^{\theta_{12}}\\
 0&1&d^{-1}\bar r_3e^{\theta_{32}}\\0&0&1
 \end{pmatrix}.
\end{align}
On \(( -\infty,k_1)\), we use
$
 \widehat V_{\mathbb R}
 =\widehat V_d^{(1)}\widehat V_e^{(1)}\widehat V_f^{(1)}
 $,
where
\begin{subequations}\label{eq:elliptic-left-first-factors}
\begin{align}
 \widehat V_d^{(1)}&=
 \begin{pmatrix}1&0&0\\0&d&0\\0&r_3e^{\theta_{23}}&1\end{pmatrix}, \qquad
  \widehat V_f^{(1)}=
 \begin{pmatrix}
 d^{-1}&0&0\\0&1&d^{-1}\bar r_3e^{\theta_{32}}\\0&0&1
 \end{pmatrix},\\
 \widehat V_e^{(1)}&=
 \begin{pmatrix}
 1&0&0\\-\bar r_1e^{\theta_{31}}&1&0\\
 (r_2+\bar r_1r_3)e^{\theta_{21}}&0&1
 \end{pmatrix}
 \begin{pmatrix}
 1&-r_1e^{\theta_{13}}&d^{-1}(\bar r_2+r_1\bar r_3)e^{\theta_{12}}\\
 0&1&0\\0&0&1
 \end{pmatrix}.
\end{align}
\end{subequations}
In addition, we introduce the following diagonal factors:
\begin{equation}
\widehat{V}_g^{(1)}
=
\operatorname{diag}(d^{-1},1,1),
\qquad
\widehat{V}_h^{(1)}
=
\operatorname{diag}(1,d,1).
\label{eq:elliptic-Vg-Vh}
\end{equation}

\begin{figure}
\centering

\begin{tikzpicture}[scale=0.7,>=Stealth]

\tikzset{
  oldsolid/.style={dashed,line width=1pt},
  kkoldsolid/.style={dashed,line width=2.2pt},
  newcontour/.style={
    black,line width=2pt,
    line cap=round,line join=round
  },
  newbranch/.style={
    black,line width=2.2pt,line cap=round
  },
  dir/.style={
    postaction={decorate},
    decoration={
      markings,
      mark=at position 0.58 with {\arrow[scale=0.8]{Stealth}}
    }
  },
  dirKA/.style={
    postaction={decorate},
    decoration={
      markings,
      mark=at position 0.72 with {\arrow[scale=0.8]{Stealth}}
    }
  }
}

\coordinate (K)    at (0,0);
\coordinate (A)    at (0.15,1.25);
\coordinate (AB)   at (0.15,-1.25);
\coordinate (I)    at (1.85,2.25);
\coordinate (MI)   at (1.85,-2.25);

\coordinate (Kone) at (-5.50,0);
\coordinate (KH)   at (-2.75,2.75);
\coordinate (KL)   at (-2.75,-2.75);

\coordinate (UR)   at (0.80,2.75);
\coordinate (DR)   at (0.80,-2.75);

\draw[kkoldsolid] (K)--(5,0);

\draw[oldsolid]
  (A) .. controls (-0.7,1.6) and (-1.5,2.4) .. (-2.3,3.4);

\draw[oldsolid]
  (A) .. controls (0.6,1.7) and (1.3,2.1) .. (I);

\draw[oldsolid]
  (AB) .. controls (-0.7,-1.6) and (-1.5,-2.4) .. (-2.3,-3.4);

\draw[oldsolid]
  (AB) .. controls (0.6,-1.7) and (1.3,-2.1) .. (MI);

\draw[newcontour,dir] (-10.8,2.75)--(KH);
\draw[newcontour,dir] (KH)--(UR);
\draw[newcontour,dir] (UR)--(5,2.75);

\draw[newcontour,dir] (-10.8,-2.75)--(KL);
\draw[newcontour,dir] (KL)--(DR);
\draw[newcontour,dir] (DR)--(5,-2.75);

\draw[newcontour,dir] (Kone)--(KH);
\draw[newcontour,dir] (Kone)--(KL);

\draw[newcontour,dirKA] (K)--(A);
\draw[newcontour,dirKA] (K)--(AB);

\draw[newcontour,dir] (-10.8,0)--(Kone);
\draw[newcontour,dir] (Kone)--(K);

\draw[newcontour] (A)--(UR);
\draw[newcontour] (AB)--(DR);

\draw[newbranch,dir] (MI)--(I);

\fill (K)    circle (1.6pt);
\fill (Kone) circle (1.6pt);
\fill (A)    circle (1.6pt);
\fill (AB)   circle (1.6pt);
\fill (I)    circle (1.6pt);
\fill (MI)   circle (1.6pt);

\node at (0.4,-0.3) {$k_0$};
\node[below] at (Kone) {$k_1$};
\node[right] at (A) {$\alpha$};
\node[right] at (AB) {$\bar{\alpha}$};


\node at (2.85,0.82)   {$\mathcal R^{(1)}_1$};
\node at (2.85,-0.82)  {$\mathcal R^{(1)}_2$};

\node at (-2.55,1.35)  {$\mathcal R^{(1)}_3$};
\node at (-2.55,-1.35) {$\mathcal R^{(1)}_4$};

\node at (-7.70,1.35)  {$\mathcal R^{(1)}_5$};
\node at (-7.70,-1.35) {$\mathcal R^{(1)}_6$};

\node at (-1.30,3.80)  {$\mathcal R^{(1)}_7$};
\node at (-1.30,-3.80) {$\mathcal R^{(1)}_8$};

\node at (-0.10,1.12)  {$1$};
\node at (-0.10,-1.12) {$2$};

\node at (-4.62,1.62)  {$3$};
\node at (-4.62,-1.62) {$4$};

\node at (-2.05,0.38)  {$5$};
\node at (-8.65,0.38)  {$6$};

\node at (2.25,-1.78) {$B$};

\end{tikzpicture}

\caption{The oriented solid curves form the jump contour
\(\Gamma^{(1)}\), while the thin dashed curves indicate portions of the
zero level set \(\Ree h_{13}=0\). The regions
\( \mathbb C= \cup_{j=1}^8  \mathcal{R}_j^{(1)} \) are also shown.}
\label{fig:Gamma1-regions}

\end{figure}
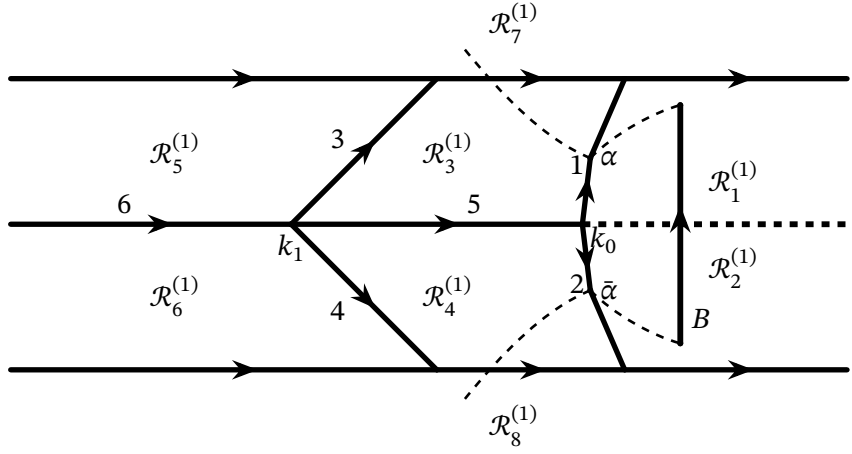

The preceding factorizations allow us to define the first lens-opening
transformation. Let \(\{\mathcal{R}_j^{(1)}\}_{j=1}^{8}\) denote the
regions shown in Figure~\ref{fig:Gamma1-regions}. We then define the
transformation as follows:
\begin{equation}
 \widehat N^{(1)}(x,t,k)=\widehat N(x,t,k)\widehat G_1(x,t,k),
 \label{eq:elliptic-first-transformation}
\end{equation}
where
\begin{equation}
 \widehat G_1=
 \begin{cases}
 (\widehat V_U^{(1)})^{-1}(\widehat V_{r,U}^{(1)})^{-1},
   &k\in\mathcal R_1^{(1)},\\
 \widehat V_L^{(1)}\widehat V_{r,L}^{(1)},
   &k\in\mathcal R_2^{(1)},\\
 (\widehat V_c^{(1)})^{-1},&k\in\mathcal R_3^{(1)},\\
 \widehat V_a^{(1)},&k\in\mathcal R_4^{(1)},\\
 (\widehat V_f^{(1)})^{-1},&k\in\mathcal R_5^{(1)},\\
 \widehat V_d^{(1)},&k\in\mathcal R_6^{(1)},\\
 (\widehat V_g^{(1)})^{-1},&k\in\mathcal R_7^{(1)},\\
 \widehat V_h^{(1)},&k\in\mathcal R_8^{(1)}.
 \end{cases}
 \label{eq:elliptic-G1}
\end{equation}

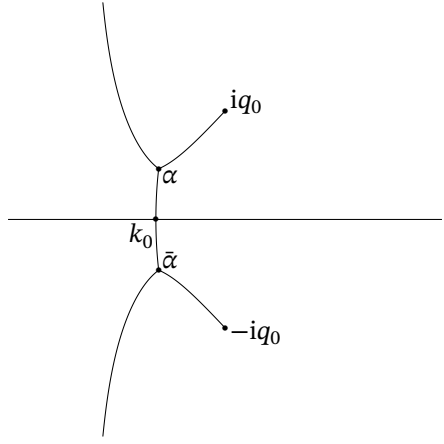
\begin{figure}
    \centering

    \begin{overpic}[width=0.36\textwidth]
        {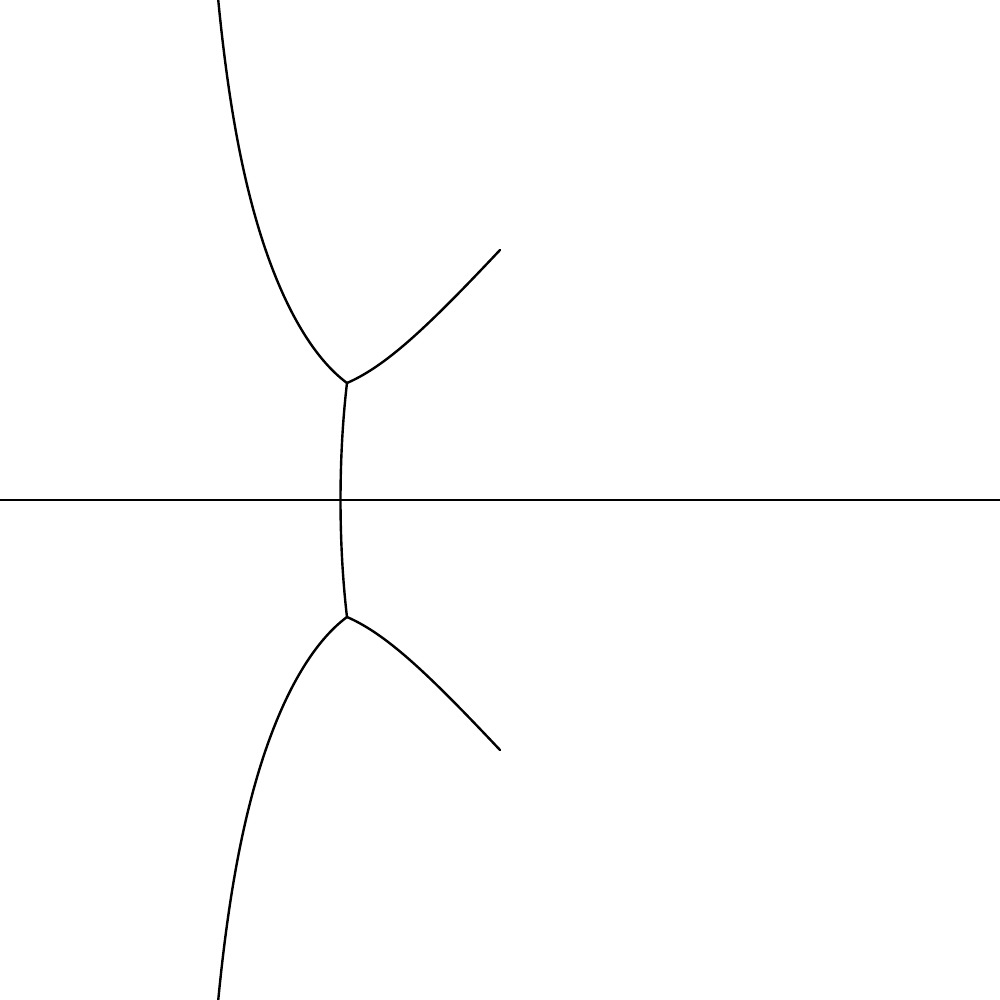}
        \put(34.70,61.70){\circle*{1.2}}
        \put(35.50,58.30){\small $\alpha$}

        \put(34.70,38.30){\circle*{1.2}}
        \put(35.50,38.70){\small $\bar{\alpha}$}
        
        \put(34.05,50.00){\circle*{1.2}}
        \put(27.80,44.80){\small $k_0$}

        \put(50.00,75.00){\circle*{1.2}}
        \put(51.20,75.80){\small $\mathrm{i}q_0$}

        \put(50.00,25.00){\circle*{1.2}}
        \put(51.20,22.50){\small $-\mathrm{i}q_0$}
    \end{overpic}

    \caption{
        The zero-level set
        $\operatorname{Re}h_{13}(\xi,k)=0$ with $q_0=1$ and $\xi=-5$.
    }
    \label{fig:h-sign-structure}
\end{figure}

\begin{remark}[On the choice of contours near $k_0$]

We briefly explain the choice of the lens boundaries near \(k_0\). When
the factorization on \((k_0,\infty)\) is extended off the real axis, the
two contour arcs connecting \(k_0\) to \(\alpha\) and
\(\bar{\alpha}\), respectively, are chosen to follow the zero level set
\(
\operatorname{Re}h_{13}(\xi,k)=0
\); see Fig. \ref{fig:h-sign-structure}.
This choice simplifies the subsequent analysis.

\end{remark}

The jump contour for \(\widehat{N}\) is denoted by
\(\Gamma^{(1)}\)~(see Fig.~\ref{fig:Gamma1-regions}), and the corresponding jump matrix by
\(\widehat{V}^{(1)}\).
We now list the nontrivial jump matrices for \(\widehat{N}^{(1)}\).
Their derivation closely parallels the corresponding calculation in the
plane-wave sector \(\mathrm{P}_{\mathrm{I}}\). We therefore omit the
details and state the resulting jump matrices directly.
\begin{subequations}\label{eq:elliptic-first-jumps}
\begin{align}
 \widehat V_1^{(1)}&=
 \begin{pmatrix}1&-r_1e^{\theta_{13}}&0\\0&1&0\\0&0&1\end{pmatrix},&
 \widehat V_2^{(1)}&=
 \begin{pmatrix}1&0&0\\-\bar r_1e^{\theta_{31}}&1&0\\0&0&1\end{pmatrix},\\[2mm]
 \widehat V_3^{(1)}&=
 \begin{pmatrix}1&0&d^{-1}(\bar r_2+r_1\bar r_3)e^{\theta_{12}}\\
 0&1&0\\0&0&1\end{pmatrix},&
 \widehat V_4^{(1)}&=
 \begin{pmatrix}1&0&0\\0&1&0\\(r_2+\bar r_1r_3)e^{\theta_{21}}&0&1\end{pmatrix},\\[2mm]
 \widehat V_5^{(1)}&=\widehat V_b^{(1)},&
 \widehat V_6^{(1)}&=\widehat V_e^{(1)},\\[2mm]
 \widehat V_B^{(1)}&=
 \begin{pmatrix}
 0&-\frac{\mathrm i(k+\lambda)}{q_0}&0\\[2mm]
 -\frac{\mathrm i(k+\lambda)}{q_0}&0&0\\0&0&1
 \end{pmatrix}.
\end{align}
\end{subequations}

\subsection{Scalar conjugations and further contour deformations}
\label{subsec:elliptic-scalar-conjugations}

The second transformation introduces a conjugating factor designed to
modify the jump matrix on \((-\infty,k_0)\).  Let \(\widehat\delta_1\) be analytic in
\(\mathbb C\setminus(-\infty,k_0)\) and satisfy
\begin{equation*}
 \widehat{\delta}_1^{+}(k)=\widehat{\delta}_1^{-}(k)
       \bigl(1+|r_1(k)|^2\bigr),\quad k\in(-\infty,k_0);
 \qquad
 \widehat\delta_1(k)\to1,\quad k\to\infty.
\end{equation*}
The Sokhotski--Plemelj formula then yields
\begin{equation}
 \widehat\delta_1(k)=
 \exp\left\{\frac{1}{2\pi\mathrm i}
 \int_{-\infty}^{k_0}
 \frac{\log(1+|r_1(s)|^2)}{s-k}\,\mathrm ds\right\}.
 \label{eq:elliptic-delta1-solution}
\end{equation}

We define the second transformation by
\begin{equation}
\widehat{N}^{(2)}(x,t,k)
=
\widehat{N}^{(1)}(x,t,k)\widehat{\Delta}_1(k),
\label{eq:elliptic-second-transformation}
\end{equation}
where
\[
\widehat{\Delta}_1(k)
=
\operatorname{diag}\left(
\widehat{\delta}_1(k),
\widehat{\delta}_1(k)^{-1},
1
\right).
\]
This transformation introduces no new contours, so the jump contour satisfies $\Gamma^{(2)}=\Gamma^{(1)}$. The corresponding jump matrix is
\begin{equation}
\widehat{V}^{(2)}
=
\bigl(\widehat{\Delta}_1^-\bigr)^{-1}
\widehat{V}^{(1)}
\widehat{\Delta}_1^+.
\label{eq:elliptic-second-jump-conjugation}
\end{equation}
The transformed jump matrices on all contour components other than
\((-\infty,k_0)\) follow directly from the conjugation formula. We
therefore focus on \((-\infty,k_0)\), where we give an explicit expression
for \(\widehat{V}^{(2)}\) together with a factorization suitable for the
next lens-opening transformation.

On the middle interval \((k_1,k_0)\), the jump matrix is given by
$
 \widehat V_5^{(2)}=\widehat V_a^{(2)}\widehat V_b^{(2)}
$,
where
\begin{align}\label{eq:elliptic-Vab-second}
 \widehat V_a^{(2)}&=
 \begin{pmatrix}
 1&-\dfrac{r_1}{1+r_1\bar r_1}
       (\widehat{\delta}_1^{-})^{-2}e^{\theta_{13}}&0\\
 0&1&0\\0&0&1
 \end{pmatrix},\qquad
 \widehat V_b^{(2)}&=
 \begin{pmatrix}
 1&0&0\\
 -\dfrac{\bar r_1}{1+r_1\bar r_1}
       (\widehat{\delta}_1^{+})^2e^{\theta_{31}}&1&0\\
 0&0&1
 \end{pmatrix}.
\end{align}
On \(( -\infty,k_1)\), a direct calculation gives
$
 \widehat V_6^{(2)}
 =\widehat V_c^{(2)}\widehat V_d^{(2)}\widehat V_e^{(2)}
 $,
with
\begin{subequations}\label{eq:elliptic-Vcde-second}
\begin{align}
 \widehat V_c^{(2)}&=
 \begin{pmatrix}
 1&-\dfrac{r_1}{1+r_1\bar r_1}
       (\widehat{\delta}_1^{-})^{-2}e^{\theta_{13}}&0\\
 0&1&0\\
 0&-\dfrac{r_1(r_2+\bar r_1r_3)}{1+r_1\bar r_1}
       (\widehat{\delta}_1^{-})^{-1}e^{\theta_{23}}&1
 \end{pmatrix},\\[3mm]
 \widehat V_d^{(2)}&=
 \begin{pmatrix}
 1&0&d^{-1}(\bar r_2+r_1\bar r_3)
                  (\widehat{\delta}_1^{+})^{-1}e^{\theta_{12}}\\
 0&1&0\\
 (r_2+\bar r_1r_3)\widehat{\delta}_1^{-}e^{\theta_{21}}&0&
 1+d^{-1}(r_2+\bar r_1r_3)(\bar r_2+r_1\bar r_3)
      \dfrac{\widehat{\delta}_1^{-}}{\widehat{\delta}_1^{+}}
 \end{pmatrix},\\[3mm]
 \widehat V_e^{(2)}&=
 \begin{pmatrix}
 1&0&0\\
 -\dfrac{\bar r_1}{1+r_1\bar r_1}
       (\widehat{\delta}_1^{+})^2e^{\theta_{31}}&1&
 -\dfrac{d^{-1}\bar r_1(\bar r_2+r_1\bar r_3)}
        {1+r_1\bar r_1}\widehat{\delta}_1^{+}e^{\theta_{32}}\\
 0&0&1
 \end{pmatrix}.
\end{align}
\end{subequations}

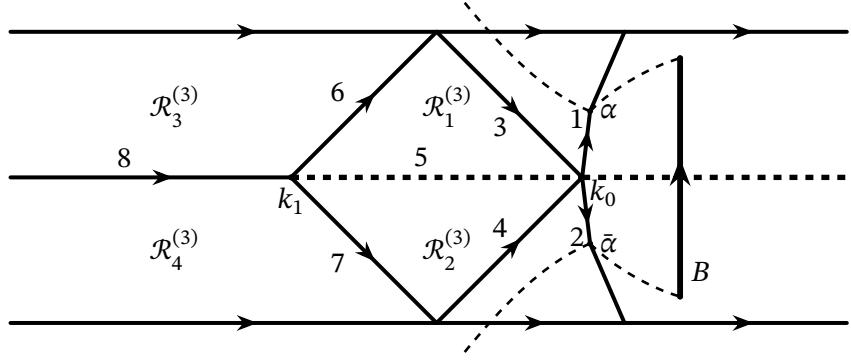
\begin{figure}[htbp]
\centering

\begin{tikzpicture}[scale=0.7,>=Stealth]

\tikzset{
  oldsolid/.style={dashed,line width=1pt},
  kkoldsolid/.style={dashed,line width=2pt},
  newcontour/.style={
    black,line width=1.5pt,
    line cap=round,line join=round
  },
  newbranch/.style={
    black,line width=2.2pt,line cap=round
  },
  dir/.style={
    postaction={decorate},
    decoration={
      markings,
      mark=at position 0.58 with {\arrow[scale=0.8]{Stealth}}
    }
  },
  dirKA/.style={
    postaction={decorate},
    decoration={
      markings,
      mark=at position 0.72 with {\arrow[scale=0.8]{Stealth}}
    }
  }
}

\coordinate (K)    at (0,0);
\coordinate (A)    at (0.15,1.25);
\coordinate (AB)   at (0.15,-1.25);
\coordinate (I)    at (1.85,2.25);
\coordinate (MI)   at (1.85,-2.25);

\coordinate (Kone) at (-5.50,0);
\coordinate (KH)   at (-2.75,2.75);
\coordinate (KL)   at (-2.75,-2.75);

\coordinate (UR)   at (0.80,2.75);
\coordinate (DR)   at (0.80,-2.75);

\draw[kkoldsolid] (K)--(5,0);

\draw[oldsolid]
  (A) .. controls (-0.7,1.6) and (-1.5,2.4) .. (-2.3,3.4);

\draw[oldsolid]
  (A) .. controls (0.6,1.7) and (1.3,2.1) .. (I);

\draw[oldsolid]
  (AB) .. controls (-0.7,-1.6) and (-1.5,-2.4) .. (-2.3,-3.4);

\draw[oldsolid]
  (AB) .. controls (0.6,-1.7) and (1.3,-2.1) .. (MI);

\draw[newcontour,dir] (-10.8,2.75)--(KH);
\draw[newcontour,dir] (KH)--(UR);
\draw[newcontour,dir] (UR)--(5,2.75);

\draw[newcontour,dir] (-10.8,-2.75)--(KL);
\draw[newcontour,dir] (KL)--(DR);
\draw[newcontour,dir] (DR)--(5,-2.75);

\draw[newcontour,dir] (Kone)--(KH);
\draw[newcontour,dir] (Kone)--(KL);

\draw[newcontour,dir] (KH)--(K);
\draw[newcontour,dir] (KL)--(K);

\draw[newcontour,dirKA] (K)--(A);
\draw[newcontour,dirKA] (K)--(AB);

\draw[newcontour,dir] (-10.8,0)--(Kone);
\draw[kkoldsolid] (Kone)--(K);

\draw[newcontour] (A)--(UR);
\draw[newcontour] (AB)--(DR);

\draw[newbranch,dir] (MI)--(I);

\fill (K)    circle (1.6pt);
\fill (Kone) circle (1.6pt);
\fill (A)    circle (1.6pt);
\fill (AB)   circle (1.6pt);
\fill (I)    circle (1.6pt);
\fill (MI)   circle (1.6pt);

\node at (0.4,-0.3) {$k_0$};
\node[below] at (Kone) {$k_1$};
\node[right] at (A) {$\alpha$};
\node[right] at (AB) {$\bar{\alpha}$};


\node at (-2.55,1.35)  {$\mathcal R^{(3)}_1$};
\node at (-2.55,-1.35) {$\mathcal R^{(3)}_2$};

\node at (-7.70,1.35)  {$\mathcal R^{(3)}_3$};
\node at (-7.70,-1.35) {$\mathcal R^{(3)}_4$};

\node at (-0.10,1.12)  {$1$};
\node at (-1.55,0.95)  {$3$};
\node at (-1.55,-0.95)  {$4$};
\node at (-0.10,-1.12) {$2$};

\node at (-4.62,1.62)  {$6$};
\node at (-4.62,-1.62) {$7$};

\node at (-3.05,0.38)  {$5$};
\node at (-8.65,0.38)  {$8$};

\node at (2.25,-1.78) {$B$};

\end{tikzpicture}

\caption{The oriented solid curves form the jump contour
\(\Gamma^{(3)}\). The regions
\(\{\mathcal{R}_j^{(3)} \}_{j=1}^4\) are also shown.}

\label{fig:Gamma3-regions}

\end{figure}

The third transformation opens the lenses associated with these two
factorizations.  Let \(\{\mathcal{R}_j^{(3)}\}_{j=1}^{4}\) denote the regions shown in Fig.~\ref{fig:Gamma3-regions}. We then define the transformation
\begin{equation}
 \widehat N^{(3)}=\widehat N^{(2)}\widehat G_2,
 \qquad
 \widehat G_2=
 \begin{cases}
 (\widehat V_b^{(2)})^{-1},&k\in\mathcal R_1^{(3)},\\
 \widehat V_a^{(2)},&k\in\mathcal R_2^{(3)},\\
 (\widehat V_e^{(2)})^{-1},&k\in\mathcal R_3^{(3)},\\
 \widehat V_c^{(2)},&k\in\mathcal R_4^{(3)},\\
 I,&\text{elsewhere}.
 \end{cases}
 \label{eq:elliptic-third-transformation}
\end{equation}
The jump contour $\Gamma^{(3)}$ is  displayed in
Fig.~\ref{fig:Gamma3-regions}.

Near \(k_0\), the jump matrices become
\begin{subequations}\label{eq:elliptic-third-jumps-k0}
\begin{align}
 \widehat V_1^{(3)}&=
 \begin{pmatrix}1&-r_1(\widehat{\delta}_1)^{-2}e^{\theta_{13}}&0\\
 0&1&0\\0&0&1\end{pmatrix},&
 \widehat V_2^{(3)}&=
 \begin{pmatrix}1&0&0\\-\bar r_1(\widehat{\delta}_1)^2e^{\theta_{31}}&1&0\\
 0&0&1\end{pmatrix},\\[2mm]
 \widehat V_3^{(3)}&=
 \begin{pmatrix}1&0&0\\
 -\dfrac{\bar r_1}{1+r_1\bar r_1}(\widehat{\delta}_1)^2e^{\theta_{31}}
 &1&0\\0&0&1\end{pmatrix},&
 \widehat V_4^{(3)}&=
 \begin{pmatrix}1&-\dfrac{r_1}{1+r_1\bar r_1}
 (\widehat{\delta}_1)^{-2}e^{\theta_{13}}&0\\
 0&1&0\\0&0&1\end{pmatrix}.
\end{align}
\end{subequations}
The jump on \((k_1,k_0)\) is eliminated. The jump matrices on the contour
segments labeled \(6\) and \(7\) are given by
\begin{align*}
 \widehat V_6^{(3)}=
 \begin{pmatrix}
 1&0&d^{-1}(\bar r_2+r_1\bar r_3)
 (\widehat{\delta}_1)^{-1}e^{\theta_{12}}\\
 0&1&0\\0&0&1
 \end{pmatrix},\qquad
 \widehat V_7^{(3)}=
 \begin{pmatrix}
 1&0&0\\0&1&0\\
 (r_2+\bar r_1r_3)\widehat{\delta}_1e^{\theta_{21}}&0&1
  \end{pmatrix}.
\end{align*}
On \((-\infty,k_1)\), the jump matrix takes the factorized form
\begin{align} \label{E:V83de}
\widehat V_8^{(3)}=\widehat V_a^{(3)}\widehat V_b^{(3)},
\end{align}
where
\begin{align*}
 \widehat V_a^{(3)}=
 \begin{pmatrix}1&0&0\\0&1&0\\
 (r_2+\bar r_1r_3)\widehat{\delta}_1^{-}e^{\theta_{21}}&0&1
 \end{pmatrix},\qquad 
 \widehat V_b^{(3)}=
 \begin{pmatrix}1&0&d^{-1}(\bar r_2+r_1\bar r_3)
 (\widehat{\delta}_1^{+})^{-1}e^{\theta_{12}}\\0&1&0\\0&0&1
 \end{pmatrix}.
\end{align*}
In addition, the jump matrix on the branch cut \(B\) is given by
\begin{equation}
 \widehat V_B^{(3)}=
 \begin{pmatrix}
 0&-\dfrac{\mathrm i(k+\lambda)}{q_0}(\widehat{\delta}_1)^{-2}&0\\[2mm]
 -\dfrac{\mathrm i(k+\lambda)}{q_0}(\widehat{\delta}_1)^2&0&0\\
 0&0&1
 \end{pmatrix}.
 \label{eq:elliptic-third-cut-jump}
\end{equation}

At this stage, only the jump on \((-\infty,k_1)\) remains to be deformed
off the real axis, which motivates the next transformation. However, the
factorization in \eqref{E:V83de} is incompatible with the corresponding
signature structure. We therefore introduce an additional conjugation to
obtain a more suitable factorization. To this end, define
\begin{equation}
 \widehat\Delta_2(k)=\diag\left( \widehat\delta_2(k),1,\widehat\delta_2^{-1}(k) \right),
 \label{eq:elliptic-Delta2}
\end{equation}
where \(\widehat\delta_2\) is analytic in
\(\mathbb C\setminus(-\infty,k_1)\) and satisfies
\begin{equation}
 \widehat{\delta}_2^{+}=\widehat{\delta}_2^{-}
 \left(1+d^{-1}\frac{|r_2+\bar r_1r_3|^2}{1+|r_1|^2}\right),
 \quad k\in(-\infty,k_1) ;
 \qquad \widehat\delta_2(k)\to1,
 \quad k\to\infty.
 \label{eq:elliptic-delta2-RHP}
\end{equation}
We then introduce the following transformation:
\begin{equation}
 \widehat N^{(4)}(x,t,k)=\widehat N^{(3)}(x,t,k)\widehat\Delta_2(k).
 \label{eq:elliptic-fourth-transformation}
\end{equation}
We next give the new factorization of \(\widehat{V}_8^{(4)}\) on
\((-\infty,k_1)\). To simplify the notation, set
\begin{equation}
 R=r_2+\bar r_1r_3,\qquad
 \bar R=\bar r_2+r_1\bar r_3,\qquad
 H=1+r_1\bar r_1+d^{-1}R\bar R.
 \label{eq:elliptic-R-H}
\end{equation}
Then
\begin{equation}
 \widehat V_8^{(4)}=\widehat V_a^{(4)}\widehat V_b^{(4)},
 \label{eq:elliptic-V8-fourth-factorization}
\end{equation}
where
\begin{align*}
 \widehat V_a^{(4)}=
 \begin{pmatrix}
 1&0&\dfrac{d^{-1}\bar R}{H}
 (\widehat{\delta}_1^{-})^{-1}(\widehat{\delta}_2^{-})^{-2}e^{\theta_{12}}\\
 0&1&0\\0&0&1
 \end{pmatrix},\qquad
 \widehat V_b^{(4)}=
 \begin{pmatrix}
 1&0&0\\0&1&0\\
 \dfrac{R}{H}\widehat{\delta}_1^{+}(\widehat{\delta}_2^{+})^2e^{\theta_{21}}
 &0&1
 \end{pmatrix}.
\end{align*}
The factorization in
\eqref{eq:elliptic-V8-fourth-factorization} immediately allows us to
deform the jump on \((-\infty,k_1)\) off the real axis.


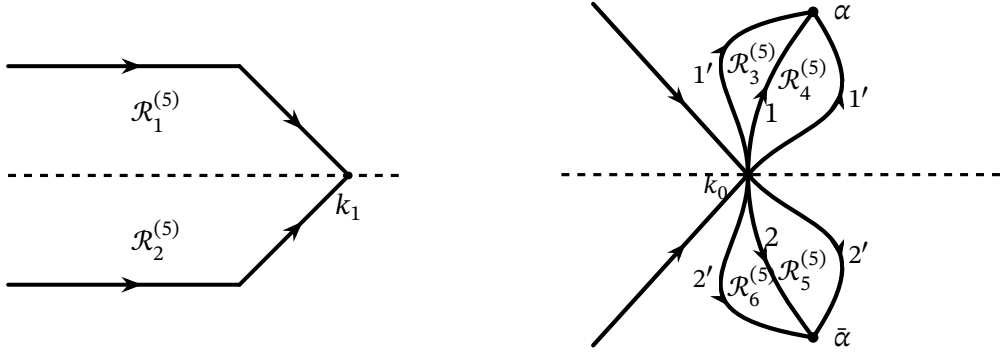
\begin{figure}
\centering
\noindent
\hspace*{0.05\textwidth}
\begin{minipage}[c]{0.34\textwidth}
\centering
\begin{tikzpicture}[scale=0.85,>=Stealth]

\tikzset{
  kkoldsolid/.style={
    dashed,
    line width=1.2pt
  },
  newcontour/.style={
    black,
    line width=1.5pt,
    line cap=round,
    line join=round
  },
  dir/.style={
    postaction={decorate},
    decoration={
      markings,
      mark=at position 0.58 with
      {\arrow[scale=0.8]{Stealth}}
    }
  }
}

\coordinate (Kone) at (0,0);
\coordinate (KUL) at (-1.70,1.70);
\coordinate (KDL) at (-1.70,-1.70);

\draw[kkoldsolid] (-5.30,0)--(0.80,0);

\draw[newcontour,dir] (-5.30,1.70)--(KUL);
\draw[newcontour,dir] (KUL)--(Kone);
\draw[newcontour,dir] (-5.30,-1.70)--(KDL);
\draw[newcontour,dir] (KDL)--(Kone);

\fill (Kone) circle (1.7pt);

\node[below=3pt] at (Kone) {$k_1$};
\node at (-3.0,1.0)  {$\mathcal R^{(5)}_1$};
\node at (-3.0,-1.0) {$\mathcal R^{(5)}_2$};

\end{tikzpicture}
\end{minipage}%
\hfill%
\begin{minipage}[c]{0.60\textwidth}
\centering
\begin{tikzpicture}[scale=1.05,>=Stealth]

\tikzset{
  kkoldsolid/.style={
    dashed,
    line width=1.2pt
  },
  newcontour/.style={
    black,
    line width=1.6pt,
    line cap=round,
    line join=round
  },
  dir/.style={
    postaction={decorate},
    decoration={
      markings,
      mark=at position 0.60 with
      {\arrow[scale=0.72]{Stealth}}
    }
  },
  dirmid/.style={
    postaction={decorate},
    decoration={
      markings,
      mark=at position 0.52 with
      {\arrow[scale=0.72]{Stealth}}
    }
  },
  raylabel/.style={
    font=\small,
    fill=white,
    inner sep=1.5pt
  }
}

\coordinate (K)  at (0,0);
\coordinate (A)  at (0.82,2.05);
\coordinate (AB) at (0.82,-2.05);
\coordinate (UL) at (-1.95,2.15);
\coordinate (LL) at (-1.95,-2.15);

\draw[kkoldsolid] (-2.35,0)--(3.20,0);

\draw[newcontour,dir] (UL)--(K);
\draw[newcontour,dir] (LL)--(K);

\draw[newcontour,dir]
  (K)
  .. controls (0,1.05) and (-1.18,1.72)
  .. (A);

\draw[newcontour,dirmid]
  (K)
  .. controls (0,1.02) and (0.55,1.68)
  .. (A);

\draw[newcontour,dir]
  (K)
  .. controls (0.62,0.72) and (1.78,0.62)
  .. (A);

\draw[newcontour,dir]
  (K)
  .. controls (0,-1.05) and (-1.18,-1.72)
  .. (AB);

\draw[newcontour,dirmid]
  (K)
  .. controls (0,-1.02) and (0.55,-1.68)
  .. (AB);

\draw[newcontour,dir]
  (K)
  .. controls (0.62,-0.72) and (1.78,-0.62)
  .. (AB);

\fill (K)  circle (1.9pt);
\fill (A)  circle (1.9pt);
\fill (AB) circle (1.9pt);

\node[font=\small] at (-0.4,-0.15) {$k_0$};
\node[right=4pt] at (A)  {$\alpha$};
\node[right=4pt] at (AB) {$\bar\alpha$};

\node[raylabel] at (-0.56,1.28)  {$1'$};
\node[raylabel] at (1.40,0.98)   {$1'$};
\node[raylabel] at (-0.56,-1.28) {$2'$};
\node[raylabel] at (1.40,-0.98)  {$2'$};

\node at (0.30,0.78)  {$1$};
\node at (0.30,-0.78) {$2$};

\node at (0.05,1.40)  {$\mathcal R^{(5)}_3$};
\node at (0.05,-1.40) {$\mathcal R^{(5)}_6$};
\node at (0.69,1.20)  {$\mathcal R^{(5)}_4$};
\node at (0.69,-1.20) {$\mathcal R^{(5)}_5$};

\end{tikzpicture}
\end{minipage}

\caption{Schematic illustration of the regions \(\{\mathcal R_j^{(5)}\}_{j=1}^6\) used in the fifth
transformation.  The left and right panels show the local contour configurations
near \(k_1\) and \(k_0\), respectively, in
Figure~\ref{fig:Gamma5-fig}. }
\label{fig:R5-local-regions}
\end{figure}

To streamline the analysis, we use the same transformation to address an
additional issue. The jumps on the contour segments connecting \(k_0\)
to \(\alpha\) and \(\bar{\alpha}\) lie in regions where
\(\theta_{31}\) has the wrong sign and may therefore exhibit exponential
growth as \(t\to\infty\). Following the strategy of
\cite{BiondiniMantzavinos2017}, we open lenses around these two segments
and extend the factors having the appropriate signature into the
corresponding decay regions. The residual jumps
that cannot be deformed remain on these segments and, after the subsequent
transformations, give rise to the jumps of the outer-model RH problem.
This lens-opening step is based on the following factorization of the
jump matrix:
\begin{equation}
\begin{aligned}
 \widehat V_1^{(4)}
 &=
 \widehat V_a^{(5)}
 \widehat V_b^{(5)}
 \widehat V_a^{(5)},\\
 \widehat V_2^{(4)}
 &=
 \widehat V_c^{(5)}
 \widehat V_d^{(5)}
 \widehat V_c^{(5)},
\end{aligned}
\label{eq:elliptic-fourth-band-factorizations}
\end{equation}
where
\begin{subequations}\label{eq:elliptic-fifth-band-factors}
\begin{align}
 \widehat V_a^{(5)}&=
 \begin{pmatrix}
 1&0&0\\
 -\dfrac{\widehat\delta_1^2\widehat\delta_2}{r_1}e^{\theta_{31}}&1&0\\
 0&0&1
 \end{pmatrix},\qquad
 \widehat V_b^{(5)}=
 \begin{pmatrix}
 0&-r_1(\widehat{\delta}_1)^{-2}(\widehat{\delta}_2)^{-1}
       e^{\theta_{13}}&0\\
 \dfrac{\widehat\delta_1^2\widehat\delta_2}{r_1}e^{\theta_{31}}&0&0\\
 0&0&1
 \end{pmatrix},\\[2mm]
 \widehat V_c^{(5)}&=
 \begin{pmatrix}
 1&-\dfrac{e^{\theta_{13}}}
       {\bar r_1\widehat\delta_1^2\widehat\delta_2}&0\\
 0&1&0\\0&0&1
 \end{pmatrix},\qquad
 \widehat V_d^{(5)}=
 \begin{pmatrix}
 0&\dfrac{e^{\theta_{13}}}
       {\bar r_1\widehat\delta_1^2\widehat\delta_2}&0\\
 -\bar r_1\widehat\delta_1^2\widehat\delta_2e^{\theta_{31}}&0&0\\
 0&0&1
 \end{pmatrix}.
\end{align}
\end{subequations}

We are now ready to define the fifth transformation. Let
\(\{\mathcal{R}_j^{(5)}\}_{j=1}^{6}\) denote the regions shown in
Figure~\ref{fig:R5-local-regions}. We then set
\begin{equation}
 \widehat N^{(5)}(x,t,k)=\widehat N^{(4)}(x,t,k)\widehat G_4(x,t,k),
 \qquad
 \widehat G_4=
 \begin{cases}
 (\widehat V_b^{(4)})^{-1},&k\in\mathcal R_1^{(5)},\\
 \widehat V_a^{(4)},&k\in\mathcal R_2^{(5)},\\
 (\widehat V_a^{(5)})^{-1},&k\in\mathcal R_3^{(5)},\\
 \widehat V_a^{(5)},&k\in\mathcal R_4^{(5)},\\
 (\widehat V_c^{(5)})^{-1},&k\in\mathcal R_5^{(5)},\\
 \widehat V_c^{(5)},&k\in\mathcal R_6^{(5)},\\
 I,&\text{elsewhere}.
 \end{cases}
 \label{eq:elliptic-fifth-transformation}
\end{equation}

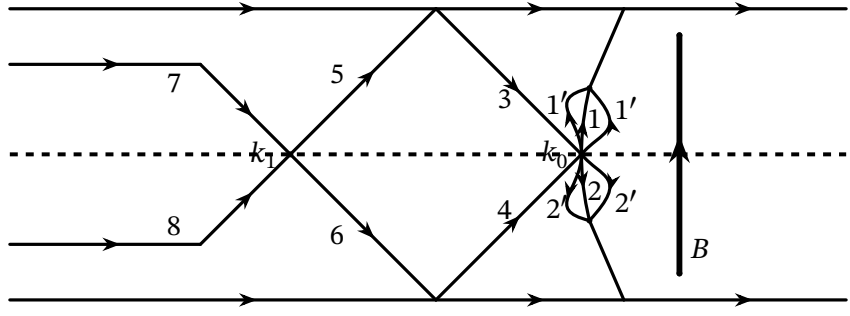
\begin{figure}
\centering 

\begin{tikzpicture}[scale=0.7,>=Stealth] 

\tikzset{ 
  newcontour/.style={ 
    black,line width=1.2pt, 
    line cap=round,line join=round 
  }, 
    kkoldsolid/.style={dashed,line width=1.5pt}, 
  newcontour/.style={ 
    black,line width=1.2pt, 
    line cap=round,line join=round 
  }, 
  newbranch/.style={ 
    black,line width=2.2pt,line cap=round 
  }, 
  dir/.style={ 
    postaction={decorate}, 
    decoration={ 
      markings, 
      mark=at position 0.58 with {\arrow[scale=0.8]{Stealth}} 
    } 
  }, 
  dirKA/.style={ 
    postaction={decorate}, 
    decoration={ 
      markings, 
      mark=at position 0.72 with {\arrow[scale=0.8]{Stealth}} 
    } 
  },
  dirmid/.style={
    postaction={decorate},
    decoration={
      markings,
      mark=at position 0.52 with {\arrow[scale=0.8]{Stealth}}
    }
  }
} 

\coordinate (K)    at (0,0); 
\coordinate (A)    at (0.15,1.25); 
\coordinate (AB)   at (0.15,-1.25); 
\coordinate (I)    at (1.85,2.25); 
\coordinate (MI)   at (1.85,-2.25); 

\coordinate (Kone) at (-5.50,0); 
\coordinate (KH)   at (-2.75,2.75); 
\coordinate (KL)   at (-2.75,-2.75); 

\coordinate (UR)   at (0.80,2.75); 
\coordinate (DR)   at (0.80,-2.75); 

\coordinate (KUL)  at (-7.20,1.70); 
\coordinate (KDL)  at (-7.20,-1.70); 

\draw[newcontour,dir] (-10.8,2.75)--(KH); 
\draw[newcontour,dir] (KH)--(UR); 
\draw[newcontour,dir] (UR)--(5,2.75); 

\draw[newcontour,dir] (-10.8,-2.75)--(KL); 
\draw[newcontour,dir] (KL)--(DR); 
\draw[newcontour,dir] (DR)--(5,-2.75); 

\draw[newcontour,dir] (Kone)--(KH); 
\draw[newcontour,dir] (Kone)--(KL); 

\draw[newcontour,dir] (-10.8,1.70)--(KUL); 
\draw[newcontour,dir] (KUL)--(Kone); 

\draw[newcontour,dir] (-10.8,-1.70)--(KDL); 
\draw[newcontour,dir] (KDL)--(Kone); 

\draw[newcontour,dir] (KH)--(K); 
\draw[newcontour,dir] (KL)--(K); 


\draw[newcontour,dirmid]
  (K) .. controls (0,0.62) and (0.10,1.02) .. (A);

\draw[newcontour,dirmid]
  (K) .. controls (0,-0.62) and (0.10,-1.02) .. (AB);

\draw[newcontour,dir] 
  (K) .. controls (0,0.66) and (-0.72,1.04) .. (A); 

\draw[newcontour,dir] 
  (K) .. controls (0.36,0.44) and (0.92,0.42) .. (A); 

\draw[newcontour,dir] 
  (K) .. controls (0,-0.66) and (-0.72,-1.04) .. (AB); 

\draw[newcontour,dir] 
  (K) .. controls (0.36,-0.44) and (0.92,-0.42) .. (AB); 

\draw[newcontour] (A)--(UR); 
\draw[newcontour] (AB)--(DR); 

\draw[newbranch,dir] (MI)--(I); 

\draw[kkoldsolid] (-10.8,0)--(5,0); 

\fill (K)    circle (1.6pt); 
\fill (Kone) circle (1.6pt); 
\fill (A)    circle (1.6pt); 
\fill (AB)   circle (1.6pt); 
\fill (I)    circle (1.6pt); 
\fill (MI)   circle (1.6pt); 

\node at (-0.5,0) {$k_0$}; 
\node at (-6.0,0) {$k_1$};


\node at (-1.45,1.05)  {$3$}; 
\node at (-1.45,-1.05) {$4$}; 

\node at (-7.70,1.35)  {$7$}; 
\node at (-7.70,-1.35) {$8$}; 

\node at (0.25,0.68)  {$1$}; 
\node at (0.25,-0.68) {$2$}; 

\node at (-0.45,0.98)  {$1'$}; 
\node at (-0.45,-0.98)  {$2'$};
\node at (0.85,0.88)  {$1'$}; 
\node at (0.85,-0.88)  {$2'$};

\node at (-4.62,1.52)  {$5$}; 
\node at (-4.62,-1.52) {$6$}; 

\node at (2.25,-1.78) {$B$}; 

\end{tikzpicture} 

\caption{The jump contour \(\Gamma^{(5)}\) for $\widehat{N}^{(5)}$.}

\label{fig:Gamma5-fig} 

\end{figure}

The resulting jump contour \(\Gamma^{(5)}\) is depicted in
Figure~\ref{fig:Gamma5-fig}. The jump matrices introduced  by the fifth transformation are
\begin{equation}
\begin{aligned}
\widehat{V}_1^{(5)}&=\widehat{V}_b^{(5)},
&\qquad
\widehat{V}_2^{(5)}&=\widehat{V}_d^{(5)},
\\
\widehat{V}_{1'}^{(5)}&=\widehat{V}_a^{(5)},
&
\widehat{V}_{2'}^{(5)}&=\widehat{V}_c^{(5)},
\\
\widehat{V}_7^{(5)}&=\widehat{V}_a^{(4)},
&
\widehat{V}_8^{(5)}&=\widehat{V}_b^{(4)}.
\end{aligned}
\label{eq:elliptic-fifth-jump-list}
\end{equation}
All remaining jump matrices are  unchanged compared with $\widehat{V}^{(4)}$.

\subsection{Modified phases and the final transformation}
\label{subsec:elliptic-modified-phases}

The choice of the preceding contour deformations may appear unmotivated
at first. In fact, the contours were selected in anticipation of the
signature structure of the modified phase functions constructed below.
The purpose of this modification is to redistribute the exponential
growth and decay so that the lens boundaries lie entirely within the
appropriate decay regions.  We will
show that, after the phase modification, all jump matrices converge
exponentially to the identity as \(t\to\infty\), except on the branch
cuts and in neighborhoods of the critical points
\(k_0\), \(k_1\), \(\alpha\), and \(\bar{\alpha}\).

The key step in modifying the phase functions is the introduction of a
\(G\)-function transformation. Remarkably, the \(G\)-function required
here is identical to that used in the scalar problem; see
\cite[Section~5]{BiondiniMantzavinos2017}. The subsequent analysis is
more involved because several phase functions are present in the
\(3\times3\) setting. Nevertheless, the scalar construction carries over
directly and considerably simplifies the analysis.

Let \(\alpha=\alpha(\xi)\) and \(k_0=k_0(\xi)\) be parameters
to be determined, and define
\begin{equation}
\label{eq:elliptic-abelian-phase}
\begin{aligned}
\hbar (\xi,k)
=
\frac{1}{2}
\left(
\int_{\mathrm{i}q_0}^{k}
+
\int_{-\mathrm{i}q_0}^{k}
\right)
\mathrm{d} \hbar (\xi,z),
\qquad
\mathrm{d}\hbar (\xi,z)
=
-4\frac{(z-k_0)(z-\alpha)(z-\bar{\alpha})}
{\gamma(\xi,z)}\,\mathrm{d}z,
\end{aligned}
\end{equation}
where
$
\gamma(\xi,k)
=
\sqrt{(k^2+q_0^2)(k-\alpha)(k-\bar{\alpha})}
$.
The branch of \(\gamma\) is chosen so that
\(
\gamma(\xi,k)=k^2+O(k)
\), 
\(
 k\to\infty,
\)
with the branch cuts specified in Remark~\ref{R:cutremark}. 
We then introduce the modified phase
\begin{align}\label{E:h13def}
h_{13}(x,t,k)=-2\mathrm{i}t\,\hbar(\xi,k).
\end{align}
The parameters \(\alpha\) and \(k_0\) are chosen to satisfy
the following two conditions:
\begin{enumerate}
\item[(i)]
the real part of \(h_{13}\) matches that of \(\theta_{13}\) as
\(k\to\infty\);

\item[(ii)]
Near \(k=\alpha\), the real part of \(h_{13}\) should be of order
\(O\bigl((k-\alpha)^{3/2}\bigr)\).
\end{enumerate}
These requirements lead precisely to the following modulation system
\begin{subequations}
\label{eq:elliptic-modulation-system}
\begin{equation}
\operatorname{Re}\alpha
=
-k_0+\frac{\xi}{4},
\qquad
\operatorname{Im}\alpha
=
\sqrt{
2k_0^2-\frac{\xi k_0}{2}+q_0^2
}.
\label{eq:elliptic-modulation-algebraic}
\end{equation}
\begin{equation}
\int_{-\mathrm{i}q_0}^{\mathrm{i}q_0}
\sqrt{
\frac{
\left(z+k_0-\frac{\xi}{4}\right)^2
+2k_0^2-\frac{\xi k_0}{2}+q_0^2
}{
z^2+q_0^2
}
}
(z-k_0)\,\mathrm{d}z
=
0.
\label{eq:elliptic-modulation-moment}
\end{equation}
\end{subequations}
This nonlinear system is the same as that derived in
\cite[Section~5]{BiondiniMantzavinos2017}.
\begin{lemma}[{\cite[Lemma~5.4]{BiondiniMantzavinos2017}}]
\label{lem:unique-k0}
For every
\(
\xi\in\left(-4\sqrt{2}\,q_0,0\right)
\),
the integral equation \eqref{eq:elliptic-modulation-moment} admits a
unique solution \(k_0=k_0(\xi)\).
\end{lemma}
This lemma ensures that \(k_0(\xi)\) is well defined for
\(
\xi\in(-4\sqrt{2}\,q_0,0)
\).
Once \(k_0\) is fixed, the algebraic relations in
\eqref{eq:elliptic-modulation-algebraic} uniquely determine
\(\alpha(\xi)\). Consequently, the Abelian phase
\(\hbar(\xi,k)\) defined by
\eqref{eq:elliptic-abelian-phase}, and hence the modified phase
\(h_{13}\), are well defined throughout
the modulated elliptic-wave sector.

\begin{remark}
\label{R:cutremark}

The branch cuts of \(\gamma(\xi,k)\) are chosen to be \(B\) and
\(\widetilde{B}\), where \(\widetilde{B}\) is the arc of the zero level
set
\[
\Ree h_{13}(\xi,k) =0,
\]
that connects \(\alpha\) to
\(\bar{\alpha}\) through \(k_0\). This arc is shown in 
Fig.~\ref{fig:h-sign-structure}. Importantly, this zero level set is
independent of the choice of branch cuts for \(\gamma\). Thus,
\(\widetilde{B}\) may first be identified from the zero level set and then
chosen as one of the branch cuts of \(\gamma\).
\end{remark}

We collect below several properties of \(\hbar(\xi,k)\) established
in \cite[Section~5]{BiondiniMantzavinos2017}. 
\begin{lemma} 
The function
\(\hbar(\xi,k)\) has the following properties:
\begin{itemize}

\item
The function \(\hbar(\xi,k)\) is analytic in
\(\mathbb{C}\setminus(B\cup\widetilde{B})\).

\item
Its boundary values satisfy
\begin{equation}
\hbar^++\hbar^-=0
\quad\text{on }B,
\qquad
\hbar^++\hbar^-=\Omega
\quad\text{on }\widetilde{B},
\label{eq:elliptic-h-jumps}
\end{equation}
where
\begin{equation}
\Omega
=
-4\left(
\int_{\mathrm{i}q_0}^{\alpha}
+
\int_{-\mathrm{i}q_0}^{\bar{\alpha}}
\right)
\frac{(z-k_0)(z-\alpha)(z-\bar{\alpha})}
{\gamma(\xi,z)}\,\mathrm{d}z
\in\mathbb{R}.
\label{eq:elliptic-Omega}
\end{equation}

\item
At each  point
\(k_\star\in\{\alpha,\bar{\alpha}\}\), the function admits the local
expansion
\begin{equation}
\hbar(\xi,k)
=
\hbar(\xi,k_\star)
+
O\bigl((k-k_\star)^{3/2}\bigr),
\qquad
\operatorname{Im}\hbar(\xi,k_\star)=0.
\label{eq:elliptic-h-branch-expansion}
\end{equation}

\item
As \(k\to\infty\),
\begin{equation}
\hbar(\xi,k)
=
-2k^2+\xi k+H_0+O(k^{-1}),
\qquad
H_0\in\mathbb{R},
\label{eq:elliptic-h-infty}
\end{equation}
where \(H_0\) has the regularized integral representation
\begin{equation}
H_0
=
-2\,
\left(
\int_{\mathrm{i}q_0}^{\infty}
+
\int_{-\mathrm{i}q_0}^{\infty}
\right)
\frac{(z-k_0)(z-\alpha)(z-\bar{\alpha})}
{\gamma(\xi,z)}-\left(z-\frac{\xi}{4}\right)\,\mathrm{d}z
-2q_0^2.
\label{eq:elliptic-H0}
\end{equation}

\end{itemize}
 \end{lemma}

\begin{figure}
\centering 

\begin{tikzpicture}[scale=0.7,>=Stealth] 

\tikzset{ 
  oldsolid/.style={dashed,line width=1pt}, 
  kkoldsolid/.style={dashed,line width=1.5pt}, 
  newcontour/.style={ 
    black,line width=1.2pt, 
    line cap=round,line join=round 
  }, 
  newbranch/.style={ 
    black,line width=2.2pt,line cap=round 
  }, 
  dir/.style={ 
    postaction={decorate}, 
    decoration={ 
      markings, 
      mark=at position 0.58 with {\arrow[scale=0.8]{Stealth}} 
    } 
  }, 
  dirKA/.style={ 
    postaction={decorate}, 
    decoration={ 
      markings, 
      mark=at position 0.72 with {\arrow[scale=0.8]{Stealth}} 
    } 
  },
  dirmid/.style={
    postaction={decorate},
    decoration={
      markings,
      mark=at position 0.52 with {\arrow[scale=0.8]{Stealth}}
    }
  }
} 

\coordinate (K)    at (0,0); 
\coordinate (A)    at (0.15,1.25); 
\coordinate (AB)   at (0.15,-1.25); 
\coordinate (I)    at (1.85,2.25); 
\coordinate (MI)   at (1.85,-2.25); 

\coordinate (Kone) at (-5.50,0); 
\coordinate (KH)   at (-2.75,2.75); 
\coordinate (KL)   at (-2.75,-2.75); 

\coordinate (UR)   at (0.80,2.75); 
\coordinate (DR)   at (0.80,-2.75); 

\coordinate (KUL)  at (-7.20,1.70); 
\coordinate (KDL)  at (-7.20,-1.70); 

\draw[oldsolid] 
  (A) .. controls (-0.7,1.6) and (-1.5,2.4) .. (-2.3,3.4); 

\draw[oldsolid] 
  (A) .. controls (0.6,1.7) and (1.3,2.1) .. (I); 

\draw[oldsolid] 
  (AB) .. controls (-0.7,-1.6) and (-1.5,-2.4) .. (-2.3,-3.4); 

\draw[oldsolid] 
  (AB) .. controls (0.6,-1.7) and (1.3,-2.1) .. (MI); 

\draw[oldsolid]
  (-6.25,3.40)
  .. controls (-6.05,2.10) and (-5.78,0.95) ..
  (Kone)
  .. controls (-5.78,-0.95) and (-6.05,-2.10) ..
  (-6.25,-3.40);

\draw[newcontour,dir] (-10.8,2.75)--(KH); 
\draw[newcontour,dir] (KH)--(UR); 
\draw[newcontour,dir] (UR)--(5,2.75); 

\draw[newcontour,dir] (-10.8,-2.75)--(KL); 
\draw[newcontour,dir] (KL)--(DR); 
\draw[newcontour,dir] (DR)--(5,-2.75); 

\draw[newcontour,dir] (Kone)--(KH); 
\draw[newcontour,dir] (Kone)--(KL); 

\draw[newcontour,dir] (-10.8,1.70)--(KUL); 
\draw[newcontour,dir] (KUL)--(Kone); 

\draw[newcontour,dir] (-10.8,-1.70)--(KDL); 
\draw[newcontour,dir] (KDL)--(Kone); 

\draw[newcontour,dir] (KH)--(K); 
\draw[newcontour,dir] (KL)--(K); 


\draw[newcontour,dirmid]
  (K) .. controls (0,0.62) and (0.10,1.02) .. (A);

\draw[newcontour,dirmid]
  (K) .. controls (0,-0.62) and (0.10,-1.02) .. (AB);

\draw[newcontour,dir] 
  (K) .. controls (0,0.66) and (-0.72,1.04) .. (A); 

\draw[newcontour,dir] 
  (K) .. controls (0.36,0.44) and (0.92,0.42) .. (A); 

\draw[newcontour,dir] 
  (K) .. controls (0,-0.66) and (-0.72,-1.04) .. (AB); 

\draw[newcontour,dir] 
  (K) .. controls (0.36,-0.44) and (0.92,-0.42) .. (AB); 

\draw[newcontour] (A)--(UR); 
\draw[newcontour] (AB)--(DR); 

\draw[newbranch,dir] (MI)--(I); 

\draw[oldsolid] (-10.8,0)--(5,0); 

\fill (K)    circle (1.6pt); 
\fill (Kone) circle (1.6pt); 
\fill (A)    circle (1.6pt); 
\fill (AB)   circle (1.6pt); 
\fill (I)    circle (1.6pt); 
\fill (MI)   circle (1.6pt); 

\node at (-0.5,0) {$k_0$}; 
\node at (-6.0,0) {$k_1$};


\node at (-1.45,1.05)  {$3$}; 
\node at (-1.45,-1.05) {$4$}; 

\node at (-7.70,1.35)  {$7$}; 
\node at (-7.70,-1.35) {$8$}; 

\node at (0.25,0.68)  {$1$}; 
\node at (0.25,-0.68) {$2$}; 

\node at (-0.45,0.98)  {$1'$}; 
\node at (-0.45,-0.98)  {$2'$};
\node at (0.85,0.88)  {$1'$}; 
\node at (0.85,-0.88)  {$2'$};

\node at (-4.62,1.52)  {$5$}; 
\node at (-4.62,-1.52) {$6$}; 

\node at (2.25,-1.78) {$B$}; 

\end{tikzpicture} 

\caption{The jump contour $\Gamma^{(6)}$ and its relative position with respect to the zero-level sets of the phase functions.}

\label{fig:Gamma6daxia-regions} 

\end{figure}
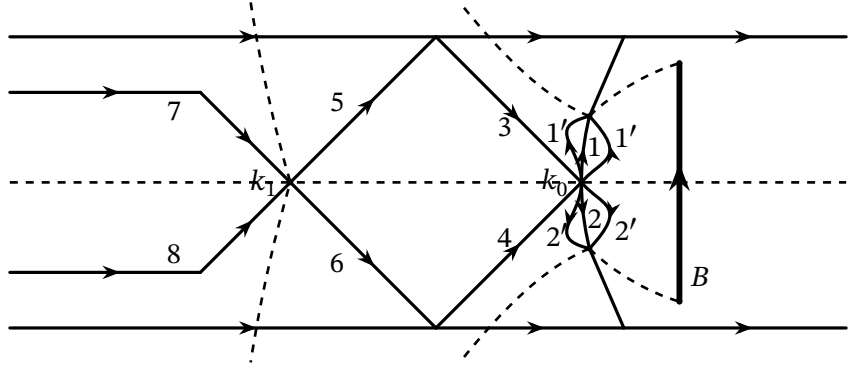

We now define
\begin{equation}
G(\xi,k)
=
\hbar(\xi,k)
-
\frac{\mathrm{i}}{t}\theta_1(x,t,k).
\label{eq:elliptic-G}
\end{equation}
It follows from \eqref{eq:elliptic-G} and the large-\(k\) expansion of
\(\theta_1\) that
\begin{equation}
 G_\infty:=\lim_{k\to\infty}G(\xi,k)=H_0+q_0^2.
 \label{eq:elliptic-G-infty}
\end{equation}
Recalling that
\(\sigma=\operatorname{diag}(1,-1,0)\), we define the sixth
transformation as follows:
\begin{equation}
 \widehat N^{(6)}(x,t,k)=e^{-\mathrm iG_\infty t\sigma}
 \widehat N^{(5)}(x,t,k)e^{\mathrm iG(\xi,k)t\sigma}.
 \label{eq:elliptic-sixth-transformation}
\end{equation}
The sixth transformation does not alter the jump contour;  namely,
\(
\Gamma^{(6)}=\Gamma^{(5)}
\).
It does, however, modify the jump matrices. In particular, the boundary
relations satisfied by \(\hbar(\xi,k)\) on \(B\) and
\(\widetilde{B}\), given in \eqref{eq:elliptic-h-jumps}, transform the
three band jumps into
\begin{subequations}
\label{eq:elliptic-sixth-band-jumps}
\begin{align}
\widehat V_1^{(6)}
&=
\begin{pmatrix}
0&
-r_1\widehat{\delta}_1^{-2}\widehat{\delta}_2^{-1}
e^{-\mathrm{i}\Omega t}
&0\\
\dfrac{\widehat{\delta}_1^2\widehat{\delta}_2}{r_1}
e^{\mathrm{i}\Omega t}
&0&0\\
0&0&1
\end{pmatrix},
\\[3mm]
\widehat V_2^{(6)}
&=
\begin{pmatrix}
0&
\dfrac{e^{-\mathrm{i}\Omega t}}
{\bar r_1\widehat{\delta}_1^2\widehat{\delta}_2}
&0\\
-\bar r_1\widehat{\delta}_1^2\widehat{\delta}_2
e^{\mathrm{i}\Omega t}
&0&0\\
0&0&1
\end{pmatrix},
\\[3mm]
\widehat V_B^{(6)}
&=
\begin{pmatrix}
0&
-\dfrac{\mathrm{i}(k+\lambda)}{q_0}
\widehat{\delta}_1^{-2}\widehat{\delta}_2^{-1}
&0\\[2mm]
-\dfrac{\mathrm{i}(k+\lambda)}{q_0}
\widehat{\delta}_1^2\widehat{\delta}_2
&0&0\\
0&0&1
\end{pmatrix}.
\end{align}
\end{subequations}
On all remaining contour segments, the jump matrices are obtained from
their fifth-transformation counterparts by replacing the original phases
\(\theta_{ij}\) in the exponential factors with the corresponding
modified phases \(h_{ij}\), where
\begin{subequations}
\label{eq:elliptic-modified-phases}
\begin{align}
h_{13}(x,t,k)
&=
\theta_{13}(x,t,k)-2\mathrm{i}G(\xi,k)t
=
-2\mathrm{i}\hbar(\xi,k)t,
\\
h_{12}(x,t,k)
&=
\theta_{12}(x,t,k)-\mathrm{i}G(\xi,k)t,
\\
h_{23}(x,t,k)
&=
\theta_{23}(x,t,k)-\mathrm{i}G(\xi,k)t.
\end{align}
\end{subequations}
The remaining modified phases are defined by
\(
h_{ji}=-h_{ij}.
\)

\begin{figure}
\centering

\begin{minipage}[c]{0.315\textwidth}
    \centering
    \includegraphics[width=\linewidth]
    {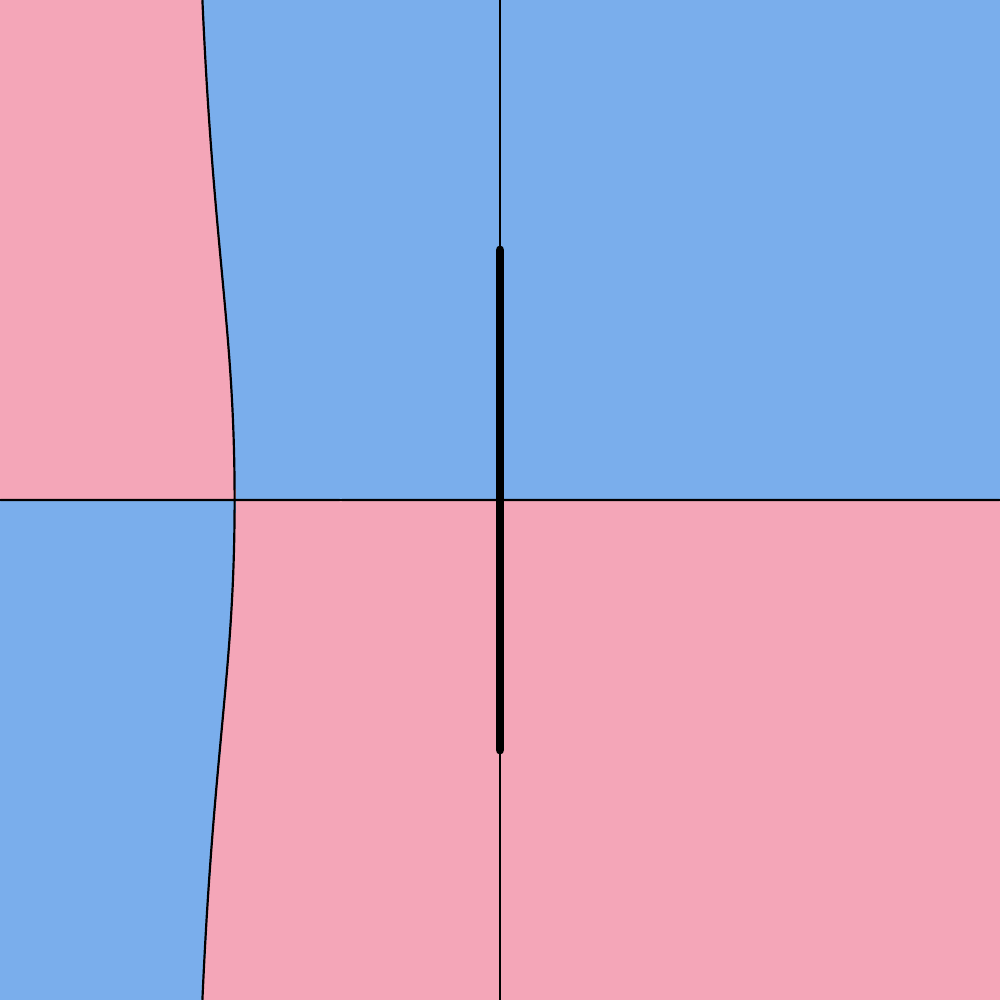}

    \vspace{2mm}
    {\small (a)}
\end{minipage}
\hfill
\begin{minipage}[c]{0.315\textwidth}
    \centering
    \includegraphics[width=\linewidth]
    {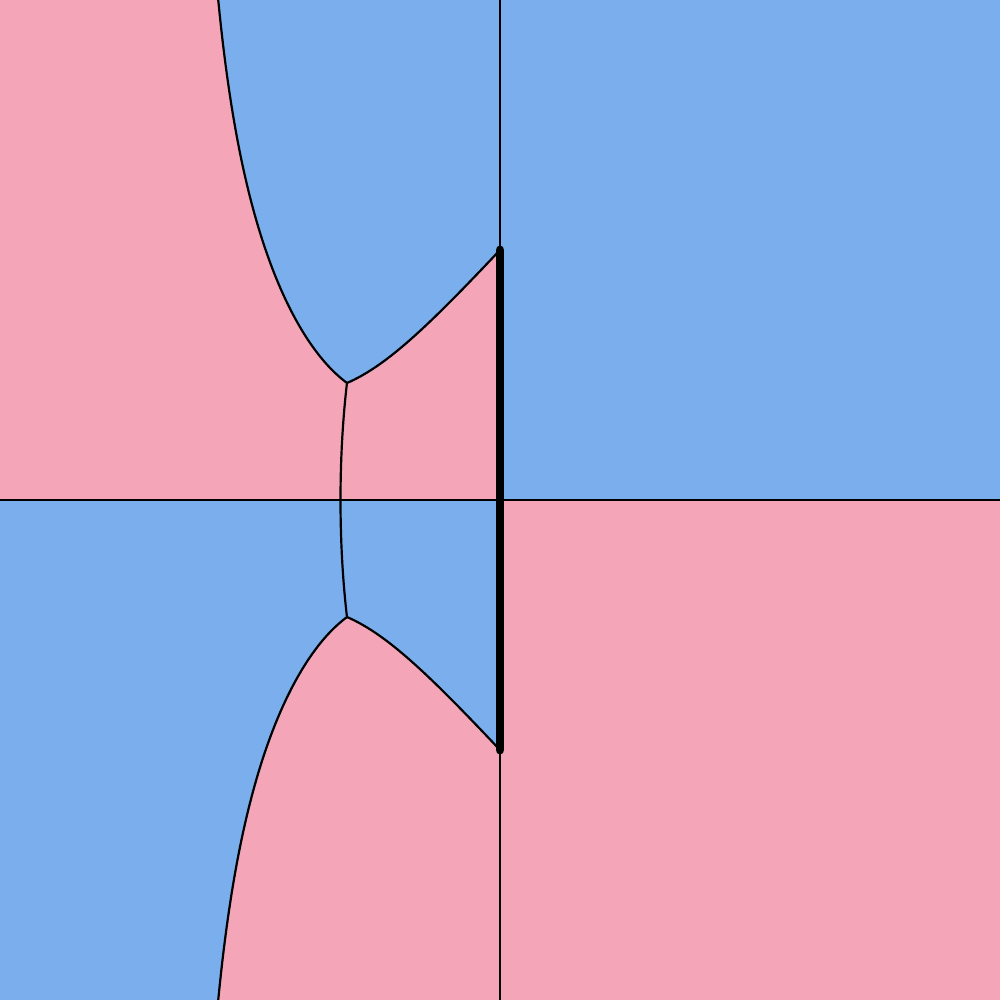}

    \vspace{2mm}
    {\small (b)}
\end{minipage}
\hfill
\begin{minipage}[c]{0.315\textwidth}
    \centering
    \includegraphics[width=\linewidth]
    {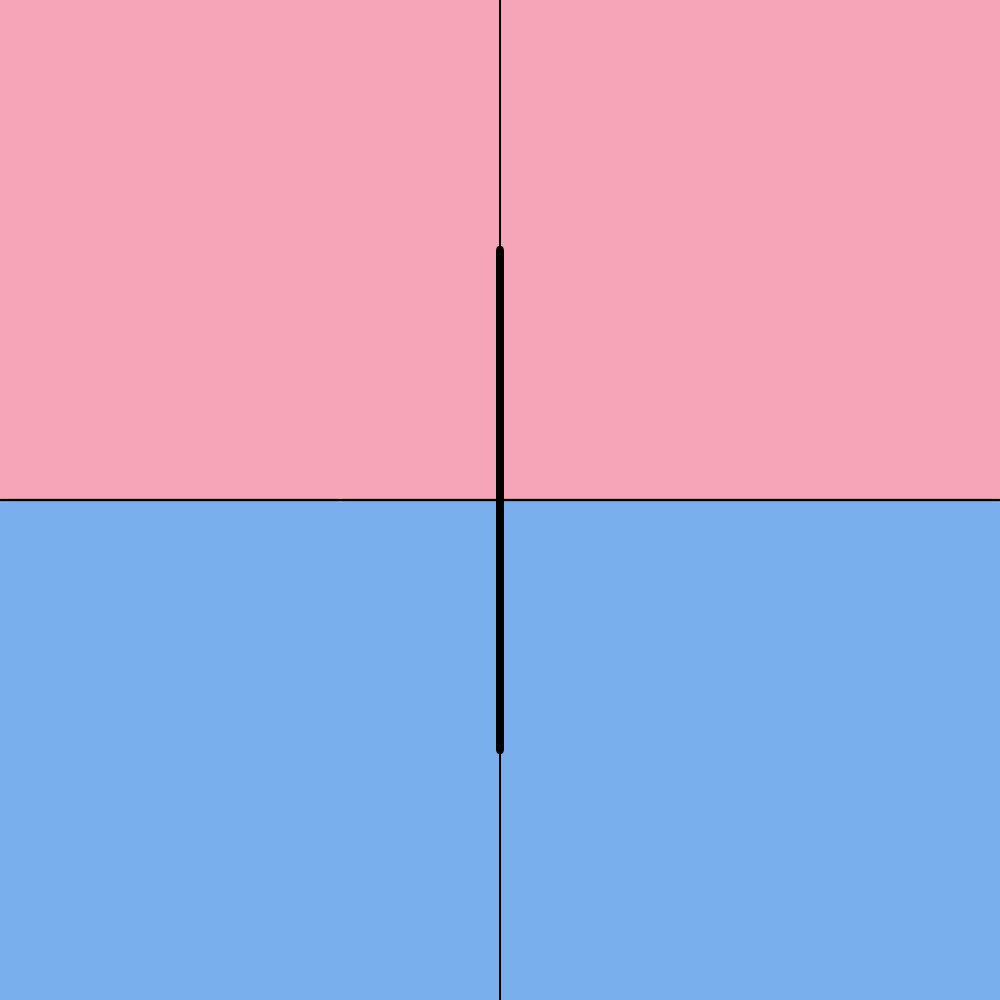}

    \vspace{2mm}
    {\small (c)}
\end{minipage}
\caption{Signature tables for
\(\operatorname{Re}h_{12}\),
\(\operatorname{Re}h_{13}\), and
\(\operatorname{Re}h_{23}\) (from left to right),
with \(q_0=1\) and \(\xi=-5\).
The pink and blue regions correspond to
\(\operatorname{Re}\theta_{ij}>0\) and
\(\operatorname{Re}\theta_{ij}<0\), respectively.}
\label{fig:modified-phase-signatures}

\end{figure}

The signature structure of the modified phases \(h_{ij}\) is shown in
Figure~\ref{fig:modified-phase-signatures}. We observe that
\(\Ree h_{23}\) has a fixed sign in each of the upper and
lower half-planes. The zero-level set of \(\Ree h_{12}\)
intersects the real axis only at \(k_1\), defined by
\eqref{eq:elliptic-k1}, whereas that of
\(\Ree h_{13}\) intersects the real axis only at \(k_0\).
This signature structure explains why the lens opening is performed
separately over the three intervals
\(
(-\infty,k_1)\), \((k_1,k_0)\) and \((k_0,\infty)
\).
Moreover, the exponentially oscillatory entries of the jump matrices
extended away from the real axis lie entirely within the corresponding
decay regions; see Figure~\ref{fig:Gamma6daxia-regions}.

We next introduce a transformation to simplify the jumps in
\eqref{eq:elliptic-sixth-band-jumps}. An analogous step also appears in
the plane-wave region $P_{\mathrm I}$. The present situation is, however, slightly
more involved, since the jumps on two bands $B$ and $\tilde{B}$ must be treated simultaneously.

Recall that $\Gamma_1^{(6)}$ and $\Gamma_2^{(6)}$ denote the oriented
upper and lower contour segments, labeled by $1$ and $2$, respectively, in
Figure~\ref{fig:Gamma6daxia-regions}. To remove the nonconstant factors
from the corresponding band jumps, we seek a scalar function
$\widehat g(\xi,k)$, analytic in
$\mathbb C\setminus(B\cup\widetilde B)$, whose boundary values satisfy
\begin{subequations}\label{eq:elliptic-ghat-RHP}
\begin{align}
 \widehat g^++\widehat g^-
 &=\mathrm i\log\!\left(\widehat\delta_1^2\widehat\delta_2\right),
 &&k\in B,\\
 \widehat g^++\widehat g^-
 &=\mathrm i\log\!\left(
   \frac{\widehat\delta_1^2\widehat\delta_2}{r_1}\right)-\omega,
 &&k\in\Gamma_1^{(6)},\\
 \widehat g^++\widehat g^-
 &=\mathrm i\log\!\left(
   \bar r_1\widehat\delta_1^2\widehat\delta_2\right)-\omega,
 &&k\in\Gamma_2^{(6)}.
\end{align}
\end{subequations}
 The real constant $\omega$ is given by
 \begin{equation*}
 \omega=\mathrm i\frac{\omega_1}{\omega_2},
\end{equation*}
where
\begin{subequations}\label{eq:elliptic-omega-periods}
\begin{align}
 \omega_1={}&
 \int_B\frac{\log(\widehat\delta_1^2\widehat\delta_2)}
                  {\gamma(\xi,s)}\,\mathrm ds
 +\int_{\Gamma_1^{(6)}}
 \frac{\log(\widehat\delta_1^2\widehat\delta_2/r_1)}
      {\gamma(\xi,s)}\,\mathrm ds
 -\int_{\Gamma_2^{(6)}}
 \frac{\log(\bar r_1\widehat\delta_1^2\widehat\delta_2)}
      {\gamma(\xi,s)}\,\mathrm ds,\\
 \omega_2={}&\int_{\widetilde B}
 \frac{\mathrm ds}{\gamma(\xi,s)}.
\end{align}
\end{subequations}
By the Plemelj formula, it is straightforward to verify that the function
\begin{align}
 \widehat g(\xi,k)
 =-\frac{\gamma(\xi,k)}{2\pi}
 \Bigg[&
 \int_B
 \frac{\log(\widehat\delta_1^2\widehat\delta_2)}
 {\gamma(\xi,s)(s-k)}\,\mathrm ds
 +\int_{\Gamma_1^{(6)}}
 \frac{\log(\widehat\delta_1^2\widehat\delta_2/r_1)+\mathrm i\omega}
 {\gamma(\xi,s)(s-k)}\,\mathrm ds  -\int_{\Gamma_2^{(6)}}
 \frac{\log(\bar r_1\widehat\delta_1^2\widehat\delta_2)+\mathrm i\omega}
 {\gamma(\xi,s)(s-k)}\,\mathrm ds
 \Bigg]
 \label{eq:elliptic-ghat-solution}
\end{align}
satisfies the jump conditions above. The constant $\omega$ is chosen so that
\(
\widehat g(\xi,k)=O(1) \),  \(k\to\infty\).
A direct calculation yields
\begin{align}
 \widehat g_\infty={}&\lim_{k\to\infty}\widehat g(\xi,k)
 \nonumber\\
 =\frac1{2\pi}\Bigg[&\int_B
 \frac{s\log(\widehat\delta_1^2\widehat\delta_2)}
      {\gamma(\xi,s)}\,\mathrm ds
 +\int_{\Gamma_1^{(6)}}
 \frac{s\left(\log(\widehat\delta_1^2\widehat\delta_2/r_1)
             +\mathrm i\omega\right)}
      {\gamma(\xi,s)}\,\mathrm ds-\int_{\Gamma_2^{(6)}}
 \frac{s\left(\log(\bar r_1\widehat\delta_1^2\widehat\delta_2)
             +\mathrm i\omega\right)}
      {\gamma(\xi,s)}\,\mathrm ds\Bigg].
 \label{eq:elliptic-ghat-infty}
\end{align}

Recall that the diagonal matrix $D$ is defined by
\begin{equation*}
 D(k)=\diag\!\bigl(d(k)^{-1/2},d(k)^{-1/2},1\bigr).
\end{equation*}
We now introduce the final transformation,
\begin{equation}
 \widehat N^{(7)}(x,t,k)
 =
 e^{-\mathrm i\widehat g_\infty\sigma}
 \widehat N^{(6)}(x,t,k)
 D(k)
 e^{\mathrm i\widehat g(\xi,k)\sigma}.
 \label{eq:elliptic-seventh-transformation}
\end{equation}
In principle, the above transformation leaves the jump contour unchanged.  For convenience in the subsequent analysis,
however, we reverse the orientation of $\Gamma_2^{(6)}$, namely,
\(
\Gamma_2^{(7)}=-\Gamma_2^{(6)}
\).  Apart from this modification, we set \(\Gamma^{(7)}=\Gamma^{(6)}\).
We next give the corresponding jump matrix
$\widehat V^{(7)}$ on the various components of the contour.

We begin with the jumps near $k_0$. We have
\begin{subequations}\label{eq:elliptic-seventh-jumps-k0}
\begin{align}
 \widehat V_{1'}^{(7)}&=
 \begin{pmatrix}
 1&0&0\\
 -\dfrac{\widehat\delta_1^2\widehat\delta_2}{r_1}
 e^{2\mathrm i\widehat g}e^{h_{31}}&1&0\\
 0&0&1
 \end{pmatrix},\qquad
 \widehat V_{2'}^{(7)}=
 \begin{pmatrix}
 1&-\dfrac{e^{-2\mathrm i\widehat g}e^{h_{13}}}
 {\bar r_1\widehat\delta_1^2\widehat\delta_2}&0\\
 0&1&0\\0&0&1
 \end{pmatrix},\\[3mm]
 \widehat V_3^{(7)}&=
 \begin{pmatrix}
 1&0&0\\
 -\dfrac{\bar r_1\widehat\delta_1^2\widehat\delta_2}
 {1+r_1\bar r_1}e^{2\mathrm i\widehat g}e^{h_{31}}&1&0\\
 0&0&1
 \end{pmatrix},\qquad
 \widehat V_4^{(7)}=
 \begin{pmatrix}
 1&-\dfrac{r_1(\widehat\delta_1)^{-2}
                   (\widehat\delta_2)^{-1}}
 {1+r_1\bar r_1}e^{-2\mathrm i\widehat g}e^{h_{13}}&0\\
 0&1&0\\0&0&1
 \end{pmatrix}.
\end{align}
\end{subequations}
Moreover, a direct calculation shows that the jump matrices
$\widehat V^{(7)}$ on $\Gamma_1^{(7)}$ and $\Gamma_2^{(7)}$ coincide.
Since
\(
\widetilde B=\Gamma_1^{(7)}\cup\Gamma_2^{(7)}
\),
the jumps on these two contour segments can be written uniformly on
$\widetilde B$. More precisely,
\begin{align}
 \widehat V_{\widetilde B}^{(7)}=
 \begin{pmatrix}
 0&-e^{-\mathrm i(\Omega t-\omega)}&0\\
 e^{\mathrm i(\Omega t-\omega)}&0&0\\
 0&0&1
 \end{pmatrix}.
\end{align}
Near \(k_1\), one obtains
\begin{subequations}\label{eq:elliptic-seventh-jumps-k1}
\begin{align}
 \widehat V_5^{(7)}&=
 \begin{pmatrix}
 1&0&d^{-1/2}\bar R\widehat\delta_1^{-1}\widehat\delta_2^{-2}
 e^{-\mathrm i\widehat g}e^{h_{12}}\\
 0&1&0\\0&0&1
 \end{pmatrix},\qquad
 \widehat V_7^{(7)}=
 \begin{pmatrix}
 1&0&0\\0&1&0\\
 \dfrac{d^{-1/2}R\widehat\delta_1\widehat\delta_2^2
 e^{\mathrm i\widehat g}}{H}e^{h_{21}}&0&1
 \end{pmatrix},\\[3mm]
 \widehat V_8^{(7)}&=
 \begin{pmatrix}
 1&0&\dfrac{d^{-1/2}\bar R\widehat\delta_1^{-1}
 \widehat\delta_2^{-2}e^{-\mathrm i\widehat g}}{H}e^{h_{12}}\\
 0&1&0\\0&0&1
 \end{pmatrix},\qquad
 \widehat V_{6}^{(7)}=
 \begin{pmatrix}
 1&0&0\\0&1&0\\
 d^{-1/2}R\widehat\delta_1\widehat\delta_2^2
 e^{\mathrm i\widehat g}e^{h_{21}}&0&1
 \end{pmatrix}.
\end{align}
\end{subequations}
Finally, the jump matrix on the branch cut \(B\) is given by
\begin{align}
 \widehat V_B^{(7)}=
 \begin{pmatrix}
 0&-\mathrm i&0\\
 -\mathrm i&0&0\\
 0&0&1
 \end{pmatrix}.
\end{align}

\subsection{Genus-one outer model}
\label{subsec:elliptic-outer-model}

After the small jumps have been removed, only the two band
jumps remain.  The leading global approximation is therefore determined by
the following model problem. 

\begin{RHP}[Genus-one outer-model RH problem]
Find a \(3\times3\) matrix-valued function
\(N^{\mathrm{out}}(\xi,k)\) with the following properties.
\begin{enumerate}
 \item[(i)] \emph{Analyticity.}
 The function \(N^{\mathrm{out}}\) is analytic in
 \(\mathbb C\setminus(B\cup\widetilde B)\) and has continuous
 nontangential boundary values on $B\cup\widetilde B$.

 \item[(ii)] \emph{Jump conditions.}
 The boundary values satisfy
 \begin{equation}
  N_+^{\mathrm{out}}(\xi,k)
  =N_-^{\mathrm{out}}(\xi,k)V^{\mathrm{out}}(\xi,k),
  \qquad k\in B\cup\widetilde B,
  \label{eq:elliptic-outer-jumps}
 \end{equation}
 where
 \begin{equation}
 V^{\mathrm{out}}(\xi,k)=
 \begin{cases}
 \begin{pmatrix}
 0&-\mathrm i&0\\
 -\mathrm i&0&0\\
 0&0&1
 \end{pmatrix},&k\in B,\\[6mm]
 \begin{pmatrix}
 0&-e^{-\mathrm i(\Omega t-\omega)}&0\\
 e^{\mathrm i(\Omega t-\omega)}&0&0\\
 0&0&1
 \end{pmatrix},&k\in\widetilde B.
 \end{cases}
 \label{eq:elliptic-outer-jump-matrices}
 \end{equation}

 \item[(iii)] \emph{Normalization.}
 As \(k\to\infty\),
 \begin{equation}
  N^{\mathrm{out}}(\xi,k)=I+O(k^{-1}).
  \label{eq:elliptic-outer-normalization-condition}
 \end{equation}

 \item[(iv)] \emph{Endpoint behavior.}
 At every endpoint
 \(k_\star\in\{\mathrm iq_0,-\mathrm iq_0,\alpha,\bar\alpha\}\),
 \begin{equation*}
  N^{\mathrm{out}}(\xi,k)
  =O\bigl(|k-k_\star|^{-1/4}\bigr),\qquad k\to k_\star.
 \end{equation*}
\end{enumerate}
\end{RHP}

\begin{figure}[htbp]
\centering
\begin{tikzpicture}[
    >=Stealth,
    scale=1.35,
    transform shape=false,
    line cap=round,
    line join=round,
    every node/.style={inner sep=2pt},
    cut/.style={black,line width=1.45pt},
    cyc/.style={black,line width=.9pt},
    dir/.style={
        postaction={decorate},
        decoration={
            markings,
            mark=at position .22 with {\arrow{>}}
        }
    },
    adir/.style={
        postaction={decorate},
        decoration={
            markings,
            mark=at position .60 with {\arrow{>}}
        }
    }
]

\draw[->,black] (-3.0,0)--(2.05,0)
    node[right=4pt] {\(\Re k\)};
\draw[->,black] (0,-2.0)--(0,2.05)
    node[above=4pt] {\(\Im k\)};

\coordinate (qplus)  at (0,1.18);
\coordinate (qminus) at (0,-1.18);

\draw[cut] (qminus)--(qplus);
\fill[black] (qplus)  circle (1.55pt);
\fill[black] (qminus) circle (1.55pt);

\coordinate (alphabar) at (-1.73,-1.02);
\coordinate (alpha)    at (-1.73,1.02);

\def\aCutT{.18}

\draw[cut] (alphabar)
    .. controls (-2.02,-.50) and (-2.02,.50) .. (alpha)
    coordinate[pos=\aCutT] (aL);

\fill[black] (alpha)    circle (1.55pt);
\fill[black] (alphabar) circle (1.55pt);

\def\aCenterX{-.60}
\def\aRadiusX{2.25}
\def\aRadiusY{.44}

\pgfmathsetmacro{\aLeftX}{
    -1.73-.87*\aCutT*(1-\aCutT)
}
\pgfmathsetmacro{\aLeftY}{
    -1.02+1.56*\aCutT+1.44*\aCutT^2-.96*\aCutT^3
}

\pgfmathsetmacro{\aAngleL}{
    acos((\aLeftX-\aCenterX)/\aRadiusX)
}
\pgfmathsetmacro{\aCenterY}{
    \aLeftY-\aRadiusY*sin(\aAngleL)
}
\coordinate (aCenter) at (\aCenterX,\aCenterY);

\pgfmathsetmacro{\aAngleR}{
    acos(-\aCenterX/\aRadiusX)
}
\pgfmathsetmacro{\aEndAngle}{\aAngleR+360}

\coordinate (aR) at
    (0,{\aCenterY+\aRadiusY*sin(\aAngleR)});

\draw[cyc,dir] (0,0)
    ellipse [x radius=.72,y radius=1.58];

\begin{scope}[shift={(aCenter)}]

    \draw[cyc,adir]
        ({\aRadiusX*cos(\aAngleL)},
         {\aRadiusY*sin(\aAngleL)})
        arc[
            start angle=\aAngleL,
            end angle=\aEndAngle,
            x radius=\aRadiusX,
            y radius=\aRadiusY
        ];

    \draw[cyc,dashed,adir]
        ({\aRadiusX*cos(\aAngleR)},
         {\aRadiusY*sin(\aAngleR)})
        arc[
            start angle=\aAngleR,
            end angle=\aAngleL,
            x radius=\aRadiusX,
            y radius=\aRadiusY
        ];

\end{scope}


\node[above left=4pt] at (alpha)
    {\(\alpha\)};
\node[below right=3pt] at (alphabar)
    {\(\bar\alpha\)};

\node[anchor=west] at (.14,.52)
    {\(B\)};
\node[anchor=east] at (-2.12,.30)
    {\(\widetilde B\)};
\node at (.90,1.57)
    {\(b\)};
\node at (1.08,-1.67)
    {\(a\)};

\end{tikzpicture}
\caption{The canonical homology basis
\(\{a,b\}\) for the genus-one surface.}
\label{fig:elliptic-homology-basis}
\end{figure}
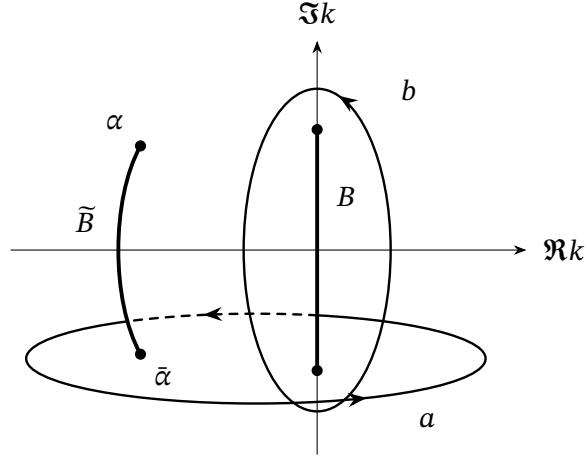

We now solve this RH problem explicitly.  The relation
\begin{equation*}
 \gamma(\xi,k)^2=(k^2+q_0^2)(k-\alpha)(k-\bar\alpha)
\end{equation*}
defines a two-sheeted genus-one Riemann surface with branch cuts
\(B\) and \(\widetilde B\).  We use the canonical homology basis
\(\{a,b\}\) shown in Figure~\ref{fig:elliptic-homology-basis}.  The
\(b\)-cycle surrounds the fixed cut \(B\) counterclockwise on the first
sheet.  The \(a\)-cycle passes through the gap between the two cuts and
returns on the second sheet.

With this choice of cycles, define the normalized Abelian differential of
the first kind by
\begin{equation}
 \mathrm d\omega_{\!A}(k)=\frac{\widehat C}{\gamma(\xi,k)}
 \,\mathrm dk,
 \qquad
 \widehat C=\left(\oint_b\frac{\mathrm dk}
 {\gamma(\xi,k)}\right)^{-1}.
 \label{eq:elliptic-normalized-differential}
\end{equation}
Thus, its \(b\)-period equals one.  Let \(\tau\) denote its \(a\)-period, namely,
\(
\tau=\oint_a\mathrm{d}\omega_{\!A}.
\)
Then, after a calculation, one obtains
\[
\tau
=
\mathrm{i}\frac{K\bigl(\sqrt{1-m^2}\bigr)}{K(m)}
\in\mathrm{i}\mathbb{R}_+,
\]
Here, \(K(\cdot)\) denotes the complete elliptic integral of the first kind.  The geometry of the cycles also gives the half-period identities
\begin{equation}
 \int_{-\mathrm iq_0}^{\mathrm iq_0}\mathrm d\omega_{\!A}=\frac12,
 \qquad
 \int_{\alpha}^{-\mathrm iq_0}\mathrm d\omega_{\!A}=\frac\tau2.
 \label{eq:elliptic-half-periods}
\end{equation}

Let the genus-one theta function be defined by
\begin{equation}
\vartheta(z)
=
\sum_{\ell\in\mathbb{Z}}
\exp\bigl(2\pi\mathrm{i}\ell z+\mathrm{i}\pi\ell^2\tau\bigr).
\label{eq:elliptic-theta-function}
\end{equation}
Since \(\operatorname{Im}\tau>0\), the series converges absolutely for
every \(z\in\mathbb{C}\). Moreover, \(\vartheta\) satisfies the following
periodicity and quasi-periodicity relations:
\begin{equation}
\vartheta(z+n)=\vartheta(z),
\qquad
\vartheta(z+n\tau)
=
e^{-2\pi\mathrm{i}nz-\mathrm{i}\pi n^2\tau}\vartheta(z),
\qquad n\in\mathbb{Z}.
\label{eq:elliptic-theta-quasiperiodicity}
\end{equation}
 We next introduce the Abel map,
\begin{equation}
 \nu(k)=\int_{\mathrm iq_0}^{k}\mathrm d\omega_{\!A}.
 \label{eq:elliptic-Abel-map}
\end{equation}
Changing sheets across either band changes the sign of the integrand.
Together with the period relations, this gives
\begin{equation}
 \nu^+(k)+\nu^-(k)=n,
 \quad k\in B;
 \qquad
 \nu^+(k)+\nu^-(k)=-\tau+n,
 \quad k\in\widetilde B, \qquad n \in  \mathbb Z.
 \label{eq:elliptic-Abel-boundary-values}
\end{equation}

Let \(c\) be the divisor constant, and define the row vector
\(\mathcal M=(\mathcal M_1,\mathcal M_2)\) by
\begin{subequations}\label{eq:elliptic-M-functions}
\begin{align}
 \mathcal M_1(\xi,k,c)
 &=e^{-\pi\mathrm i/4}
 \frac{\vartheta\!\left(
  \dfrac{\Omega t}{2\pi}-\dfrac{\omega}{2\pi}+\dfrac14
  +\nu(k)+c\right)}{\vartheta(\nu(k)+c)},\\[2mm]
 \mathcal M_2(\xi,k,c)
 &=e^{\pi\mathrm i/4}
 \frac{\vartheta\!\left(
  \dfrac{\Omega t}{2\pi}-\dfrac{\omega}{2\pi}+\dfrac14
  -\nu(k)+c\right)}{\vartheta(-\nu(k)+c)}.
\end{align}
\end{subequations}
It follows immediately from
\eqref{eq:elliptic-theta-quasiperiodicity} that the row vector satisfies
the following jump conditions:
\begin{equation}
 \mathcal M^+=\mathcal M^-
 \begin{cases}
 \begin{pmatrix}0&\mathrm i\\-\mathrm i&0\end{pmatrix},&k\in B,\\[4mm]
 \begin{pmatrix}
 0&e^{-\mathrm i(\Omega t-\omega)}\\
 e^{\mathrm i(\Omega t-\omega)}&0
 \end{pmatrix},&k\in\widetilde B.
 \end{cases}
 \label{eq:elliptic-M-jumps}
\end{equation}
The jump conditions in \eqref{eq:elliptic-M-jumps} do not yet coincide
with those of the outer model in
\eqref{eq:elliptic-outer-jump-matrices}. To account for this difference,
we introduce the fourth-root function
\begin{equation}
p(k)
=
\left[
\frac{(k-\mathrm{i}q_0)(k-\alpha)}
{(k+\mathrm{i}q_0)(k-\bar{\alpha})}
\right]^{1/4},
\qquad
p(k)\to1
\quad\text{as}\quad
k\to\infty, 
\label{eq:elliptic-p}
\end{equation}
where the cuts are chosen along \(B\cup\widetilde B\).  Then
its boundary values satisfy
\begin{equation}
p^+(k)=\mathrm{i}\,p^-(k),
\qquad
k\in B\cup\widetilde B.
\label{eq:elliptic-p-jump}
\end{equation}
We now define the function
\begin{equation}
 \widetilde N^{\mathrm{out}}(\xi,k,c)=
 \begin{pmatrix}
 \dfrac12(p+p^{-1})\mathcal M_1(\xi,k,c)&
 \dfrac{\mathrm i}{2}(p-p^{-1})\mathcal M_2(\xi,k,c)&0\\[3mm]
 -\dfrac{\mathrm i}{2}(p-p^{-1})\mathcal M_1(\xi,k,-c)&
 \dfrac12(p+p^{-1})\mathcal M_2(\xi,k,-c)&0\\[3mm]
 0&0&1
 \end{pmatrix}.
 \label{eq:elliptic-outer-unnormalized}
\end{equation}
It is straightforward to verify that
\(\widetilde N^{\mathrm{out}}(\xi,k,c)\) satisfies the jump conditions
in \eqref{eq:elliptic-outer-jump-matrices}. However, it does not satisfy
the required normalization at infinity. Since \(p(k)\to1\) as
\(k\to\infty\), we have
\begin{equation}
\widetilde N_\infty^{\mathrm{out}}
:=
\lim_{k\to\infty}\widetilde N^{\mathrm{out}}(\xi,k,c)
=
\operatorname{diag}\left(
\mathcal M_1(\xi,\infty,c),
\mathcal M_2(\xi,\infty,-c),
1
\right).
\label{eq:elliptic-outer-limit}
\end{equation}
It is therefore natural to expect that
\begin{equation}
N^{\mathrm{out}}(\xi,k,c)
=
\bigl(\widetilde N_\infty^{\mathrm{out}}\bigr)^{-1}
\widetilde N^{\mathrm{out}}(\xi,k,c)
\label{eq:elliptic-outer-normalized}
\end{equation}
solves the genus-one outer-model RH problem.
To verify this claim, it remains only to establish that
\(N^{\mathrm{out}}(\xi,k,c)\), defined by
\eqref{eq:elliptic-outer-normalized}, is analytic for
\(
k\in\mathbb{C}\setminus\bigl(B\cup\widetilde B\bigr).
\)
We now choose the divisor constant \(c\) so that
\(\widetilde N^{\mathrm{out}}\) has no poles away from the two bands.

The  poles of $\widetilde N^{\mathrm{out}}$ can occur only
at zeros of \(\vartheta(\nu(k)\pm c)\).  The zeros of \(\vartheta\) form
the lattice
$
 \frac{1+\tau}{2}+\mathbb Z+\tau\mathbb Z
$.
On the other hand, \(p-p^{-1}\) has one finite simple zero, located at
\begin{equation}
 k_{q}=\frac{q_0\Ree\alpha}{q_0+\Imm\alpha}. 
 \label{eq:elliptic-divisor-point}
\end{equation}
To place the zero of the corresponding theta-function factor at
\(k_{ q}\), we choose
\begin{equation}
c
=
\nu(k_{q})+\frac{1+\tau}{2}.
\label{eq:elliptic-divisor-choice}
\end{equation}
The zero of \(p-p^{-1}\) then cancels the possible theta-function pole.
Thus \(\widetilde N^{\mathrm{out}}\) is analytic away from the two bands,
with only the allowed fourth-root behavior at their endpoints.   Thus   \(\widetilde N^{\mathrm{out}}\)  is the solution of the outer RH problem.

To record the coefficient needed in the reconstruction formula, write
\begin{equation}
 N_1^{\mathrm{out}}:=\lim_{k\to\infty}
 k\bigl(N^{\mathrm{out}}(\xi,k)-I\bigr)\quad  \text{and} 
 \quad \nu_\infty:=\nu(\infty).
 \label{eq:elliptic-outer-coefficient-definition}
\end{equation}
A direct expansion of \eqref{eq:elliptic-outer-normalized}
shows that the  \((2,1)\)-entry and  \((3,1)\)-entry of  $N_1^{\mathrm{out}}$ are
\begin{equation}
(N_1^{\mathrm{out}})_{31}=0, \qquad 
 (N_1^{\mathrm{out}})_{21}
 =\frac{\mathrm i(q_0+\Imm\alpha)}2
 \frac{\vartheta(\frac{\Omega t-\omega}{2\pi}+\frac14+\nu_\infty-c)
       \vartheta(-\nu_\infty-c)}
      {\vartheta(\nu_\infty-c)
       \vartheta(\frac{\Omega t-\omega}{2\pi}+\frac14-\nu_\infty-c)}.
 \label{eq:elliptic-outer-21-unsimplified}
\end{equation}
This formula can be simplified by expressing the divisor constant in terms
of \(\nu_\infty\).  To justify this relation, introduce the meromorphic function
\begin{equation}
 f(k):=p(k)\bigl[p^{-1}(k)-p(k)\bigr]
 =1-p(k)^2
 =1-\frac{(k-\mathrm iq_0)(k-\alpha)}
          {\gamma(\xi,k)}
\end{equation}
on the genus-one Riemann surface. Its divisor is
\begin{equation}
 (f)=k_{q}+\infty_1-(-\mathrm iq_0)-\bar\alpha,
\end{equation}
where \(\infty_1\) denotes the point at infinity on the first sheet.
Since this divisor is principal, Abel's theorem gives
\begin{equation}
 \nu(k_{q})
 \equiv
 \nu(-\mathrm iq_0)+\nu(\bar\alpha)-\nu_\infty
 \pmod{\mathbb Z+\tau\mathbb Z}.
 \label{eq:elliptic-divisor-Abel-identities}
\end{equation}
On the other hand, recalling the choice of cycles shown in
Fig.~\ref{fig:elliptic-homology-basis} and using the relations in
\eqref{eq:elliptic-half-periods}, we find that
\[
\nu(-\mathrm{i}q_0)\in\frac{1}{2}+\mathbb{Z},
\qquad
\nu(\bar{\alpha})\in\frac{1}{2}-\frac{\tau}{2}+\mathbb{Z}.
\]
Combining these identities with \eqref{eq:elliptic-divisor-choice} yields
\begin{equation*}
 c\equiv\frac12-\nu_\infty
 \pmod{\mathbb Z+\tau\mathbb Z}.
\end{equation*}
We may therefore use the representative
\(c=\frac12-\nu_\infty\).  Substitution into
\eqref{eq:elliptic-outer-21-unsimplified}, followed by the periodicity of
\(\vartheta\), gives the final expression
\begin{equation}
 (N_1^{\mathrm{out}})_{21}
 =\frac{\mathrm i(q_0+\Imm\alpha)}2
 \frac{\vartheta\!\left(\frac{\Omega t-\omega}{2\pi}
                  +2\nu_\infty-\frac14\right)\vartheta(-\frac12)}
      {\vartheta\!\left(\frac{\Omega t-\omega}{2\pi}-\frac14\right)
       \vartheta(2\nu_\infty-\frac12)}.
 \label{eq:elliptic-outer-21}
\end{equation}

\subsection{Local parametrices}
\label{subsec:elliptic-local-parametrices}
After the preceding transformations, the original RH problem has been
reduced to the deformed problem for \(\widehat N^{(7)}\). For \(k\) outside fixed neighborhoods of the points
\(\alpha\), \(\bar\alpha\), \(k_0\), and \(k_1\), and of
the bands \(B\cup\widetilde B\), there exists a constant
\(c>0\) such that
\[
\widehat V^{(7)}(k)
=
I+O(e^{-ct}),
\qquad t\to\infty.
\]
The nondecaying jumps on \(B\cup\widetilde B\) are described
by the genus-one outer model \(N^{\mathrm{out}}\).  It therefore remains to construct local parametrices near
\(\alpha\), \(\bar\alpha\), \(k_0\), and \(k_1\).

\subsubsection{The Airy parametrix near \(\alpha\) and \( \bar \alpha\)}
\label{subsubsec:elliptic-Airy-alpha}

\begin{figure}
\centering
\begin{tikzpicture}[
    >=Stealth,
    scale=0.95,
    contour/.style={
        black,
        line width=0.95pt,
        line cap=round,
        line join=round
    },
    band/.style={
        black,
        line width=1.35pt,
        line cap=round
    },
    directed/.style={
        postaction={decorate},
        decoration={
            markings,
            mark=at position 0.58 with {\arrow{>}}
        }
    },
    boundary/.style={
        black,
        dashed,
        line width=0.85pt,
        line cap=round
    }
]

\begin{scope}

    \coordinate (A)  at (-0.05,0.12);
    \coordinate (L)  at (-2.08,-0.72);
    \coordinate (R)  at (2.00,1.15);
    \coordinate (BL) at (-0.86,-2.02);
    \coordinate (BR) at (0.90,-2.04);

    \draw[boundary] (0,0) circle[radius=2.25];

    \draw[contour,directed]
        (L)
        .. controls (-1.52,-0.58) and (-0.72,0.02) ..
        (A);

    \draw[contour,directed]
        (A)
        .. controls (0.42,0.58) and (1.18,0.98) ..
        (R);

    \draw[band,directed]
        (BL)
        .. controls (-0.86,-1.42) and (-0.40,-0.42) ..
        (A);

    \draw[contour,directed]
        (BR)
        .. controls (0.73,-1.28) and (0.40,-0.36) ..
        (A);

    \fill (A) circle (2pt);
    \node at (-0.20,0.42) {\(\alpha\)};

    \node at (-1.44,0.08) {\(\Gamma_2^{\alpha}\)};
    \node at (1.00,1.30)  {\(\Gamma_1^{\alpha}\)};
    \node at (-1.08,-1.08) {\(\Gamma_3^{\alpha}\)};
    \node at (1.15,-1.02) {\(\Gamma_4^{\alpha}\)};


\end{scope}

\end{tikzpicture}

\caption{The local contour
$
\Gamma^{\alpha}=\bigcup_{j=1}^4 \Gamma_j^{\alpha}
$
in a neighborhood of \(\alpha\).}
\label{fig:local-alpha-contour}
\end{figure}
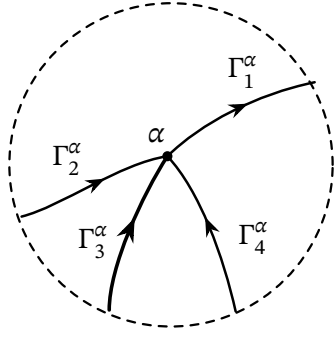
The outer model captures the nondecaying band jumps away from their
endpoints.  Its approximation is not uniform, however, when \(k\)
approaches the branch points \(\alpha\) and \(\bar\alpha\).   We therefore replace the outer
approximation by local Airy corrections.  We first carry out the
construction near \(\alpha\); the correction near \(\bar\alpha\) then
follows from the Schwarz symmetry of the RH problem.

Recall that the Abelian phase has the local expansion 
\begin{equation}
 \hbar (k)=\hbar (\alpha)
  +C_\alpha(k-\alpha)^{3/2}
  +O((k-\alpha)^{5/2}),  \quad C_\alpha\ne0.
 \label{eq:elliptic-h-alpha-refined}
\end{equation}
It is straightforward to verify that
\(
\hbar (\xi,\alpha)=\frac{\Omega}{2}\in\mathbb{R}
\),
where \(\Omega\) is defined in \eqref{eq:elliptic-Omega}.
The important feature of
\eqref{eq:elliptic-h-alpha-refined} is the leading
\((k-\alpha)^{3/2}\) term.  This term suggests a
\(2/3\)-power change of variables and hence a standard Airy model.
 Choose a sufficiently small disk
\begin{equation*}
 D_\epsilon(\alpha)=\{k:|k-\alpha|<\epsilon\}.
\end{equation*}
For convenience, we denote the portion of \(\Gamma^{(7)}\) inside
\(D_\epsilon(\alpha)\) by
\(
\Gamma^\alpha
\),
see Figure~\ref{fig:local-alpha-contour}.
Then we introduce the conformal coordinate in the disk:
\begin{equation}
 \zeta_\alpha(k)=
 \left[\frac{3\mathrm it}{2}
       \left(\hbar(k)-\frac{\Omega}{2}\right)\right]^{2/3}
 =t^{2/3}(k-\alpha)\psi_\alpha(k),\qquad
 \psi_\alpha(\alpha)\ne0.
 \label{eq:elliptic-Airy-coordinate}
\end{equation}
Here \(\psi_\alpha\) is analytic in \(D_\epsilon(\alpha)\).  After
decreasing \(\epsilon\), if necessary, we may assume that
\(\psi_\alpha\) has no zeros in the disk.  The map
\(k\mapsto\zeta_\alpha(k)\) is then biholomorphic from
\(D_\epsilon(\alpha)\) onto a neighborhood of the origin in the
\(\zeta_\alpha\)-plane.  
We select the branch so that
\(\Gamma^{\alpha}_3\) is mapped to the negative real
axis.  If necessary, the remaining local contour arcs can be deformed within
the disk so that their images coincide with the other three Airy rays
shown in Figure~\ref{fig:app-Airy-contour}.   With this choice of \(\zeta_\alpha\), we have
\begin{equation}
 \frac{4}{3}\zeta_{\alpha}(k)^{3/2}
 =2\mathrm it
 \left(\hbar(k)-\frac{\Omega}{2}\right).
 \label{eq:elliptic-Airy-phase-identity}
\end{equation}
Consequently, the local phase factors are transformed into the standard
Airy exponential factors.

The jump matrices still contain nonconstant scalar prefactors, which
must be removed to obtain the standard Airy jumps. To absorb these
factors, we introduce the diagonal matrix
\begin{equation}
 Y_\alpha(k)=
 \begin{pmatrix}
 e^{-\mathrm i\Omega t/2}\widehat\delta_1^{-1}
       \widehat\delta_2^{-1}r_1e^{-\mathrm i\widehat g}&0&0\\
 0&-e^{\mathrm i\Omega t/2}\widehat\delta_1e^{\mathrm i\widehat g}&0\\
 0&0&1
 \end{pmatrix}.
 \label{eq:elliptic-Y-alpha}
\end{equation}
The assumption \(r_1(\alpha)\ne0\) ensures that
\(Y_\alpha(k)\) is well defined for sufficiently small
\(\epsilon>0\).
 We next verify that conjugation by \(Y_\alpha\) transforms all the local
jumps into the standard Airy form. Let \(V_j^\alpha\),
\(j=1,\ldots,4\), denote the jump matrix on the contour arc
\(\Gamma_j^\alpha\). A direct calculation yields
\begin{equation} \label{eq:elliptic-Airy-local-jump-one}
\begin{aligned}
  Y_\alpha^{-1}  V_1^{\alpha}
 Y_\alpha
 &=\begin{pmatrix}1&e^{h_{13}+\mathrm i\Omega t}&0\\0&1&0\\0&0&1\end{pmatrix}
 =\begin{pmatrix}1&e^{-\frac43\zeta_\alpha^{3/2}}&0\\0&1&0\\0&0&1\end{pmatrix},\\
 Y_\alpha^{-1}  V_2^{\alpha}  Y_\alpha
 &= Y_\alpha^{-1}  V_4^{\alpha}  Y_\alpha
 =\begin{pmatrix}1&0&0\\e^{\frac43\zeta_\alpha^{3/2}}&1&0\\0&0&1\end{pmatrix}.
 \end{aligned}
\end{equation}
On the band \(\Gamma_3^\alpha\), both boundary values of \(Y_\alpha\)
are involved. Accordingly, we have
\begin{align}
 \bigl( Y_\alpha^-(k)\bigr)^{-1}
  V_3^{\alpha}(k)  Y_\alpha^+(k)
 &=
 \begin{pmatrix}
 0&
 e^{\mathrm i\omega}
 e^{\mathrm i(\widehat g^+(k)+\widehat g^-(k))}
 \frac{\widehat\delta_1(k)^2\widehat\delta_2(k)}{r_1(k)}
 &0\\[2mm]
 -e^{-\mathrm i\omega}
 e^{-\mathrm i(\widehat g^+(k)+\widehat g^-(k))}
 \frac{r_1(k)}
 {\widehat\delta_1(k)^2\widehat\delta_2(k)}
 &0&0\\
 0&0&1
 \end{pmatrix}
 \nonumber\\
 &=
 \begin{pmatrix}
 0&1&0\\
 -1&0&0\\
 0&0&1
 \end{pmatrix}.
 \label{eq:elliptic-Airy-band-calculation}
\end{align}
Next, we construct a \(3\times3\) matrix-valued function from the
solution of the standard Airy model RH problem and show that it satisfies
the jump relations in
\eqref{eq:elliptic-Airy-local-jump-one} and
\eqref{eq:elliptic-Airy-band-calculation}.

Let \(N^{\mathrm{Ai}}(\zeta)\) denote the solution of the Airy model RH
problem in Appendix~\ref{App:BBB}, and define its \(3\times3\) embedding by
\[
N^{\mathrm{Ai},\alpha}(\zeta)
:=
\begin{pmatrix}
N^{\mathrm{Ai}}(\zeta)&\mathbf{0}_{2\times1}\\
\mathbf{0}_{1\times2}&1
\end{pmatrix}.
\]
It is straightforward to verify that
\(N^{\mathrm{Ai},\alpha}(\zeta_\alpha)\) satisfies precisely the jump
relations in \eqref{eq:elliptic-Airy-local-jump-one} and
\eqref{eq:elliptic-Airy-band-calculation}.
Similarly, define
\[
N_0^{\mathrm{Ai},\alpha}(\zeta)
:=
\begin{pmatrix}
N_0^{\mathrm{Ai}}(\zeta)&\mathbf{0}_{2\times1}\\
\mathbf{0}_{1\times2}&1
\end{pmatrix},
\]
where \(N_0^{\mathrm{Ai}}(\zeta)\) is the leading factor in the
large-\(\zeta\) expansion of \(N^{\mathrm{Ai}}(\zeta)\). The local
correction near \(\alpha\) is then defined by
\begin{equation}
N^{\mathrm{loc},\alpha}(x,t,k)
=
N^{\mathrm{out}}(x,t,k)Y_\alpha(k)
\bigl[N_0^{\mathrm{Ai},\alpha}(\zeta_\alpha(k))\bigr]^{-1}
N^{\mathrm{Ai},\alpha}(\zeta_\alpha(k))
Y_\alpha(k)^{-1},
\qquad
k\in D_\epsilon(\alpha).
\label{eq:elliptic-local-alpha}
\end{equation}
By construction, \(N^{\mathrm{loc},\alpha}\) satisfies the required local
jump condition
\[
N_+^{\mathrm{loc},\alpha}(x,t,k)
=
N_-^{\mathrm{loc},\alpha}(x,t,k)
V^{(7)}(x,t,k),
\qquad
k\in\Gamma^{(7)} \cap D_\epsilon(\alpha).
\]
Moreover, \(\zeta_\alpha\) tends to infinity as \(t\to\infty\).
Thus, the normalization
\eqref{eq:app-Airy-normalization} implies that, as \(t\to\infty\),
\begin{equation}
N^{\mathrm{out}}(x,t,k) \bigl(N^{\mathrm{loc},\alpha}(x,t,k)\bigr)^{-1}
-I
=
O(t^{-1}),
\qquad
k\in\partial D_\epsilon(\alpha).
\label{E:asyofNaialpha}
\end{equation}

The local contours and jump matrices near \(\alpha\) and
\(\bar\alpha\) are related by Schwarz symmetry.  We can therefore obtain the local  parametrix near
\(\bar\alpha\) directly from the one constructed near \(\alpha\), without
solving another Airy model problem.  More precisely, we define
\begin{equation}
N^{\mathrm{loc},\bar\alpha}(x,t,k)
=
\widehat\sigma_2
\bigl[N^{\mathrm{loc},\alpha}(x,t,k^*)\bigr]^*
\widehat\sigma_2,
\qquad
\widehat\sigma_2
=
\begin{pmatrix}
0&-\mathrm{i}&0\\
\mathrm{i}&0&0\\
0&0&1
\end{pmatrix}.
\label{eq:elliptic-local-alphabar}
\end{equation}
It also inherits the same matching
estimate as \(N^{\mathrm{loc},\alpha}\); namely,
\begin{align} \label{E:asyofNaibaralpha}
N^{\mathrm{out}}(x,t,k) \bigl(N^{\mathrm{loc},\bar\alpha}(x,t,k)\bigr)^{-1}
-I
=
O(t^{-1}),
\qquad
k\in\partial D_\epsilon(\bar\alpha),
\end{align}
as \(t\to\infty\).

\subsubsection{The local model near \(k_0\)}
\label{subsubsec:elliptic-local-k0}

\begin{figure}
\centering

\begin{tikzpicture}[
  scale=1.05,
  >=Stealth,
  c/.style={
    line width=1.05pt
  },
  dir/.style={
    postaction={decorate},
    decoration={
      markings,
      mark=at position .58 with {\arrow{>}}
    }
  }
]

\draw[dashed,gray] (0,0) circle (2.0);


\draw[c,dir]
  (0,0)
  .. controls (.30,.06) and (.92,1.15) ..
  (1.66,1.43);

\draw[c,dir]
  (0,0)
  .. controls (.16,.90) and (.34,1.58) ..
  (.48,1.94);

\draw[c,dir]
  (0,0)
  .. controls (-.22,.79) and (-.29,1.52) ..
  (-.23,1.98);


\draw[c,dir]
  (-1.414,1.414)--(0,0);     

\draw[c,dir]
  (-1.414,-1.414)--(0,0);    


\draw[c,dir]
  (0,0)
  .. controls (-.22,-.79) and (-.29,-1.52) ..
  (-.23,-1.98);

\draw[c,dir]
  (.48,-1.94)
  .. controls (.34,-1.58) and (.16,-.90) ..
  (0,0);

\draw[c,dir]
  (0,0)
  .. controls (.30,-.06) and (.92,-1.15) ..
  (1.66,-1.43);


\foreach \p/\lab in {
  {(1.34,.98)}/1,
  {(.62,1.46)}/2,
  {(-.08,1.76)}/3,
  {(-1.02,0.72)}/4,
  {(-1.02,-0.72)}/5,
  {(-.08,-1.76)}/6,
  {(.62,-1.46)}/7,
  {(1.34,-.98)}/8
}
\node at \p {\(\lab\)};

\fill (0,0) circle (1.8pt);
\node at (-0.44,0) {\(k_0\)};

\end{tikzpicture}

\caption{The eight local arcs \(\Gamma_j^{(\ell)}\) near \(k_0\).}
\label{fig:elliptic-k0-eight-rays}

\end{figure}
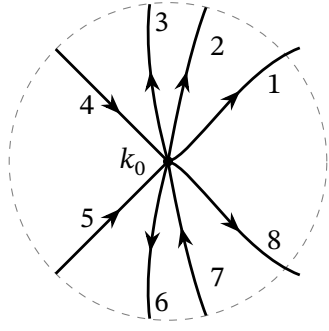

First, choose a sufficiently small disk \(D_\epsilon(k_0)\) centered at
\(k_0\). For an appropriate choice of \(\epsilon\), the intersection of
the jump contour with this disk consists of eight local arcs. We relabel
these arcs and write
\(
\Gamma^{(\ell)}
=
\bigcup_{j=1}^{8}\Gamma_j^{(\ell)}
\);
see Figure~\ref{fig:elliptic-k0-eight-rays}. For convenience, let
\(V^{(\ell)}\) denote the restriction of \(\widehat V^{(7)}\) to
\(\Gamma^{(\ell)}\), and let \(V_j^{(\ell)}\) denote its restriction to
\(\Gamma_j^{(\ell)}\). For later reference, we list the corresponding
jump matrices explicitly:
\begin{equation}\label{eq:elliptic-k0-eight-jumps}
\begin{aligned}
 V_1^{(\ell)}=V_3^{(\ell)}
 &=\begin{pmatrix}
 1&0&0\\
 -\dfrac{\widehat\delta_1^2\widehat\delta_2}{r_1}
       e^{2\mathrm i\widehat g}e^{h_{31}}&1&0\\0&0&1
 \end{pmatrix},\qquad
 V_6^{(\ell)}=V_8^{(\ell)}
 =\begin{pmatrix}
 1&-\dfrac{e^{-2\mathrm i\widehat g}e^{h_{13}}}
               {\bar r_1\widehat\delta_1^2\widehat\delta_2}&0\\
 0&1&0\\0&0&1
 \end{pmatrix},\\[2mm]
 V_2^{(\ell)}=V_7^{(\ell)}
 &=\begin{pmatrix}
 0&-e^{-\mathrm i(\Omega t-\omega)}&0\\
 e^{\mathrm i(\Omega t-\omega)}&0&0\\0&0&1
 \end{pmatrix},\qquad
 V_4^{(\ell)}
 =\begin{pmatrix}
 1&0&0\\
 -\dfrac{\bar r_1}{1+r_1\bar r_1}
   \widehat\delta_1^2\widehat\delta_2e^{2\mathrm i\widehat g}e^{h_{31}}
   &1&0\\0&0&1
 \end{pmatrix},\\[2mm]
 V_5^{(\ell)}
 &=\begin{pmatrix}
 1&-\dfrac{r_1}{1+r_1\bar r_1}
   \widehat\delta_1^{-2}\widehat\delta_2^{-1}
   e^{-2\mathrm i\widehat g}e^{h_{13}}&0\\
 0&1&0\\0&0&1
 \end{pmatrix}.
\end{aligned}
\end{equation}
We reduce the local jumps listed above to those of a solvable
parabolic-cylinder model in several steps.

\begin{figure}
\centering

\begin{tikzpicture}[
  scale=1.05,
  >=Stealth,
  c/.style={
    line width=1.05pt
  },
  dir/.style={
    postaction={decorate},
    decoration={
      markings,
      mark=at position .58 with {\arrow{>}}
    }
  }
]

\draw[dashed,gray] (0,0) circle (2.0);


\draw[c,dir]
  (0,0)
  .. controls (.30,.06) and (.92,1.15) ..
  (1.66,1.43);

\draw[c,dashed]
  (0,0)
  .. controls (.16,.90) and (.34,1.58) ..
  (.48,1.94);

\draw[c,dir]
  (0,0)
  .. controls (-.22,.79) and (-.29,1.52) ..
  (-.23,1.98);


\draw[c,dir]
  (-1.414,1.414)--(0,0);     

\draw[c,dir]
  (-1.414,-1.414)--(0,0);    


\draw[c,dir]
  (0,0)
  .. controls (-.22,-.79) and (-.29,-1.52) ..
  (-.23,-1.98);

\draw[c,dashed]
  (.48,-1.94)
  .. controls (.34,-1.58) and (.16,-.90) ..
  (0,0);

\draw[c,dir]
  (0,0)
  .. controls (.30,-.06) and (.92,-1.15) ..
  (1.66,-1.43);


\foreach \p/\lab in {
  {(1.34,.98)}/1,
  {(-.08,1.76)}/3,
  {(-1.02,0.72)}/4,
  {(-1.02,-0.72)}/5,
  {(-.08,-1.76)}/6,
  {(1.34,-.98)}/8
}
\node at \p {\(\lab\)};

\fill (0,0) circle (1.8pt);
\node at (-0.44,0) {\(k_0\)};

\end{tikzpicture}

\caption{The six local arcs \(\widetilde{\Gamma}_j^{(\ell)}\) at \(k_0\).}
\label{fig:elliptic-k0-six-rays}

\end{figure}
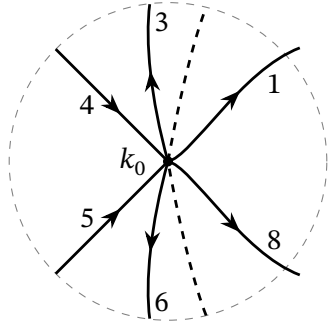

First, we eliminate the jumps on
\(\Gamma_2^{(\ell)}\) and \(\Gamma_7^{(\ell)}\). These arcs are the
local portions of the band \(\widetilde B\), and hence their jumps
are not part of the standard parabolic-cylinder model.  We introduce the following piecewise transformation
\begin{equation}
 \widetilde N^{(\ell)}(x,t,k)
       =\widehat N^{(7)}(x,t,k)C_{\mathrm r}(k),\qquad
 C_{\mathrm r}(k)=
 \begin{cases}
 I,&\text{on the \(+\) side of the band $\widetilde B$ in }D_\epsilon(k_0),\\
 V_2^{(\ell)},&\text{on the \(-\) side of the band $\widetilde B$ in }D_\epsilon(k_0).
 \end{cases}
 \label{eq:elliptic-k0-band-removal}
\end{equation}
This transformation removes the jumps on
\(\Gamma_2^{(\ell)}\) and \(\Gamma_7^{(\ell)}\). Consequently,
\(\widetilde N^{(\ell)}\) has jumps only on the six remaining local arcs
inside \(D_\epsilon(k_0)\). We denote their union by
\(\widetilde\Gamma^{(\ell)}\); see
Figure~\ref{fig:elliptic-k0-six-rays}, and denote the corresponding jump
matrix by \(\widetilde V^{(\ell)}\).
In addition to removing the jumps on
\(\Gamma_2^{(\ell)}\) and \(\Gamma_7^{(\ell)}\), the transformation
modifies the jumps on \(\Gamma_1^{(\ell)}\) and
\(\Gamma_8^{(\ell)}\). More precisely, the jump matrices on these two
arcs become
\begin{subequations}\label{eq:elliptic-k0-conjugated-jumps}
\begin{align}
 \widetilde V_1^{(\ell)}
 &=(V_2^{(\ell)})^{-1}V_1^{(\ell)}V_2^{(\ell)}
 =\begin{pmatrix}
 1&\dfrac{\widehat\delta_1^2\widehat\delta_2}{r_1}
 e^{2\mathrm i\widehat g}e^{h_{31}}
 e^{-2\mathrm i(\Omega t-\omega)}&0\\
 0&1&0\\0&0&1
 \end{pmatrix},\\[2mm]
 \widetilde V_8^{(\ell)}
 &=(V_2^{(\ell)})^{-1}V_8^{(\ell)}V_2^{(\ell)}
 =\begin{pmatrix}
 1&0&0\\
 \dfrac{e^{-2\mathrm i\widehat g}}
 {\bar r_1\widehat\delta_1^2\widehat\delta_2}
 e^{h_{13}}e^{2\mathrm i(\Omega t-\omega)}&1&0\\
 0&0&1
 \end{pmatrix}.
\end{align}
\end{subequations}
The other jumps are unchanged 

\begin{figure}
\centering

\begin{tikzpicture}[
  scale=1.05,
  >=Stealth,
  c/.style={
    line width=1.05pt
  },
  dir/.style={
    postaction={decorate},
    decoration={
      markings,
      mark=at position 0.58 with {\arrow{>}}
    }
  }
]

\draw[dashed,gray] (0,0) circle (2.0);

\draw[c,dir] (-2.0,0)--(0,0);

\draw[c,dir]
  (0,0)
  .. controls (0,0.92) and (0.18,1.56) ..
  (0.48,1.94);

\draw[c,dir]
  (0.48,-1.94)
  .. controls (0.18,-1.56) and (0,-0.92) ..
  (0,0);

\node at (.68,2.22) {\(\Gamma^{(\ell)}_2\)};
\node at (.68,-2.22) {\(\Gamma^{(\ell)}_7\)};

\node at (.68,0.02) {\(\mathcal D_R\)};
\node at (-.68,0.92) {\(\mathcal D_U\)};
\node at (-.68,-0.92) {\(\mathcal D_L\)};
\fill (0,0) circle (1.8pt);
\node at (-0.34,-0.3) {\(k_0\)};

\end{tikzpicture}

\caption{ The regions \(\mathcal D_U\), \(\mathcal D_L\) and \(\mathcal D_R\).}
\label{fig:elliptic-k0-two-arcs}

\end{figure}
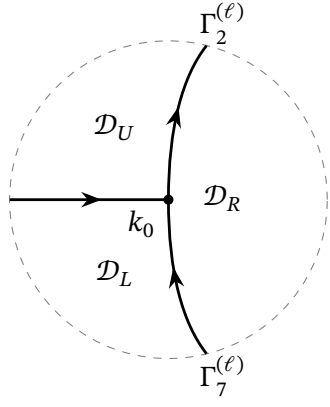

The next step is to introduce a local coordinate in
\(D_\epsilon(k_0)\) and then use a suitable conjugation to reduce the
local jumps to a solvable model RH problem formulated in this coordinate.
 To this end,  introduce the scaled coordinate
\begin{equation}
 \zeta_3=\sqrt{-2\phi_0''(k_0)}\,\sqrt t\,(k-k_0)
   =:\sqrt t\,\widehat\gamma_1(k-k_0),
 \qquad k\in D_\epsilon(k_0),
 \label{eq:elliptic-k0-scaled-coordinate}
\end{equation}
where
\begin{equation}
 \phi_0(k)=
 \begin{cases}
 \hbar(k)-\hbar^+(k_0),&k\in\mathcal D_U\cup\mathcal D_L,\\
 -\bigl(\hbar(k)-\hbar^-(k_0)\bigr),&k\in\mathcal D_R.
 \end{cases}
 \label{eq:elliptic-k0-glued-phase}
\end{equation}
Here $\widehat\gamma_1>0$ and
the regions \(\mathcal D_U\cup\mathcal D_L\) and \(\mathcal D_R\) are
shown in Figure~\ref{fig:elliptic-k0-two-arcs}. They represent the
portions of \(D_\epsilon(k_0)\) lying on the two sides of the band
\(\widetilde B\).
In terms of the scaled coordinate
\eqref{eq:elliptic-k0-scaled-coordinate}, the local expansion of the
phase $h_{31}$ takes the form
\begin{equation}
 h_{31}(k)
 =\begin{cases}
 h_{31}^+(k_0)-\frac{\mathrm i \zeta_3^2}{2}+O(t|k-k_0|^3), \qquad k \in \mathcal D_U\cup\mathcal D_L.\\[2mm]
  h_{31}^-(k_0)+\frac{\mathrm i \zeta_3^2}{2}+O(t|k-k_0|^3), \qquad k \in \mathcal D_R.
 \end{cases}
 \label{eq:elliptic-k0-quadratic-phase}
\end{equation}

As in the previous local
constructions, we introduce a conjugating matrix \(Y_{k_0}\) to reduce
these jumps to those of a solvable  model problem. The construction of
\(Y_{k_0}\) relies on a local factorization of
\(\widehat\delta_1e^{\mathrm{i}\widehat g}\), or equivalently, on a
suitable logarithmic representation of this product. Some care is
required because both \(\widehat\delta_1\) and
\(e^{\mathrm{i}\widehat g}\) may exhibit oscillatory behavior near
\(k_0\). We now present a constructive lemma that provides a suitable local
factorization of
\(\widehat\delta_1e^{\mathrm{i}\widehat g}\).

Define $\nu_3=-\frac1{2\pi}\log\!\left(1+\frac1{|r_1(k_0)|^2}\right)$<0, and
\begin{equation}
  \mathcal S_0(k)=
 \exp\left\{-\frac1{2\pi\mathrm i}
 \int_{-\infty}^{k_0}\log\mathcal G_0(s)
 \left(\frac1{s-k}-\frac1{s-k_0}\right)\,\mathrm ds\right\},
 \label{eq:elliptic-S-zero}
\end{equation}
where
\begin{equation*}
 \mathcal G_0(k)=
 \left(\frac{1+r_1(k)\bar r_1(k)}{r_1(k)\bar r_1(k)}\right)
 \frac{|r_1(k_0)|^2}{1+|r_1(k_0)|^2}.
\end{equation*}
The assumption \(r_1(k_0)\ne0\) is important here, as it
allows us to choose \(\epsilon>0\) sufficiently small
to ensure that \(\mathcal G_0(k)\) is well defined and nonzero.
Then we have the following lemma.
\begin{lemma}
There exists a function \(\mathcal F_0(k)\), analytic in
\(D_\epsilon(k_0)\), such that
\(\widehat\delta_1e^{\mathrm i\widehat g}\) admits the following local
factorization:
\begin{equation}
 \widehat\delta_1e^{\mathrm i\widehat g}=
 \begin{cases}
 (r_1 (k)\bar r_1(k))^{1/2} (k-k_0)^{\mathrm i\nu_3}
 \mathcal F_0 (k) \mathcal S_0^{-1}(k),
 &k\in\mathcal D_U,\\[3mm]
 (r_1(k)\bar r_1(k))^{-1/2} (k-k_0)^{\mathrm i\nu_3}
 \mathcal F_0 (k)\mathcal S_0^{-1} (k),
 &k\in\mathcal D_L,\\[3mm]
 \left(\dfrac{r_1(k)}{\bar r_1(k)}\right)^{1/2}
 (k-k_0)^{-\mathrm i\nu_3}  e^{-\mathrm i\omega}
 \mathcal F_0^{-1}(k)\mathcal S_0(k)\widehat\delta_2^{-1}(k),
 &k\in\mathcal D_R,
 \end{cases}
 \label{eq:elliptic-delta-g-local-factorization}
\end{equation}
where the regions \(\mathcal D_U\), \(\mathcal D_L\) and \(\mathcal D_R\) are
shown in Figure~\ref{fig:elliptic-k0-two-arcs}.
\end{lemma}

\begin{proof}
We prove the lemma by explicitly constructing \(\mathcal F_0\).
We first define
\begin{equation}
 \mathcal Q_0(k)=
 \begin{cases}
 \dfrac{\widehat\delta_1e^{\mathrm i\widehat g}}
       {(r_1\bar r_1)^{1/2}(k-k_0)^{\mathrm i\nu_3}},
 &k\in\mathcal D_U,\\[4mm]
 \dfrac{(r_1\bar r_1)^{1/2}\widehat\delta_1e^{\mathrm i\widehat g}}
       {(k-k_0)^{\mathrm i\nu_3}},
 &k\in\mathcal D_L,\\[4mm]
 \left(\dfrac{r_1}{\bar r_1}\right)^{1/2}
 e^{-\mathrm i\omega}\widehat\delta_1^{-1}e^{-\mathrm i\widehat g}
 (k-k_0)^{-\mathrm i\nu_3}\widehat\delta_2^{-1},
 &k\in\mathcal D_R.
 \end{cases}
 \label{eq:elliptic-Q-zero}
\end{equation}
The jump relations for \(\widehat g\) in
\eqref{eq:elliptic-ghat-RHP} imply that
\(\mathcal Q_0^+=\mathcal Q_0^-\) for
\(k\in\Gamma^{(\ell)}_2\cup\Gamma^{(\ell)}_7\).
Moreover, combining the jump relation for \(\widehat\delta_1\) with
the definition of \(\nu_3\), we obtain
\begin{equation}
 \mathcal Q_0^+=\mathcal Q_0^-\mathcal G_0,
 \qquad k\in(-\infty,k_0)\cap D_\epsilon(k_0).
 \label{eq:elliptic-Q-zero-real-jump}
\end{equation}
The integral representation \eqref{eq:elliptic-S-zero} shows that
\(\mathcal S_0\) is analytic in
\(\mathbb C\setminus(-\infty,k_0]\) and satisfies
\begin{equation}
 \mathcal S_0^+(k)=\mathcal S_0^-(k)\mathcal G_0(k)^{-1},
 \qquad k\in(-\infty,k_0).
 \label{eq:elliptic-S-zero-properties}
\end{equation}
This suggests defining
\begin{equation}
 \mathcal F_0(k)=\mathcal Q_0(k)\mathcal S_0(k),
 \qquad k\in D_\epsilon(k_0).
 \label{eq:elliptic-F-zero}
\end{equation}
The jumps of \(\mathcal Q_0\) and \(\mathcal S_0\) cancel on
\((-\infty,k_0)\cap D_\epsilon(k_0)\).
Since both functions are bounded near \(k_0\), the possible singularity
of \(\mathcal F_0\) at \(k_0\) is removable.
Thus \(\mathcal F_0\) is analytic in \(D_\epsilon(k_0)\).
Finally, substituting
\(\mathcal Q_0=\mathcal F_0\mathcal S_0^{-1}\) into
\eqref{eq:elliptic-Q-zero} and rearranging yields the local
factorization \eqref{eq:elliptic-delta-g-local-factorization}.
\end{proof}

\begin{remark}
The local factorization
\eqref{eq:elliptic-delta-g-local-factorization}
separates the oscillatory factors from those with well-defined limits.
More precisely, the factors \((k-k_0)^{\pm\mathrm i\nu_3}\) oscillate
near \(k_0\), whereas all remaining factors have well-defined limits
as \(k\to k_0\) nontangentially to the real axis.
This is immediate for all factors except \(\mathcal S_0(k)\).
For this function, the integral representation
\eqref{eq:elliptic-S-zero} and the endpoint behavior of Cauchy-type
integrals yield
\begin{equation}  \label{E:estofmathcals0}
 \bigl|\mathcal S_0(k)-1\bigr|
 \leq C|k-k_0|\,\bigl|\log|k-k_0|\bigr|
\end{equation}
as \(k\to k_0\) nontangentially.
In particular, \(\mathcal S_0(k)\to1\).
\end{remark}

Substituting the local factorization
\eqref{eq:elliptic-delta-g-local-factorization} and the scaled coordinate
\eqref{eq:elliptic-k0-scaled-coordinate} into the jump
matrix \(\widetilde V^{(\ell)}\), we can rewrite it as follows:
\begin{subequations}\label{eq:elliptic-six-ray-scalar-jumps}
\begin{align}
\widetilde V_3^{(\ell)}&=
 \begin{pmatrix}1&0&0\\
 -\bar r_1 \zeta_3^{2\mathrm i\nu_3}
 (\widehat\gamma_1\sqrt t)^{-2\mathrm i\nu_3}
 \mathcal F_0^2\mathcal S_0^{-2}\widehat\delta_2e^{h_{31}}&1&0\\
 0&0&1\end{pmatrix},\\
 \widetilde V_4^{(\ell)}&=
 \begin{pmatrix}1&0&0\\
 -\dfrac{r_1\bar r_1^2}{1+r_1\bar r_1} \zeta_3^{2\mathrm i\nu_3}
 (\widehat\gamma_1\sqrt t)^{-2\mathrm i\nu_3}
 \mathcal F_0^2\mathcal S_0^{-2}\widehat\delta_2e^{h_{31}}&1&0\\
 0&0&1\end{pmatrix},\\
 \widetilde V_5^{(\ell)}&=
 \begin{pmatrix}
 1&-\dfrac{r_1^2\bar r_1}{1+r_1\bar r_1}\zeta_3^{-2\mathrm i\nu_3}
 (\widehat\gamma_1\sqrt t)^{2\mathrm i\nu_3}
 \mathcal F_0^{-2}\mathcal S_0^2\widehat\delta_2^{-1}e^{h_{13}}&0\\
 0&1&0\\0&0&1
 \end{pmatrix},\\
\widetilde V_6^{(\ell)}&=
 \begin{pmatrix}
 1&-r_1 \zeta_3^{-2\mathrm i\nu_3}
 (\widehat\gamma_1\sqrt t)^{2\mathrm i\nu_3}
 \mathcal F_0^{-2}\mathcal S_0^2\widehat\delta_2^{-1}e^{h_{13}}&0\\
 0&1&0\\0&0&1
 \end{pmatrix},\\
\widetilde  V_1^{(\ell)}&=
 \begin{pmatrix}
 1&\dfrac1{\bar r_1}\zeta_3^{-2\mathrm i\nu_3}
 (\widehat\gamma_1\sqrt t)^{2\mathrm i\nu_3}
 \mathcal F_0^{-2}\mathcal S_0^2\widehat\delta_2^{-1}
 e^{h_{31}-2\mathrm i\Omega t}&0\\
 0&1&0\\0&0&1
 \end{pmatrix},\\
 \widetilde V_8^{(\ell)}&=
 \begin{pmatrix}1&0&0\\
 \dfrac1{r_1}\zeta_3^{2\mathrm i\nu_3}
 (\widehat\gamma_1\sqrt t)^{-2\mathrm i\nu_3}
 \mathcal F_0^2\mathcal S_0^{-2}\widehat\delta_2
 e^{h_{13}+2\mathrm i\Omega t}&1&0\\
 0&0&1
 \end{pmatrix}.
\end{align}
\end{subequations}


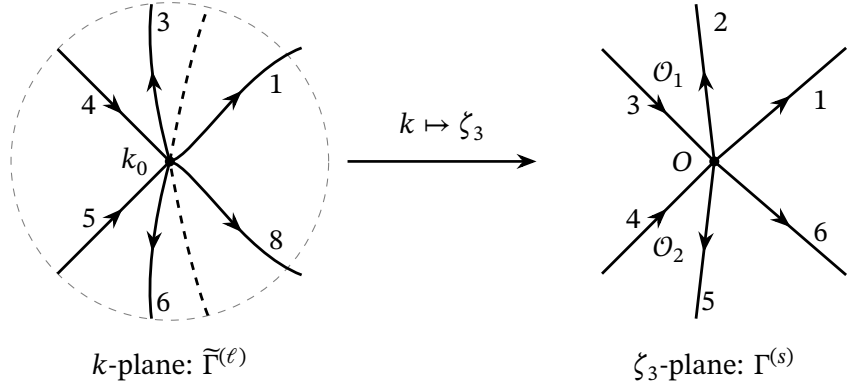
\begin{figure}
\centering

\begin{tikzpicture}[
  >=Stealth,
  c/.style={
    line width=1.05pt
  },
  dir/.style={
    postaction={decorate},
    decoration={
      markings,
      mark=at position .58 with {\arrow{>}}
    }
  }
]

\begin{scope}[xshift=-3.6cm,scale=1.05]

\draw[dashed,gray] (0,0) circle (2.0);


\draw[c,dir]
  (0,0)
  .. controls (.30,.06) and (.92,1.15) ..
  (1.66,1.43);

\draw[c,dashed]
  (0,0)
  .. controls (.16,.90) and (.34,1.58) ..
  (.48,1.94);

\draw[c,dir]
  (0,0)
  .. controls (-.22,.79) and (-.29,1.52) ..
  (-.23,1.98);


\draw[c,dir]
  (-1.414,1.414)--(0,0);     

\draw[c,dir]
  (-1.414,-1.414)--(0,0);    


\draw[c,dir]
  (0,0)
  .. controls (-.22,-.79) and (-.29,-1.52) ..
  (-.23,-1.98);

\draw[c,dashed]
  (.48,-1.94)
  .. controls (.34,-1.58) and (.16,-.90) ..
  (0,0);

\draw[c,dir]
  (0,0)
  .. controls (.30,-.06) and (.92,-1.15) ..
  (1.66,-1.43);


\foreach \p/\lab in {
  {(1.34,.98)}/1,
  {(-.08,1.76)}/3,
  {(-1.02,0.72)}/4,
  {(-1.02,-0.72)}/5,
  {(-.08,-1.76)}/6,
  {(1.34,-.98)}/8
}
\node at \p {\(\lab\)};

\fill (0,0) circle (1.8pt);
\node at (-0.44,0) {\(k_0\)};

\node at (0,-2.55)
  {\(k\)-plane: $\widetilde{\Gamma}^{(\ell)}$};

\end{scope}


\draw[->,line width=1pt]
  (-1.25,0)--(1.25,0)
  node[midway,above=4pt] {\(k\mapsto \zeta_3\)};

\begin{scope}[xshift=3.6cm,scale=1.05]


\draw[c,dir]
  (0,0)--(1.66,1.43);


\draw[c,dir]
  (0,0)--(-.23,1.98);


\draw[c,dir]
  (-1.414,1.414)--(0,0);     

\draw[c,dir]
  (-1.414,-1.414)--(0,0);    


\draw[c,dir]
  (0,0)--(-.23,-1.98);


\draw[c,dir]
  (0,0)--(1.66,-1.43);


\foreach \p/\lab in {
  {(1.34,.78)}/1,
  {(-0.4,0)}/O,
  {(.08,1.76)}/2,
  {(-1.02,0.72)}/3,
  {(-1.02,-0.72)}/4,
  {(-.08,-1.76)}/5,
  {(1.34,-.88)}/6
}
\node at \p {\(\lab\)};
\node at (-0.58,1.1) {\(\mathcal O_1\)};
\node at (-0.58,-1.1) {\(\mathcal O_2\)};
\fill (0,0) circle (1.8pt);

\node at (0,-2.55)
  {\(\zeta_3\)-plane: $\Gamma^{(s)}$};

\end{scope}

\end{tikzpicture}

\caption{
The local contour $\widetilde{\Gamma}^{(\ell)}$ near \(k_0\) in the \(k\)-plane and the corresponding
six-ray contour $\Gamma^{(s)}$  in the \(\zeta_3\)-plane.
}
\label{fig:elliptic-k0-six-ray-map-to}

\end{figure}

The expressions above suggest defining the conjugating matrix
\(Y_{k_0}\) by
\begin{equation}
 \begin{aligned}
 Y_{k_0}
 &=\diag\bigl(\mathfrak D_0,\mathfrak D_0^{-1},1\bigr),\\
 \mathfrak D_0
 &=(\widehat\gamma_1\sqrt t)^{\mathrm i\nu_0}
 \mathcal F_0(k_0)^{-1}\widehat\delta_2(k_0)^{-1/2}
 e^{-\frac{1}{2}h_{31}^+(k_0)}.
 \end{aligned}
 \label{eq:elliptic-k0-Y-zero-MS}
\end{equation}
A straightforward algebraic calculation then shows that
\(Y_{k_0}^{-1}\widetilde V_j^{(\ell)}Y_{k_0}
\to\widetilde V^{\mathrm{PC}}_{\pi(j)}\)
as \(t\to\infty\), where
\begin{equation*}
 (\pi(1),\pi(3),\pi(4),\pi(5),\pi(6),\pi(8))=(1,2,3,4,5,6).
\end{equation*}
Here, the jump matrix \(\widetilde V^{\mathrm{PC}}\) is given by
\begin{equation}\label{eq:elliptic-k0-PC-frozen-jumps}
\begin{aligned}
 \widetilde  V_1^{\mathrm{PC}}(\zeta_3)&=
 \begin{pmatrix}
 1&\dfrac1{\bar r_1(k_0)}\zeta_3^{-2\mathrm i\nu_3}e^{\mathrm i \zeta_3^2/2}&0\\
 0&1&0\\0&0&1\end{pmatrix}, \qquad
 \widetilde  V_2^{\mathrm{PC}}(\zeta_3)=
 \begin{pmatrix}1&0&0\\
 -\bar r_1(k_0)\zeta_3^{2\mathrm i\nu_3}e^{-\mathrm i\zeta_3^2/2}&1&0\\
 0&0&1\end{pmatrix},\\[2mm]
  \widetilde  V_3^{\mathrm{PC}}(\zeta_3)&=
 \begin{pmatrix}1&0&0\\
 -\dfrac{\bar r_1(k_0)|r_1(k_0)|^2}{1+|r_1(k_0)|^2}
 \zeta_3^{2\mathrm i\nu_3}e^{-\mathrm i \zeta_3^2/2}&1&0\\
 0&0&1\end{pmatrix},\quad
  \widetilde  V_4^{\mathrm{PC}}(\zeta_3)=
 \begin{pmatrix}
 1&-\dfrac{r_1(k_0)|r_1(k_0)|^2}{1+|r_1(k_0)|^2}
 \zeta_3^{-2\mathrm i\nu_3}e^{\mathrm i\zeta_3^2/2}&0\\
 0&1&0\\0&0&1\end{pmatrix},\\[2mm]
  \widetilde  V_5^{\mathrm{PC}}(\zeta_3)&=
 \begin{pmatrix}
 1&-r_1(k_0) \zeta_3^{-2\mathrm i\nu_3}e^{\mathrm i \zeta_3^2/2}&0\\
 0&1&0\\0&0&1\end{pmatrix},\qquad
 \widetilde   V_6^{\mathrm{PC}}(\zeta_3)=
 \begin{pmatrix}1&0&0\\
 \dfrac1{r_1(k_0)} \zeta_3^{2\mathrm i\nu_3}e^{-\mathrm i \zeta_3^2/2}&1&0\\
 0&0&1\end{pmatrix}.
\end{aligned}
\end{equation}

For each fixed \(k\in\widetilde{\Gamma}^{(\ell)}\setminus\{k_0\}\), we have
\(|\zeta_3(k)|\to\infty\) as \(t\to\infty\).
After a simple deformation if necessary, the image of the local
contour \(\widetilde \Gamma^{(\ell)}\) in the \(\zeta_3\)-plane may be taken
to consist of straight rays. We denote this contour by
\(\Gamma^{(s)}=\bigcup_{j=1}^6\Gamma_j^{(s)}\); see Fig. \ref{fig:elliptic-k0-six-ray-map-to}.
The preceding discussion motivates the following model RH problem
in the \(\zeta_3\)-plane:

\begin{RHP}[Model RH problem in the $\zeta_3$-plane]
Given the oriented contour $\Gamma^{(s)}$ and the jump matrix
$\widetilde V^{\mathrm{PC}}(\zeta_3)$, find a $3\times3$ matrix-valued function
$\mathcal P_0(\zeta_3)$ satisfying the following conditions:

\begin{enumerate}

\item[(i)] \emph{Analyticity.}
The function $\mathcal P_0(\zeta_3)$ is analytic in
\(
\mathbb C\setminus\Gamma^{(s)}
\),
and admits continuous non-tangential boundary values
$\mathcal P_0^\pm(\zeta_3)$ on 
$\Gamma^{(s)}$.

\item[(ii)] \emph{Jump condition.}
The boundary values satisfy
\begin{equation}
\mathcal P_0^+(\zeta_3)
=
\mathcal P_0^-(\zeta_3)
\widetilde V^{\mathrm{PC}}(\zeta_3),
\qquad
\zeta_3\in\Gamma^{(s)}.
\label{eq:elliptic-k0-PC-jump}
\end{equation}

 \item[(iii)] \emph{Behavior at the origin.}
 The function \(\mathcal P_0(\zeta_3)\) remains bounded as
 \(\zeta_3\to0\).

\item[(iv)] \emph{Normalization.}
As $\zeta_3\to\infty$,
\begin{equation}
\mathcal P_0(\zeta_3)
=
I+O(\zeta_3^{-1}).
\label{eq:elliptic-k0-PC-normalization}
\end{equation}

\end{enumerate}
\end{RHP}

\begin{figure}
\centering
\begin{tikzpicture}[>=Stealth,
 dir/.style={postaction={decorate},decoration={markings,
 mark=at position .60 with {\arrow{>}}}}]
 \foreach \ang/\lab in {45/1,315/4}{
  \draw[thick,dir] (0,0)--(\ang:1.65);
  \node at (\ang:1.9) {\(\lab\)};
 }
 \foreach \ang/\lab in {135/2,225/3}{
  \draw[thick,dir] (\ang:1.65)--(0,0);
  \node at (\ang:1.9) {\(\lab\)};
 }
\end{tikzpicture}
\caption{The jump contour for \(\widehat{\mathcal P}_0\).}
\label{fig:elliptic-k0-PC-four-rays}
\end{figure}
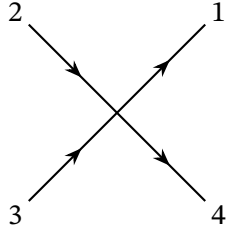

A key observation is that a simple transformation reduces the six-ray
model RH problem to a four-ray problem whose solution can be expressed
in terms of the standard parabolic cylinder model.
Let \(\mathcal O_1\) and \(\mathcal O_2\) denote the regions marked
in the right panel of Figure~\ref{fig:elliptic-k0-six-ray-map-to}. We define the transformation as follows:

\begin{equation}
 \widehat{\mathcal P}_0(\zeta_3)=\mathcal P_0(\zeta_3)
 \begin{cases}
  \bigl(\widetilde V_2^{\mathrm{PC}}(\zeta_3)  \bigl)^{-1},&\zeta_3\in\mathcal O_1,\\
 \widetilde V_5^{\mathrm{PC}}(\zeta_3),
 &\zeta_3\in\mathcal O_2,\\
 I,&\text{elsewhere}.
 \end{cases}
 \label{eq:elliptic-k0-PC-four-ray-reduction}
\end{equation}
The jump contour for \(\widehat{\mathcal P}_0\) is shown in
Figure~\ref{fig:elliptic-k0-PC-four-rays}.
The corresponding jump matrices
\(\{\widehat V_j^{\mathrm{PC}}\}_{j=1}^4\) are given by
\begin{equation}\label{eq:elliptic-k0-PC-four-jumps}
\begin{aligned}
 \widehat V_1^{\mathrm{PC}}&=\widetilde V_1^{\mathrm{PC}},
 \qquad \widehat V_4^{\mathrm{PC}}=\widetilde V_6^{\mathrm{PC}},\\
 \widehat V_2^{\mathrm{PC}}&=
 \begin{pmatrix}
 1&0&0\\
 \dfrac{\bar r_1(k_0)}{1+|r_1(k_0)|^2}
 \zeta_3^{2\mathrm i\nu_3}e^{-\mathrm i\zeta_3^2/2}&1&0\\
 0&0&1
 \end{pmatrix},\\[2mm]
 \widehat V_3^{\mathrm{PC}}&=
 \begin{pmatrix}
 1&\dfrac{r_1(k_0)}{1+|r_1(k_0)|^2}
 \zeta_3^{-2\mathrm i\nu_3}e^{\mathrm i\zeta_3^2/2}&0\\
 0&1&0\\
 0&0&1
 \end{pmatrix}.
\end{aligned}
\end{equation}
\par
\par

Let \(N^{\mathrm{PC}}(\zeta;p,q)\) denote the solution of the standard
parabolic cylinder model RH problem in Appendix~\ref{App:AAA}.
A direct calculation shows that the block-diagonal matrix
\begin{equation}
 \widehat{\mathcal P}_0(\zeta_3)=
 \begin{pmatrix}
 N^{\mathrm{PC}}\!
 \left(\zeta_3;\dfrac{1}{\bar r_1(k_0)},-\dfrac{1}{r_1(k_0)}\right)
 &\mathbf 0_{2\times1}\\[3mm]
 \mathbf 0_{1\times2}&1
 \end{pmatrix}
 \label{eq:elliptic-k0-PC-standard-embedding}
\end{equation}
satisfies the jump relations \eqref{eq:elliptic-k0-PC-four-jumps}.
Therefore, the solution of the six-ray model RH problem is given by
\begin{equation*}
 \mathcal P_0(\zeta_3)=
 \begin{pmatrix}
 N^{\mathrm{PC}}\!
 \left(\zeta_3;\dfrac{1}{\bar r_1(k_0)},-\dfrac{1}{r_1(k_0)}\right)
 &\mathbf 0_{2\times1}\\[3mm]
 \mathbf 0_{1\times2}&1
 \end{pmatrix} \times
 \begin{cases}
 V_2^{\mathrm{PC}}(\zeta_3),
 &\zeta_3\in\mathcal O_1,\\[2mm]
 \bigl(V_5^{\mathrm{PC}}(\zeta_3)\bigr)^{-1},
 &\zeta_3\in\mathcal O_2,\\[2mm]
 I,&\text{elsewhere}.
 \end{cases}
\end{equation*}

It remains to determine the asymptotic expansion of
\(\mathcal P_0(\zeta_3)\), the unique solution of the six-ray model
RH problem, as \(\zeta_3\to\infty\).
Since the factors relating \(\mathcal P_0\) and
\(\widehat{\mathcal P}_0\) tend exponentially to the identity
in their respective sectors, the expansion \eqref{E:mXasymoxia}
of the standard parabolic cylinder model yields
\begin{equation}
 \mathcal P_0(\zeta_3)=I+\frac{\mathcal C_0}{\zeta_3}+O(\zeta_3^{-2}),
 \qquad  \zeta_3 \to\infty,
 \label{eq:elliptic-k0-PC-expansion}
\end{equation}
uniformly for \(\arg \zeta_3\in[0,2\pi]\), where the  coefficient matrix $\mathcal C_0$ is
\begin{equation}
 \mathcal C_0=
 \begin{pmatrix}
 0&-\dfrac{\sqrt{2\pi}\,r_1(k_0)e^{-\pi\nu_3/2}e^{\pi\mathrm i/4}}
             {\Gamma(\mathrm i\nu_3)}&0\\[2mm]
 \dfrac{\sqrt{2\pi}\,\bar r_1(k_0)e^{-\pi\nu_3/2}
                     e^{-\pi\mathrm i/4}}{\Gamma(-\mathrm i\nu_3)}
 &0&0\\
 0&0&0
 \end{pmatrix}.
 \label{eq:elliptic-k0-PC-first-coefficient-MS}
\end{equation}

Recall that the local analysis in \(D_\epsilon(k_0)\) led to a
solvable model RH problem for \(\mathcal P_0\).
This  suggests defining the local parametrix near \(k_0\) by
\begin{equation}
 N^{\mathrm{loc},k_0}(x,t,k)
 =N^{\mathrm{out}}(x,t,k)C_{\mathrm r}(k) Y_{k_0}\mathcal P_0(\zeta_3(k))
 Y_{k_0}^{-1}  C_{\mathrm r}^{-1}(k),\qquad k \in D_\epsilon(k_0).
 \label{eq:elliptic-local-k0}
\end{equation}

\begin{remark}
\label{rem:elliptic-outer-Cr-analyticity}
The product \(N^{\mathrm{out}}(k)C_{\mathrm r}(k)\) is analytic and
invertible in \(D_\epsilon(k_0)\).
\end{remark}
\subsubsection{The local model near \(k_1\)}
\label{subsubsec:elliptic-local-k1}
We now construct the local parametrix near \(k_1\). We begin by
rewriting the scalar factor \(\widehat{\delta}_2\). Recall that
\(\widehat{\delta}_2\) admits the integral representation
\begin{equation}
\widehat{\delta}_2(k)
=
\exp\left\{
\frac{1}{2\pi\mathrm{i}}
\int_{-\infty}^{k_1}
\frac{
\log\left(
1+d(s)^{-1}\dfrac{|R(s)|^2}{1+|r_1(s)|^2}
\right)
}{s-k}\,\mathrm{d}s
\right\}.
\label{eq:elliptic-delta2-Cauchy-explicit}
\end{equation}
As before, integration by parts yields
\begin{equation}
\widehat{\delta}_2(k)
=
(k-k_1)^{\mathrm{i}\nu_4}e^{-\eta_4(k)},
\label{eq:elliptic-delta2-k1-endpoint}
\end{equation}
where
\begin{align}
\nu_4
&=
-\frac{1}{2\pi}
\log\left(
1+d(k_1)^{-1}
\frac{|R(k_1)|^2}{1+|r_1(k_1)|^2}
\right)
<0,
\\
\eta_4(k)
&=
\frac{1}{2\pi\mathrm{i}}
\int_{-\infty}^{k_1}
\log(k-s)\,
\mathrm{d}\log\left(
1+d(s)^{-1}
\frac{|R(s)|^2}{1+|r_1(s)|^2}
\right).
\label{eq:elliptic-nu3-eta3}
\end{align}
Moreover, as \(k\to k_1\), the function \(\eta_4\) satisfies
\begin{align} \label{E:estofeth4k}
\left|\eta_4(k)-\eta_4(k_1)\right|
\leq
C|k-k_1|\,\bigl|\log|k-k_1|\bigr|.
\end{align}

Let \(D_\epsilon(k_1)\) be a sufficiently small disk centered at
\(k_1\). We need to introduce a local coordinate in \(D_\epsilon(k_1)\). To this
end, we first expand the phase function \(h_{12}(k)\) near \(k_1\)
\begin{equation*}
h_{12}(x,t,k)
=
h_{12}(x,t,k_1)
+\frac{1}{2}\partial_k^2h_{12}(x,t,k_1)(k-k_1)^2
+O\bigl(t|k-k_1|^3\bigr).
\end{equation*}
For convenience, define \(\widehat \gamma_2=\widehat \gamma_2(\xi)\) by
\begin{align*}
\widehat{\gamma}_2
=-\frac{\mathrm{i}}{t}
\partial_k^2h_{12}(x,t,k_1)=
\frac{
8\bigl(q_0^2+k_0^2-k_0\operatorname{Re}\alpha\bigr)^2
}{
\bigl(k_0^2+(\operatorname{Im}\alpha)^2\bigr)
\bigl(k_1^2+q_0^2\bigr)
}
>0,
\qquad
-4\sqrt{2}\,q_0<\xi<0.
\end{align*}
We then introduce the local coordinate
\begin{equation}
\zeta_4
=
\sqrt{\widehat \gamma_2 t}\,(k-k_1).
\label{eq:elliptic-zeta4}
\end{equation}
In terms of \(\zeta_4\), the local expansion of the phase becomes
\begin{equation}
h_{12}(x,t,k)
=
h_{12}(x,t,k_1)
+\frac{\mathrm{i}}{2}\zeta_4^2
+O\bigl(t|k-k_1|^3\bigr).
\label{eq:elliptic-h12-zeta4}
\end{equation}

We next analyze the scalar factor appearing in the local jump matrices
near \(k_1\). Using \eqref{eq:elliptic-zeta4}, the factor
\(\widehat{\delta}_1\widehat{\delta}_2^2e^{\mathrm{i}\widehat g}\)
can be written as
\begin{equation}
\widehat{\delta}_1(k)\widehat{\delta}_2(k)^2
e^{\mathrm{i}\widehat g(k)}
=
e^{2\mathrm{i}\nu_4\log\zeta_4}\delta_{4,a}
\begin{cases}
\bigl(1+|r_1(k_1)|^2\bigr)\delta_{4,b}^+(k),
&k\in\mathbb{C}_+,\\
\delta_{4,b}^-(k),
&k\in\mathbb{C}_-,
\end{cases}
\label{eq:elliptic-delta4-factorization}
\end{equation}
where
\begin{align}
\delta_{4,a}
&=
\bigl(\sqrt{\widehat \gamma_2 t}\bigr)^{-2\mathrm{i}\nu_4}
e^{-2\eta_4(k_1)}
\widehat{\delta}_1^-(k_1)
e^{\mathrm{i}\widehat g(k_1)},
\\
\delta_{4,b}^{\pm}(k)
&=
\frac{\widehat{\delta}_1(k)}
{\widehat{\delta}_1^\pm(k_1)}
e^{\mathrm{i}(\widehat g(k)-\widehat g(k_1))}
e^{-2(\eta_4(k)-\eta_4(k_1))}.
\label{eq:elliptic-delta4-ab}
\end{align}
Using the estimate \eqref{E:estofeth4k}, we immediately obtain
\begin{equation}
|\delta_{4,b}^{\pm}(k)-1|
\leq
C|k-k_1|\,\bigl|\log|k-k_1|\bigr|,
\qquad
k\in\mathbb{C}_\pm,\quad k\to k_1.
\label{eq:elliptic-delta4-remainder-bound}
\end{equation}
We are now ready to introduce a conjugating matrix that removes the
nonessential scalar factors from the local jumps. Define
\begin{equation}
 Y_{k_1}
=
\begin{pmatrix}
\delta_{4,a}^{-1/2}e^{1/2 h_{12}(k_1)}&0&0\\
0&1&0\\
0&0&\delta_{4,a}^{1/2}e^{1/2 h_{21}(k_1)}
\end{pmatrix}.
\label{eq:elliptic-Y-k1-MS}
\end{equation}
Then, as \(t\to\infty\) with \(\zeta_4\) fixed, the four conjugated
local jump matrices satisfy
\begin{equation}
( Y_{k_1})^{-1}
\widehat V_{j+6}^{(7)}(k)
 Y_{k_1}
\longrightarrow
V_j^{\mathrm{PC},k_1}(\zeta_4),
\qquad
j=1,2,3,4,
\label{eq:elliptic-k1-jump-limits}
\end{equation}
where $V_j^{\mathrm{PC},k_1}(\zeta_4)$ is given by
\begin{subequations}\label{eq:elliptic-k1-four-frozen-jumps}
\begin{align}
 V_1^{\mathrm{PC},k_1}(\zeta_4)&=
 \begin{pmatrix}
 1&0&\dfrac{d(k_1)^{-1/2}\bar R(k_1)}{1+|r_1(k_1)|^2}
       \zeta_4^{-2\mathrm i\nu_4}e^{\mathrm i\zeta_4^2/2}\\
 0&1&0\\0&0&1
 \end{pmatrix},\\[2mm]
 V_2^{\mathrm{PC},k_1}(\zeta_4)&=
 \begin{pmatrix}1&0&0\\0&1&0\\
 \dfrac{d(k_1)^{-1/2}R(k_1)(1+|r_1(k_1)|^2)}
 {1+|r_1(k_1)|^2+d(k_1)^{-1}|R(k_1)|^2}
 \zeta_4^{2\mathrm i\nu_4}e^{-\mathrm i\zeta_4^2/2}&0&1
 \end{pmatrix},\\[2mm]
 V_3^{\mathrm{PC},k_1}(\zeta_4)&=
 \begin{pmatrix}
 1&0&\dfrac{d(k_1)^{-1/2}\bar R(k_1)}
 {1+|r_1(k_1)|^2+d(k_1)^{-1}|R(k_1)|^2}
 \zeta_4^{-2\mathrm i\nu_4}e^{\mathrm i\zeta_4^2/2}\\
 0&1&0\\0&0&1
 \end{pmatrix},\\[2mm]
 V_4^{\mathrm{PC},k_1}(\zeta_4)&=
 \begin{pmatrix}1&0&0\\0&1&0\\
 d(k_1)^{-1/2}R(k_1)\zeta_4^{2\mathrm i\nu_4}
 e^{-\mathrm i\zeta_4^2/2}&0&1
 \end{pmatrix}.
\end{align}
\end{subequations}
We next show that a \(3\times3\) matrix-valued function satisfying
precisely the jump relations in
\eqref{eq:elliptic-k1-four-frozen-jumps} on the \(\zeta_4\)-plane can
be constructed from the solution of the standard parabolic-cylinder
model RH problem in Appendix~\ref{App:AAA}.

Let \(N^{\mathrm{PC}}(\zeta;p,q)\) denote the solution of the standard
parabolic-cylinder model RH problem in Appendix~\ref{App:AAA}. Set
\begin{equation}
\widehat X(\zeta_4)
=
N^{\mathrm{PC}}\left(
\zeta_4;
\frac{d(k_1)^{-1/2}\bar R(k_1)}
{1+|r_1(k_1)|^2},
-d(k_1)^{-1/2}R(k_1)
\right).
\label{eq:elliptic-k1-PC-parameters}
\end{equation}
We then define its \(3\times3\) embedding by
\begin{equation}
\mathcal P_{k_1}(\zeta_4)
=
\begin{pmatrix}
\widehat X_{11}&0&\widehat X_{12}\\
0&1&0\\
\widehat X_{21}&0&\widehat X_{22}
\end{pmatrix},
\label{eq:elliptic-k1-PC-embedding}
\end{equation}
where \(\widehat X_{ij}=\widehat X_{ij}(\zeta_4)\) denotes the
\((i,j)\) entry of \(\widehat X(\zeta_4)\). It is straightforward to
verify that \(\mathcal P_{k_1}(\zeta_4)\) satisfies precisely the jump
relations in \eqref{eq:elliptic-k1-four-frozen-jumps}.
This suggests defining the local parametrix near \(k_1\) as follows:
\begin{equation}
 N^{\mathrm{loc},k_1}(x,t,k)
 =N^{\mathrm{out}}(x,t,k)  Y_{k_1}\mathcal P_{k_1}(\zeta_4(k))
        Y_{k_1}^{-1},
 \qquad k\in D_\epsilon(k_1).
 \label{eq:elliptic-local-k1}
\end{equation}

Finally, we record the large-\(\zeta_4\) behavior of
\(\mathcal P_{k_1}(\zeta_4)\). It follows from
\eqref{E:mXasymoxia} and the definition
\eqref{eq:elliptic-k1-PC-embedding} that, as
\(\zeta_4\to\infty\),
\begin{equation}
\mathcal P_{k_1}(\zeta_4)
=
I+\frac{\mathcal C_1}{\zeta_4}
+O(\zeta_4^{-2}),
\label{eq:elliptic-k1-PC-expansion}
\end{equation}
where
\begin{equation}
\mathcal C_1
=
\begin{pmatrix}
0&0&
-\dfrac{
\sqrt{2\pi}\,e^{-\pi\nu_4/2}\sqrt{d(k_1)}
e^{\pi\mathrm{i}/4}
}{
R(k_1)\Gamma(\mathrm{i}\nu_4)
}
\\[3mm]
0&0&0
\\[1mm]
\dfrac{
\sqrt{2\pi}\,e^{-\pi\nu_4/2}e^{-\pi\mathrm{i}/4}
\sqrt{d(k_1)}(1+|r_1(k_1)|^2)
}{
\bar R(k_1)\Gamma(-\mathrm{i}\nu_4)
}
&0&0
\end{pmatrix}.
\label{eq:elliptic-k1-PC-first-coefficient}
\end{equation}

\subsection{Matching and the small norm  RH problem}
\label{subsec:elliptic-error-RHP}
In the preceding subsection, we constructed local parametrices near
\(\alpha\), \(\bar\alpha\), \(k_0\), and \(k_1\).
We now combine them into the piecewise-defined matrix
\begin{equation}
 \widehat N^{\mathrm{loc}}(k)=
 \begin{cases}
 N^{\mathrm{loc},\alpha}(k),&k\in D_\epsilon(\alpha),\\
 N^{\mathrm{loc},\bar\alpha}(k),&k\in D_\epsilon(\bar\alpha),\\
 N^{\mathrm{loc},k_0}(k),&k\in D_\epsilon(k_0),\\
 N^{\mathrm{loc},k_1}(k),&k\in D_\epsilon(k_1).
 \end{cases}
 \label{eq:elliptic-combined-local-corrections}
\end{equation}
For convenience, set
\begin{equation}
 \widehat{\mathcal D}
 =\bigcup_{a\in\{\alpha,\bar\alpha,k_0,k_1\}} D_\epsilon(a),
 \qquad
 \partial\widehat{\mathcal D}
 =\bigcup_{a\in\{\alpha,\bar\alpha,k_0,k_1\}}\partial D_\epsilon(a).
 \label{eq:elliptic-four-disks}
\end{equation}
Each boundary circle is oriented counterclockwise.
With this notation, we have the following lemma.

\begin{lemma}
\label{lem:elliptic-matching}
The combined local matrix
\(\widehat N^{\mathrm{loc}}\) is analytic in
\(\widehat{\mathcal D}\setminus\Gamma^{(7)}\) and satisfies
\begin{equation*}
 (\widehat N^{\mathrm{loc}})^+
 =(\widehat N^{\mathrm{loc}})^- \widehat V^{\mathrm{loc}},
 \qquad k\in\Gamma^{(7)}\cap\widehat{\mathcal D},
\end{equation*}
where the jump matrix \(V^{\mathrm{loc}}\) satisfies the estimates
\begin{equation}  \label{eq:elliptic-matching-estimates}
\begin{aligned}
 \|\widehat V^{(7)}-\widehat V^{\mathrm{loc}}\|_
 {L^\infty(\Gamma^{(7)}\cap\widehat{\mathcal D})}
 &\lesssim t^{-1/2}\log t,\\
 \|\widehat V^{(7)}- \widehat V^{\mathrm{loc}}\|_
 {L^1(\Gamma^{(7)}\cap\widehat{\mathcal D})}
 &\lesssim t^{-1}\log t.
\end{aligned}
\end{equation}
Moreover, as \(t\to\infty\),
\begin{equation} \label{eq:elliptic-matching-estimates-1}
 \bigl\|(N^{\mathrm{out}}(\widehat N^{\mathrm{loc}})^{-1}-I\bigr\|_
 {L^1\cap L^\infty(\partial\widehat{\mathcal D})}
 =O(t^{-1/2}).
\end{equation}
More precisely, the following expansions hold uniformly on $\partial\widehat{\mathcal D}$:
\begin{equation} \label{eq:elliptic-matching-estimates-2}
 N^{\mathrm{out}}(k) \bigl(\widehat N^{\mathrm{loc}}(k)\bigr)^{-1}-I=
 \begin{cases}
 O(t^{-1}),
 &k\in\partial D_\epsilon(\alpha)\cup
       \partial D_\epsilon(\bar\alpha),\\[4mm]
 -\dfrac{N^{\mathrm{out}}(k)  C_{\mathrm r}Y_{k_0}\mathcal C_0
          Y_{k_0}^{-1}C_{\mathrm r}^{-1} \left( N^{\mathrm{out}}(k)\right)^{-1} }
        {\widehat\gamma_1\sqrt t\,(k-k_0)}
 +O(t^{-1}),
 &k\in\partial D_\epsilon(k_0),\\[4mm]
 -\dfrac{ N^{\mathrm{out}}(k)   Y_{k_1}\mathcal C_1Y_{k_1}^{-1} \left( N^{\mathrm{out}}(k)\right)^{-1}}
        {\sqrt{\widehat\gamma_2t}\,(k-k_1)}
 +O(t^{-1}),
 &k\in\partial D_\epsilon(k_1).
 \end{cases}
\end{equation}
\end{lemma}

\begin{proof}
We first prove \eqref{eq:elliptic-matching-estimates}.
On
\(
 \Gamma^{(7)}\cap
 \bigl(D_\epsilon(\alpha)\cup D_\epsilon(\bar\alpha)\bigr)
\),
the local jumps agree exactly with the original jumps:
\(\widehat V^{(7)}=V^{\mathrm{loc}}\).
Thus the required estimates hold on these portions of the contour.
It remains to establish the estimates on the portions of the contour
inside the disks centered at \(k_0\) and \(k_1\).
We give the details near \(k_0\), where the argument is more involved.
The estimates near \(k_1\) follow in the same way.

For \(k\in\Gamma_1^{(\ell)}\)~(see Fig.\ref{fig:elliptic-k0-eight-rays}), we have
\begin{equation*}
 \widehat V^{(7)}-V^{\mathrm{loc}}
 =C_{\mathrm r}Y_{k_0}
 \begin{pmatrix}
 0&\widehat f(k)&0\\
 0&0&0\\
 0&0&0
 \end{pmatrix}
 Y_{k_0}^{-1}C_{\mathrm r}^{-1},
\end{equation*}
where
\begin{align*}
 \widehat f(k)
 =\frac{\zeta_3(k)^{-2\mathrm i\nu_3(\xi)}}{\bar r_1(k)}
 \frac{\mathcal F_0(k)^{-2}\widehat\delta_2(k_0)}
      {\mathcal F_0(k_0)^{-2}\widehat\delta_2(k)}
 \mathcal S_0(k)^2
 e^{h_{31}(k)-h_{31}^-(k_0)}-\frac{\zeta_3(k)^{-2\mathrm i\nu_3(\xi)}}{\bar r_1(k_0)}
 e^{\mathrm i\zeta_3(k)^2/2}.
\end{align*}
The matrix \(C_{\mathrm r}Y_{k_0}\) and its inverse are uniformly
bounded for \((x,t)\in\mathrm E_{\mathrm I}\).
It therefore suffices to estimate \(\widehat f(k)\).

Set \(u=|k-k_0|\).
The functions \(\mathcal F_0\), \(\widehat\delta_2\), and \(r_1\)
are locally Lipschitz near \(k_0\), and their values at \(k_0\)
are nonzero. After reducing \(\epsilon\) if necessary, these
functions remain bounded away from zero in the disk.
Together with \eqref{E:estofmathcals0}, these properties give
\begin{align*}
 \left|
 \frac{\mathcal F_0(k)^{-2}\widehat\delta_2(k_0)}
      {\mathcal F_0(k_0)^{-2}\widehat\delta_2(k)}
 \mathcal S_0(k)^2-1
 \right|
 \lesssim u\bigl(1+|\log u|\bigr),\qquad
 \left|\frac{1}{\bar r_1(k)}-\frac{1}{\bar r_1(k_0)}\right|
 \lesssim u.
\end{align*}
On the other hand, by \eqref{eq:elliptic-k0-quadratic-phase} and the sign structure
of \(h_{31}\) on \(\Gamma_1^{(\ell)}\),
we may choose \(\epsilon>0\) sufficiently small so that
\begin{equation*}
 \left|e^{h_{31}(k)-h_{31}^-(k_0)}\right|
 +\left|e^{\mathrm i\zeta_3(k)^2/2}\right|
 \lesssim e^{-ctu^2}
\end{equation*}
for some \(c>0\).
The remainder estimate in \eqref{eq:elliptic-k0-quadratic-phase}
then yields
\begin{equation*}
 \left|
 e^{h_{31}(k)-h_{31}^-(k_0)}
 -e^{\mathrm i\zeta_3(k)^2/2}
 \right|
 \lesssim tu^3e^{-ctu^2}, \qquad k \in \Gamma_1^{(\ell)}.
\end{equation*}
Combining all the estimates obtained above, we conclude that
\begin{equation}
 |\widehat f(k)|
 \lesssim
 \bigl[u(1+|\log u|)+tu^3\bigr]e^{-ctu^2}
 \lesssim t^{-1/2}\log t.
  \label{eq:elliptic-f-Linfty}
\end{equation}
This therefore gives the required \(L^\infty\) estimate for
\(\widehat V^{(7)}-V^{\mathrm{loc}}\) on \(\Gamma_1^{(\ell)}\).
Integrating the same bound along \(\Gamma_1^{(\ell)}\), we obtain
\begin{align*}
 \|\widehat f\|_{L^1(\Gamma_1^{(\ell)}\cap D_\epsilon(k_0))}
 \lesssim
 \int_0^\infty u(1+|\log u|)e^{-ctu^2}\,\mathrm du
 +t\int_0^\infty u^3e^{-ctu^2}\,\mathrm du \lesssim t^{-1}\log t.
\end{align*}
This proves the corresponding \(L^1\) estimate.
The remaining local arcs near \(k_0\) can be  treated in a similar way. Together with the analogous
argument near \(k_1\), this proves
\eqref{eq:elliptic-matching-estimates}.

It remains to prove \eqref{eq:elliptic-matching-estimates-1}
and \eqref{eq:elliptic-matching-estimates-2}.
On each component of \(\partial\widehat{\mathcal D}\),
the modulus of the corresponding scaled local coordinate tends
uniformly to infinity as \(t\to\infty\).
\eqref{eq:elliptic-matching-estimates-2} therefore follow from
the definition of \(\widehat N^{\mathrm{loc}}\) and the
expansions
\eqref{E:asyofNaialpha},
\eqref{E:asyofNaibaralpha},
\eqref{eq:elliptic-k0-PC-expansion}, and
\eqref{eq:elliptic-k1-PC-expansion}.
Finally, the norm estimate \eqref{eq:elliptic-matching-estimates-1}
is a direct consequence of these boundary expansions.
\end{proof}

We are now ready to carry out the error analysis.
Define the error matrix by
\begin{equation}
 \widehat{\mathcal E}(x,t,k)=
 \begin{cases}
 \widehat N^{(7)}(x,t,k)(N^{\mathrm{out}}(k))^{-1},
          &k\in\mathbb C\setminus\widehat{\mathcal D},\\[1mm]
 \widehat N^{(7)}(x,t,k)(\widehat N^{\mathrm{loc}}(k))^{-1},&k\in\widehat{\mathcal D}.
 \end{cases}
 \label{eq:elliptic-error-definition}
\end{equation}
The jump contour for \(\widehat{\mathcal E}(x,t,k)\) is
\[
 \begin{aligned}
 \Gamma^{\mathcal E}
 ={}&
 \left(
 \Gamma^{(7)}\setminus \overline{D_\epsilon(\alpha)\cup D_\epsilon(\bar \alpha) \cup B \cup \widetilde B}
 \right)  
 \cup\partial\widehat{\mathcal D}.
 \end{aligned}
\]
The corresponding jump matrix is given by

\begin{equation}
 \widehat V^{\mathcal E}=
 \begin{cases}
 N^{\mathrm{out}}\widehat V^{(7)}(N^{\mathrm{out}})^{-1},
                   &k\in\Gamma^{\mathcal E}\setminus \left(\overline{\widehat{\mathcal D}} \cup B \cup \widetilde B \right),\\
 \widehat N^{\mathrm{loc},-}  
       \widehat V^{(7)} (\widehat N^{\mathrm{loc},+})^{-1},
                   &k\in\Gamma^{\mathcal E}\cap \widehat{\mathcal D},\\
 N^{\mathrm{out}}(\widehat N^{\mathrm{loc}})^{-1},
                   &k\in\partial\widehat{\mathcal D}.
 \end{cases}
 \label{eq:elliptic-error-jumps-MS}
\end{equation}
\begin{remark}
We clarify two points about the jump matrix
\(\widehat V^{\mathcal E}\).
$\mathrm {(i)}$. On \(\Gamma^{(7)}\cap
\bigl(D_\epsilon(\alpha)\cup D_\epsilon(\bar\alpha)\bigr)\),
we have \(\widehat V^{\mathrm{loc}}=\widehat V^{(7)}\).
Hence \(\widehat V^{\mathcal E}=I\) on this part of the contour.
$\mathrm {(ii)}$. The matrices \(N^{\mathrm{out}}\) and \(N^{(7)}\) satisfy the same
jump relations on \((B\cup\widetilde B)\setminus\widehat{\mathcal D}\).
Moreover, \(\widehat N^{\mathrm{loc}}\) and \(N^{(7)}\) satisfy the
same jump relation on \(\widetilde B\cap D_\epsilon(k_0)\).
Thus, the jumps on the bands \(B\cup\widetilde B\) cancel completely
in the error RH problem.
\end{remark}

Using the sign structure of the phase functions \(h_{ij}\) and
the estimates \eqref{eq:elliptic-matching-estimates} and
\eqref{eq:elliptic-matching-estimates-1}, we obtain the following lemma.
We omit the proof, which is entirely analogous to that of
Lemma~\ref{L:estiofWe}.
\begin{lemma}
Let
\(
\widehat w^{\mathcal{E}}(k)=\widehat V^{\mathcal{E}}(k)-I
\).
Then, as \((x,t)\in\mathrm{E}_{\mathrm{I}}\) and \(t\to\infty\), there exists a constant \(c>0\) such that
\begin{subequations}
\label{eq:w-error-estimates-ew}
\begin{align}
\|\widehat w^{\mathcal{E}}\|_{L^1\cap L^\infty(\Gamma^{\mathcal{E}}\setminus \overline{\widehat{\mathcal{D}}})}
&\lesssim e^{-ct},
\label{eq:w-error-outside-1}\\
\|\widehat w^{\mathcal{E}}\|_{L^\infty(\partial\widehat{\mathcal{D}})}
&\lesssim t^{-1/2},
\label{eq:w-error-boundary-2}\\
\|\widehat w^{\mathcal{E}}\|_{L^\infty(\Gamma^{\mathcal{E}}\cap \widehat{\mathcal{D}})}
&\lesssim t^{-1/2}\log t,
\label{eq:w-error-local-Linfty-3}\\
\|\widehat w^{\mathcal{E}}\|_{L^1(\Gamma^{\mathcal{E}}\cap \widehat{\mathcal{D}})}
&\lesssim t^{-1}\log t.
\label{eq:w-error-local-L1-4}
\end{align}
\end{subequations}
\end{lemma}
This lemma gives
 \begin{align} \label{eq:elliptic-error-global-norms-1}
 \|\widehat w^{\mathcal E}\|_{L^\infty(\Gamma^{\mathcal E})}
 \lesssim t^{-1/2}\log t,\qquad
 \|\widehat w^{\mathcal E}\|_{L^1(\Gamma^{\mathcal E})}
 \lesssim t^{-1/2}.
 \end{align}
By interpolation, we also obtain
\begin{align} \label{eq:elliptic-error-global-norms}
 \|\widehat w^{\mathcal E}\|_{L^2(\Gamma^{\mathcal E})}
 \leq
 \|\widehat w^{\mathcal E}\|_{L^1(\Gamma^{\mathcal E})}^{1/2}
 \|\widehat w^{\mathcal E}\|_{L^\infty(\Gamma^{\mathcal E})}^{1/2}
 \lesssim t^{-1/2}(\log t)^{1/2}.
\end{align}
Thus, \(\widehat{\mathcal E}(x,t,k)\) satisfies a small-norm RH problem.
We now apply standard RH theory to derive its Cauchy integral
representation.  Define the operator \(C_{\widehat w^{\mathcal E}}\) by
\[
 C_{\widehat w^{\mathcal E}}f
 =C_-(f\widehat w^{\mathcal E}).
\]
By \eqref{eq:elliptic-error-global-norms-1} and \eqref{eq:elliptic-error-global-norms}, its operator norm on
\(L^2(\Gamma^{\mathcal E})\) tends to zero as \(t\to\infty\).
Hence \(I-C_{\widehat w^{\mathcal E}}\) is invertible for all
sufficiently large \(t\).
Then standard Beals--Coifman theory gives
\begin{equation}
 \widehat{\mathcal E}(x,t,k)=I+\frac1{2\pi\mathrm i}
 \int_{\Gamma^{\mathcal E}}
 \frac{\widehat\mu^{\mathcal E}(x,t,s)\widehat w^{\mathcal E}(x,t,s)}
      {s-k}\,\mathrm ds, \qquad k \in \mathbb C \setminus \Gamma^{\mathcal{E}},
 \label{eq:elliptic-error-Cauchy}
\end{equation}
where
\begin{equation}
 \widehat\mu^{\mathcal E}
 =I+(I-C_{\widehat w^{\mathcal E}})^{-1}
                      C_{\widehat w^{\mathcal E}}I.
 \label{eq:elliptic-error-density}
\end{equation}
Expanding the Cauchy kernel in \eqref{eq:elliptic-error-Cauchy} as \(k\to\infty\), we obtain
\begin{equation}
 \widehat{\mathcal E}(x,t,k)
 =I+\frac{\widehat{\mathcal E}^{(1)}(x,t)}k+O(k^{-2}),
 \qquad
 \widehat{\mathcal E}^{(1)}(x,t)
 =-\frac1{2\pi\mathrm i}\int_{\Gamma^{\mathcal E}}
       \widehat\mu^{\mathcal E}(x,t,s)\widehat w^{\mathcal E}(x,t,s)\,\mathrm ds.
 \label{eq:elliptic-error-first-coefficient}
\end{equation}
We now estimate the  coefficient $\widehat{\mathcal E}^{(1)}$.

\begin{lemma}
As \(t\to\infty\), the coefficient \(\widehat \mathcal{E}^{(1)}\) admits the
asymptotic expansion
\begin{align}
 \widehat{\mathcal E}^{(1)}(x,t)
 =\frac{[N^{\mathrm{out}}C_{\mathrm r}](k_0)
       Y_{k_0}\mathcal C_0Y_{k_0}^{-1}
       [N^{\mathrm{out}}C_{\mathrm r}](k_0)^{-1}}
          {\widehat\gamma_1\sqrt t}+\frac{N^{\mathrm{out}}(k_1)Y_{k_1}
        \mathcal C_1Y_{k_1}^{-1}N^{\mathrm{out}}(k_1)^{-1}}
          {\sqrt{\widehat\gamma_2 t}}
 +O(t^{-1}\log t),
 \label{eq:elliptic-error-residue-formula}
\end{align}
where the error term is uniform for \(\xi\in \mathcal K_-\).
\end{lemma}

\begin{proof}
The $L^2$ bounds in \eqref{eq:elliptic-error-global-norms} imply
$\|\widehat{\mu}^{\mathcal{E}}-I\|_{L^2(\Gamma^{\mathcal{E}})}
=O(t^{-1/2}(\log t)^{1/2})$.  Cauchy--Schwarz, together with
\eqref{eq:w-error-estimates-ew}, shows that
\begin{equation}
\begin{aligned}
 \widehat\mathcal E^{(1)}=&-\frac{1}{2\pi\mathrm i}
 \int_{\partial \widehat {\mathcal D}}\widehat{w}^{\mathcal E}(s)\,\mathrm ds -\frac{1}{2\pi\mathrm i}
 \int_{\Gamma^{\mathcal E} \setminus \partial\widehat {\mathcal D}} \widehat{w}^{\mathcal E}(s)\,\mathrm ds-\frac{1}{2\pi\mathrm i}
 \int_{\Gamma^{\mathcal E} }(\widehat{\mu}^{\mathcal{E}}(s)-I) \widehat{w}^{\mathcal E}(s),\mathrm ds \\
 =&
 -\frac{1}{2\pi\mathrm i}
 \int_{\partial \widehat{\mathcal D}} \widehat{w}^{\mathcal E}(s)\,\mathrm ds
 +O(t^{-1}\log t)\\
 =&-\frac{1}{2\pi\mathrm i}
 \int_{\partial \widehat{\mathcal D}} \left[ N^{\mathrm{out}}(s) \bigl(\widehat N^{\mathrm{loc}}(s)\bigr)^{-1}-I \right]\,\mathrm ds
 +O(t^{-1}\log t), \qquad  t \to \infty.
 \label{eq:E1-circle-integral-E}
 \end{aligned}
\end{equation}
Substituting
\eqref{eq:elliptic-matching-estimates-2} into the remaining circle integrals and
evaluating them by residues gives \eqref{eq:elliptic-error-residue-formula}.
\end{proof}

\subsection{Proof of Theorem \ref{thm:long-EW}}
\label{subsec:elliptic-reconstruction}
We now reverse the transformations to recover the potential.
Recall the reconstruction formula
\(\mathbf q=-2\mathrm i(M_1)_{\mathrm{rc}}\), where
\begin{equation}
 M_1(x,t)=\lim_{k\to\infty}k\bigl(M(x,t,k)-M_\infty\bigr),
 \qquad
 (M_1)_{\mathrm{rc}}=
 \begin{pmatrix}
 (M_1)_{21}\\
 (M_1)_{31}
 \end{pmatrix}.
 \label{eq:elliptic-reconstruction-column}
\end{equation}
To compute \(M_1\), we let \(k\to\infty\) within either of the
regions \(\mathcal R_7^{(1)}\) and \(\mathcal R_8^{(1)}\) shown in
Figure~\ref{fig:Gamma1-regions}.
In these regions, the inverse transformations give
\begin{align}
 M(x,t,k)
 ={}&[M_\infty\tilde\sigma S_\infty\tilde\sigma^{-1}]
 \Bigl[\tilde\sigma
 e^{\mathrm i(G_\infty t+\widehat g_\infty)\sigma}
 \widehat{\mathcal E}N^{\mathrm{out}}
 e^{-\mathrm i\widehat g\sigma}D^{-1}
 e^{-\mathrm iGt\sigma}
 \widehat\Delta_2^{-1}\widehat\Delta_1^{-1}
 \widehat G_1^{-1}S^{-1}\tilde\sigma^{-1}\Bigr].
 \label{eq:elliptic-undo-transformations}
\end{align}

First, a direct calculation using the boundary relation
\(\mathbf q_+=e^{\mathrm i(\vartheta_+-\vartheta_-)}\mathbf q_-\)
gives
\begin{equation}
 M_\infty\tilde\sigma S_\infty\tilde\sigma^{-1}
 =\begin{pmatrix}
 1&0&0\\
 \mathbf0&\mathbf q_-^\perp/q_0&\mathbf q_-/q_0
 \end{pmatrix} =: \widehat M_\infty.
 \label{eq:elliptic-Mhat-infinity}
\end{equation}
Second, in the regions \(\mathcal R_7^{(1)}\) and
\(\mathcal R_8^{(1)}\), all matrix factors on the right-hand side of
\eqref{eq:elliptic-undo-transformations} are diagonal except for
\(\widehat M_\infty\), \(\tilde\sigma^{\pm1}\),
\(\widehat{\mathcal E}\), and \(N^{\mathrm{out}}\).
Expanding both sides of \eqref{eq:elliptic-undo-transformations}
as \(k\to\infty\) and comparing the coefficients of \(k^{-1}\),
we obtain
\begin{equation}
 (M_1)_{\mathrm{rc}}
 =
 \left[
 \widehat M_\infty\tilde\sigma
 e^{\mathrm i(G_\infty t+\widehat g_\infty)\sigma}
 (\widehat{\mathcal E}^{(1)}+N_1^{\mathrm{out}})
 e^{-\mathrm i(G_\infty t+\widehat g_\infty)\sigma}
 \tilde\sigma^{-1}
 \right]_{\{2,3\}\times\{1\}}.
 \label{eq:elliptic-root-column-expansion}
\end{equation}
Here, \([A]_{\{2,3\}\times\{1\}}\) denotes the column vector
formed by the \((2,1)\) and \((3,1)\) entries of \(A\).  Next, we define
\begin{equation}
\begin{aligned}
 \widecheck{\mathcal E}(x,t)
 ={}&\frac{[N^{\mathrm{out}}C_{\mathrm r}](k_0)
       Y_{k_0}\mathcal C_0Y_{k_0}^{-1}
       [N^{\mathrm{out}}C_{\mathrm r}](k_0)^{-1}}
          {\widehat\gamma_1}+\frac{N^{\mathrm{out}}(k_1)Y_{k_1}
        \mathcal C_1Y_{k_1}^{-1}N^{\mathrm{out}}(k_1)^{-1}}
          {\sqrt{\widehat\gamma_2}}.
\end{aligned}
 \label{eq:elliptic-scaled-error-explicit}
\end{equation}
Then by \eqref{eq:elliptic-error-residue-formula}, we  may write
\begin{equation}
 \widehat{\mathcal E}^{(1)}(x,t)
 =t^{-1/2}\widecheck{\mathcal E}(x,t)+O(t^{-1}\log t).
 \label{eq:elliptic-scaled-error-coefficient}
\end{equation}
On the other hand, by \eqref{eq:elliptic-outer-21-unsimplified}
and \eqref{eq:elliptic-outer-21} we have
\begin{equation}
 \begin{aligned}
 (N_1^{\mathrm{out}})_{31}=0,\qquad
 (N_1^{\mathrm{out}})_{21}
 =\frac{\mathrm i(q_0+\Imm\alpha)}2
 \frac{\vartheta\!\left(\frac{\Omega t-\omega}{2\pi}
                  +2\nu_\infty-\frac14\right)\vartheta(-\frac12)}
      {\vartheta\!\left(\frac{\Omega t-\omega}{2\pi}-\frac14\right)
       \vartheta(2\nu_\infty-\frac12)}.
 \end{aligned}
 \label{eq:elliptic-outer-21-31}
\end{equation}
Substituting
\eqref{eq:elliptic-scaled-error-coefficient} and
\eqref{eq:elliptic-outer-21-31} into
\eqref{eq:elliptic-root-column-expansion}, and then using
\eqref{eq:reconstruction}, gives
\begin{align}
 \mathbf q(x,t)
 =\frac{q_0+\Im\alpha}{q_0}
 e^{-2\mathrm i(G_\infty t+\widehat g_\infty)}
 \frac{\vartheta\!\left(\frac{\Omega t-\omega}{2\pi}
                    +2\nu_\infty-\frac14\right)\vartheta(-\frac12)}
      {\vartheta\!\left(\frac{\Omega t-\omega}{2\pi}-\frac14\right)
       \vartheta(2\nu_\infty-\frac12)}
 \mathbf q_- +t^{-1/2}\mathbf q_{\mathrm{rad}}^{\mathrm{ell}}(x,t)
       +O(t^{-1}\log t),
 \label{eq:elliptic-final-asymptotics-MS}
\end{align}
where the radiative coefficient is given by
\begin{equation}
 \mathbf q_{\mathrm{rad}}^{\mathrm{ell}}(x,t)
 =-\frac{2\mathrm i}{q_0}\left(
 \widecheck{\mathcal E}_{21}
 e^{-2\mathrm i(G_\infty t+\widehat g_\infty)}\mathbf q_-
 +\widecheck{\mathcal E}_{31}
 e^{-\mathrm i(G_\infty t+\widehat g_\infty)}\mathbf q_-^\perp
 \right).
 \label{eq:elliptic-radiation-MS}
\end{equation}
The remainder is uniform for \(\xi\in \mathcal K_-\).

\section{Long--time asymptotics in  the Painlev\'e region}
\label{sec:positive-transition}


\begin{figure}[htbp]
\centering

\begin{minipage}[c]{0.32\textwidth}
    \centering
    \includegraphics[width=\linewidth]
    {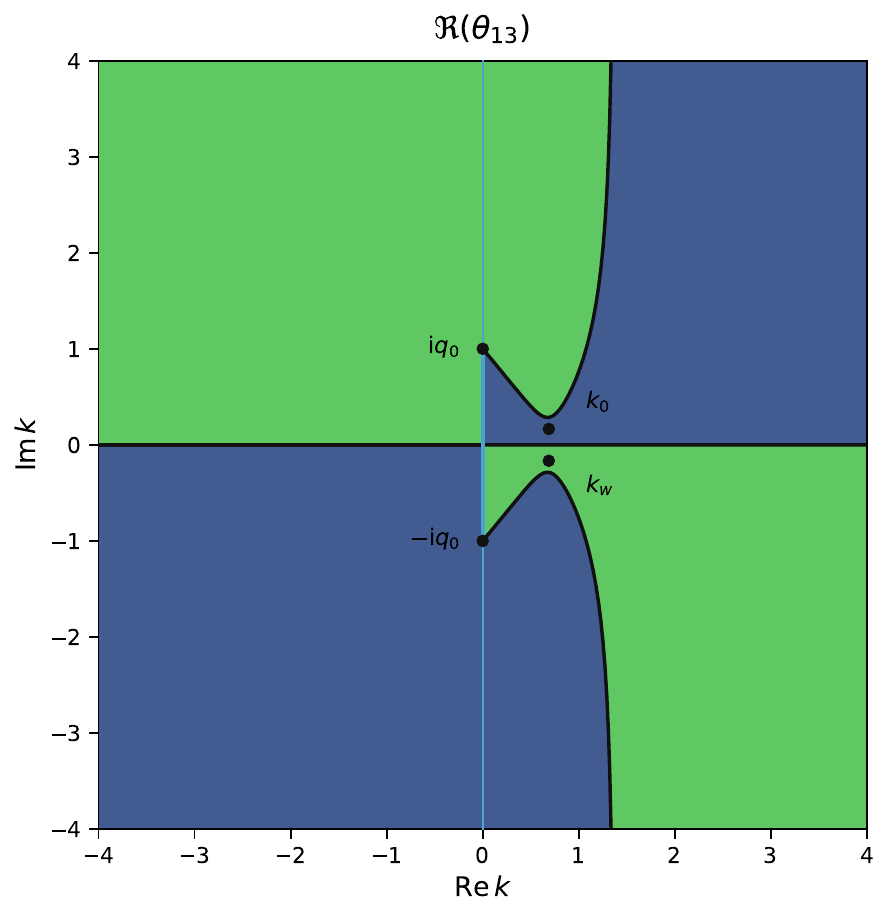}

    \vspace{1mm}
    {\small (a) $0<\xi<\xi_c$}
\end{minipage}
\hfill
\begin{minipage}[c]{0.32\textwidth}
    \centering
    \includegraphics[width=\linewidth]
    {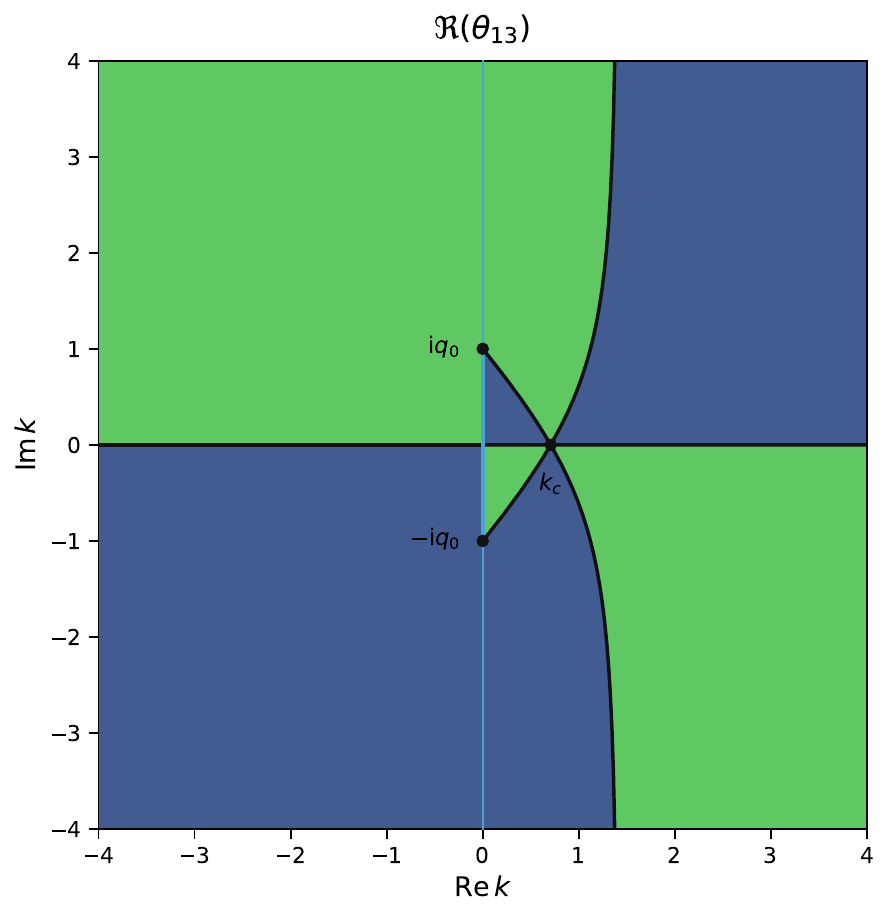}

    \vspace{1mm}
    {\small (b) $\xi=\xi_c$}
\end{minipage}
\hfill
\begin{minipage}[c]{0.32\textwidth}
    \centering
    \includegraphics[width=\linewidth]
    {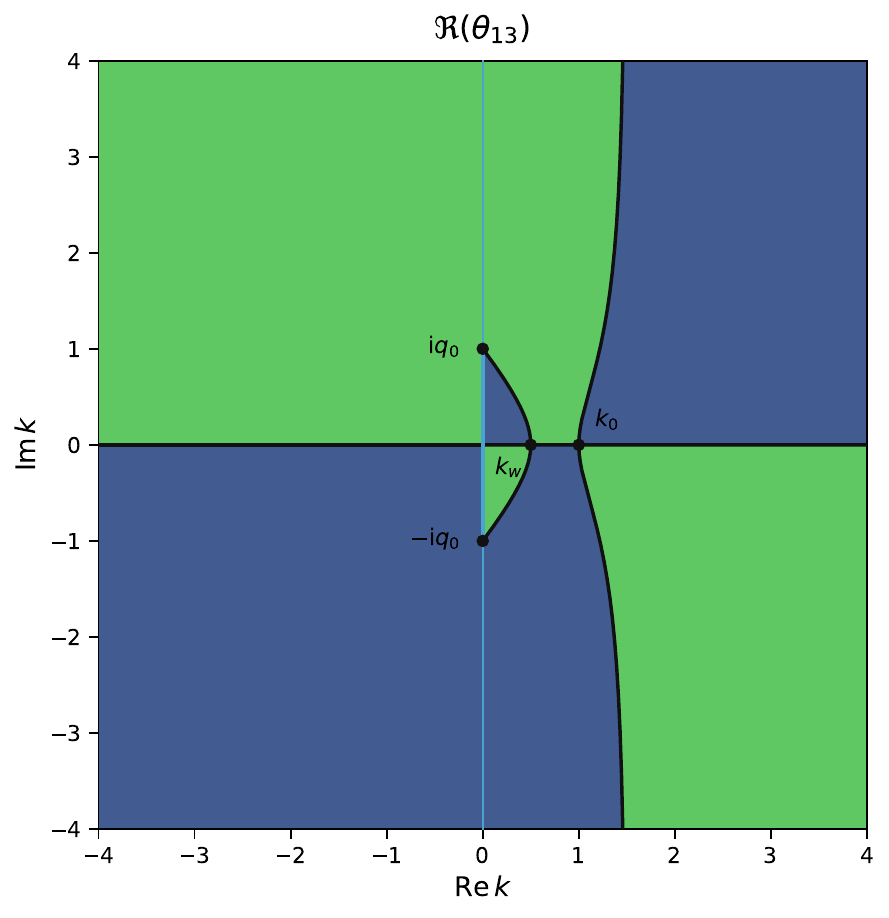}

    \vspace{1mm}
    {\small (c) $\xi>\xi_c$}
\end{minipage}

\caption{
Stationary points of $\theta_{13}$ and the sign distribution of
$\operatorname{Re}\theta_{13}$ for
$0<\xi<\xi_c$, $\xi=\xi_c$, and $\xi>\xi_c$, respectively, from left to
right. For $0<\xi<\xi_c$, the two stationary points form a
complex-conjugate pair, whereas for $\xi>\xi_c$ they are both
real. At the critical value $\xi=\xi_c$, the two stationary points
coalesce at $k_c$.
}
\label{fig:theta13-stationary-sign-structure}

\end{figure}

The asymptotic formula derived in Section~\ref{sec:steepest-descent}
is valid in the interior of the plane-wave region, but is not uniform
as \(\xi\) approaches the critical value \(\xi_c=4\sqrt2\,q_0\).
The same loss of uniformity occurs when the critical ray is approached
from the elliptic-wave region. Our goal in this section is therefore
to obtain a uniform asymptotic description of the transition near
\(\xi=\xi_c\).
More precisely, we study the long-time behavior of solutions
in the transition region:
\begin{equation}
 \mathcal T_{\mathrm I}=\left\{(x,t):
 |\xi-\xi_c|\leq Ct^{-2/3}\right\}.
 \label{eq:tr-region}
\end{equation}
We write \(\mathcal T_+=\mathcal T\cap\{\xi \ge \xi_c\}\) and
\(\mathcal T_-=\mathcal T\cap\{\xi\le \xi_c\}\).

We need to examine how the stationary points of the phase functions
behave as \(t\to\infty\) with \((x,t)\in\mathcal T\).
We begin with \(\theta_{13}(x,t,k)\), whose stationary points are $k_w$ and $k_0$; see \eqref{E:expreofkwk0}.
For \((x,t)\in\mathcal T_+\), both points are real and approach
\[
 k_c=\frac{q_0}{\sqrt2}
\]
 at a rate of \(O(t^{-1/3})\).
For \((x,t)\in\mathcal T_-\), they form a complex conjugate pair
and coalesce at \(k_c\) on the same scale.
Thus, for all sufficiently large \(t\) with \((x,t)\in\mathcal T\),
both stationary points lie in a fixed disk centered at \(k_c\).   Figure~\ref{fig:theta13-stationary-sign-structure} shows the evolution of the stationary points and the
sign structure of \(\theta_{13}\) for
\(\xi<\xi_c\), \(\xi=\xi_c\), and
\(\xi>\xi_c\).  Moreover, at \(\xi=\xi_c\), the quadratic term in the local expansion of
\(\theta_{13}\) at \(k_c\) vanishes. This means that the quadratic approximation near the stationary
point \(k_0\) no longer provides a uniform description of the
transition. We therefore carry out a double-scaling analysis near
\(k_c\), using a Painlev\'e-II model in place of the quadratic
approximation.

We next consider \(\theta_{12}\) and \(\theta_{23}\).
Since \(\theta_{23}\) has no real stationary points for
\((x,t) \in \mathcal T\), it remains to analyze \(\theta_{12}\).
As \(t\to\infty\) with \((x,t)\in\mathcal T\), its unique real
stationary point $k_1$ approaches
\[
 k_d=\frac{7q_0}{4\sqrt2}.
\]
The quadratic term in the local expansion of \(\theta_{12}\)
at this point remains nonzero.
Hence the parabolic-cylinder model used in Section~\ref{sec:steepest-descent} remains valid near \(k_d\)
and yields a contribution of order \(t^{-1/2}\) to the final
asymptotic formula.

For \((x,t)\in\mathcal T_+\), the sign structures of
\(\theta_{12}\) and \(\theta_{23}\) agree with those shown in
Figure~\ref{fig:phase-signature-charts}. They remain qualitatively unchanged  for \((x,t)\in\mathcal T_-\). However,
the position of the real stationary point \(k_1\) of \(\theta_{12}\)
relative to the fixed point \(k_d\) changes as \(\xi\) passes
through \(\xi_c\). Since
\[
 k_1(\xi)=\frac{\xi}{4}-\frac{q_0^2}{\xi}
\]
is strictly increasing for \(\xi>0\) and \(k_1(\xi_c)=k_d\),
the point \(k_1\) lies to the right of \(k_d\) when
\((x,t)\in\mathcal T_+\), and to its left when
\((x,t)\in\mathcal T_-\). At \(\xi=\xi_c\), the two points coincide.  Moreover, we have
\begin{align} \label{E:estk1kd}
 k_1-k_d=\frac9{32}(\xi-\xi_c)+O((\xi-\xi_c)^2)
 =O(t^{-2/3}), \qquad t \to \infty \quad \text{with} \quad (x,t) \in \mathcal{T}.
\end{align}

\begin{figure}
\centering
\begin{tikzpicture}[
    x=0.92cm,
    y=0.92cm,
    >={Stealth[length=2.5mm,width=1.7mm]},
    contour/.style={
        black,
        line width=1.7pt,
        line cap=round,
        line join=round
    },
    axis/.style={
        black,
        dashed,
        line width=1.00pt,
        line cap=round
    },
    branch/.style={
        black,
        line width=2.2pt,
        line cap=round
    },
    raylabel/.style={
        font=\small,
        fill=white,
        inner sep=2.5pt
    },
    dir/.style={
        postaction={decorate},
        decoration={
            markings,
            mark=at position 0.56 with {
                \arrow{Stealth[length=3mm,width=2mm]}
            }
        }
    }
]

\coordinate (kc)    at (-2.0,0);
\coordinate (kzero) at (-0.5,0);
\coordinate (kd)    at (3.1,0);
\coordinate (kone)  at (4.2,0);

\coordinate (kcup)   at (-2.0,1.8);
\coordinate (kcdown) at (-2.0,-1.8);

\coordinate (kup)   at (1.3,1.8);
\coordinate (kdown) at (1.3,-1.8);

\coordinate (krup)   at (4.92,0.72);
\coordinate (krdown) at (4.92,-0.72);

\draw[axis] (-6,0)--(6.4,0);

\draw[contour,dir] (-6,1.8)--(kcup);
\draw[contour,dir] (kcup)--(kup);
\draw[contour,dir] (kup)--(6.4,1.8);

\draw[contour,dir] (-6,-1.8)--(kcdown);
\draw[contour,dir] (kcdown)--(kdown);
\draw[contour,dir] (kdown)--(6.4,-1.8);

\draw[branch,dir] (-4.0,-1.05)--(-4.0,1.05);

\draw[contour,dir] (kcup)--(kc);
\draw[contour,dir] (kcdown)--(kc);

\draw[contour,dir] (kc)--(kzero);

\draw[contour,dir] (kzero)--(kup);
\draw[contour,dir] (kzero)--(kdown);
\draw[contour,dir] (kup)--(kd);
\draw[contour,dir] (kdown)--(kd);

\draw[contour,dir] (kd)--(kone);

\draw[contour] (kone)--(krup);
\draw[contour] (kone)--(krdown);
\draw[contour,dir] (krup)--(6.4,0.72);
\draw[contour,dir] (krdown)--(6.4,-0.72);

\fill[black] (kc)    circle (2.8pt);
\fill[black] (kzero) circle (2.8pt);
\fill[black] (kd)    circle (2.8pt);
\fill[black] (kone)  circle (2.8pt);

\node[below=0pt,xshift=-8pt]  at (kc)    {$k_c$};
\node[below=0pt,xshift=-2pt] at (kzero) {$k_0$};
\node[below=1pt,xshift=3pt]  at (kd)    {$k_d$};
\node[below=1pt,xshift=-2pt] at (kone)  {$k_1$};

\node[raylabel] at (-4.38,-0.90) {$B$};

\node[raylabel] at (-1.66,1.02)  {$1$};
\node[raylabel] at (-1.66,-1.02) {$2$};
\node[raylabel] at (0.00,1.02)   {$3$};
\node[raylabel] at (0.00,-1.12)  {$4$};
\node[raylabel] at (-1.15,0.32)  {$5$};

\end{tikzpicture}
\caption{The jump contour $\mathcal{L}^{(1)}$ for $\widetilde N^{(1)}$ in the case $(x,t) \in \mathcal T_+$.}
\label{fig:final-lenses-tran}
\end{figure}
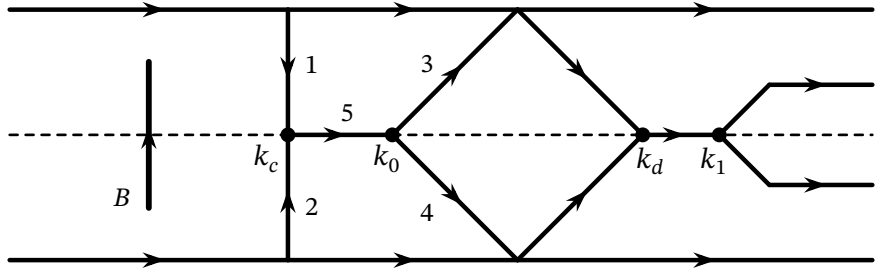

\subsection{Opening of lenses}
\label{S:Tranopenning}

We now describe the lens-opening strategy for the nonlinear
steepest-descent analysis in the transition region.
We first consider \((x,t)\in\mathcal T_+\).
The construction is similar to that used in the plane-wave region
\(\mathrm P_{\mathrm I}\) in Section~\ref{sec:steepest-descent},
with some adjustments near the critical points $k_c$ and $k_d$.

Recall that, in Section~\ref{sec:steepest-descent}, we divided
the real axis into the three intervals:
\(
 (-\infty,k_0)\), \((k_0,k_1)\), 
 \((k_1,\infty)\),
and extended the jump factors off the real axis on each interval.
For \((x,t)\in\mathcal T_+\), we have four relevant points, which are ordered as
\(
 k_c<k_0<k_d<k_1
\).
We  modify the construction by opening lenses along
\(
 (-\infty,k_c) \), \( (k_0,k_d) \), \((k_1,\infty) \)
in the same way as in Section~\ref{sec:steepest-descent},
while leaving the jumps on \((k_c,k_0)\) and \((k_d,k_1)\)
on the real axis.
The resulting jump contour is denoted by \(\mathcal L^{(1)}\);
see Figure~\ref{fig:final-lenses-tran}. Let \(\widetilde N^{(1)}\) denote the RH unknown after these
deformations, and let \(\widetilde V^{(1)}\) denote the
corresponding jump matrix. It is easy to see that, as \(t\to\infty\), all jump matrices away
from the branch cut \(B\) and fixed neighborhoods of \(k_c\) and
\(k_d\) approach the identity exponentially. The jumps near \(k_c\) and
\(k_d\) therefore require separate treatment.
To write down the exact jump matrices, we first modify the  conjugating functions.

In the transition region $\mathcal T$, we replace the conjugation matrices
\(\Delta_1\) and \(\Delta_2\) introduced in
Section~\ref{sec:steepest-descent} by
\(\widetilde\Delta_1\) and \(\widetilde\Delta_2\), defined by
\begin{equation}
 \widetilde\Delta_1
 =\diag(\widetilde\delta_1,\widetilde\delta_1^{-1},1),
 \qquad
 \widetilde\Delta_2
 =\diag(\widetilde\delta_2,1,\widetilde\delta_2^{-1}).
 \label{eq:tr-Deltas}
\end{equation}
The scalar functions \(\widetilde\delta_1\) and
\(\widetilde\delta_2\) are defined by the same integral formulas
as \(\delta_1\) and \(\delta_2\), with the integration endpoints
chosen for the transition region. More precisely,
\begin{equation}
 \begin{aligned}
 \widetilde\delta_1(k)
 &=\exp\left\{-\frac1{2\pi\ii}\int_{k_c}^{\infty}
 \frac{\log(1+|\rho_2(s)|^2)}{s-k}\dd s\right\},\\
 \widetilde\delta_2(\xi,k)
 &=\exp\left\{-\frac1{2\pi\ii}\int_{k_d}^{\infty}
 \frac{\log h(s)}{s-k}\dd s\right\},\qquad
 h(s)
 =1+\frac{d(s)^{-1}|\rho_1(s)|^2}{1+|\rho_2(s)|^2}.
 \end{aligned}
 \label{eq:tr-delta12}
\end{equation}
With these scalar factors, the function \(g(\xi,k)\) defined in
\eqref{eq:g-solution} is replaced by
\begin{equation}
 \widetilde g(\xi,k)
 =-\frac{\lambda(k)}{2\pi\ii}\int_B
 \frac{\ii\log(\widetilde\delta_1(s)^2\widetilde\delta_2(s))}
      {\lambda(s)(s-k)}\dd s.
 \label{eq:tr-g}
\end{equation}
The corresponding limit at infinity is
\begin{equation}
 \widetilde g_\infty(\xi)
 =\frac1{2\pi\ii}\int_B
 \frac{\ii\log(\widetilde\delta_1(s)^2\widetilde\delta_2(s))}
      {\lambda(s)}\dd s.
 \label{eq:tr-g-infty}
\end{equation}

 We describe only the jumps on \(B\) and near \(k_c\).
The jumps near \(k_d\) are omitted for the reason explained
in Remark~\ref{R:explaink1}.
The jump matrices on the subcontours
labeled \(1,2,3,4\) in Figure~\ref{fig:final-lenses-tran} can be written as
\begin{equation}
 \widetilde V_j^{(1)}
 =V_j^{(6)}\big|_{\delta_1\to\widetilde\delta_1,\,
                  \delta_2\to\widetilde\delta_2,\,
                  g\to\widetilde g},
 \qquad j=1,2,3,4,B,
 \label{eq:tr-unchanged-jumps}
\end{equation}
where \(V_j^{(6)}\) are the jump matrices given in
Section~\ref{sec:steepest-descent}.
Here, \(a\to b\) means that \(a\) is replaced by \(b\). More precisely, we have
\begin{align}\label{eq:tr-local-prejumps}
 \widetilde V_1^{(1)}&=
 \begin{pmatrix}1&0&0\\
 \rho_2\widetilde\delta_1^2\widetilde\delta_2
 e^{2\ii\widetilde g}e^{\theta_{31}}&1&0\\0&0&1\end{pmatrix},&
 \widetilde V_2^{(1)}&=
 \begin{pmatrix}1&\bar\rho_2\widetilde\delta_1^{-2}
 \widetilde\delta_2^{-1}e^{-2\ii\widetilde g}e^{\theta_{13}}&0\\
 0&1&0\\0&0&1\end{pmatrix},\notag\\
 \widetilde V_3^{(1)}&=
 \begin{pmatrix}1&\dfrac{\bar\rho_2}{1+\rho_2\bar\rho_2}
 \widetilde\delta_1^{-2}\widetilde\delta_2^{-1}
 e^{-2\ii\widetilde g}e^{\theta_{13}}&0\\0&1&0\\0&0&1\end{pmatrix},&
 \widetilde V_4^{(1)}&=
 \begin{pmatrix}1&0&0\\
 \dfrac{\rho_2}{1+\rho_2\bar\rho_2}
 \widetilde\delta_1^2\widetilde\delta_2
 e^{2\ii\widetilde g}e^{\theta_{31}}&1&0\\0&0&1\end{pmatrix},\notag\\
  \widetilde V_B^{(1)}&=
  \begin{pmatrix}
  0&-\mathrm i&0\\
  -\mathrm i&0&0\\
  0&0&1
 \end{pmatrix}.
\end{align}
The jump matrix on $\mathcal L^{(1)}_5$ is given by
\begin{equation}
 \widetilde V_5^{(1)}=
 \begin{pmatrix}
 (1+\rho_2\bar\rho_2)
 \dfrac{\widetilde\delta_1^+}{\widetilde\delta_1^-}
 &
 \dfrac{\bar\rho_2e^{-2\ii\widetilde g}e^{\theta_{13}}}
       {\widetilde\delta_2\widetilde\delta_1^-\widetilde\delta_1^+}
 &0\\[3mm]
 \rho_2\widetilde\delta_1^-\widetilde\delta_1^+
 \widetilde\delta_2e^{2\ii\widetilde g}e^{\theta_{31}}
 &
 \dfrac{\widetilde\delta_1^-}{\widetilde\delta_1^+}
 &0\\
 0&0&1
 \end{pmatrix}.
 \label{eq:tr-V5-explicit}
\end{equation}

\begin{remark}
\label{R:explaink1}
The explicit jump matrices near \(k_d\) can be obtained in the
same way. However, we omit them here for brevity.
We explain the reason below.
The preceding analysis of \(\theta_{12}\) shows that a
parabolic-cylinder local model applies near \(k_d\), just as in
the local analysis near \(k_1\) in
Section~\ref{sec:steepest-descent}. This model contributes
\(O(t^{-1/2})\) to the long-time asymptotics.
As we will see, the Painlev\'e-II local parametrix near \(k_c\)
produces a correction of order \(t^{-1/3}\) to the leading term.
We therefore include the smaller contribution from the
neighborhood of \(k_d\) in the remainder and omit the explicit
jump matrices there.
\end{remark}

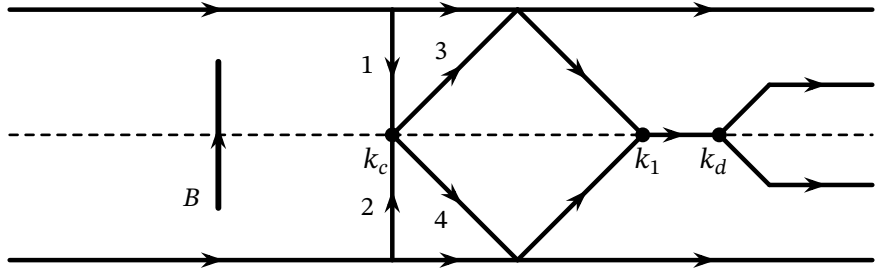
\begin{figure}
\centering
\begin{tikzpicture}[
    x=0.92cm,
    y=0.92cm,
    >={Stealth[length=2.5mm,width=1.7mm]},
    contour/.style={
        black,
        line width=1.7pt,
        line cap=round,
        line join=round
    },
    axis/.style={
        black,
        dashed,
        line width=1.00pt,
        line cap=round
    },
    branch/.style={
        black,
        line width=2.2pt,
        line cap=round
    },
    raylabel/.style={
        font=\small,
        fill=white,
        inner sep=2.5pt
    },
    dir/.style={
        postaction={decorate},
        decoration={
            markings,
            mark=at position 0.56 with {
                \arrow{Stealth[length=3mm,width=2mm]}
            }
        }
    }
]

\coordinate (kc)   at (-0.5,0);
\coordinate (kone) at (3.1,0);
\coordinate (kd)   at (4.2,0);

\coordinate (kcup)   at (-0.5,1.8);
\coordinate (kcdown) at (-0.5,-1.8);

\coordinate (kup)   at (1.3,1.8);
\coordinate (kdown) at (1.3,-1.8);

\coordinate (krup)   at (4.92,0.72);
\coordinate (krdown) at (4.92,-0.72);

\draw[axis] (-6,0)--(6.4,0);

\draw[contour,dir] (-6,1.8)--(kcup);
\draw[contour,dir] (kcup)--(kup);
\draw[contour,dir] (kup)--(6.4,1.8);

\draw[contour,dir] (-6,-1.8)--(kcdown);
\draw[contour,dir] (kcdown)--(kdown);
\draw[contour,dir] (kdown)--(6.4,-1.8);

\draw[branch,dir] (-3.0,-1.05)--(-3.0,1.05);

\draw[contour,dir] (kcup)--(kc);
\draw[contour,dir] (kcdown)--(kc);

\draw[contour,dir] (kc)--(kup);
\draw[contour,dir] (kc)--(kdown);
\draw[contour,dir] (kup)--(kone);
\draw[contour,dir] (kdown)--(kone);

\draw[contour,dir] (kone)--(kd);

\draw[contour] (kd)--(krup);
\draw[contour] (kd)--(krdown);
\draw[contour,dir] (krup)--(6.4,0.72);
\draw[contour,dir] (krdown)--(6.4,-0.72);

\fill[black] (kc)   circle (2.8pt);
\fill[black] (kone) circle (2.8pt);
\fill[black] (kd)   circle (2.8pt);

\node[below=1pt,xshift=-6pt]  at (kc)   {$k_c$};
\node[below=1pt,xshift=2pt] at (kone) {$k_1$};
\node[below=1pt,xshift=-2pt] at (kd)   {$k_d$};

\node[raylabel] at (-3.38,-0.90) {$B$};


\node[raylabel] at (-0.86,1.02)  {$1$};
\node[raylabel] at (-0.86,-1.02) {$2$};

\node[raylabel] at (0.20,1.22)  {$3$};
\node[raylabel] at (0.20,-1.22) {$4$};

\end{tikzpicture}

\caption{The jump contour $\mathcal{L}^{(1)}$ for $\widetilde N^{(1)}$ in the case $(x,t) \in \mathcal T_-$.}
\label{fig:final-lenses-tranew}
\end{figure}

We now describe the lens opening for \((x,t)\in\mathcal T_-\),
the part of the transition region on the elliptic-wave side.
Unlike the analysis in the interior of the elliptic-wave region
in Section~\ref{sec:elliptic-wave}, the present construction does not require a
\(G\)-function transformation.
A key observation is that, when \((x,t)\in\mathcal T_-\) and the
lenses are chosen as in Figure~\ref{fig:final-lenses-tranew},
the nondecaying jumps near \(k_c\) are confined to a small fixed
disk \(D_\epsilon(k_c)\).
More precisely, the sign structure of \(\Ree\theta_{13}\) shows
that the jump matrices on the local lens contours outside this
disk are exponentially close to the identity as \(t\to\infty\).
Inside this disk, we construct a Painlev\'e-II local parametrix
to approximate the solution of the deformed RH problem. Thus we can construct a small-norm RH problem without further modifying the phase.

The lens opening therefore follows the same basic strategy as
in \(\mathcal T_+\), with a small adjustment to the intervals
used in the deformation.
For $(x,t) \in \mathcal T_-$, the point \(k_0\) is no longer real,
and the remaining real points satisfy
\(
 k_c<k_1<k_d
\).
We open lenses along
\(
 (-\infty,k_c) \),  \( (k_c,k_1)\), \( (k_d,\infty)
\),
while leaving the jump on \((k_1,k_d)\) on the real axis.
We continue to denote the resulting jump contour by
\(\mathcal L^{(1)}\)~(see Fig.~\ref{fig:final-lenses-tranew}), and the transformed RH unknown and its
jump matrix by \(\widetilde N^{(1)}\) and
\(\widetilde V^{(1)}\), respectively.  Moreover, for \(j=1,2,3,4,B\), the jump matrices
\(\widetilde V_j^{(1)}\) are given by the same formulas
for \((x,t)\in\mathcal T_-\) as for \((x,t)\in\mathcal T_+\).

\subsection{Outer model and local model}
Since the jump  $\widetilde V^{(1)}$ on the branch cut \(B\) is the same as $V^{(6)}_B$ in
Section~\ref{sec:steepest-descent}, we can use the same outer model,
namely,
\begin{align}
 N_B(k)=\frac12
 \begin{pmatrix}
 \Lambda+\Lambda^{-1}&-(\Lambda-\Lambda^{-1})&0\\
 -(\Lambda-\Lambda^{-1})&\Lambda+\Lambda^{-1}&0\\
 0&0&2
 \end{pmatrix},
  \label{eq:tr-MB-explicit}
\end{align}
where  $\Lambda(k)=
\left(\frac{k-\mathrm{i}q_0}{k+\mathrm{i}q_0}\right)^{1/4}$ is normalized to one at infinity, as in
\eqref{eq:Lambda-outer}.  As before, \(N_B\) has the following large-\(k\) expansion:
\begin{align}
 N_B(k)=I+\frac{N_B^{(1)}}k+O(k^{-2}),\qquad
 N_B^{(1)}=
 \begin{pmatrix}0&\ii q_0/2&0\\\ii q_0/2&0&0\\0&0&0\end{pmatrix}.
 \label{eq:tr-MB-expansion}
\end{align}

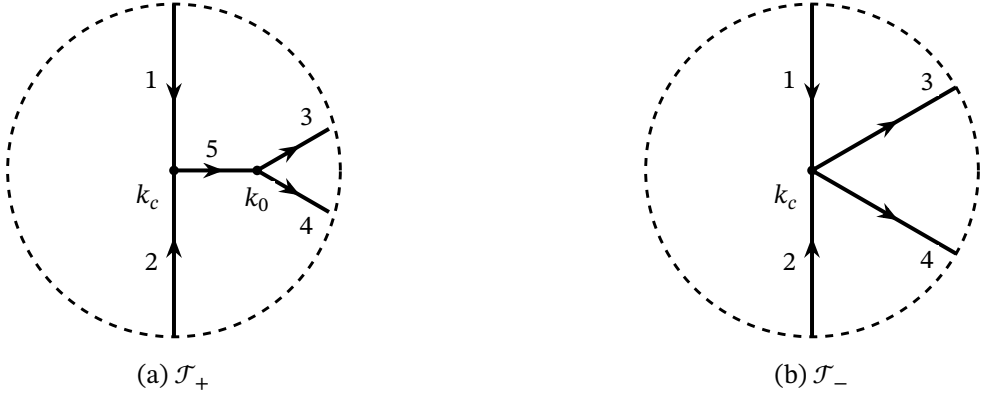
\begin{figure}
\centering

\begin{minipage}[t]{0.47\textwidth}
\centering
\begin{tikzpicture}[
 >=Stealth,
 ray/.style={
     black,
     line width=1.5pt,
     postaction={decorate},
     decoration={
         markings,
         mark=at position .60 with {\arrow[scale=0.8]{>}}
     }
 },
 lab/.style={font=\small,fill=white,inner sep=2pt}
]
\coordinate (o) at (0,0);
\coordinate (b) at (1.1,0);

\draw[black,dashed,line width=1.1pt] (o) circle (2.2);

\draw[ray] (0,2.2)--(o);
\draw[ray] (0,-2.2)--(o);
\draw[ray] (o)--(b);
\draw[ray] (b)--(2.05,.55);
\draw[ray] (b)--(2.05,-.55);

\fill[black] (o) circle (1.8pt);
\fill[black] (b) circle (1.8pt);

\node[lab,below left=3pt] at (o) {$k_c$};
\node[lab,below=4pt] at (b) {$k_0$};

\node[lab] at (-.30,1.2)  {$1$};
\node[lab] at (-.30,-1.2) {$2$};
\node[lab] at (1.75,.72)  {$3$};
\node[lab] at (1.75,-.72) {$4$};
\node[lab] at (.50,.28)   {$5$};

\end{tikzpicture}

\par\smallskip
{\small (a) $\mathcal T_+$}
\end{minipage}
\hfill
\begin{minipage}[t]{0.47\textwidth}
\centering
\begin{tikzpicture}[
 >=Stealth,
 ray/.style={
     black,
     line width=1.5pt,
     postaction={decorate},
     decoration={
         markings,
         mark=at position .60 with {\arrow[scale=0.8]{>}}
     }
 },
 lab/.style={font=\small,fill=white,inner sep=2pt}
]
\coordinate (o) at (0,0);

\draw[black,dashed,line width=1.1pt] (o) circle (2.2);

\draw[ray] (0,2.2)--(o);
\draw[ray] (0,-2.2)--(o);
\draw[ray] (o)--(30:2.2);
\draw[ray] (o)--(-30:2.2);

\fill[black] (o) circle (1.8pt);
\node[lab,below left=3pt] at (o) {$k_c$};

\node[lab] at (-.30,1.2)  {$1$};
\node[lab] at (-.30,-1.2) {$2$};
\node[lab] at (1.52,1.18)  {$3$};
\node[lab] at (1.52,-1.18) {$4$};

\end{tikzpicture}

\par\smallskip
{\small (b) $\mathcal T_-$}
\end{minipage}

\caption{Local contours in $D_\epsilon(k_c)$:
(a) for $(x,t)\in\mathcal T_+$; (b) for $(x,t)\in\mathcal T_-$.}
\label{fig:tr-local-plus}
\end{figure}

We now construct the local model near \(k_c\).
Let \(D_\epsilon(k_c)\) be a sufficiently small disk centered at \(k_c\); see Fig. \ref{fig:tr-local-plus}.
The next step is to  introduce a suitable local coordinate in this disk. We first compute the Taylor expansion of the phase \(\theta_{13}\) about \(k_c\).
Writing \(\theta_{13}=\ii t\vartheta_{13}\), then we have
\begin{align}
 \vartheta_{13}(\xi,k)=\vartheta_{13}(\xi,k_c)
 -\frac{2}{\sqrt3}(\xi-\xi_c)(k-k_c)
+\frac{8\sqrt6}{9q_0}(k-k_c)^3 +S(\xi,k),
 \label{eq:tr-phase-Taylor}
\end{align}
where
$$
 S(\xi,k)=-\frac{2\sqrt6}{9q_0}(\xi-\xi_c)(k-k_c)^2
 +\frac{4\sqrt3}{27q_0^2}(\xi-\xi_c)(k-k_c)^3
 +O((k-k_c)^4).
$$
Motivated by the above expansion, we introduce the local variable
\(z\) and the bounded parameter \(y\) as follows:
\begin{equation}
 z=\left(\frac{8\sqrt6}{9q_0}\right)^{1/3} t^{1/3}(k-k_c), \qquad
 y=-\frac{2}{\sqrt3}\left(\frac{8\sqrt6}{9q_0}\right)^{-1/3}
 t^{2/3}(\xi-4\sqrt2q_0).
 \label{eq:tr-y-z}
\end{equation}
Consequently, near $k_c$ we have
\begin{equation}
 \theta_{13}(x,t,k)=\theta_{13}(x,t,k_c)
 +\ii(yz+z^3)+\ii tS(\xi,k).
 \label{eq:tr-phase-normal-form}
\end{equation}

 We now isolate the endpoint behavior of \(\widetilde\delta_1\).  Integration by parts yields the factorization
\begin{equation}
 \widetilde\delta_1(k)
 =e^{\ii\widetilde\nu_1\log_\pi(k_c-k)}
  e^{\widetilde\eta_1(k)},
\end{equation}
where
\[
\begin{aligned}
 \widetilde\nu_1
 =-\frac{1}{2\pi}\log\bigl(1+|\rho_2(k_c)|^2\bigr)\leq0,\qquad
 \widetilde\eta_1(k)
 =\frac{1}{2\pi\ii}\int_{k_c}^{\infty}
   \log_\pi(s-k)\,\dd\log\bigl(1+|\rho_2(s)|^2\bigr).
\end{aligned}
\]
In particular, \(\widetilde\eta_1(k)\) has a finite limit as
\(k\to k_c\), and satisfies
\[
 \widetilde\eta_1(k)-\widetilde\eta_1(k_c)
 =O\bigl(|k-k_c|(1+|\log|k-k_c||)\bigr),
 \qquad k\to k_c.
\]
 In terms of the scaled variable in
\eqref{eq:tr-y-z}, \(\widetilde\delta_1\) can also be rewrite as
\begin{equation}
 \widetilde\delta_1(k)=
 \exp\left\{\ii\widetilde\nu_1
 \log\left( \left(\frac{8\sqrt6}{9q_0} t \right )^{-1/3} \right)+\ii\widetilde\nu_1\log_\pi(-z)\right\}
 e^{\widetilde\eta_1(k)}.
 \label{eq:tr-delta-z}
\end{equation}

Examining the jump matrices in equation~\eqref{eq:tr-local-prejumps}, we see that we
need to analyze the scalar factor \(\widetilde\delta_1^2\widetilde\delta_2
 e^{2\ii\widetilde g}\).
It follows from \eqref{eq:tr-delta-z} that
\begin{equation}
 \widetilde\delta_1(k)^2\widetilde\delta_2(k)
 e^{2\ii\widetilde g(k)}
 =(-z)^{2\ii\widetilde\nu_1}\widetilde d_0\widetilde d_1(k),
 \label{eq:tr-d0d1-factor}
\end{equation}
where
\begin{align*}  
 \widetilde d_0=
 \left(\frac{8\sqrt6}{9q_0}\right)^{-2\ii\widetilde\nu_1/3}
 t^{-2\ii\widetilde\nu_1/3}
 e^{2(\widetilde\eta_1(k_c)+\ii\widetilde g(k_c))}
 \widetilde\delta_2(k_c),
\qquad
 \widetilde d_1(k)= e^{2(\widetilde\eta_1(k)-\widetilde\eta_1(k_c))}
 e^{2\ii(\widetilde g(k)-\widetilde g(k_c))}
 \frac{\widetilde\delta_2(k)}{\widetilde\delta_2(k_c)}.
\end{align*}
It follows that $|\widetilde d_0|=1$ and 
\(\widetilde d_1(k)=1+O(|k-k_c|(1+|\log|k-k_c||))\).
Write
\begin{equation}
 \rho_2(k_c)=\rho_0e^{\ii\arg\rho_2(k_c)},
 \qquad \rho_0=|\rho_2(k_c)|,
 \label{eq:tr-rho0}
\end{equation}
and introduce the diagonal conjugator
\begin{equation}
 \widetilde Y=\diag\left(
 \widetilde d_0^{-1/2}e^{\theta_{13}(k_c)/2}
 e^{-\ii\arg\rho_2(k_c)/2},
 \widetilde d_0^{1/2}e^{\theta_{31}(k_c)/2}
 e^{\ii\arg\rho_2(k_c)/2},1\right).
 \label{eq:tr-Y}
\end{equation}
Thus, the local jump matrices in \eqref{eq:tr-local-prejumps}
satisfy
\[
 \widetilde Y^{-1}\widetilde V_j^{(1)}\widetilde Y
 \longrightarrow V_j^Y,
 \qquad k\to k_c,\qquad j=1,2,3,4,5,
\]
where the limiting jump matrices \(\{V_j^Y\}_{j=1}^5\) are given by \eqref{eq:tr-VY}.

The analysis above motivates the following definition of the
local parametrix near \(k_c\):
\begin{equation}
 N^{\mathrm{loc},k_c}(x,t,k)=
 \widetilde Y M^Y(z(k);y)\widetilde Y^{-1},
 \qquad k\in D_\epsilon(k_c).
 \label{eq:tr-local-kc}
\end{equation}
For convenience, we define
\(
 \mathcal L^\epsilon=D_\epsilon(k_c)\cap\mathcal L^{(1)}
\).
We then have the following lemma.
\begin{lemma}
\label{L:Nlockc}
For each \((x,t)\), the matrix
\(N^{\mathrm{loc},k_c}(x,t,k)\) defined in
\eqref{eq:tr-local-kc} is analytic and bounded in
\( D_\epsilon(k_c)\setminus \mathcal L^\epsilon\).
Its boundary values satisfy
\[
\left( N^{\mathrm{loc},k_c} \right)^+(x,t,k)
 =\left( N^{\mathrm{loc},k_c} \right)^-(x,t,k)
  V^{\mathrm{loc},k_c}(x,t,k),
 \qquad k\in \mathcal L^\epsilon.
\]
Moreover, for all sufficiently large \(t\) and
\((x,t)\in\mathcal T\), we have
\begin{equation}
\begin{aligned}
 \|\widetilde V^{(1)}-V^{\mathrm{loc},k_c}\|_
 {L^\infty(\mathcal L^\epsilon)}
 &\leq Ct^{-1/3}\log t,\\
 \|\widetilde V^{(1)}-V^{\mathrm{loc},k_c}\|_
 {L^1(\mathcal L^\epsilon)}
 &\leq Ct^{-2/3}\log t.
\end{aligned}
\label{E:estVlockc}
\end{equation}
Furthermore, as \(t\to\infty\),
\begin{align}
 (N^{\mathrm{loc},k_c})^{-1}(x,t,k)-I
 =- \frac{\widetilde Y M_1^Y(y)\widetilde Y^{-1}}
 {\left(\frac{8\sqrt6}{9q_0}\right)^{1/3}
  t^{1/3}(k-k_c)}
 +O(t^{-2/3}),
 \qquad k\in\partial  D_\epsilon(k_c),
 \label{E:bzsy1}
\end{align}
where \(M_1^Y\) is given by \eqref{E:defMX1dd}.
\end{lemma}

Before proving this lemma, we first establish the following useful
estimates for the relevant phase functions.
\begin{lemma}\label{lem:uniform-local-exponentials}
For sufficiently small $\epsilon>0$, there exist constants
$C,c,T>0$ such that, for all $(x,t)\in\mathcal{T}$ with
$t\ge T$,
\begin{align}
 \left|e^{\theta_{31}(x,t,k)}\right|
 &\le C e^{-ct|k-k_c|^3},
 && k\in\mathcal{L}_1^\epsilon\cup\mathcal{L}_4^\epsilon,
 \label{eq:uniform-exp-plus}\\
 \left|e^{\theta_{13}(x,t,k)}\right|
 &\le C e^{-ct|k-k_c|^3},
 && k\in\mathcal{L}_2^\epsilon\cup\mathcal{L}_3^\epsilon.
 \label{eq:uniform-exp-minus}
\end{align}
\end{lemma}

\begin{proof}
We prove only \eqref{eq:uniform-exp-plus}, since the proof of \eqref{eq:uniform-exp-minus} is similar.
We let $\theta_{31}=\ii t \vartheta_{31}$. By \eqref{eq:tr-phase-Taylor}, we have
\begin{align} \label{E:Taytheta31}
\vartheta_{31}(\xi,k)=\vartheta_{31}(\xi,k_c)
 +\frac{2}{\sqrt3}(\xi-\xi_c)(k-k_c)
-\frac{8\sqrt6}{9q_0}(k-k_c)^3 -S(\xi,k),
\end{align}
where $\vartheta_{31}(\xi,k_c)\in\mathbb{R}$ and
\begin{align} \label{E:estogS}
 |S(\xi,k)|
 \le C_1\left(
 |\xi-\xi_c|\,|k-k_c|^2+|k-k_c|^4
 \right).
\end{align}
This estimate holds uniformly for $\xi \in \mathcal T$ and $k\in D_\epsilon(k_c)$.

We first consider $\mathcal{L}_1^\epsilon$.
For $k=k_c+\mathrm{i}s$, with $0<s<\epsilon$, the expansion~\eqref{E:Taytheta31} gives
\[
 -\operatorname{Im}\vartheta_{31}(\xi,k)
 \le
 \frac{2|\xi-\xi_c|}{\sqrt{3}}s
 -\frac{8\sqrt{6}}{9q_o}s^3
 +C_1\left(|\xi-\xi_c|s^2+s^4\right).
\]
Since $s<\epsilon$, the term
$|\xi-\xi_c|s^2$ is bounded by
$\epsilon|\xi-\xi_c|s$.
After decreasing $\epsilon$ to absorb the quartic term,
we obtain
\[
- \operatorname{Im}\vartheta_{31}(\xi,k)
 \le -c_1s^3+C_2|\xi-\xi_c|s
\]
for some $c_1,C_2>0$.
Young's inequality gives
\[
 C_2|\xi-\xi_c|s
 \le \frac{c_1}{2}s^3+C_3|\xi-\xi_c|^{3/2}.
\]
Consequently, for $(x,t) \in \mathcal T$ we have
\[
 -\operatorname{Im}\vartheta_{31}(\xi,k)
 \le -\frac{c_1}{2}|k-k_c|^3
      +C_3|\xi-\xi_c|^{3/2} \le -\frac{c_1}{2}|k-k_c|^3
      +C_4 t^{-1},
 \qquad k\in\mathcal{L}_1^\epsilon.
\]
Since \(\theta_{31}=\mathrm{i}t\vartheta_{31}\), the inequality
above immediately yields the estimate \eqref{eq:uniform-exp-plus}
for \(k\in\mathcal L_1^\epsilon\).

We next consider $k \in \mathcal{L}_4^\epsilon$. For \((x,t)\in\mathcal T_-\), the proof proceeds in the same way
as on \(\mathcal L_1^\epsilon\).
We therefore focus on \((x,t)\in\mathcal T_+\).
In this case, \(\mathcal L_4^\epsilon\) has its endpoint on the
real axis at \(k_0\), which makes the corresponding local
estimates more involved. We first establish the following estimate for \((x,t)\in\mathcal T_+\).
\[
 0\le k_0(\xi)-k_c
 =\frac{\xi_c-\xi+\sqrt{\xi^2-\xi_c^2}}{8}
 \le C_4|\xi-\xi_c|^{1/2}.
\]
Using this bound, we immediately obtain
\begin{align} \label{E:estkkc}
 |k-k_c| \le |k-k_0|+|k_0-k_c| \le C_5\left(|\operatorname{Im}k|+
 |\xi-\xi_c|^{1/2}
 \right),
 \qquad k\in\mathcal{L}_4^\epsilon.
\end{align}
Using elementary inequalities, we arrive at
\begin{align} \label{E:estkkc3}
 |k-k_c|^3
 \le C_6\left(
 |\xi-\xi_c|^{3/2}+|\operatorname{Im}k|^3
 \right).
\end{align}
It follows from \eqref{E:estkkc} and \eqref{E:estkkc3} that for $k \in \mathcal{L}^{\epsilon}_4$,
\begin{align} \label{E:estiofxixic}
 |\xi-\xi_c|\,|k-k_c|^2
\lesssim  \epsilon \left(
 |\xi-\xi_c|^{3/2}
 +|\xi-\xi_c| |\operatorname{Im}k|  \right), \qquad
 |k-k_c|^4
\lesssim   \epsilon \left(
 |\xi-\xi_c|^{3/2}+|\operatorname{Im}k|^3
 \right).
\end{align}
Moreover, since \(\arg(k-k_c)\in(-\pi/6,0)\) for
\(k\in\mathcal L_4^\epsilon\), we have
\begin{align}
 \operatorname{Im}\bigl((k-k_c)^3\bigr)
&=
 \operatorname{Im}k
 \left(
 3(\operatorname{Re}k-k_c)^2-(\operatorname{Im}k)^2
 \right) \nonumber\\
 &=\operatorname{Im}k  \left(
 3(\operatorname{Re}(k-k_c))^2-(\operatorname{Im}k)^2 \right)
 \le -2|\operatorname{Im}k|^3.  \label{E:estImkkc3}
\end{align}

Substituting the estimates \eqref{E:estogS}, \eqref{E:estiofxixic},
and \eqref{E:estImkkc3} into the phase expansion \eqref{E:Taytheta31}
and decreasing \(\epsilon\) further if necessary, we obtain
\begin{align} \label{E:zjgd}
 -\operatorname{Im}\vartheta_{31}(\xi,k)
 \le -c_3|\operatorname{Im}k|^3
      +C_7|\xi-\xi_c| |\operatorname{Im}k|
      +C_7|\xi-\xi_c|^{3/2}
\end{align}
for some constants \(c_3,C_7>0\).
Using  Young's inequality again gives
\[
 C_7|\xi-\xi_c||\operatorname{Im}k|
 \le \frac{c_3}{2}|\operatorname{Im}k|^3
      +C_8|\xi-\xi_c|^{3/2}.
\]
Finally, substituting the estimate above into \eqref{E:zjgd}
and using \eqref{E:estkkc3}, we conclude that
\[
-\operatorname{Im}\vartheta_{31}(\xi,k)
 \le -c_4|k-k_c|^3+C_9|\xi-\xi_c|^{3/2},
 \qquad k\in\mathcal{L}_4^\epsilon,
\]
for some $c_4, C_9>0$. Using the boundedness of \(t|\xi-\xi_c|^{3/2}\) for
\((x,t)\in\mathcal T_+\), together with the identity
\(\theta_{31}=\mathrm{i}t\vartheta_{31}\), we immediately obtain
the estimate \eqref{eq:uniform-exp-plus} for
\(k\in\mathcal L_4^\epsilon\).

\end{proof}

\begin{remark}\label{R:mpedphseest}
The proof of Lemma~\ref{lem:uniform-local-exponentials} also gives the corresponding
estimates for the model exponential factors:
\begin{subequations}\label{E:model-exp-est}
\begin{align}
\left|
\exp\{-\ii yz-\ii z^3\}
\right|
&\le
C e^{-ct|k-k_c|^3},
\qquad
k\in \mathcal{L}_1^\epsilon \cup \mathcal{L}_4^\epsilon,
\label{E:model-exp-est-b}
\\
\left|
\exp\{\ii yz+\ii z^3\}
\right|
&\le
C e^{-ct|k-k_c|^3},
\qquad
k\in \mathcal{L}_2^\epsilon\cup \mathcal{L}_3^\epsilon.
\label{E:model-exp-est-a}
\end{align}
\end{subequations}
\end{remark}

We are now ready to prove Lemma~\ref{L:Nlockc}.

\begin{proof}[Proof of Lemma~\ref{L:Nlockc}]
We illustrate the estimate \eqref{E:estVlockc} on \(\mathcal{L}_1^{\epsilon}\).
By the definition of \(N^{\mathrm{loc},k_c}\), its jump matrix
\(V^{\mathrm{loc},k_c}\) satisfies
\[
\widetilde V^{(1)}(k)-V^{\mathrm{loc},k_c}(k)
=
\widetilde Y
\begin{pmatrix}
0&0&0\\
f(k)&0&0\\
0&0&0
\end{pmatrix}
\widetilde Y^{-1},
\]
where
\begin{align}
f(k)=\left(\rho_2(k)  e^{-\ii \mathrm{arg} \rho_2(k_c)}\right)e^{\theta_{31}-\theta_{31}(k_c)}(-z)^{2\ii \widetilde{\nu}_1} \widetilde{d}_1(k)
-\rho_0(-z)^{2\ii\widetilde\nu_1}e^{-\ii \left(yz+z^3  \right)}.
\end{align}
Since \(\widetilde Y\) and \(\widetilde Y^{-1}\) are uniformly bounded in
\(D_{\epsilon}(k_c)\), it is enough to estimate $f(k)$.
We first note the estimates
$$
\left|\rho_2(k)  e^{-\ii \mathrm{arg}\rho_2(k_c)}-\rho_0 \right| \lesssim |k-k_c|, \qquad
\left| \widetilde{d}_1(k)-1 \right|  \lesssim |k-k_c|(1+|\log|k-k_c||).
$$
On the other hand, using
\eqref{eq:uniform-exp-plus}, \eqref{E:model-exp-est-b}, the elementary inequality
\[ 
|e^{\omega_1}-e^{\omega_2}| \le \left(|e^{\omega_1}|+|e^{\omega_2}|\right) |\omega_1-\omega_2|, \qquad \omega_1,\omega_2\in \mathbb C,
 \] and  the remainder estimate
\eqref{E:estogS}, we obtain
\[ 
\left| e^{\theta_{31}(x,t,k)-\theta_{31}(x,t,k_c))} - e^{-\ii(yz+z^3)} \right| \le Ct|S(\xi,k)|e^{-ct|k-k_c|^3}. 
\]
Therefore, we conclude that
\[
|f(k)|
\le
C|k-k_c|e^{-ct|k-k_c|^3}
+
Ct|S(\xi,k)|e^{-ct|k-k_c|^3}
+
C|\widetilde d_1(k)-1|e^{-ct|k-k_c|^3} .
\]
Hence, putting \(u=|k-k_c|\), and using the  scaling
\(
|\xi-\xi_c|\le Ct^{-2/3}
\),
we get
\[
\begin{aligned}
|f(k)|
\le{}&
C\sup_{u>0} u e^{-ctu^3}
+
Ct|\xi-\xi_c|\sup_{u>0} u^2 e^{-ctu^3} + Ct|\xi-\xi_c|\sup_{u>0} u^3 e^{-ctu^3}\\
&+
Ct\sup_{u>0} u^4 e^{-ctu^3}
+
C\sup_{u>0}
\left(1+|\log u|\right)u e^{-ctu^3}.
\end{aligned}
\]
Consequently,
\(
|f(k)|
\le
C t^{-1/3}\log t 
\).
Similarly, by direct integration along \(\mathcal{L}_1^{\epsilon}\), we obtain
\[
\int_{\mathcal{L}_1^{\epsilon}}|f(s)|\,|ds|
\le
C t^{-2/3}\log t .
\]
The estimates on the other components of the local contour are obtained in the
same way. We thus proved \eqref{E:estVlockc}.

It remains to prove \eqref{E:bzsy1}.
For \(k\in\partial D_\epsilon(k_c)\), the local coordinate satisfies
\(z\to\infty\) as \(t\to\infty\).
Applying the expansion \eqref{E:asyMX}, we obtain \eqref{E:bzsy1}.

\end{proof}

We next consider the local parametrix near \(k_d\).
Its construction is entirely analogous to that of the parabolic
cylinder parametrix near \(k_1\) in
Section~\ref{subsec:local-parametrices}.
Its contribution to the long-time asymptotics is \(O(t^{-1/2})\)
and therefore does not affect the leading correction of order
\(t^{-1/3}\). We thus omit the details of the construction and
record only the estimates needed below.

Let \(D_\epsilon(k_d)\) be a sufficiently small fixed disk
centered at \(k_d\), and set
\(
 \widetilde{\mathcal L}^\epsilon
 =\mathcal L^{(1)}\cap D_\epsilon(k_d)
\).
Denote the corresponding local parametrix by
\(N^{\mathrm{loc},k_d}\). It satisfies the jump relation
\[
 \bigl(N^{\mathrm{loc},k_d}\bigr)^+(x,t,k)
 =
 \bigl(N^{\mathrm{loc},k_d}\bigr)^-(x,t,k)
 V^{\mathrm{loc},k_d}(x,t,k),
 \qquad k\in\widetilde{\mathcal L}^\epsilon.
\]
As in the analysis near \(k_1\)~(see Section~\ref{subsec:local-parametrices} ), the difference between the exact
and model jump matrices satisfies, for sufficiently large \(t\),
\begin{equation}\label{eq:local-jump-Linfty}
\begin{aligned}
 \left\|\widetilde V^{(1)}-V^{\mathrm{loc},k_d}\right\|_
 {L^\infty(\widetilde{\mathcal L}^\epsilon)}
 &\lesssim t^{-1/2}\log t,
 \\
 \left\|\widetilde V^{(1)}-V^{\mathrm{loc},k_d}\right\|_
 {L^1(\widetilde{\mathcal L}^\epsilon)}
 &\lesssim t^{-1}\log t.
\end{aligned}
\end{equation}
Moreover, the local parametrix satisfies the boundary matching estimate
\begin{equation}
 \left\|N^{\mathrm{loc},k_d}-I\right\|_
 {L^1 \cap L^\infty (\partial D_\epsilon(k_d))}
 \lesssim t^{-1/2}.
 \label{eq:tr-local-kd-matching}
\end{equation}

For convenience, we combine the local parametrices near \(k_c\)
and \(k_d\) into a single matrix \(N^{\mathrm{loc}}\).
Set
\(
 \widetilde{\mathcal D}
 =D_\epsilon(k_c)\cup D_\epsilon(k_d)
\)
and define \(N^{\mathrm{loc}}\) on \(\widetilde{\mathcal D}\) by
\begin{equation}
 N^{\mathrm{loc}}
 =
 \begin{cases}
 N^{\mathrm{loc},k_c},&k\in D_\epsilon(k_c),\\
 N^{\mathrm{loc},k_d},&k\in D_\epsilon(k_d).
 \end{cases}
 \label{eq:tr-local-combined}
\end{equation}
This combined local parametrix will be used in the next subsection
to construct a small-norm RH problem.

\subsection{The small norm RH problem}
We define the final transformation to obtain a small-norm RH problem as follows:
\begin{equation}
 \mathcal E_{\mathrm{tr}}(x,t,k)=
 \begin{cases}
 \widetilde N(x,t,k)N^{\mathrm{loc}}(x,t,k)^{-1}
 N_B(k)^{-1},&k\in\widetilde{\mathcal D},\\
 \widetilde N(x,t,k)N_B(k)^{-1},
 &k\notin\widetilde{\mathcal D}.
 \end{cases}
 \label{eq:tr-error-definition}
\end{equation}
The jump contour of \( \mathcal E_{\mathrm{tr}} \) is $\mathcal{L}_{\mathrm{tr}}= \left(\mathcal{L}^{(1)} \setminus B \right) \cup \partial \widetilde{\mathcal D}$, where \(\partial \widetilde{\mathcal D}\) is oriented counterclockwise.  The corresponding jump matrix \(V_{\mathrm{tr}}\) is given by
\begin{equation}
V_{\mathrm{tr}}(k)
=
\begin{cases}
N_B(k)
\bigl(N^{\mathrm{loc}}(k)\bigr)^{-1}
\bigl(N_B(k)\bigr)^{-1},
&
k\in \partial \widetilde{\mathcal D},
\\[1.4ex]
N_B (k)
\widetilde V^{(1)}(k)
N_B(k)^{-1},
&
k\in \mathcal{L}^{(1)}
\setminus\bigl(\overline{ \widetilde{\mathcal D} }\cup B\bigr),
\\[1.4ex]
N_B(k)
\left[
N^{\mathrm{loc},-}(k)
\widetilde V^{(1)}(k)
\bigl(N^{\mathrm{loc},+}(k)\bigr)^{-1}
\right]
N_B(k)^{-1},
&
k\in \mathcal{L}^{(1)} \cap \widetilde{\mathcal D}.
\end{cases}  \label{E:VEdef}
\end{equation}
Let \(w_{\mathrm{tr}}=V_{\mathrm{tr}}-I\).  Using the sign structure of the phase functions \(\theta_{ij}\) and
the estimates \eqref{E:estVlockc}, \eqref{E:bzsy1}, \eqref{eq:local-jump-Linfty} and
\eqref{eq:tr-local-kd-matching}, we obtain the following lemma.
We omit the proof, which is entirely analogous to that of
Lemma~\ref{L:estiofWe}.

\begin{lemma} \label{L:mxy}
As \(t\to\infty\), the following estimates hold uniformly for
\((x,t)\in\mathcal T\):
\begin{subequations}\label{E:wE-estimates}
\begin{align}
&\left\|w_{\mathrm{tr}}\right\|_{L^1\cap L^\infty
\left(\mathcal{L}_{\mathrm{tr}} \setminus \overline{\widetilde{\mathcal D}}\right)}
\le
C e^{-ct},
\label{E:wE-estimates-a}
\\
&\left\|w_{\mathrm{tr}}\right\|_{L^1\cap L^\infty(\partial\widetilde{\mathcal D})}
\le
C t^{-1/3},
\label{E:wE-estimates-b}
\\
&\left\|w_{\mathrm{tr}}\right\|_{L^1\left(\mathcal{L}_{\mathrm{tr}} \cap \widetilde{\mathcal D}\right)}
\le
C t^{-2/3}\log t,
\label{E:wE-estimates-c}
\\
&\left\|w_{\mathrm{tr}}\right\|_{L^\infty\left(\mathcal{L}_{\mathrm{tr}} \cap \widetilde{\mathcal D}\right)}
\le
C t^{-1/3}\log t.
\label{E:wE-estimates-d}
\end{align}
\end{subequations}

\end{lemma}

The estimates in the lemma above show that 
\begin{equation}
 \|w_{\mathrm{tr}}\|_{L^1(\mathcal{L}_{\mathrm{tr}})}
 \lesssim t^{-1/3},
 \qquad
 \|w_{\mathrm{tr}}\|_{L^\infty(\mathcal{L}_{\mathrm{tr}})}
 \lesssim t^{-1/3}\log t.
 \label{eq:tr-error-bounds}
\end{equation}
In particular, by interpolation,
\begin{align} \label{E:wmathrntr}
\left\| w_{\mathrm{tr}} \right\|_{L^2(\mathcal{L}_{\mathrm{tr}})}
\le
C t^{-1/3}(\log t)^{1/2}.
\end{align}
Define the Cauchy operator
\(
 C_{w_{\mathrm{tr}}}h
 =C_-\bigl(hw_{\mathrm{tr}}\bigr)
\).
Then \(\|C_{w_{\mathrm{tr}}}\|_{\mathcal B(L^2 (\mathcal{L}_{\mathrm{tr}}) )}\to0\), so
\(I-C_{w_{\mathrm{tr}}^{\mathcal E}}\) is invertible for all sufficiently
large \(t\).  Thus we may define
\begin{equation*}
 \mu_{\mathrm{tr}}=I+
 (I-C_{w_{\mathrm{tr}}})^{-1}
 C_{w_{\mathrm{tr}}}I  \in I+ L^2 (\mathcal{L}_{\mathrm{tr}}).
\end{equation*}
In particular, one can verify that 
\begin{align} \label{E:extmutr}
\left\|\mu_{\mathrm{tr}}-I\right\|_{L^2(\mathcal{L}_{\mathrm{tr}})} \lesssim   \left\| w_{\mathrm{tr}} \right\|_{L^2(\mathcal{L}_{\mathrm{tr}})}  \lesssim  
 t^{-1/3}(\log t)^{1/2}.
\end{align}
By the standard theory for small-norm RH problems, the error
function \( \mathcal E_{\mathrm{tr}} \)  can be represented as
\begin{equation}
 \mathcal E_{\mathrm{tr}}(x,t,k)=I+
 \frac{1}{2\pi\ii}\int_{\mathcal{L}_{\mathrm{tr}}}
 \frac{\mu_{\mathrm{tr}}(x,t,s)
 w_{\mathrm{tr}}(x,t,s)}{s-k}\dd s, \qquad k \in \mathbb C \setminus  \mathcal{L}_{\mathrm{tr}}.
 \label{eq:tr-error-Cauchy}
\end{equation}
Therefore, the coefficient of \(k^{-1}\) in the expansion of \( \mathcal E_{\mathrm{tr}} \) at infinity
is
\begin{equation}
 \mathcal E_{\mathrm{tr}}^{(1)}(x,t)
 :=\lim_{k\to\infty}k(\mathcal E_{\mathrm{tr}}(x,t,k)-I)
 =-\frac{1}{2\pi\ii}\int_{\mathcal{L}_{\mathrm{tr}}}
 \mu_{\mathrm{tr}}(x,t,s) w_{\mathrm{tr}}(x,t,s) \dd s.
 \label{eq:tr-E1-definition}
\end{equation}
\begin{lemma}
As \(t\to\infty\),
\begin{equation}
 \mathcal E_{\mathrm{tr}}^{(1)}(x,t)
 =-\frac{1}{2\pi\ii}\int_{\partial D_\epsilon(k_c)}
 w_{\mathrm{tr}}(s)\dd s+O(t^{-1/2}).
 \label{eq:tr-E1-circle}
\end{equation}
\end{lemma}
\begin{proof}
From the Cauchy representation \eqref{eq:tr-E1-definition}, we have
\[
 \mathcal E_{\mathrm{tr}}^{(1)}(x,t)
=
-\frac{1}{2\pi\mathrm{i}}
\int_{\partial D_\epsilon(k_c)}
w_{\mathrm{tr}}(x,t,s)\,ds
+
F_1(x,t)+F_2(x,t)+F_3(x,t),
\]
where
\[
\begin{aligned}
F_1(x,t)
&=
-\frac{1}{2\pi\mathrm{i}}
\int_{\partial D_\epsilon(k_d)}
w^E(x,t,s)\,ds, \qquad
F_2(x,t)
=
-\frac{1}{2\pi\mathrm{i}}
\int_{\mathcal{L}_{\mathrm{tr} }\setminus\partial \widetilde{\mathcal D}}
w^E(x,t,s)\,ds, \\
F_3(x,t)
&=
-\frac{1}{2\pi\mathrm{i}}
\int_{\mathcal{L}_{\mathrm{tr} } }(\mu_{\mathrm{tr}}-I)
w_{\mathrm{tr}}(x,t,s)\,ds.
\end{aligned}
\]
The matching estimate \eqref{eq:tr-local-kd-matching} gives
\(|F_1(x,t)|\lesssim t^{-1/2}\), while \eqref{E:wE-estimates}
yields \(|F_2(x,t)|\lesssim t^{-2/3}\log t\).
Combining \eqref{E:wmathrntr} and \eqref{E:extmutr}
with H\"older's inequality, we obtain
\(|F_3(x,t)|\lesssim t^{-2/3}\log t\).
Together, these bounds establish \eqref{eq:tr-E1-circle}.
\end{proof}

Note that, for \(k\in\partial D_\epsilon(k_c)\), we have
\begin{equation*}
 w_{\mathrm{tr}} (k)=
 N_B(k)\bigl(N^{\mathrm{loc},k_c}(k)^{-1}-I\bigr)N_B(k)^{-1}.
\end{equation*}
Since \(\widetilde Y\), \(N_B(k)\), and their inverses are analytic
in \(D_\epsilon(k_c)\), applying the residue theorem to the
asymptotic expansion \eqref{E:bzsy1} gives
\begin{align}
 -\frac{1}{2\pi\mathrm{i}}
\int_{\partial D_\epsilon(k_c)}
w_{\mathrm{tr}}(x,t,s)\,ds
 =\frac{N_B(k_c)\widetilde Y M_1^Y(y)\widetilde Y^{-1}N_B(k_c)^{-1}}
 {\left(\frac{8\sqrt6}{9q_0}\right)^{1/3} t^{1/3}}+O(t^{-2/3}), \qquad t \to \infty.
 \label{eq:tr-E1-residue}
\end{align}
For convenience, we define
\(\widetilde{\mathcal E}_{\mathrm{tr}}^{(1)}(x,t)\) by
\begin{equation}
 \widetilde{\mathcal E}_{\mathrm{tr}}^{(1)}(x,t)= \left(\frac{8\sqrt6}{9q_0}\right)^{-1/3} \left[ N_B(k_c)\widetilde Y M_1^Y(y)\widetilde Y^{-1}N_B(k_c)^{-1} \right].
 \label{eq:tr-Etilde}
\end{equation}
Combining \eqref{eq:tr-E1-circle} and \eqref{eq:tr-E1-residue},
we obtain
\begin{align} \label{E:estoooo}
 \mathcal E_{\mathrm{tr}}^{(1)}(x,t)= t^{-1/3}  \widetilde{\mathcal E}_{\mathrm{tr}}^{(1)}(x,t)+O(t^{-1/2}), \qquad
t \to \infty.
\end{align}

\subsection{Proof of Theorem \ref{thm:tr-final}} \label{S:poggP}
As before, let \(M_1\) denote the coefficient of \(k^{-1}\) in
the large-\(k\) expansion of the matrix-valued function \(M\),
and let \((M_1)_{\mathrm{rc}}\) denote the column vector formed
by the \((2,1)\) and \((3,1)\) entries of \(M_1\).
Reversing all the transformations in this section, as in the
previous two sections, we obtain a formula analogous to
\eqref{eq:M1-root-column}:
\begin{equation}
 (M_1)_{\mathrm{rc}}=
 \left[M_\infty\widetilde\sigma
 e^{\ii\widetilde g_\infty\sigma}
 \bigl(\mathcal E_{\mathrm{tr}}^{(1)}+N_B^{(1)}\bigr)
 e^{-\ii\widetilde g_\infty\sigma}\widetilde\sigma^{-1}
 \right]_{\{2,3\}\times\{1\}},
 \label{eq:tr-root-column}
\end{equation}
where \([A]_{\{2,3\}\times\{1\}}\) denotes the column vector
formed by the \((2,1)\) and \((3,1)\) entries of \(A\).  

We substitute the explicit expressions for the constant matrices
\(M_\infty\), \(\sigma\), and \(\widetilde\sigma\), the expression for
\(N_B^{(1)}\) in \eqref{eq:tr-MB-expansion}, and the asymptotic
expansion of \(\mathcal E_{\mathrm{tr}}^{(1)}(x,t)\) in
\eqref{E:estoooo} into \eqref{eq:tr-root-column}.
Using the reconstruction formula \eqref{eq:reconstruction},
we then obtain by direct calculation
\begin{equation*}
 \mathbf q(x,t)
 =\mathbf q_+e^{-2\mathrm{i}\widetilde g_\infty}
 +t^{-1/3}\mathbf q_P(x,t)+O(t^{-1/2}),
 \qquad t\to\infty,
\end{equation*}
where
\begin{equation}
 \mathbf q_P(x,t)
 =-\frac{2\mathrm{i}}{q_0}
 \left(
 \bigl(\widetilde{\mathcal E}_{\mathrm{tr}}^{(1)}\bigr)_{21}
 e^{-2\mathrm{i}\widetilde g_\infty}\mathbf q_+
 +
 \bigl(\widetilde{\mathcal E}_{\mathrm{tr}}^{(1)}\bigr)_{31}
 e^{-\mathrm{i}\widetilde g_\infty}\mathbf q_+^\perp
 \right).
 \label{eq:tr-qP}
\end{equation}
Here $\widetilde{\mathcal E}_{\mathrm{tr}}^{(1)}$ is given by \eqref{eq:tr-Etilde}. We now give the explicit expressions for the \((2,1)\) and
\((3,1)\) entries of this matrix.
Using the expressions for \(N_B(k)\), \(M_1^Y\), and
\(\widetilde Y\) in \eqref{eq:tr-MB-expansion},
\eqref{E:defMX1dd}, and \eqref{eq:tr-Y}, respectively, we find
\(
 \bigl(\widetilde{\mathcal E}_{\mathrm{tr}}^{(1)}\bigr)_{31}=0
\)
and
$$
\bigl(\widetilde{\mathcal E}_{\mathrm{tr}}^{(1)}\bigr)_{21}=\left(\frac{8\sqrt6}{9q_0}\right)^{-1/3} \left(\kappa_1 \mathcal{V}^*(y)+\kappa_2 \mathcal{V}(y)+\kappa_3 \widehat{\mathcal{V}}(y) \right).
$$
 The coefficients $\kappa_j$, $j=1,2,3$, are given by
\begin{align} \label{kappaj123}
\kappa_1=\frac{k_c+\lambda(k_c)}{2 \lambda(k_c)} (\widetilde Y_{11})^{-2},\qquad
\kappa_2=\frac{k_c-\lambda(k_c)}{2 \lambda(k_c)} (\widetilde Y_{11})^{2},\qquad 
\kappa_3=\frac{\ii q_0}{\lambda(k_c)}.
\end{align}
The functions \(\mathcal V(y)\) and \(\widehat{\mathcal V}(y)\)
can be expressed in terms of the increasing tritronqu\'ee solution
\(Q(y)\) of an inhomogeneous Painlev\'e-II equation, as given in
\eqref{E:defmathcalV} and \eqref{E:hatmathcalV}, respectively.
Thus, \(\mathbf q_P(x,t)\) defined by \eqref{eq:tr-qP} simplifies to
\begin{equation}
 \mathbf q_P(x,t)   
 =-
\frac{\ii \left( \kappa_1 \mathcal{V}^*(y)+\kappa_2 \mathcal{V}(y)+\kappa_3 \widehat{\mathcal{V}}(y) \right)}{q_0 \left(\frac{\sqrt6}{9q_0}\right)^{1/3}}
 e^{-2\mathrm{i}\widetilde g_\infty}\mathbf q_+.
 \label{eq:tr-qP-re}
\end{equation}
This completes the proof of the theorem \ref{thm:tr-final}.

\appendix
\section{Parabolic-cylinder model}\label{App:AAA}
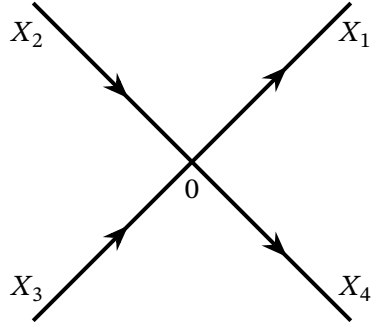
\begin{figure}
\centering
\begin{tikzpicture}[
    scale=0.7,
    >=Stealth,
    ray/.style={
        black,
        line width=1.5pt,
        postaction={decorate},
        decoration={
            markings,
            mark=at position .60 with {\arrow{>}}
        }
    }
]

    \draw[ray] (-3,-3)--(0,0);
    \draw[ray] (0,0)--(3,3);
    \draw[ray] (-3,3)--(0,0);
    \draw[ray] (0,0)--(3,-3);

    \node at (0,-0.5) {$0$};
    \node at (3.1,2.4) {$X_1$};
    \node at (-3.1,2.4) {$X_2$};
    \node at (-3.1,-2.4) {$X_3$};
    \node at (3.1,-2.4) {$X_4$};

\end{tikzpicture}
\caption{The jump contour
$X=\bigcup_{j=1}^{4}X_j$ for
$N^{\mathrm{PC}}(\zeta;p,q)$.}
\label{fig:B}
\end{figure}

In this appendix, we collect some basic results for the model RH problem. Its jump contour is
\(
    X=\bigcup_{j=1}^{4}X_j
\),
where, as shown in Figure~\ref{fig:B},
\begin{equation}
\label{E:Xj}
\begin{aligned}
X_1&=\left\{\zeta\in\mathbb{C}:
\zeta=r\mathrm{e}^{\pi\mathrm{i}/4},\ 0\leq r<\infty\right\},\qquad 
X_2=\left\{\zeta\in\mathbb{C}:
\zeta=r\mathrm{e}^{3\pi\mathrm{i}/4},\ 0\leq r<\infty\right\},\\
X_3&=\left\{\zeta\in\mathbb{C}:
\zeta=r\mathrm{e}^{5\pi\mathrm{i}/4},\ 0\leq r<\infty\right\},\qquad  
X_4=\left\{\zeta\in\mathbb{C}:
\zeta=r\mathrm{e}^{7\pi\mathrm{i}/4},\ 0\leq r<\infty\right\}.
\end{aligned}
\end{equation}
Let $p$ and $q$ belong to a subset $\mathbb{D}$ of the complex plane, and define
\(
    \nu=\nu(p,q)
    =
    -\frac{1}{2\pi}\log(1-pq)
\), $ pq<0$.

\begin{RHP}
Find a $2\times2$ matrix-valued function
$N^{\mathrm{PC}}(\zeta;p,q)$ satisfying the following conditions:

\begin{enumerate}

\item
$N^{\mathrm{PC}}(\cdot;p,q)$ is analytic in $\mathbb{C}\setminus X$.

\item
$N^{\mathrm{PC}}$ admits continuous boundary values on
$X\setminus\{0\}$ satisfying
\[
    N^{\mathrm{PC}+}(\zeta;p,q)
    =
    N^{\mathrm{PC}-}(\zeta;p,q)
    V^{\mathrm{PC}}(\zeta;p,q),
    \qquad
    \zeta\in X\setminus\{0\},
\]
where
\begin{align*}
V_1^{\mathrm{PC}}
&=
\begin{pmatrix}
1 & p\zeta^{-2\mathrm{i}\nu}
      \mathrm{e}^{\mathrm{i}\zeta^2/2}\\
0 & 1
\end{pmatrix},
&
V_2^{\mathrm{PC}}
&=
\begin{pmatrix}
1 & 0\\
-\dfrac{q}{1-pq}\zeta^{2\mathrm{i}\nu}
      \mathrm{e}^{-\mathrm{i}\zeta^2/2} & 1
\end{pmatrix},
\\[2mm]
V_3^{\mathrm{PC}}
&=
\begin{pmatrix}
1 & \dfrac{p}{1-pq}\zeta^{-2\mathrm{i}\nu}
      \mathrm{e}^{\mathrm{i}\zeta^2/2}\\
0 & 1
\end{pmatrix},
&
V_4^{\mathrm{PC}}
&=
\begin{pmatrix}
1 & 0\\
-q\zeta^{2\mathrm{i}\nu}
      \mathrm{e}^{-\mathrm{i}\zeta^2/2} & 1
\end{pmatrix}.
\end{align*}
Here,
\(
    \zeta^{2\mathrm{i}\nu}
    =
    \mathrm{e}^{2\mathrm{i}\nu\log\zeta}
\).

\item
As $\zeta\to\infty$,
\(
    N^{\mathrm{PC}}(\zeta;p,q)\to I
\).

\item
As $\zeta\to0$,
\(
    N^{\mathrm{PC}}(\zeta;p,q)=O(1)
\).

\end{enumerate}
\end{RHP}

The model RH problem can be solved using the standard parabolic-cylinder construction; see, for example~\cite{pz93}. We omit the details and record only the large-$\zeta$ asymptotics of its solution.

\begin{lemma} \label{L:asyofLemmaPC}

The solution $N^{\mathrm{PC}}(\zeta;p,q)$ has the asymptotic expansion
\begin{equation}
\label{E:mXasymoxia}
N^{\mathrm{PC}}(\zeta;p,q)
=
I
+
\frac{N_{\infty}^{\mathrm{PC}}(p,q)}{\zeta}
+
O\left(\frac{1}{\zeta^2}\right),
\qquad
\zeta\to\infty,
\end{equation}
where the error term is uniform with respect to
$\arg\zeta\in[0,2\pi]$ and for $p$ and $q$ in compact subsets of
$\mathbb{D}$. The coefficient
$N_{\infty}^{\mathrm{PC}}(p,q)$ is given by
\begin{equation}
\label{E:m1Xdefbuzbz}
N_{\infty}^{\mathrm{PC}}(p,q)
=
\begin{pmatrix}
0 & \mathrm{i}\beta_{12}^{\mathrm{PC}}\\
-\mathrm{i}\beta_{21}^{\mathrm{PC}} & 0
\end{pmatrix},
\end{equation}
where
\begin{equation}
\label{E:beta1232}
\beta_{12}^{\mathrm{PC}}
=
\frac{
\sqrt{2\pi}\,
\mathrm{e}^{-\pi\nu/2}
\mathrm{e}^{-\pi\mathrm{i}/4}
}{
q\,\Gamma(\mathrm{i}\nu)
},
\qquad
\beta_{21}^{\mathrm{PC}}
=
\frac{
\sqrt{2\pi}\,
\mathrm{e}^{-\pi\nu/2}
\mathrm{e}^{\pi\mathrm{i}/4}
}{
p\,\Gamma(-\mathrm{i}\nu)
}.
\end{equation}
Here, $\Gamma$ denotes the Gamma function. Moreover, for every compact subset
$K\subset\mathbb{D}$,
\[
    \sup_{p,q\in K}
    \sup_{\zeta\in\mathbb{C}\setminus X}
    \left\|N^{\mathrm{PC}}(\zeta;p,q)\right\|
    <\infty.
\]
 \end{lemma}

 \section{Airy model}\label{App:BBB}
This appendix presents the Airy model used to construct the local
parametrices near \(\alpha\) and \(\bar{\alpha}\) in
Section~\ref{subsubsec:elliptic-Airy-alpha}.   The Airy model arises frequently in the study of integrable systems and the results presented here are standard. For completeness, this appendix provides a concise and self-contained account
of the facts needed in the present work. Further details may be found,
for example, in \cite{BoutetDeMonvelLenellsShepelsky2022}.

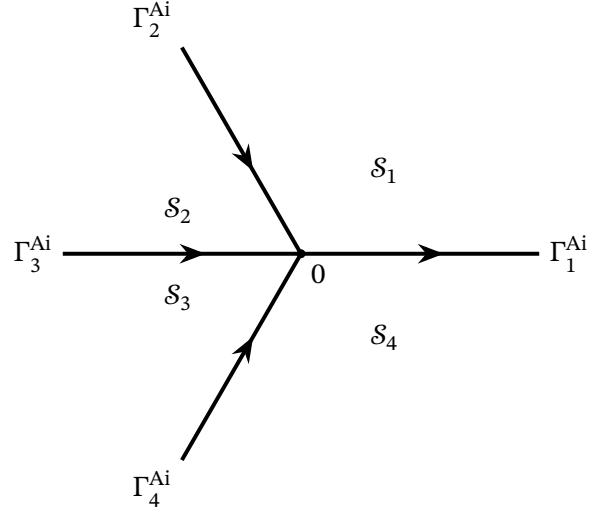
\begin{figure}
\centering
\begin{tikzpicture}[
    >=Stealth,
    scale=1.05,
    ray/.style={
        black,
        line width=1.5pt,
        postaction={decorate},
        decoration={
            markings,
            mark=at position .60 with {\arrow{>}}
        }
    }
]
    \coordinate (O) at (0,0);

    \draw[ray] (O)--(3,0);
    \draw[ray] (120:3)--(O);
    \draw[ray] (-3,0)--(O);
    \draw[ray] (240:3)--(O);

    \fill (O) circle (1.6pt)
        node[below right] {\(0\)};

    \node[right] at (3,0)
        {\(\Gamma_1^{\mathrm{Ai}}\)};
    \node[above left] at (120:3)
        {\(\Gamma_2^{\mathrm{Ai}}\)};
    \node[left] at (-3,0)
        {\(\Gamma_3^{\mathrm{Ai}}\)};
    \node[below left] at (240:3)
        {\(\Gamma_4^{\mathrm{Ai}}\)};

    \node at (1.05,1.05)   {\(\mathcal S_1\)};
    \node at (-1.55,.55)   {\(\mathcal S_2\)};
    \node at (-1.55,-.55)  {\(\mathcal S_3\)};
    \node at (1.05,-1.05)  {\(\mathcal S_4\)};

\end{tikzpicture}
\caption{The oriented Airy contour
\(\Gamma^{\mathrm{Ai}}=\bigcup_{j=1}^4\Gamma_j^{\mathrm{Ai}}\)
and the four sectors \(\mathcal S_j\).}
\label{fig:app-Airy-contour}
\end{figure}

We state the associated
RH problem first and then give its explicit solution in
terms of Airy functions.  
Let $\Gamma^{\mathrm{Ai}}
 :=\bigcup_{j=1}^4\Gamma_j^{\mathrm{Ai}}$, where
\begin{equation}
 \begin{aligned}
 \Gamma_1^{\mathrm{Ai}}=\mathbb R_+,\qquad
 \Gamma_2^{\mathrm{Ai}}=e^{2\pi\mathrm i/3}\mathbb R_+,\qquad
 \Gamma_3^{\mathrm{Ai}}=\mathbb R_-, \qquad
 \Gamma_4^{\mathrm{Ai}}=e^{-2\pi\mathrm i/3}\mathbb R_+
 \end{aligned}.
 \label{eq:app-Airy-rays}
\end{equation}
The first ray is oriented away from the origin, whereas the other three
rays are oriented toward the origin, as shown in
Figure~\ref{fig:app-Airy-contour}.   The rays divide the
plane into the four open sectors
 \begin{align*}
 \mathcal S_1&=\{\zeta:0<\arg\zeta<2\pi/3\},&
 \mathcal S_2&=\{\zeta:2\pi/3<\arg\zeta<\pi\},\\
 \mathcal S_3&=\{\zeta:-\pi<\arg\zeta<-2\pi/3\},&
 \mathcal S_4&=\{\zeta:-2\pi/3<\arg\zeta<0\}.
 \end{align*}
Set
\begin{equation}
 \sigma_3:=\operatorname{diag}(1,-1),\qquad
 \varpi_{\mathrm A}:=e^{2\pi\mathrm i/3},\qquad
 J_{\mathrm A}:=
 \begin{pmatrix}0&1\\-1&0\end{pmatrix},
 \label{eq:app-Airy-notation}
\end{equation}
and introduce the following   function
\begin{equation}
 N_0^{\mathrm{Ai}}(\zeta)
 :=\frac{e^{\pi\mathrm i/12}}{2\sqrt\pi}
 \begin{pmatrix}
 \zeta^{-1/4}&0\\
 0&\zeta^{1/4}
 \end{pmatrix}
 \begin{pmatrix}
 1&1\\
 -1&1
 \end{pmatrix}
 e^{-\pi\mathrm i\sigma_3/4}.
 \label{eq:app-Airy-leading-factor}
\end{equation}

We can now formulate the model problem precisely.

\begin{RHP}
Find a \(2\times2\) matrix-valued function
\(N^{\mathrm{Ai}}(\zeta)\) with the following properties.
\begin{enumerate}
 \item[(i)] \emph{Analyticity.}
 The function \(N^{\mathrm{Ai}}\) is analytic in
 \(\mathbb C\setminus\Gamma^{\mathrm{Ai}}\) and has continuous
 nontangential boundary values on each open ray
 \(\Gamma_j^{\mathrm{Ai}}\setminus\{0\}\).

 \item[(ii)] \emph{Jump conditions.}
 The boundary values satisfy
 \begin{equation}
  N^{\mathrm{Ai},+}(\zeta)
  =N^{\mathrm{Ai},-}(\zeta)V^{\mathrm{Ai}}(\zeta),
  \qquad \zeta\in\Gamma^{\mathrm{Ai}}\setminus\{0\},
  \label{eq:app-Airy-RHP-jump}
 \end{equation}
 where
 \begin{equation}
 V^{\mathrm{Ai}}(\zeta)=
 \begin{cases}
 \begin{pmatrix}
 1&e^{-\frac43\zeta^{3/2}}\\
 0&1
 \end{pmatrix},
 &\zeta\in\Gamma_1^{\mathrm{Ai}},\\[5mm]
 \begin{pmatrix}
 1&0\\
 e^{\frac43\zeta^{3/2}}&1
 \end{pmatrix},
 &\zeta\in\Gamma_2^{\mathrm{Ai}}\cup\Gamma_4^{\mathrm{Ai}},\\[5mm]
 J_{\mathrm A},
 &\zeta\in\Gamma_3^{\mathrm{Ai}}.
 \end{cases}
 \label{eq:app-Airy-jump-matrices}
 \end{equation}

 \item[(iii)] \emph{Behavior at the origin.}
 The function \(N^{\mathrm{Ai}}(\zeta)\) remains bounded as
 \(\zeta\to0\) within each sector \(\mathcal S_j\).

 \item[(iv)] \emph{Normalization at infinity.}
 As \(\zeta\to\infty\),
 \begin{equation}
  \bigl(N_0^{\mathrm{Ai}}(\zeta)\bigr)^{-1}
  N^{\mathrm{Ai}}(\zeta)
  =I+O(\zeta^{-3/2}),
  \label{eq:app-Airy-normalization}
 \end{equation}
 uniformly with respect to $\mathrm{arg} \zeta \in [0, 2 \pi]$.
\end{enumerate}
\end{RHP}

We next construct the solution.  In terms of the Airy function \(\operatorname{Ai}(\cdot)\).  Define
\begin{equation}
 \mathcal A(\zeta):=
 \begin{cases}
 \begin{pmatrix}
 \operatorname{Ai}(\zeta)&
 \operatorname{Ai}(\varpi_{\mathrm A}^2\zeta)\\
 \operatorname{Ai}'(\zeta)&
 \varpi_{\mathrm A}^2
 \operatorname{Ai}'(\varpi_{\mathrm A}^2\zeta)
 \end{pmatrix}
 e^{-\pi\mathrm i\sigma_3/6},
 &\Im\zeta>0,\\[6mm]
 \begin{pmatrix}
 \operatorname{Ai}(\zeta)&
 -\varpi_{\mathrm A}^2
 \operatorname{Ai}(\varpi_{\mathrm A}\zeta)\\
 \operatorname{Ai}'(\zeta)&
 -\operatorname{Ai}'(\varpi_{\mathrm A}\zeta)
 \end{pmatrix}
 e^{-\pi\mathrm i\sigma_3/6},
 &\Im\zeta<0.
 \end{cases}
 \label{eq:app-Airy-fundamental-matrix}
\end{equation} 
The solution of the model problem can then be expressed in terms of
\(\mathcal A(\zeta)\) as follows:
\begin{equation}
N^{\mathrm{Ai}}(\zeta)
 =\mathcal A(\zeta)
 \begin{cases}
 e^{\frac23\zeta^{3/2}\sigma_3},
 &\zeta\in\mathcal S_1\cup\mathcal S_4,\\[2mm]
 \begin{pmatrix}
 1&0\\
 -1&1
 \end{pmatrix}
 e^{\frac23\zeta^{3/2}\sigma_3},
 &\zeta\in\mathcal S_2,\\[5mm]
 \begin{pmatrix}
 1&0\\
 1&1
 \end{pmatrix}
 e^{\frac23\zeta^{3/2}\sigma_3},
 &\zeta\in\mathcal S_3.
 \end{cases}
 \label{eq:app-Airy-explicit-solution}
\end{equation}
We now verify that the matrix \(N^{\mathrm{Ai}}\) defined by
\eqref{eq:app-Airy-explicit-solution} satisfies all the conditions of
the Airy model RH problem.
\begin{theorem}
\label{thm:app-Airy-model}
The matrix \(N^{\mathrm{Ai}}\) defined by
\eqref{eq:app-Airy-explicit-solution} is the unique solution of the
Airy-model RH problem.  
\end{theorem}

\begin{proof}
We first establish the uniqueness of the solution to the Airy model RH
problem.  Since
\(\det V^{\mathrm{Ai}}=1\), the determinant of any solution has no jump.
Its behavior at the origin and at infinity, together with Liouville's
theorem, gives
\[
\det N^{\mathrm{Ai}}(\zeta)
=
\frac{e^{\pi\mathrm{i}/6}}{2\pi}
\neq0.
\]
Thus, every solution is invertible. If \(N_1^{\mathrm{Ai}}\) and
\(N_2^{\mathrm{Ai}}\) are two solutions, then
\(N_1^{\mathrm{Ai}}(N_2^{\mathrm{Ai}})^{-1}\) extends to an entire
function and tends to \(I\) at infinity. Liouville's theorem therefore
implies that it is identically \(I\), proving uniqueness.

Since the Airy function is entire,
\(N^{\mathrm{Ai}}(\zeta)\) is analytic in
\(\mathbb{C}\setminus\Gamma^{\mathrm{Ai}}\). The Airy connection formula
\[
\operatorname{Ai}(z)
+\varpi_{\mathrm A}\operatorname{Ai}(\varpi_{\mathrm A}z)
+\varpi_{\mathrm A}^2
\operatorname{Ai}(\varpi_{\mathrm A}^2z)
=0
\]
implies that \( N^{\mathrm{Ai}}(\zeta)\) satisfies the jump
condition \eqref{eq:app-Airy-RHP-jump}. Moreover, the boundedness of
\(\operatorname{Ai}\) and \(\operatorname{Ai}'\) near the origin,
together with \eqref{eq:app-Airy-explicit-solution}, gives the required
behavior at \(\zeta=0\).

Finally, the standard large-\(z\) asymptotic expansions
\[
\operatorname{Ai}(z)
=
\frac{z^{-1/4}}{2\sqrt{\pi}}
e^{-\frac{2}{3}z^{3/2}}
\left(1+O(z^{-3/2})\right),
\qquad
\operatorname{Ai}'(z)
=
-\frac{z^{1/4}}{2\sqrt{\pi}}
e^{-\frac{2}{3}z^{3/2}}
\left(1+O(z^{-3/2})\right),
\]
 yield the normalization
\eqref{eq:app-Airy-normalization}. Therefore,
\( N^{\mathrm{Ai}}(\zeta)\) satisfies all the conditions of
the Airy model RH problem and, by uniqueness, is its unique solution.
\end{proof}


\section{Painlev\'e-II model}
\label{app:PII}
\begin{figure}
\centering
\begin{tikzpicture}[>=Stealth,
 ray/.style={black,line width=1.5pt,postaction={decorate},
 decoration={markings,mark=at position .60 with {\arrow{>}}}},
 lab/.style={font=\small,fill=white,inner sep=2pt}]
 \begin{scope}[shift={(-3.4,0)}]
  \draw[ray] (0,1.65)--(0,0);
  \draw[ray] (0,-1.65)--(0,0);
  \draw[ray] (0,0)--(1,0);
  \draw[ray] (1,0)--(2.6,.924);
  \draw[ray] (1,0)--(2.6,-.924);
  \node[lab] at (-.28,.85) {1};
  \node[lab] at (-.28,-.85) {2};
  \node[lab] at (1.95,.85) {3};
  \node[lab] at (1.95,-.85) {4};
  \node[lab] at (.50,.28) {5};
  \node[lab] at (1,-.30) {$b$};
  \node at (.90,-2.0) {$(x,t) \in \mathcal T_+$};
 \end{scope}
 \begin{scope}[shift={(3.0,0)}]
  \draw[ray] (0,1.65)--(0,0);
  \draw[ray] (0,-1.65)--(0,0);
  \draw[ray] (0,0)--(2,1.154);
  \draw[ray] (0,0)--(2,-1.154);
  \node[lab] at (-.28,.85) {1};
  \node[lab] at (-.28,-.85) {2};
  \node[lab] at (1.25,1.0) {3};
  \node[lab] at (1.25,-1.0) {4};
  \node at (.80,-2.0) {$(x,t) \in \mathcal T_-$};
 \end{scope}
\end{tikzpicture}
\caption{The oriented model contours \(\Sigma^Y\).}
\label{fig:tr-Y-contours}
\end{figure}
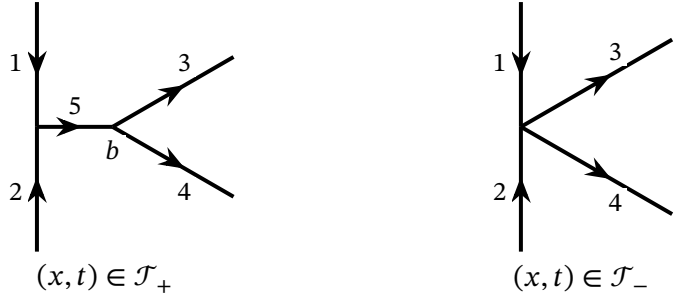

We are now ready to formulate a Painlev\'e-II model RH problem
in the \(z\)-plane. We first specify its jump contour.
For \((x,t)\in\mathcal T_+\), we set
\(b=\left(\frac{8\sqrt6}{9q_0}\right)^{1/3} t^{1/3}(k_0-k_c)\ge 0\), whereas for
\((x,t)\in\mathcal T_-\), we set \(b=0\).
It is easy to see that \(b=O(1)\) in the former case.
Define
\[
\begin{aligned}
 \Sigma_1^Y=\ii\mathbb R_+,
 \quad \Sigma_2^Y=-\ii\mathbb R_+,\quad
 \Sigma_3^Y=b+e^{\ii\pi/6}\mathbb R_+,
 \quad \Sigma_4^Y=b+e^{-\ii\pi/6}\mathbb R_+,\quad
 \Sigma_5^Y=[0,b].
\end{aligned}
\]
The model jump contour~(see Fig. \ref{fig:tr-Y-contours}) is then given by
\begin{equation}
 \Sigma^Y=
 \begin{cases}
 \displaystyle\bigcup_{j=1}^5\Sigma_j^Y,
 &(x,t)\in\mathcal T_+,\\[2mm]
 \displaystyle\bigcup_{j=1}^4\Sigma_j^Y,
 &(x,t)\in\mathcal T_-.
 \end{cases}
 \label{eq:tr-SigmaY}
\end{equation}

We then formulate the following RH problem in the \(z\)-plane.

\begin{RHP}\label{prop:tr-Y-RHP}
Find a \(3\times3\) matrix \(M^Y(z;y)\) with the following properties:
\begin{enumerate}
 \item[(i)] \(M^Y\) is analytic in \(\mathbb C\setminus\Sigma^Y\)
 and has nontangential boundary values on $\Sigma^Y$.
 \item[(ii)] The boundary values satisfy the jump condition
 \begin{equation}
  M_+^Y(z;y)=M_-^Y(z;y)V_j^Y(z;y),\qquad z\in\Sigma_j^Y,
  \label{eq:tr-Y-jump-RHP}
 \end{equation}
 where the matrix $V_j^Y$  is given by 
\begin{align}\label{eq:tr-VY}
 V_1^Y&=\begin{pmatrix}1&0&0\\
 \rho_0(-z)^{2\ii\widetilde\nu_1}e^{-\ii\Phi}&1&0\\0&0&1
 \end{pmatrix},
 &V_2^Y&=\begin{pmatrix}1&
 \rho_0(-z)^{-2\ii\widetilde\nu_1}e^{\ii\Phi}&0\\
 0&1&0\\0&0&1\end{pmatrix},\notag\\
 V_3^Y&=\begin{pmatrix}1&
 \dfrac{\rho_0}{1+\rho_0^2}(-z)^{-2\ii\widetilde\nu_1}
 e^{\ii\Phi}&0\\0&1&0\\0&0&1\end{pmatrix},
 &V_4^Y&=\begin{pmatrix}1&0&0\\
 \dfrac{\rho_0}{1+\rho_0^2}(-z)^{2\ii\widetilde\nu_1}
 e^{-\ii\Phi}&1&0\\0&0&1\end{pmatrix}, \notag\\
 V_5^Y&=\begin{pmatrix}
 1&\dfrac{\rho_0}{1+\rho_0^2}(-z)_+^{-2\ii\widetilde\nu_1}
 e^{\ii\Phi}&0\\[2mm]
 \dfrac{\rho_0}{1+\rho_0^2}(-z)_-^{2\ii\widetilde\nu_1}
 e^{-\ii\Phi}&1+\rho_0^2&0\\
 0&0&1
 \end{pmatrix}.
 \end{align}
Here $\Phi=yz+z^3$.
 \item[(iii)] \(M^Y(z;y)=I+O(z^{-1})\) as \(z\to\infty\).
 \item[(iv)] \(M^Y\) is bounded at \(z=0\) and, when
\((x,t)\in\mathcal T_+\), also at \(z=b\).
\end{enumerate}
\end{RHP}

When \(\rho_0=0\), the solution of the RH problem above
is the identity matrix. We therefore assume
\(\rho_0\ne0\) in what follows.
We note that this RH problem can be solved using the solution of 
Jimbo--Miwa Painlev\'e-II model RH problem studied in \cite{Miller}.
We first recall this Painlev\'e-II model and collect the properties
of its solution that we will need. Throughout this appendix, let
\[
 \sigma=\operatorname{diag}(1,-1,0),\qquad
 \sigma_3=\operatorname{diag}(1,-1),\qquad
 \sigma_1=\begin{pmatrix}0&1\\1&0\end{pmatrix}.
\]
Then we define the RH problem as follows.
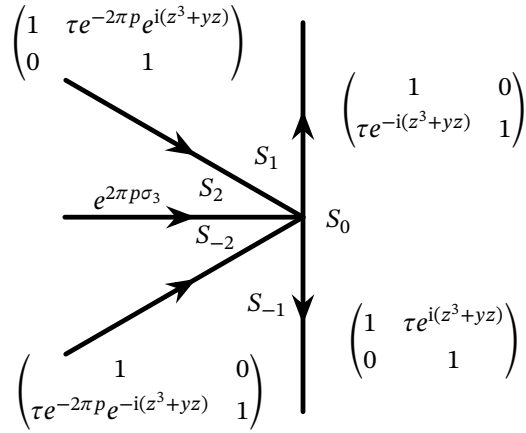
\begin{figure}[htbp]
\centering
\begin{tikzpicture}[
 scale=1.05,
 >=Stealth,
 line cap=round,
 line join=round,
 every node/.style={font=\small},
 contour/.style={draw=black,line width=1.8pt},
 arrowcontour/.style={
  contour,
  postaction={decorate},
  decoration={markings,
   mark=at position 0.55 with {\arrow{Stealth}}}
 }
]
 \coordinate (O) at (0,0);

 \draw[arrowcontour] (150:3.45)--(O);
 \draw[arrowcontour] (210:3.45)--(O);
 \draw[arrowcontour] (-3,0)--(O);
 \draw[arrowcontour] (O)--(0,2.45);
 \draw[arrowcontour] (O)--(0,-2.45);

 \node at (-0.45,0.75) {$S_1$};
 \node at (-1.15,0.35) {$S_2$};
 \node at (-1.10,-0.25) {$S_{-2}$};
 \node at (-0.45,-1.10) {$S_{-1}$};
 \node at (0.45,-0.05) {$S_0$};

 \node at (-2.20,0.25) {$e^{2\pi p\sigma_3}$};

 \node at (-2.20,2.20) {$
  \begin{pmatrix}
   1&\tau e^{-2\pi p}e^{\mathrm{i}(z^3+yz)}\\[0.4ex]
   0&1
  \end{pmatrix}$};

 \node at (1.65,1.35) {$
  \begin{pmatrix}
   1&0\\[0.4ex]
   \tau e^{-\mathrm{i}(z^3+yz)}&1
  \end{pmatrix}$};

 \node at (1.65,-1.55) {$
  \begin{pmatrix}
   1&\tau e^{\mathrm{i}(z^3+yz)}\\[0.4ex]
   0&1
  \end{pmatrix}$};

 \node at (-2.05,-2.20) {$
  \begin{pmatrix}
   1&0\\[0.4ex]
   \tau e^{-2\pi p}e^{-\mathrm{i}(z^3+yz)}&1
  \end{pmatrix}$};
\end{tikzpicture}
\caption{The oriented jump contour and jump matrices for \(W(z;y)\).}
\label{fig:zeta-contour-jumps}
\end{figure}

\begin{RHP}[Jimbo--Miwa Painlev\'e-II problem \cite{Miller}]
\label{rhp:PII-generalized}
Let \(y,p,\tau\in\mathbb{C}\) be related by
\(
\tau^2=e^{2\pi p}-1
\).
Seek a \(2\times 2\) matrix-valued function
\(
W(z;y)=W(z;y,p,\tau)
\)
with the following properties.

\begin{itemize}
\item[]\textbf{Analyticity.}
The matrix \(W(z;y)\) is analytic for \(z\) in the five sectors
 $S_0$: $|\arg(z)|<\tfrac{1}{2}\pi$, $S_1$: $\tfrac{1}{2}\pi<\arg(z)<\tfrac{5}{6}\pi$, $S_{-1}$: $-\tfrac{5}{6}\pi<\arg(z)<-\tfrac{1}{2}\pi$, $S_2$: $\tfrac{5}{6}\pi<\arg(z)<\pi$, and $S_{-2}$: $-\pi<\arg(z)<-\tfrac{5}{6}\pi$.
It takes continuous boundary values on the excluded rays and at the origin from
each sector.

\item[] \textbf{Jump conditions.}
The boundary values satisfy
\[
W_+(z;y)
=
W_-(z;y)V^{\mathrm{PII}}(z;y),
\]
where \(V^{\mathrm{PII}}(z;y)\) is the matrix defined on the jump
contour shown in Figure~\ref{fig:zeta-contour-jumps}.

\item[]\textbf{Normalization.}
As \(z\to\infty\), uniformly in all directions,
\(
W (z;y)z^{\mathrm{i}p\sigma_3}\to I
\).
\end{itemize}
\end{RHP}

The RH problem above was studied systematically in \cite{Miller}.
Throughout this appendix, we take \(p=-\widetilde\nu_1>0\) and
\(\tau=-\rho_0<0\). For these parameter values, the RH problem
has a unique solution for every \(y\in\mathbb R\).
Moreover, the product \(W(z;y)z^{\mathrm{i}p\sigma_3}\) admits
the complete asymptotic expansion
\begin{equation} 
 W(z;y)z^{\mathrm{i}p\sigma_3}
 \sim I+\sum_{j=1}^{\infty}W_j(y)z^{-j},
 \qquad z\to\infty,
 \label{E:Wasy}
\end{equation}
uniformly in all directions of the \(z\)-plane.
For \(p=-\widetilde\nu_1>0\), \(\tau=-\rho_0<0\), and
\(y\in\mathbb R\), the first coefficient has the form
\begin{align} \label{E:W1asy}
 W_1(y)=
 \begin{pmatrix}
  \widehat{\mathcal V}(y)&-\mathcal V^*(y)\\[0.6ex]
  \mathcal V(y)&\widehat{\mathcal V}^*(y)
 \end{pmatrix},
\end{align}
where the superscript \(*\) denotes complex conjugation.
The scalar functions \(\mathcal V\) and \(\widehat{\mathcal V}\)
can be expressed in terms of an increasing tritronqu\'ee solution
of an inhomogeneous Painlev\'e-II equation.
More precisely, let \(Q(y)\) be the solution of
\begin{equation}
 Q''(y)+\frac23yQ(y)-2Q(y)^3
 +\frac23\mathrm{i} \widetilde{\nu}_1-\frac13=0,
 \label{eq:PII-Q-equation}
\end{equation}
characterized by
\begin{equation}
 Q(y)=\mathrm{i}\left(-\frac y3\right)^{1/2}
 -\left(\frac14-\frac{\mathrm{i} \widetilde{\nu}_1}{2}\right)\frac1y
 +O(|y|^{-5/2}),
 \qquad |y|\to\infty, \qquad |\arg(-y)|<2\pi/3.
 \label{eq:PII-Q-asymptotics}
\end{equation}
Then the function \(\mathcal V(y)\) is determined by the logarithmic
derivative relation
\[
 \frac{\mathcal V'(y)}{\mathcal V(y)}=Q(y)
\]
together with the asymptotic condition
\begin{equation}
 \mathcal V(y)
 =\frac{\rho_0  \widetilde{\nu}_1 \Gamma(-\mathrm{i} \widetilde{\nu}_1 )}{2\sqrt{\pi}}\,
e^{-\frac{3\pi\mathrm{i}}{4}}
e^{\frac{\pi \widetilde{\nu}_1}{2}}
2^{\mathrm{i} \widetilde{\nu}_1}
 e^{-2\mathrm{i}(-y/3)^{3/2}}
 (-3y)^{-1/4+\mathrm{i}\widetilde{\nu}_1/2}
 \bigl[1+O(|y|^{-5/4})\bigr],
 \qquad y\to-\infty.
 \label{eq:PII-V-negative-asymptotics}
\end{equation}
Thus,  \(\mathcal V(y)\) can be written in the following form:
\begin{equation} \label{E:defmathcalV}
\mathcal V(y)=
\begin{cases}
\displaystyle
\alpha_0e^{\mathrm{i}\phi_0}
e^{-\frac{2}{9} \sqrt{3}\ii \left(- y\right)^{3/2}}
(- 3 y)^{-\frac14+\frac{\ii \widetilde{\nu}_1}{2}}
\\[1mm]
\displaystyle\qquad\qquad\qquad
{}\times
\exp\left\{
\int_{-\infty}^{y}
\left[
Q(s)-\ii\left(-\frac{s}{3}\right)^{1/2}
+
\left(\frac14-\frac{\ii \widetilde{\nu}_1}{2}\right)\frac{1}{s}
\right]\mathrm ds
\right\},
& y<0,
\\[5mm]
\displaystyle
\mathcal V(-1)\exp\left\{
\int_{-1}^{y}Q(s)\,\mathrm ds
\right\},
& y\ge 0,
\end{cases}
\end{equation}
where
\[
 \alpha_0=|C_{p,\tau}|
 =\sqrt{-\frac{\widetilde\nu_1}{2}},
 \qquad
 \phi_0=\frac{\pi}{4}
 +\widetilde\nu_1\log2
 +\arg\Gamma(-\mathrm{i}\widetilde\nu_1).
\]
Finally, \(\widehat{\mathcal V}\) is recovered from
\(\mathcal V\) through
\begin{align} \label{E:hatmathcalV}
 \widehat{\mathcal V}(y)=
 \begin{cases}
 \displaystyle
 \mathrm{i}\widetilde\nu_1\left(-\frac y3\right)^{1/2}
 +\mathrm{i}\int_{-\infty}^{y}
 \left[
  |\mathcal V(s)|^2
  +\frac{\widetilde\nu_1}{2\sqrt3}(-s)^{-1/2}
 \right]\mathrm{d}s,
 &y<0,\\[5mm]
 \displaystyle
 \widehat{\mathcal V}(-1)
 +\mathrm{i}\int_{-1}^{y}|\mathcal V(s)|^2\,\mathrm{d}s,
 &y\geq0.
 \end{cases}
\end{align}
The results above are obtained by specializing the corresponding
results in \cite{Miller} to the parameter values used here.
We therefore omit their proofs. The reader may also refer to the appendix of \cite{GengZhangWang2026}.


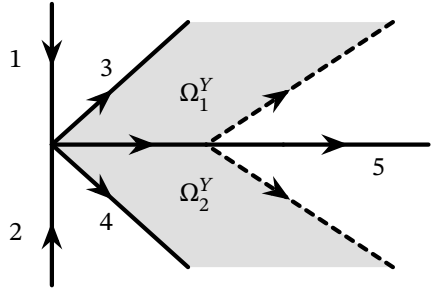
\begin{figure}[htbp]
\centering
\begin{tikzpicture}[
    scale=1.2,
    >=latex,
    line cap=round,
    line join=round,
    every node/.style={font=\small},
    contour/.style={draw=black, line width=1.6pt},
    arrowcontour/.style={
        contour,
        postaction={decorate},
        decoration={
            markings,
            mark=at position 0.55 with {\arrowreversed{Stealth}}
        }
    }
]

    \coordinate (C) at (3.8,0);
    \coordinate (D) at (1.25,0);

    \coordinate (S1) at (2.75, 1.35);
    \coordinate (S2) at (2.75,-1.35);

    \coordinate (U2) at (2.95,0);
    \coordinate (L2) at (2.95,0);

    \coordinate (U1) at (5.007, 1.35);
    \coordinate (L1) at (5.007,-1.35);

    \fill[gray!25] (U1)--(S1)--(D)--(U2)--cycle;
    \fill[gray!25] (L1)--(S2)--(D)--(L2)--cycle;

    \draw[arrowcontour,dashed] (U1)--(U2);
    \draw[arrowcontour,dashed] (L1)--(L2);

    \draw[arrowcontour] (5.4,0)--(C);
    \draw[arrowcontour] (C)--(D);

    \draw[arrowcontour] (S1)--(D);
    \draw[arrowcontour] (S2)--(D);

    \draw[arrowcontour] (D)--(1.25, 1.55);
    \draw[arrowcontour] (D)--(1.25,-1.55);

    \node at (2.85, 0.55) {\(\Omega_1^Y\)};
    \node at (2.85,-0.55) {\(\Omega_2^Y\)};
    
        \node at (0.85, 0.95) {\(1\)};
         \node at (0.85, -0.95) {\(2\)};
          \node at (1.85, 0.85) {\(3\)};
          \node at (1.85, -0.85) {\(4\)};
          \node at (4.85, -0.25) {\(5\)};

\end{tikzpicture}
\caption{The shaded regions: \(\{ \Omega_j^Y\}_{j=1}^2\).
The solid black curves form the jump contour for \(\widehat{M}^Y\).}
\label{fig:RX-regions}
\end{figure}

We next show how to solve the local model RH
problem~\ref{prop:tr-Y-RHP} using the solution \(W(z;y)\) of
the Painlev\'e-II model above.
Let \(\Omega_1^Y\) and \(\Omega_2^Y\) denote the two sectors
shown in Figure~\ref{fig:RX-regions}.
We first introduce the following transformation:
\begin{equation}
 \widehat M^Y(z;y)=M^Y(z;y)(-z)^{-\ii\widetilde\nu_1\sigma}
 \begin{cases}
  G^Y(z;y),&(x,t)\in\mathcal T_+,\\
  I,&(x,t)\in\mathcal T_-,
 \end{cases}
 \label{eq:tr-MhatY}
\end{equation}
where 
\begin{equation}
 G^Y(z;y)=
 \begin{cases}
 \begin{pmatrix}
 1&-\dfrac{\rho_0}{1+\rho_0^2}e^{\ii\Phi(z;y)}&0\\
 0&1&0\\0&0&1
 \end{pmatrix},&z\in\Omega_1^Y,\\[5mm]
 \begin{pmatrix}
 1&0&0\\
 \dfrac{\rho_0}{1+\rho_0^2}e^{-\ii\Phi(z;y)}&1&0\\
 0&0&1
 \end{pmatrix},&z\in\Omega_2^Y,\\[5mm]
 I,&z\notin\Omega_1^Y\cup\Omega_2^Y.
 \end{cases}
 \label{eq:tr-GY}
\end{equation}
Let \(\widehat{\Sigma}^Y\) denote the jump contour for
\(\widehat M^Y(z;y)\), shown in Figure~\ref{fig:RX-regions}.
The corresponding jump matrix \(\widehat V^Y(z;y)\) is given by
\begin{align}\label{eq:tr-hatVY}
\widehat V_1^Y&=\begin{pmatrix}1&0&0\\
 \rho_0 e^{-\ii\Phi}&1&0\\0&0&1
 \end{pmatrix},
 &\widehat V_2^Y&=\begin{pmatrix}1&
 \rho_0 e^{\ii\Phi}&0\\
 0&1&0\\0&0&1\end{pmatrix},\notag\\
\widehat V_3^Y&=\begin{pmatrix}1&
 \dfrac{\rho_0}{1+\rho_0^2}
 e^{\ii\Phi}&0\\0&1&0\\0&0&1\end{pmatrix},
 &\widehat V_4^Y&=\begin{pmatrix}1&0&0\\
 \dfrac{\rho_0}{1+\rho_0^2}
 e^{-\ii\Phi}&1&0\\0&0&1\end{pmatrix}, \notag\\
 \widehat V_5^Y&=e^{-2\pi \widetilde{\nu}_1 \sigma}.
 \end{align}
Let \(W(z;y)=W(z;y,p,\tau)\) denote the solution of the
Jimbo--Miwa Painlev\'e-II problem~\ref{rhp:PII-generalized}.
A direct algebraic calculation shows that the matrix
\[
 \begin{pmatrix}
  \sigma_1 W(-z;y,-\widetilde\nu_1,-\rho_0)\sigma_1
  &\mathbf 0_{2\times1}\\
  \mathbf 0_{1\times2}&1
 \end{pmatrix}
\]
satisfies the jump relations \eqref{eq:tr-hatVY}.

A standard application of Liouville's theorem shows that local model RH
problem~\ref{prop:tr-Y-RHP}  has at most one solution.
Its unique solution is therefore given by
\begin{align} \label{E:deMY}
 M^Y(z;y)=
 \begin{pmatrix}
  \sigma_1 W(-z;y,-\widetilde\nu_1,-\rho_0)\sigma_1
  &\mathbf 0_{2\times1}\\
  \mathbf 0_{1\times2}&1
 \end{pmatrix} \times
 \widehat G^Y(z;y)
 (-z)^{\mathrm{i}\widetilde\nu_1\sigma}.
\end{align}
where
$$
\widehat G^Y(z;y)=
\begin{cases}
  \left(G^Y(z;y)\right)^{-1},&(x,t)\in\mathcal T_+,\\
  I,&(x,t)\in\mathcal T_-.
 \end{cases}
$$
Indeed, it is straightforward to verify that the function defined
above satisfies all the conditions of the local model RH
problem~\ref{prop:tr-Y-RHP}.

 In the relevant sectors,
the factor \( \widehat G^{Y}(z;y)\)   approaches the identity exponentially
fast as \(z\to\infty\) and therefore does not affect any algebraic coefficient
in the large-\(z\) expansion. Consequently, the local solutions in the two
cases have identical expansion coefficients.  Using \eqref{E:Wasy}, \eqref{E:W1asy}, and \eqref{E:deMY},
we then obtain the following lemma.

\begin{lemma}
The  unique solution \(M^Y(z;y)\) of local model RH
problem~\ref{prop:tr-Y-RHP}  admits the following expansion as \(z\to\infty\):
\begin{align} \label{E:asyMX}
M^Y(z;y)
=
I+\frac{M_1^Y(y)}{z}
+
\mathcal O\left(\frac{1}{z^2}\right),
\end{align}
uniformly for \(\arg z\in[0,2\pi]\).  Moreover,
\begin{align} \label{E:defMX1dd}
M_1^Y(y)
=
\begin{pmatrix}
-\widehat{ \mathcal V}^*(y) & -\mathcal V(y)&0
\\[0.6ex]
\mathcal{V}^*(y) & -\widehat{ \mathcal V}(y)&0\\
0&0&0
\end{pmatrix},
\end{align}
where \(\mathcal V(y)\) and  \(\widehat{\mathcal V}(y)\)  are defined by~\eqref{E:defmathcalV} and \eqref{E:hatmathcalV}, respectively.
\end{lemma}



\begin{thebibliography}{99}

\bibitem{Manakov1974}
S.~V. Manakov,
\newblock On the theory of two-dimensional stationary self-focusing of
electromagnetic waves,
\newblock \emph{Sov. Phys. JETP} \textbf{38} (1974), 248--253.

\bibitem{BZ1970}
A.L. Berkhoer, V.E. Zakharov,
Self excitation of waves with different polarizations in nonlinear
media, Sov. Phys. JETP 31 (1970) 486--490.

\bibitem{CP1999}
C. Sulem, P.L. Sulem,
\emph{The Nonlinear Schr\"odinger Equation: Self-Focusing and Wave Collapse},
Springer, New York, 1999.

\bibitem{A1987}
G.P. Agrawal,
Modulation instability induced by cross-phase modulation,
Phys. Rev. Lett. 59 (1987) 880--883.

\bibitem{A1989}
G.P. Agrawal,
Nonlinear Fiber Optics,
Academic Press, New York, 1989.

\bibitem{BDW1987}
K.J. Blow, N.J. Doran, D. Wood,
Polarization instabilities for solitons in birefringent fibers,
Opt. Lett. 12 (1987) 202.

\bibitem{QZL2019}
Y.-H. Qin, L.-C. Zhao, L. Ling,
Nondegenerate bound-state solitons in multicomponent Bose-Einstein
condensates, Phys. Rev. E 100 (2019) 022212.

\bibitem{KrausBiondiniKovacic2015}
D.~Kraus, G.~Biondini, and G.~Kova\v{c}i\v{c},
\newblock The focusing Manakov system with nonzero boundary conditions,
\newblock \emph{Nonlinearity} \textbf{28} (2015), 3101--3151.

\bibitem{BF1967}
T.B. Benjamin, J.E. Feir,
The disintegration of wave trains in deep water,
J. Fluid Mech. 27 (1967) 417--430.

\bibitem{GelashRaskovalov2023}
A.~Gelash and A.~Raskovalov,
\newblock Vector breathers in the Manakov system,
\newblock \emph{Stud. Appl. Math.} \textbf{150} (2023), 841--882.

\bibitem{pz93}
Deift, P., Zhou, X.:
A steepest descent method for oscillatory Riemann--Hilbert problems.
Asymptotics for the mKdV equation.
\emph{Ann. of Math. (2)} 137(2), 295--368 (1993).

\bibitem{BLM2021}
Biondini, G., Li, S., Mantzavinos, D.:
Long-time asymptotics for the focusing nonlinear Schr\"odinger equation
with nonzero boundary conditions in the presence of a discrete spectrum.
\emph{Comm. Math. Phys.} 382(3), 1495--1577 (2021).

\bibitem{BiondiniFagerstrom2015}
G.~Biondini and E.~Fagerstrom,
\newblock The integrable nature of modulational instability,
\newblock \emph{SIAM J. Appl. Math.} \textbf{75} (2015), 136--163.

\bibitem{BiondiniMantzavinos2016}
G.~Biondini and D.~Mantzavinos,
\newblock Universal nature of the nonlinear stage of modulational instability,
\newblock \emph{Phys. Rev. Lett.} \textbf{116} (2016), 043902.

\bibitem{BiondiniMantzavinos2017}
G.~Biondini and D.~Mantzavinos,
\newblock Long-time asymptotics for the focusing nonlinear Schr\"odinger
equation with nonzero boundary conditions at infinity and asymptotic stage of
modulational instability,
\newblock \emph{Commun. Pure Appl. Math.} \textbf{70} (2017), 2300--2365.

\bibitem{BiondiniLiMantzavinosTrillo2018}
G.~Biondini, S.~Li, D.~Mantzavinos, and S.~Trillo,
\newblock Universal behavior of modulationally unstable media,
\newblock \emph{SIAM Rev.} \textbf{60} (2018), 888--908.

\bibitem{FrisquetEtAl2015}
B.~Frisquet, B.~Kibler, J.~Fatome, P.~Morin, F.~Baronio, M.~Conforti,
G.~Millot, and S.~Wabnitz,
\newblock Polarization modulation instability in a Manakov fiber system,
\newblock \emph{Phys. Rev. A} \textbf{92} (2015), 053854.

\bibitem{OnoratoOsborneSerio2006}
M.~Onorato, A.~R. Osborne, and M.~Serio,
\newblock Modulational instability in crossing sea states: A possible
mechanism for the formation of freak waves,
\newblock \emph{Phys. Rev. Lett.} \textbf{96} (2006), 014503.

\bibitem{BJM2018}
M. Borghese, R. Jenkins, K.T.-R. McLaughlin,
Long time asymptotic behavior of the focusing nonlinear Schr\"odinger equation,
Ann. Inst. H. Poincar\'e C Anal. Non Lin\'eaire 35 (2018) 887--920.

\bibitem{Miller}
P.~D. Miller,
\newblock On the increasing tritronqu\'ee solutions of the Painlev\'e-II
equation,
\newblock \emph{SIGMA} \textbf{14} (2018), Paper No.~125, 38~pp.

\bibitem{BilmanLingMiller2020}
D.~Bilman, L.~Ling, and P.~D. Miller,
\newblock Extreme superposition: Rogue waves of infinite order and the
Painlev\'e-III hierarchy,
\newblock \emph{Duke Math. J.} \textbf{169} (2020), 671--760.

\bibitem{GengZhangWang2026}
H.~Zhang, X.~Geng, and K.~Wang,
\newblock Tritronqu\'ee Painlev\'e-II asymptotics for the focusing nonlinear
Schr\"odinger equation on a modulationally unstable background,
\newblock arXiv:2606.29156v2, 2026.

\bibitem{RadhakrishnanLakshmananHietarinta1997}
R. Radhakrishnan, M. Lakshmanan, and J. Hietarinta,
Inelastic collision and switching of coupled bright solitons in
optical fibers,
\emph{Phys. Rev. E} \textbf{56} (1997), 2213--2216.

\bibitem{KannaLakshmanan2001}
T. Kanna and M. Lakshmanan,
Exact soliton solutions, shape changing collisions, and partially
coherent solitons in coupled nonlinear Schr\"odinger equations,
\emph{Phys. Rev. Lett.} \textbf{86} (2001), 5043--5046.

\bibitem{Wright2003}
O.~C. Wright,
The Darboux transformation of some Manakov systems,
\emph{Appl. Math. Lett.} \textbf{16} (2003), 647--652.

\bibitem{LingGuoZhao2014}
L. Ling, B. Guo, and L.-C. Zhao,
High-order rogue waves in vector nonlinear Schr\"odinger equations,
\emph{Phys. Rev. E} \textbf{89} (2014), 041201(R).

\bibitem{BaronioDegasperisConfortiWabnitz2012}
F. Baronio, A. Degasperis, M. Conforti, and S. Wabnitz,
Solutions of the vector nonlinear Schr\"odinger equations:
Evidence for deterministic rogue waves,
\emph{Phys. Rev. Lett.} \textbf{109} (2012), 044102.

\bibitem{ElginEnolskiIts2007}
J.~N. Elgin, V.~Z. Enolski, and A.~R. Its,
Effective integration of the nonlinear vector Schr\"odinger equation,
\emph{Physica D} \textbf{225} (2007), 127--152.

\bibitem{MM2013}
N. V. Priya, M. Senthilvelan, and M. Lakshmanan,
Physical Review E 88, 022918 (2013).

\bibitem{Kalla2011}
C. Kalla,
Breathers and solitons of generalized nonlinear Schr\"odinger equations
as degenerations of algebro-geometric solutions,
\emph{J. Phys. A: Math. Theor.} \textbf{44} (2011), 335210.

\bibitem{YangYang2023}
B. Yang and J. Yang,
Rogue wave patterns associated with Okamoto polynomial hierarchies,
\emph{Stud. Appl. Math.} \textbf{151} (2023), 60--115.

\bibitem{MuQinGrimshaw2015}
G. Mu, Z. Qin, and R. Grimshaw,
Dynamics of rogue waves on a multisoliton background in a vector
nonlinear Schr\"odinger equation,
\emph{SIAM J. Appl. Math.} \textbf{75} (2015), 1--20.

\bibitem{LingZhaoGuo2015}
L. Ling, L.-C. Zhao, and B. Guo,
Darboux transformation and multi-dark soliton for $N$-component
nonlinear Schr\"odinger equations,
\emph{Nonlinearity} \textbf{28} (2015), 3243--3261.

\bibitem{LingPelinovskySu2026}
L. Ling, D.~E. Pelinovsky, and H. Su,
Orbital stability of breathers for coupled nonlinear
Schr\"odinger equations,
\emph{Adv. Math.} \textbf{495} (2026), 110961.

\bibitem{LingSu2027}
L. Ling and H. Su,
Orbital stability of vector multi-solitons in coupled NLS and
modified KdV equations,
\emph{J. Differential Equations} \textbf{486} (2027), 114803.

\bibitem{Liu2019}
J. Liu,
$L^2$-Sobolev space bijectivity of the inverse scattering of a
$3\times3$ AKNS system,
\emph{Nonlinearity} \textbf{32} (2019), 1012--1043.

\bibitem{P2023}
B. Prinari,
Inverse scattering transform for nonlinear Schr\"odinger systems
on a nontrivial background: A survey of classical results, new developments
and future directions,
\emph{J. Nonlinear Math. Phys.}, 30 (2023), pp. 317--383.

\bibitem{EgorovaMichorTeschl2018}
I.~Egorova, J.~Michor, and G.~Teschl,
Long-time asymptotics for the Toda shock problem:
non-overlapping spectra,
\emph{J. Math. Phys. Anal. Geom.}
\textbf{14} (2018), 406--451.

\bibitem{EgorovaMichorPryimakTeschl2023}
I.~Egorova, J.~Michor, A.~Pryimak, and G.~Teschl,
Long-time asymptotics for Toda shock waves in the modulation region,
\emph{J. Math. Phys. Anal. Geom.}
\textbf{19} (2023), 396--442.

\bibitem{EgorovaGladkaKotlyarovTeschl2013}
I.~Egorova, Z.~Gladka, V.~Kotlyarov, and G.~Teschl,
Long-time asymptotics for the Korteweg--de Vries equation
with step-like initial data,
\emph{Nonlinearity}
\textbf{26} (2013), 1839--1864.

\bibitem{EgorovaPiorkowskiTeschl2024}
I.~Egorova, M.~Piorkowski, and G.~Teschl,
Asymptotics of KdV shock waves via the Riemann--Hilbert approach,
\emph{Indiana Univ. Math. J.}
\textbf{73} (2024), 645--690.

\bibitem{BoutetDeMonvelLenellsShepelsky2021}
A.~Boutet de Monvel, J.~Lenells, and D.~Shepelsky,
The focusing NLS equation with step-like oscillating background:
scenarios of long-time asymptotics,
\emph{Comm. Math. Phys.}
\textbf{383} (2021), 893--952.

\bibitem{BoutetDeMonvelLenellsShepelsky2022}
A.~Boutet de Monvel, J.~Lenells, and D.~Shepelsky,
The focusing NLS equation with step-like oscillating background:
the genus 3 sector,
\emph{Comm. Math. Phys.}
\textbf{390} (2022), 1081--1148.


\end{thebibliography}
\end{document}